\documentstyle[prd,aps,preprint,tighten,epsf,amstex]{revtex}

\newcommand{\DSone}{\Delta S\!=\!1}
\newcommand{\DStwo}{\Delta S\!=\!2}
\newcommand{\epe}{\epsilon^{\prime}/\epsilon}
\newcommand{\repe}{{\rm Re}(\epsilon^{\prime}/\epsilon)}
\newcommand{\ep}{\epsilon^{\prime}}
\newcommand{\DDmone}{\Delta D\!=\!-1}
\newcommand{\DIhalf}{\Delta I=1/2}
\newcommand{\DIthalf}{\Delta I=3/2}

\newcommand{\repezero}{[\omega/ (\sqrt{2} | \epsilon |)]_{\rm exp}P_0}
\newcommand{\repetwo}{ [\omega/ (\sqrt{2} | \epsilon |)]_{\rm exp}P_2}
\newcommand{\repezeroi}{[\omega/ (\sqrt{2}|\epsilon|)]_{\rm exp}P_0^i}
\newcommand{\repetwoi}{ [\omega/ (\sqrt{2}|\epsilon|)]_{\rm exp}P_2^i}

\newcommand{\qbar}{\overline{q}}
\newcommand{\Kbar}{\overline{K}}

\def\cancel#1#2{\ooalign{$\hfil#1\mkern1mu/\hfil$\crcr$#1#2$}}
\def\slash#1{\mathpalette\cancel{#1}}

\newcommand{\DSoneHc}{{\cal H}_c^{(\Delta S=1)}}
\newcommand{\DSoneH}{{\cal H}^{(\Delta S=1)}}
\newcommand{\DStwoH}{{\cal H}^{(\Delta S=2)}}

\newcommand{\DStwoQ}{Q^{(\Delta S=2)}}

\newcommand{\sutlr}{SU(3)_L \otimes SU(3)_R}

\newcommand{\Lcpt}{\Lambda_{\chi PT}}
\newcommand{\Lqcpt}{\Lambda_{Q \chi PT}}
\newcommand{\cpt}{\chi{\rm PT}}

\def\mpi2{m_\pi^2}
\def\mK2{m_K^2}

\def\mres{m_{\rm res}}

\newcommand{\bea}{\begin{eqnarray}}
\newcommand{\eea}{\end{eqnarray}}
\newcommand{\be}{\begin{equation}}
\newcommand{\ee}{\end{equation}}

\def\lvec#1{\setbox0=\hbox{$#1$}
    \setbox1=\hbox{$\scriptstyle\leftarrow$}
    #1\kern-\wd0\smash{
    \raise\ht0\hbox{$\raise1pt\hbox{$\scriptstyle\leftarrow$}$}}
    \kern-\wd1\kern\wd0}
\def\rvec#1{\setbox0=\hbox{$#1$}
    \setbox1=\hbox{$\scriptstyle\rightarrow$}
    #1\kern-\wd0\smash{
    \raise\ht0\hbox{$\raise1pt\hbox{$\scriptstyle\rightarrow$}$}}
    \kern-\wd1\kern\wd0}

\begin{document}
\bibliographystyle{apsrev}
\epsfclipon


\newcounter{Intro}
\setcounter{Intro}{1}

\newcounter{Analytic}
\setcounter{Analytic}{1}

\newcounter{ContChiralPert}
\setcounter{ContChiralPert}{1}

\newcounter{DwfChiralMod}
\setcounter{DwfChiralMod}{1}

\newcounter{RunBasics}
\setcounter{RunBasics}{1}

\newcounter{TestDwfChiral}
\setcounter{TestDwfChiral}{1}

\newcounter{WilsonCoef}
\setcounter{WilsonCoef}{1}

\newcounter{Npr}
\setcounter{Npr}{1}

\newcounter{LatMe}
\setcounter{LatMe}{1}

\newcounter{IThreeOverTwoMe}
\setcounter{IThreeOverTwoMe}{1}

\newcounter{IOneOverTwoMe}
\setcounter{IOneOverTwoMe}{1}

\newcounter{PhysME}
\setcounter{PhysME}{1}

\newcounter{RealAZeroATwo}
\setcounter{RealAZeroATwo}{1}

\newcounter{ImagAZeroATwo}
\setcounter{ImagAZeroATwo}{1}

\newcounter{Conclusions}
\setcounter{Conclusions}{1}

\newcounter{Acknowledgments}
\setcounter{Acknowledgments}{1}

\newcounter{Appendix}
\setcounter{Appendix}{1}

\newcounter{Tables}
\setcounter{Tables}{1}

\newcounter{Figures}
\setcounter{Figures}{1}


\draft

\preprint{CU-TP-1028, BNL-HET-01/30, RBRC-194}

\title{Kaon Matrix Elements and CP-violation from Quenched Lattice QCD:
(I) the 3-flavor case}

\author{
T.~Blum$^a$,
P.~Chen$^b$,
N.~Christ$^b$,
C.~Cristian$^b$,
C.~Dawson$^c$,
G.~Fleming$^b$\footnote{Current address:  Physics Dept., The Ohio
State University, Columbus, OH 43210}
R.~Mawhinney$^b$,
S.~Ohta$^{ad}$,
G.~Siegert$^b$,
A.~Soni$^c$,
P.~Vranas$^e$,
M.~Wingate$^{a*}$,
L.~Wu$^b$,
Y.~Zhestkov$^b$}

\address{
\vspace{0.5in}
$^a$RIKEN-BNL Research Center,
Brookhaven National Laboratory,
Upton, NY 11973}

\address{$^b$Physics Department,
Columbia University,
New York, NY 10027}

\address{$^c$Physics Department,
Brookhaven National Laboratory,
Upton, NY 11973}

\address{$^d$Institute for Particle and Nuclear Studies,
KEK,
Tsukuba, Ibaraki, 305-0801, Japan }

\address{$^e$IBM Research,
Yorktown Heights, New York, 10598 } 

\date{\today}
\maketitle

\begin{abstract}
We report the results of a calculation of the $K \to \pi \pi$ matrix
elements relevant for the $\DIhalf$ rule and $\epe$ in quenched lattice
QCD using domain wall fermions at a fixed lattice spacing $a^{-1}
\sim 2$ GeV.  Working in the three-quark effective theory, where only
the $u$, $d$ and $s$ quarks enter and which is known perturbatively to
next-to-leading order, we calculate the lattice $K \to \pi$ and $K \to
|0\rangle$ matrix elements of dimension six, four-fermion operators.
Through lowest order chiral perturbation theory these yield $K \to \pi
\pi$ matrix elements, which we then normalize to continuum values
through a non-perturbative renormalization technique.  For the ratio of
isospin amplitudes $|A_0|/|A_2|$ we find a value of $25.3 \pm 1.8$
(statistical error only) compared to the experimental value of 22.2,
with individual isospin amplitudes $10-20$\% below the experimental
values.  For $\epe$, using known central values for standard model
parameters, we calculate $(-4.0 \pm 2.3) \times 10^{-4}$ (statistical
error only) compared to the current experimental average of $(17.2 \pm
1.8) \times 10^{-4}$.  Because we find a large cancellation between the
$I = 0$ and $I = 2$ contributions to $\epe$, the result may be very
sensitive to the approximations employed.  Among these are the use of:
quenched QCD, lowest order chiral perturbation theory and continuum
perturbation theory below 1.3 GeV.  We have also calculated the kaon
$B$ parameter, $B_K$ and find $B_{K,\overline{MS}}(2 \, {\rm  GeV}) =
0.532(11)$.  Although currently unable to give a reliable systematic
error, we have control over statistical errors and more simulations
will yield information about the effects of the approximations on this
first-principles determination of these important quantities.
\end{abstract}


\newpage


\section{Introduction}
\label{sec:intro}

\ifnum\theIntro=1
%
%

The experimental observation of CP violation in kaon decays
\cite{Christenson:1964fg,Alavi-Harati:1999xp,Alavi-Harati:web,Fanti:1999nm,Fanti:web}
presents a continuing challenge to theoretical calculations within the
Standard Model and its possible extensions.  The Standard Model allows
CP violation through the single avenue set down by Kobayashi and
Maskawa almost 30 years ago \cite{Kobayashi:1973fv}, but a quantitative
comparison between theory and experiment requires the calculation of
well-defined electroweak interactions involving quarks, when the quarks
are bound into kaons and pions.  These ``weak matrix elements'' can be
calculated from first principles using the techniques of lattice QCD,
although many technical difficulties have impeded the realization of
this goal.  A large number of analytical and phenomenological
techniques have also been employed to estimate these matrix elements
and these are reviewed in \cite{Bertolini:1998vd}.  The work
described in this paper represents a complete calculation of the
matrix elements, using the approximations described below, that
determines the amplitudes $A_0$ and $A_2$ which describe two pion decays
of kaons, both their magnitudes and their CP-violating phases.
We also calculate the kaon $B$ parameter, $B_K$, which enters
Standard-Model predictions for the CP violation effects first
seen by Cronin and Fitch \cite{Christenson:1964fg}.

A major approximation made in this work is the use of quenched lattice
QCD in the evaluation of the matrix elements and the determination of
their normalizations.  This truncation of the full theory reduces the
required computer power markedly, but is an uncontrolled
approximation.  In most cases where quenched results are compared with
experimental values, agreement is at or better than the $\sim 25$\%
level, but there is no convincing argument that such agreement must be
uniformly good for all low-energy hadronic phenomena.  It should be
stressed that, if the necessary computer power were available to
generate an ensemble of dynamical fermion lattices, the numerical work
and analysis in this paper could be easily redone, yielding values
without the approximation of quenching.

Almost all attempts to calculate the matrix elements needed for CP
violation using lattice QCD have been done in the quenched
approximation.  The first lattice calculations using Wilson fermions
were unsuccessful \cite{Bernard:1985tm,Bochicchio:1985xa}, primarily
due to the lack of chiral symmetry on the lattice.  Staggered fermions
do provide a remnant chiral symmetry on the lattice and a calculation
of the matrix elements studied here has been done
\cite{Pekurovsky:1998jd}.  To match continuum and lattice operators for
staggered fermions, perturbation theory was used \cite{Sharpe:1994ur}.
Due to the large size of the one-loop perturbative corrections for
unimproved staggered fermions, the matching introduces large
uncertainties.  The current calculation uses domain wall fermions,
which have controllable chiral symmetry breaking at finite lattice
spacing, and a non-perturbative renormalization technique to relate
lattice quantities to the continuum.

The electroweak physics responsible for $K \to \pi \pi$ decays is
readily described by an effective weak Hamiltonian, valid for low
energy processes, which is given by four-quark operators multiplied by
perturbatively calculable Wilson coefficients.  In Section
\ref{sec:analytic}, we give our notation for the effective Hamiltonian
and the operator basis we will use.  We discuss both the three-quark
effective Hamiltonian, where $u$, $d$, and $s$ quarks can appear, and
the four-quark Hamiltonian, which includes the $c$ quark.  The Wilson
coefficients are known in both cases, although the three-quark case
requires using continuum perturbation theory down to a scale below
the charm quark mass, $m_c \approx 1.3$ GeV.  The $\sutlr$ quantum
numbers of the operators are given, since these determine their mixing
under renormalization and their behavior in the chiral limit.  In this
section we also give the relations between the matrix elements we
calculate and the quantities $\epsilon^\prime$ and $\epsilon$.

A second approximation made in this work is the use of primarily lowest
order chiral perturbation theory in the determination of the desired $K
\to \pi \pi$ matrix elements \cite{Bernard:1985wf}.  We evaluate $K \to
\pi$ and $K \to |0\rangle$ matrix elements in quenched lattice QCD and
then use lowest-order, full QCD chiral perturbation theory to determine
$K \to \pi\pi$ matrix elements.  This is reviewed in Section
\ref{sec:cont_chiral_pert}.  Thus, our calculation is strictly an
evaluation of the relevant matrix elements for small quark masses.  The
effects of quenching on lowest order full QCD chiral perturbation
theory and the chiral limit of quenched QCD are still subjects where
analytic understanding is limited.  We address quenching effects in our
results where analytic calculations offer guidance as to the mass
dependence expected in quenched amplitudes.  However, in general, such
phenomena are neglected in the quenched approximation and their
presence serves as a measure of the size of systematic error.  Once we
have determined values for the $K \to \pi\pi$ matrix elements valid in
the region of small quark mass, we then use the known chiral logarithms
in full QCD to extrapolate to the physical kaon mass.  The size of
these next-to-leading-order, chiral logarithms provides an indication
of the importance of the other next-to-leading-order terms which we do
not include in our extrapolation. Terms of this type, {\em i.e.} $m^2
\ln( m^2)$ where $m$ is a pseudoscalar mass, we will refer to as
conventional chiral logarithms.  Similar $m^2 \ln (m^2)$ terms also
occur in the quenched theory, along with the more singular quenched
chiral logarithms \cite{Morel:1987xk,Sharpe:1990me,Bernard:1992mk}
discussed in Section \ref{sec:cont_chiral_pert}.

To employ chiral perturbation theory as discussed in the previous
paragraph, it is important to use a lattice fermion formulation which
preserves chiral symmetry for the low energy physics.  (The presence of
chiral symmetry also simplifies operator mixing and renormalization,
which we discuss shortly.) \ \ A major theoretical advance in this area
\cite{Kaplan:1992bt} is provided by the domain wall
\cite{Kaplan:1992bt,Shamir:1993zy,Furman:1995ky} and overlap fermion
\cite{Narayanan:1993wx,Narayanan:1993ss} formulations of lattice
fermions.  Here we use the domain wall fermion formulation, which has
been shown, even for the quenched theory, to have small chiral symmetry
breaking effects for currently accessible values for the length of the
introduced fifth dimension \cite{Blum:2000kn,AliKhan:2000iv}.  In
Section \ref{sec:dwf_chiral_mod} we discuss the features of domain wall
fermions relevant for this calculation, paying particular attention to
the non-universal character of the chiral symmetry breaking for power
divergent operators and the topological near-zero modes present in
quenched calculations at finite volume.  This discussion will be
important for understanding the chiral limit of our matrix elements and
in the subtraction of power divergent terms from them.

In Section \ref{sec:run_basics}, we discuss the basic parameters of our
numerical calculations.  Then in Section \ref{sec:test_dwf_chiral} we
present further tests of the chiral properties of domain wall fermions,
in particular extending the results of \cite{Blum:2000kn} to the case
of Ward-Takahashi identities involving power divergent operators.  Here
we also determine the size of quenched chiral logarithm effects in our
simulations.  The numerical examples in this section complement the
theoretical explanations in Section \ref{sec:dwf_chiral_mod}.

The continuum perturbation theory calculations of the Wilson
coefficients for the low energy effective Hamiltonian have been done to
next-to-leading order \cite{Buras:1993dy,Ciuchini:1995cd}.  Using the
results from these calculations, we must evolve the Wilson coefficients
to the scale where we have renormalized our lattice operators.  This is
discussed in Section \ref{sec:wilson_coef} and involves some subtlety
due to the matching between the Wilson coefficients calculated in full
QCD and our quenched operators.  In addition, we must also incorporate
perturbatively calculated matching factors to move from the modified
minimal subtraction ($\overline{\rm MS}$) scheme used in the continuum
to the regularization independent scheme used for our lattice
operators.

To handle the renormalization of lattice operators, we employ another
major theoretical advance of recent years, the non-perturbative
renormalization (NPR) technique.  In this method one adopts a
renormalization scheme for defining renormalized operators that is
independent of the regularization.  Such a scheme can then be
implemented in both perturbation theory (where dimensional
regularization is typically used) and in a non-perturbative lattice
calculation.  This NPR approach avoids the use of lattice perturbation
theory and the attendant worries about its accuracy.  In principle, NPR
permits the use of perturbation theory to be restricted to short
distances where its validity is more certain.  Of the two most
developed approaches to NPR, the Schroedinger
functional\cite{Luscher:1997jn} and momentum-space based RI method
\cite{Martinelli:1995ty}, we have adopted the latter method since much
important analytical work for the kaon system has already been done
supporting this approach.  In Section \ref{sec:npr}, we discuss in some
detail how we have implemented this technique for the $\DSone$
operators of primary interest in this report.  This represents one of
the most complicated cases where this technique has been used to date
and we have only removed mixings with the dominant lower-dimensional
operators.  It is worth noting that this technique is particularly well
suited for use with domain wall fermions, since the definition of the
regularization independent scheme involves off-shell quark fields.  For
domain wall fermions the suppression of explicit chiral symmetry
breaking and the consequent elimination of order $a$ lattice spacing
errors occurs both on- and off-shell.

In Section \ref{sec:lat_me}, we discuss the precise quantities that we
measure on the lattice to determine $K \to \pi$ and $K \to |0 \rangle$
matrix elements.  We have used standard ratios of lattice Green's
functions to measure these matrix elements, but the presence of
topological near-zero modes leads to preferred choices for the factors
in the ratio to minimize the effects of zero modes.  The tables referred
to in Section \ref{sec:lat_me} report our bare lattice values
for these quantities.

We can now use our lattice results for the bare $K \to \pi$ and $K \to
|0 \rangle$ matrix elements to evaluate the chiral perturbation theory
constants which determine $K \to \pi \pi$ matrix elements.  In Section
\ref{sec:i_3over2_me} we discuss the $\DIthalf$ matrix elements, where
the chiral perturbation theory constants come directly from $K \to \pi$
matrix elements.  Depending on the operator involved, these operators
can vanish or be non-zero in the chiral limit.  We find that it is
important to know the coefficients of the conventional chiral logarithm
terms from analytic calculations in order to determine the chiral
perturbation theory constants.

In Section \ref{sec:i_1over2_me}, we perform a similar analysis of our
lattice data to determine the chiral perturbation theory constants for
$\DIhalf$ matrix elements.  This case is more subtle numerically, since
it involves the cancellation of unphysical, power divergent effects
between $K \to \pi$ and $K \to |0 \rangle$ matrix elements in the
determination of the desired physical chiral perturbation theory
constants.  For one group of operators, we can check this cancellation
by using the Wigner-Eckart theorem to relate $\DIhalf$ constants, which
involve subtractions, to $\DIthalf$ constants, which do not.  We find
the agreement expected.  The end result of our numerical determinations
are the values given in Table \ref{tab:final_alpha_charm_out}.  These
are lattice values from a quenched calculation, using the formulae from
chiral perturbation theory for full QCD.

In Section \ref{sec:phys_me} we discuss how to take these final lattice
values and calculate physical quantities.  In the spirit of the
quenched approximation we take these quenched results as an
approximation for the desired full QCD quantities.  In particular, for
$K \to \pi \pi$ matrix elements which vanish in the chiral limit, we
take our quenched values for the slope with respect to quark mass of
these matrix elements as the value for the slope for the full QCD
matrix elements.  For $K \to \pi \pi$ matrix elements which are
non-zero in the chiral limit, the chiral limit value in the quenched
theory is used as the chiral limit value in the full theory.  We can
then determine physical matrix elements at the kaon mass by
extrapolating in lowest order chiral perturbation theory.  Since the
chiral logarithms are known, we can also extrapolate including the
effects of the logarithms.  This is not a complete higher order chiral
perturbation theory calculation, but gives an indication of the size of
the effects entering at next order.

In Section \ref{sec:real_a0_a2} we combine the matrix elements, Wilson
coefficients, non-perturbative renormalization and central values for
Standard-Model parameters to give physical values for Re($A_0$),
Re($A_2$) and their ratio, which reflects the $\Delta I= 1/2$ rule.
Figures \ref{fig:Re_A0}, \ref{fig:Re_A2} and \ref{fig:ReA0_over_ReA2}
show our results for the various extrapolations, along with the
physical values.  The general agreement with the experimental values is
quite good, in spite of the many approximations in the calculation.
We also report our results for the kaon $B$ parameter, $B_K$, at
the end of this section.

Section \ref{sec:imag_a0_a2} also combines matrix elements, Wilson
coefficients, non-perturbative renormalization and central values for
Standard-Model parameters, but now the values for Im($A_0$), Im($A_2$)
and Re($\epe$) are the focus.  Figures \ref{fig:Im_A0} and
\ref{fig:Im_A2} show Im($A_0$) and Im($A_2$) and Figure
\ref{fig:epe_all} shows Re($\epe$).  For Re($\epe$), a large
cancellation is occurring between individual isospin contributions, as
can be seen in Figure \ref{fig:epe_choice2}.  It is important to note
that the magnitudes of each of the two individual isospin terms are
very similar to the experimental value for Re($\epe$).  We would like
to point out that even though in principle lattice techniques allow a
calculation of $\epe$, here we have used experimental information
regarding the phases of $\epsilon^{\prime}$ and $\epsilon$ in our
calculation of Re($\epe$).

Table \ref{tab:choice2_mu_2_13_charm_out_vs_exp} gives our final values
for the physical quantities Re($A_0$), Re($A_2$), Re($A_0$)/Re($A_2$)
and Re($\epe$).  Our conclusions are given in Section
\ref{sec:conclusions} and the five Appendices contain further details
about our conventions, the decomposition of operators into irreducible
representations of $\sutlr$ and other definitions used in the text.

\fi


\section{General Analytic Framework}
\label{sec:analytic}

\ifnum\theAnalytic=1
%
%


\subsection{$K \rightarrow \pi \pi $ in the Standard Model}
\label{subsec:kpp_std_model}

At energies below the electroweak scale, the weak interactions can be
described by local four-fermion operators due to the essentially
point-like character of the vector boson interactions for low
energies.  Simple charged vector boson exchange produces current-current
operators, with both currents left-handed, of the form
$ (\bar{q} \, q^{\prime})_{(V-A)} \;
  (\bar{q}^{\prime \prime} \, q^{\prime \prime \prime} )_{(V-A)}.
$
Additional low-energy four-fermion operators arise from more
complicated Standard-Model processes involving loops with heavy
particles, including the vector bosons and the top quark.  The naive
suppression of these non-exchange operators, due to the large masses in
the loop propagators and additional powers of the couplings, is offset
somewhat by the large phase space for the loop integrals and the large
logarithms which appear due to the disparity between GeV scale hadronic
physics and these heavy masses. The operator product expansion and the
renormalization group provide the framework for understanding such
logarithmic enhancements and, coupled with continuum perturbation
theory, provide a way to calculate these logarithmic effects.  Such
calculations yield the the low-energy four-fermion operators' Wilson
coefficients, which encapsulate the high energy physics in the
low-energy effective theory.

Thus, for energies well below the electroweak scale but above the
bottom quark mass, we have an effective weak Hamiltonian with
four-fermion interactions, where the coefficients of a given operator
depend on $\mu$, $m_t$, $m_W$, $m_Z$, $\alpha_s$, $\alpha$ and the
elements of the Cabibbo-Kobayashi-Maskawa (CKM) matrix, $V_{lm}$.  The
four-fermion interactions can involve all quark fields, except the top,
giving the Hamiltonian the generic form
\begin{equation}
  {\cal H_{\rm eff}}
           = \frac{G_F}{\sqrt{2}} \sum_{i} A_i(\mu, m_t, m_W, m_Z,
	     \alpha_s, \alpha, V_{lm}) \;
	     ( \bar{q}_i \, \Gamma_i \, q^{\prime}_i) \;
	     ( \bar{q}^{\prime \prime}_i \, \Gamma_i^{\prime} \,
	       q^{\prime \prime \prime}_i )
  \label{eq:Heff}
\end{equation}
The scale $\mu$ which appears in this equation is introduced through
the normalization condition required to define the composite
four-fermion operators, whose dependence on $\mu$ is not shown.  The
explicit $\mu$ dependence of the coefficients $A_i$ cancels the $\mu$
dependence implicit in these operators.  In studying physics at energy
scales well below the bottom quark mass, we can remove the bottom quark
from the operators that appear in $\cal H_{\rm eff}$, renormalizing
at a scale $\mu$ which is generally chosen near the scale of the
physics under consideration.  Of course, the Wilson coefficients $A_i$
must now depend explicitly on the bottom quark mass, $m_b$.  A similar
elimination of the charm degrees of freedom can be achieved if $\cal
H_{\rm eff}$ is specialized to a form valid for energies well below
the charm quark mass.

Following the general discussion above, one can determine the terms in
the low-energy effective Hamiltonian relevant to particular processes,
such as the $\DSone$, $\DDmone$ case of primary interest in this
study.  The terms arising from simple vector boson exchange, which
should play a dominant role in the $\DIhalf$ rule because of their
large Wilson coefficients, were first discussed in
\cite{Altarelli:1974ex,Gaillard:1974nj}, where it was also found that
the $A_i$ coefficients for these terms could explain some of the
enhancement given by the $\DIhalf$ rule.  Subsequently, additional
low-energy terms arising from Standard-Model graphs involving loops
were identified \cite{Vainshtein:1975sv,Shifman:1977tn} and
their importance for CP violation in the full six-quark Standard Model
emphasized in Refs.\ \cite{Gilman:1979wm,Gilman:1979bc,Bijnens:1984ye}.
These additional low-energy four-quark operators are referred to as
penguin operators and are further refined into QCD and electroweak
penguin operators.  Historically attention was first focused on the QCD
penguins, since the electroweak penguins are suppressed by a power of
the electroweak coupling $\alpha$.  However, as reviewed below, the
electroweak penguins are important for CP violation in the Standard
Model since they are non-zero to lowest order in the light quark
masses, are enhanced by the $\DIhalf$ rule and enter with coefficients
that increase with the top quark mass.

For our calculations, the energy scale that can be used in the
effective theory must be well below $m_b$, since we will work on a
lattice with $a^{-1} \sim 2$ GeV.  We do, however, have the ability to
work both with an effective theory valid for energies at or above $m_c$
(a four-flavor theory) and with a three-flavor theory that is only
valid for energies below $m_c$.  Thus, we will actually deal with two
effective Hamiltonians for $\DSone$ processes.  For clarity, we will
denote the four-flavor $\DSone$ effective Hamiltonian valid for
energies below $m_b$ by $\DSoneHc$ and use $\DSoneH$ for the
three-flavor theory valid only for energies below $m_c$.  Note, the
renormalization scale $\mu$ that appears in $\DSoneHc$ is
conventionally chosen well above $m_c$ while the $\mu$ that appears in
$\DSoneH$ should be chosen above $m_s$.  (Of course, in both cases we
would like to choose $\mu$ in a region where perturbation theory can be
used.)  In the notation of Ref.\ \cite{Gilman:1979bc}, operators in the
effective theory are given by $O_i$ for the five-quark theory which
includes the up, down, strange, charm and bottom quarks, by $P_i$ for
the effective four-quark theory and by $Q_i$ for the effective
three-quark theory including only the up, down and strange quarks
explicitly.  We follow this notation, but since we will not deal with
the effective five-quark theory, we also use $O_i$ to represent a
generic operator.  Using the operator basis defined below and following
Refs.\ \cite{Gilman:1979bc,Buras:1993dy,Ciuchini:1995cd} the effective
Hamiltonians can be written as:
\begin{align}
    \DSoneHc =
  & \frac{G_F}{\sqrt{2}} V_{ud}V_{us}^* \left\{
    \sum_{i=1}^{2} C_i(\mu) \left[ P_i + ( \tau - 1 ) P_i^c \right]
    + \tau \sum_{i=3}^{10} C_i(\mu) P_i \right\} \label{eq:DSone_ham_c}
  \\
    \DSoneH =
  & \frac{G_F}{\sqrt{2}} V_{ud}V_{us}^* \left\{
    \sum_{i=1}^{10} \left[ z_i(\mu) + \tau y_i(\mu) \right] Q_i
    \right\} \label{eq:DSone_ham}
\end{align}
Here $G_F$ is the Fermi coupling constant, $V_{kl}$ are elements of the
CKM matrix, $\lambda_k \equiv V_{kd}V^*_{ks}$ for $k = u, c, t$ and
$\tau = - \lambda_t/\lambda_u$.  For the four-flavor theory , we denote
the Wilson coefficients by real numbers $C_i(\mu)$ and the four-quark
operators by $P_i$ and $P_i^c$.  In general, charm quark fields will
appear in the operators $P_i$ as well as $P_i^c$.  For the three-flavor
theory, we denote the Wilson coefficients by the real numbers
$y_i(\mu)$ and $z_i(\mu)$ and use $Q_i$ to represent the four-quark
operators, which are made of up, down and strange quark fields only.
The dependence of the Wilson coefficients on the other parameters shown
in Eq.\ \ref{eq:Heff} is suppressed.

Before describing the operator basis in detail, a few important
features of the effective $\DSone$ Hamiltonians should be noted.
\begin{enumerate}
\item
  In these Hamiltonians, CP violation enters entirely through the
  parameter $\tau$, since we choose the standard representation of the
  CKM matrix of Ref.\ \cite{PDBook} where $V_{td}$, and thus $\tau$, is
  complex.
\item
  Of the 12 operators entering $\DSoneHc$, only 9 are linearly
  independent in a regularization that preserves Fierz
  transformations.  Similarly, for the 10 operators entering $\DSoneH$,
  only 7 are linearly independent.  The calculations of the Wilson
  coefficients most commonly use an overcomplete basis, since this
  allows one to transparently see how the original physics is inherited
  by the operators in the low energy effective theory.
\item
  The Wilson coefficients, which can be thought of as the couplings
  for the low-energy theory, vary markedly in size.  The Wilson
  coefficient for the vector boson exchange term is of ${\cal O}(1)$.  The
  QCD penguin terms are naively of ${\cal O}(\alpha_s)$ while the electroweak
  penguins are naively of ${\cal O}(\alpha)$.  This simple counting is
  influenced by the large logarithms generated from QCD running,
  which we will discuss further in \ref{sec:wilson_coef}.
\end{enumerate}
The numerical results reported here are for the three-flavor theory,
where the charm quark mass has been integrated out.  In the remainder
of this section we summarize the relevant low-energy four-fermion
operators for the three- and four-flavor theories and establish
notation for both cases.

As mentioned above, charged vector boson exchange gives rise to
left-left current interactions, with a particular color trace structure
($Q_2$, $P_2$ and $P_2^c$ below).  Mixing under renormalization
produces a left-left operator with the other possible color trace
($Q_1$, $P_1$ and $P_1^c$ below).  Letting $(L,R)$ denote the $\sutlr$
representation of an operator and $I$ its isospin, we give the quantum
numbers of the operators as $(L,R) \; I$.  Then with $\alpha$ and
$\beta$ denoting color indices, the charged vector boson
exchange operators in our basis are
\cite{Gilman:1979bc,Ciuchini:1995cd}
\newline \newline \indent {\bf Current-current operators:}
\begin{xalignat}{4}
    Q_1 \equiv P_1 & = (\bar{s}_\alpha d_\alpha)_{V-A} \;
	    (\bar{u}_\beta  u_\beta )_{V-A}
  & (8,1)  & \quad 1/2
  & (27,1) & \quad 1/2
  & (27,1) & \quad 3/2 \label{eq:Q1} \label{eq:P1}
  \\
    P_1^c & = (\bar{s}_\alpha d_\alpha)_{V-A} \;
	      (\bar{c}_\beta  c_\beta )_{V-A}
  & (8,1)  & \quad 1/2
  &        &
  &        & \label{eq:P1c}
  \\
    &
  & &
  & &
  & &
  \nonumber \\
    Q_2 \equiv P_2 & = (\bar{s}_\alpha d_\beta)_{V-A} \;
	    (\bar{u}_\beta  u_\alpha)_{V-A}
  & (8,1)  & \quad 1/2
  & (27,1) & \quad 1/2
  & (27,1) & \quad 3/2 \label{eq:Q2} \label{eq:P2}
  \\
    P_2^c & = (\bar{s}_\alpha d_\beta)_{V-A} \;
	      (\bar{c}_\beta  c_\alpha)_{V-A}
  & (8,1)  & \quad 1/2
  &        &
  &        & \label{eq:P2c} 
\end{xalignat}
Here the subscript $(V-A)$ refers to a quark bilinear of the form
$\qbar \gamma_\mu (1 - \gamma_5) q$.  Operators with color trace
structure similar to $Q_1$, are referred to as color diagonal operators
while $Q_2$ is an example of a color mixed operator.  Note that the
exchange operators in Eq.\ \ref{eq:Q1} and Eq.\ \ref{eq:Q2} get a
contribution from more than one representation of $\sutlr$ and contain
both $I=1/2$ and 3/2 parts.

In addition to the simple exchange diagrams which lead to the operators
of Eqs.\ \ref{eq:P1}-\ref{eq:P2c}, loop diagrams in the Standard Model
(the penguin diagrams) produce additional four-fermion terms in the
effective theory.  In the penguin diagrams relevant to this paper, a
top quark loop appears in the full electroweak theory.  QCD penguins
involve gluon exchange with this top quark loop, while electroweak
penguins involve $Z^0$ and photon exchange with the top quark loop.
The resulting four-fermion operators in the effective theory include
interactions between left-handed and right-handed currents and both
color diagonal and color mixed operators arise.  For effective
operators generated by the QCD penguin diagrams, all quarks which are
present in the effective theory enter with equal weight, since the
strong interactions couple equally to each flavor.
\newline \newline
\indent {\bf QCD penguin operators:}
\begin{xalignat}{4}
    Q_3 & = (\bar{s}_\alpha d_\alpha)_{V-A} \;
            \sum_{q=u,d,s} (\bar{q}_\beta  q_\beta )_{V-A}
  & (8,1)  & \quad 1/2
  &   &
  &   & \label{eq:Q3}
  \\ 
    P_3 & = (\bar{s}_\alpha d_\alpha)_{V-A} \;
             \sum_{q=u,d,s,c} (\bar{q}_\beta  q_\beta )_{V-A}
  & (8,1)  & \quad 1/2
  &   &
  &   & \label{eq:P3}
  \\
    &
  &   &
  &   &
  &   & \nonumber
  \\
    Q_4 & = (\bar{s}_\alpha d_\beta )_{V-A} \;
            \sum_{q=u,d,s} (\bar{q}_\beta  q_\alpha)_{V-A}
  & (8,1)  & \quad 1/2
  &   &
  &   & \label{eq:Q4}
  \\
    P_4 & = (\bar{s}_\alpha d_\beta )_{V-A} \;
             \sum_{q=u,d,s,c} (\bar{q}_\beta  q_\alpha)_{V-A}
  & (8,1)  & \quad 1/2
  &   &
  &   & \label{eq:P4}
  \\
    &
  &   &
  &   &
  &   & \nonumber
  \\
    Q_5 & = (\bar{s}_\alpha d_\alpha)_{V-A} \;
            \sum_{q=u,d,s} (\bar{q}_\beta  q_\beta )_{V+A}
  & (8,1)  & \quad 1/2
  &   &
  &   & \label{eq:Q5}
  \\
    P_5 &  = (\bar{s}_\alpha d_\alpha)_{V-A} \;
              \sum_{q=u,d,s,c} (\bar{q}_\beta  q_\beta )_{V+A}
  & (8,1)  & \quad 1/2
  &   &
  &   & \label{eq:P5}
  \\
    &
  &   &
  &   &
  &   & \nonumber
  \\
    Q_6 & = (\bar{s}_\alpha d_\beta )_{V-A} \;
            \sum_{q=u,d,s} (\bar{q}_\beta  q_\alpha)_{V+A}
  & (8,1)  & \quad 1/2
  &   &
  &   & \label{eq:Q6}
  \\
    P_6 & = (\bar{s}_\alpha d_\beta )_{V-A} \;
             \sum_{q=u,d,s,c} (\bar{q}_\beta  q_\alpha)_{V+A}
  & (8,1)  & \quad 1/2
  &   &
  &   & \label{eq:P6}
\end{xalignat}
Here the subscript $(V+A)$ refers to a quark bilinear of the form
$\qbar \gamma_\mu (1 + \gamma_5) q$.  As the list above shows, the QCD
penguin operators all have $I = 1/2$ and are singlets under
$SU(3)_R$, even though they contain right-handed quark fields.

The electroweak penguin operators have the same quark flavors as the
QCD penguins, but each quark bilinear is multiplied by its electric
charge, $e_q$.
\newline \newline \indent {\bf Electroweak penguin operators:}
\begin{alignat}{4}
    Q_7 & = \frac{3}{2} (\bar{s}_\alpha d_\alpha)_{V-A} \;
            \sum_{q=u,d,s} e_q (\bar{q}_\beta  q_\beta )_{V+A}
  & \quad (8,8)  & \; 1/2
  & \qquad (8,8)  & \; 3/2
  &  & \label{eq:Q7}
  \\
    P_7 & = \frac{3}{2} (\bar{s}_\alpha d_\alpha)_{V-A} \;
                \sum_{q=u,d,s,c} e_q (\bar{q}_\beta  q_\beta )_{V+A}
  & \quad (8,8)  & \; 1/2
  & \qquad (8,8)  & \; 3/2
  & \qquad (8,1)  & \; 1/2 \label{eq:P7}
  \\
    &
  &  &
  &  &
  &  & \nonumber
  \\
    Q_8 & = \frac{3}{2} (\bar{s}_\alpha d_\beta )_{V-A} \;
            \sum_{q=u,d,s} e_q (\bar{q}_\beta  q_\alpha)_{V+A}
  & (8,8)  & \; 1/2
  & (8,8)  & \; 3/2
  &  & \label{eq:Q8}
  \\
    P_8 & = \frac{3}{2} (\bar{s}_\alpha d_\beta )_{V-A} \;
                \sum_{q=u,d,s,c} e_q (\bar{q}_\beta  q_\alpha)_{V+A}
  & (8,8)  & \; 1/2
  & (8,8)  & \; 3/2
  & (8,1)  & \; 1/2 \label{eq:P8}
  \\
    &
  &  &
  &  &
  &  & \nonumber
  \\
    Q_9 & = \frac{3}{2} (\bar{s}_\alpha d_\alpha)_{V-A} \;
            \sum_{q=u,d,s} e_q (\bar{q}_\beta  q_\beta )_{V-A}
  & (8,1)  & \; 1/2
  & (27,1) & \; 1/2
  & (27,1) & \; 3/2 \label{eq:Q9}
  \\
    P_9 & = \frac{3}{2} (\bar{s}_\alpha d_\alpha)_{V-A} \;
                \sum_{q=u,d,s,c} e_q (\bar{q}_\beta  q_\beta )_{V-A}
  & (8,1)  & \; 1/2
  & (27,1) & \; 1/2
  & (27,1) & \; 3/2 \label{eq:P9}
  \\
    &
  &  &
  &  &
  &  & \nonumber
  \\ 
    Q_{10} & = \frac{3}{2} (\bar{s}_\alpha d_\beta )_{V-A} \;
               \sum_{q=u,d,s} e_q (\bar{q}_\beta  q_\alpha)_{V-A}
  & (8,1)  & \; 1/2
  & (27,1) & \; 1/2
  & (27,1) & \; 3/2 \label{eq:Q10}
  \\
    P_{10} & = \frac{3}{2} (\bar{s}_\alpha d_\beta )_{V-A} \;
                   \sum_{q=u,d,s,c} e_q (\bar{q}_\beta  q_\alpha)_{V-A}
  & (8,1)  & \; 1/2
  & (27,1) & \; 1/2
  & (27,1) & \; 3/2 \label{eq:P10}
\end{alignat}
Note that $Q_7$ and $Q_8$ are in a single representation of $\sutlr$,
so their $I=1/2$ and 3/2 matrix elements can be related by the
Wigner-Eckert theorem.  This is not true for $P_7$ and $P_8$,
since the addition of the charm quark brings in a contribution from a
different $\sutlr$ representation.

These operators can also be decomposed into irreducible representations
of isospin and $\sutlr$ and the details are given in
the appendix.  For the left-left operators made of $u,d$ and $s$
quarks, there is a single (27,1) and two (8,1) irreducible
representations.  Thus, there are only 3 matrix elements needed to
determine $Q_1$, $Q_2$, $Q_3$, $Q_4$, $Q_9$ and $Q_{10}$.

With these definitions and knowledge of the Wilson coefficients, $K
\rightarrow \pi \pi$ processes in the Standard Model can be expressed
in terms of the matrix elements $\langle \pi \pi | P_i(\mu) | K
\rangle$ defined in the four-quark effective theory or the three-quark
effective theory matrix elements $\langle \pi \pi | Q_i(\mu) | K
\rangle$.  Notice that here we have shown explicitly the dependence of
the operator on the scale $\mu$, which cancels the $\mu$ dependence of
the Wilson coefficients.  Since the Wilson coefficients are calculated
in continuum perturbation theory using dimensional regularization and
we will calculate the hadronic matrix elements using a lattice
regularization, we must relate, or match, operators normalized on the
lattice and the continuum operators.  This matching will also involve
operator mixing, so in general one has
\begin{equation}
  O_i^{\rm cont}(\mu) = Z_{ij}(\mu,a) O_j^{\rm lat}(a)
\end{equation}
where $a$ is the lattice spacing.
In this work, we employ a relatively new technique, non-perturbative
renormalization, as part of the calculation of the $Z_{ij}$'s.  This is
explained in detail in Section \ref{sec:npr}.  Before turning to our lattice
determination of $\langle \pi \pi | O_i(\mu) | K \rangle$ matrix
elements, we summarize the effective Hamiltonian for $\DStwo$
transitions in the Standard Model.


\subsection{$K^0$-$\Kbar^0$ Mixing in the Standard Model}
\label{subsec:kkbar_std_model}

In the development of the Standard Model, the $K^0$-$\Kbar^0$ system
has played an important role.  The GIM mechanism \cite{Glashow:1970gm}
provided a natural theoretical explanation for the small mass
difference between the $K_L$ and $K_S$ and was subsequently used to
give an estimate for the charm quark mass \cite{Gaillard:1974hs}.
These calculations were done for the case of only four quarks, where
there is no imaginary part to the $K^0$-$\Kbar^0$ mass matrix and no CP
violation.  For the six-quark Standard Model, this system should in
general exhibit CP violation and the low energy theory describing these
effects, including QCD corrections to leading logarithm order, was
first given in \cite{Gilman:1980di,Gilman:1983ap}.  Subsequent
work has determined the Wilson coefficients to next-to-leading order
\cite{Buras:1990fn,Herrlich:1996vf}.

We write the $\DStwo$ Hamiltonian for the effective three-flavor
theory to NLO as
\cite{Buras:1990fn}
\begin{align}
    \DStwoH
  & 
    =
    \frac{G_F^2}{16 \pi^2} M_W^2
    \left[
        \lambda_c^2 \eta_1 S_0(x_c) 
      + \lambda_t^2 \eta_2 S_0(x_t) 
      + 2 \lambda_c \lambda_t \eta_3 S_0(x_c, x_t)
    \right] \nonumber
  \\
  &
    \times
    \left[ \alpha_s^{(3)} \right]^{-2/9}
    \left[ 1 + \frac{\alpha_S^{(3)}(\mu)}{4 \pi} J_3 \right] \DStwoQ
      + h.c.
  \label{eq:DStwo_ham}
\end{align}  
where
\begin{xalignat}{4}
    \DStwoQ & = (\bar{s}_\alpha d_\alpha )_{V-A} \;
	        (\bar{s}_\beta  d_\beta)_{V-A}
  & (27,1)  & \quad 1
  &   &
  &   & \label{eq:Qll}
\end{xalignat}
$x_q = m_q^2/M_W^2$ and the functions $S_0(x_i)$ and $S_0(x_i, x_j)$
are the Inami-Lim functions \cite{Inami:1981fz}.  $J_3$ is defined
as
\begin{equation}
  J_3 \equiv \frac{ \gamma^{(0)} \beta_1}{2 \beta_0^2}
  - \frac{\gamma^{(1)}}{2 \beta_0}
\end{equation}
where $\gamma^{(i)}$ is the $i$th-order contribution to the anomalous
dimension for $\DStwoQ$ and $\beta_j$ are the $j$th order coefficients
for the QCD beta function in a three flavor theory.  In
addition, $\alpha_S^{(3)}(\mu)$ is the QCD running coupling for
a three flavor theory.

The coefficients $\eta_i$ are known to NLO
\cite{Buras:1990fn,Herrlich:1996vf} and have the values
\begin{equation}
  \eta_1 = 1.38 \pm 0.20, \qquad
  \eta_2 = 0.57 \pm 0.01, \qquad
  \eta_3 = 0.37 \pm 0.04.
\end{equation}
CP violating processes involving $K^0 - \Kbar^0$ mixing in the Standard
Model are then known if the CKM matrix elements are known and the
matrix element $\langle \Kbar^0 | \DStwoQ | K^0 \rangle$ is known.
Since for three degenerate quarks, $\DStwoQ$ is part of the same (27,1)
irreducible representation as $Q_1$ and $Q_2$, one can relate the
$\langle \Kbar^0 | \DStwoQ | K^0 \rangle$ matrix element to $\langle
\pi^+ | Q_1 | K^+ \rangle$ and $\langle \pi^+ | Q_2 | K^+ \rangle$.


\subsection{Connecting Experiment and Theory}
\label{subsec:theory_to_exp}

The previous two subsections have given the $\DSone$ and $\DStwo$
effective Hamiltonians in the notation we will use in this paper.
To further establish our notation and conventions, we now collect the 
relevant formulae to connect these Hamiltonians with the experimentally
measured quantities.  For a more comprehensive review, the reader
is referred to \cite{Winstein:1993sx,Buras:1998ra}.

Considering only the strong Hamiltonian, a neutral kaon, the $K^0$,
containing an anti-strange and down quark and its anti-particle, the
$\overline{K}^0$, containing an anti-down and strange quark are energy
eigenstates.  We adopt the conventional definitions of parity, $P$ and
charge conjugation, $C$, for quark fields in the Standard Model, giving
$CP| K^0 \rangle = - | \Kbar^0 \rangle$.  While charge conjugation and
parity are valid symmetries of the strong interactions, they are
violated by the weak interactions.  Allowing for the weak interactions
to also violate $CP$, for the neutral kaons seen in nature one writes
\begin{eqnarray}
  | K_S \rangle & = & p| K^0 \rangle - q | \overline{K}^0 \rangle \\
  | K_L \rangle & = & p| K^0 \rangle + q | \overline{K}^0 \rangle
\end{eqnarray}
with $p^2 + q^2 = 1$.  $CP$ is not a valid symmetry if the resulting
physical states have $p \neq q$.  Provided $CP$ violating effects are
small, $K_S$, being predominantly $CP$ even, has a much shorter
lifetime than $K_L$, since $K_S$ decay to two pions, where more phase
space is available, conserves $CP$.

The quantities measured experimentally to determine $CP$ violation are
\begin{alignat}{2}
  \eta_{+-} & = |\eta_{+-}| e^{i \phi_{+-}} & = &
    \frac{ A(K_L \rightarrow \pi^+ \pi^- )}
         { A(K_S \rightarrow \pi^+ \pi^- )}
    \\
  \eta_{00} &= |\eta_{00}| e^{i \phi_{00}} & = &
    \frac{ A(K_L \rightarrow \pi^0 \pi^0 )}
         { A(K_S \rightarrow \pi^0 \pi^0 )}
\end{alignat}
The current values for these quantities are \cite{PDBook} $|\eta_{+-}|
\approx | \eta_{00} | = 2.28 \times 10^{-3}$ and $|\phi_{+-}| \approx |
\phi_{00} | = 44^0$.

It is important to distinguish between $CP$ violation due to mixing,
also known as indirect $CP$ violation, and $CP$ violation in decays,
also referred to as direct $CP$ violation.  $CP$ violation due to
mixing refers to $K_L \leftrightarrow K_S$ transitions (or alternately
$K^0 \leftrightarrow \overline{K}^0$) and if all $CP$ violation came
from this source, one would find $\eta_{+-} = \eta_{00}$.  The initial
states would mix and the decay processes would preserve $CP$.  Allowing
for $CP$ violation in decays, one defines
\begin{equation}
  \eta_{+-} = \epsilon + \ep, \qquad
  \eta_{00} = \epsilon - 2 \ep
\end{equation}
and a non-zero value for $\ep$ signals $CP$ violation in decays.  The
current value for $\epsilon$ is $(2.271 \pm 0.017) \times 10^{-3}$ and
for $\epe$ is $(2.1 \pm 0.5) \times 10^{-3}$ \cite{PDBook}.

To relate the experimental quantities to the theoretical matrix
elements calculated here, it is conventional to define the isospin
amplitudes by
\begin{eqnarray}
  A \left( K^0 \rightarrow \pi \pi(I) \right)
& =
& A_I e^{i \delta_I} \\
  A \left( \overline{K}^0 \rightarrow \pi \pi(I) \right)
& =
& -A_I^* e^{i \delta_I}
\end{eqnarray}
where $I$ gives the isospin state of the pions and $\delta_I$
is the final-state phase shift determined from $\pi\pi$ scattering.
In general, $ A \left( K^0 \rightarrow \pi \pi(I) \right)
 =  \langle \pi \pi(I) | -i {\cal{H}} | K^0 \rangle$.
Defining $\bar{\epsilon}$ through
\begin{equation}
  \frac{p}{q} = \frac{( 1 + \bar{\epsilon})}{( 1 - \bar{\epsilon})}
\end{equation}
and using the isospin decomposition
\begin{alignat}{2}
    | \pi^0 \pi^0 \rangle
  &
    =
  &
      \sqrt{\frac{2}{3}} \; | \pi \pi(I=2) \rangle
  &
    - \sqrt{\frac{1}{3}} \; | \pi \pi(I=0) \rangle
  \\
    \sqrt{\frac{1}{2}} \; \left( | \pi^+ \pi^- \rangle
    + | \pi^- \pi^+ \rangle \right)
  &
    =
  &
      \sqrt{\frac{1}{3}} \; | \pi \pi(I=2) \rangle
  &
    + \sqrt{\frac{2}{3}} \; | \pi \pi(I=0) \rangle
\end{alignat}
one can show \cite{Winstein:1993sx}
\begin{equation}
  \epsilon = \bar{\epsilon}
	   + i \left( \frac{ {\rm Im} A_0 }{ {\rm Re} A_0} \right)
  \label{eq:e_from_a0_a2}
\end{equation}
\begin{equation}
  \epsilon^\prime = \frac{ i e^{i(\delta_2 - \delta_0)}}{ \sqrt{2} }
		    \frac{ {\rm Re} A_2 }{ {\rm Re} A_0 }
	     \left[ \frac{ {\rm Im} A_2 }{ {\rm Re} A_2 } -
	            \frac{ {\rm Im} A_0 }{ {\rm Re} A_0 } \right]
  \label{eq:ep_from_a0_a2}
\end{equation}
We define
\begin{eqnarray}
  \omega & \equiv & \frac{ {\rm Re} A_2 }{ {\rm Re} A_0 } \\
  P_0 & \equiv & \frac{ {\rm Im} A_0 }{ {\rm Re} A_0 } \\
  P_2 & \equiv & \frac{ {\rm Im} A_2 }{ {\rm Re} A_2 }
\end{eqnarray}
and simplify Eqs.\ \ref{eq:e_from_a0_a2} and \ref{eq:ep_from_a0_a2}
to
\begin{equation}
  \epsilon = \bar{\epsilon} + i P_0
  \label{eq:e_from_p0_p2}
\end{equation}
\begin{equation}
  \epsilon^\prime = \frac{ i e^{i(\delta_2 - \delta_0)}}{ \sqrt{2} }
		    \omega \left[ P_2 - P_0 \right]
  \label{eq:ep_from_p0_p2}
\end{equation}
The equations above assume that both $\bar{\epsilon}$ and $\omega$
are small quantities, which is true for the physical values of
quark masses.  In particular, the small value of $\omega$ (0.045)
is the quantitative expression of the $\DIhalf$ rule.  For our
quenched QCD simulations, we must be careful to only use these
formula for situations where both $\bar{\epsilon}$ and $\omega$
are small.

There are corrections to Eq.\ \ref{eq:ep_from_p0_p2} from isospin
violations.  These will not be included in our current calculation
but have been estimated by \cite{Buras:1987wc,Ecker:2000zr}.

From Eq.\ \ref{eq:ep_from_p0_p2} one sees that $CP$ violation in decays
comes from a non-zero value of $P_2 - P_0$.  This in turn arises
through isospin-dependent imaginary parts of $A_I$.  In the Standard
Model, CP-violating imaginary contributions to $A_0$ and $A_2$ enter
only through the CKM matrix element $V_{td}$.  The effects of $V_{td}$
enter through the penguin operators and in particular, the major
contribution to ${\rm Im} A_2$ is expected to come from the electroweak
penguin operators, while the QCD penguin operators should produce most
of ${\rm Im} A_0$.  Given that $P_2 - P_0$ determines the size of direct
$CP$ violation effects, estimates of the generic size of $P_0$ and
$P_2$ do not tightly constrain $\ep$.

Since a non-perturbative lattice calculation of $K \rightarrow \pi \pi$
matrix elements yields $A_0$ and $A_2$, the calculation also produces a
value for $\omega$.  The value of $\omega$ is an interesting quantity
in its own right and because of its dependence only on the real parts
of the amplitudes, it probes Standard-Model physics that is quite
different from CP violation.

To determine $\epsilon$, one needs the value for $\bar{\epsilon}$
which in turn comes from a determination of the off-diagonal
elements of the two by two matrix governing the evolution of
the $K^0$-$\Kbar^0$ system \cite{Buras:1998ra}.  These off-diagonal
contributions are commonly parameterized by defining $B_K(\mu)$
through
\begin{equation}
  \langle \Kbar^0 | \DStwoQ(\mu) | K^0 | \rangle \equiv \frac{8}{3}
  B_K(\mu) f_K^2 m_K^2
  \label{eq:bk_def}
\end{equation}
and the renormalization group invariant parameter $\hat{B}_K$
by
\begin{equation}
  \hat{B}_K \equiv B_K(\mu) \left[ \alpha_s^{(3)}(\mu) \right]^{-2/9}
  \left[ 1 + \frac{\alpha_s^{(3)}(\mu)}{4 \pi} J_3 \right].
\end{equation}
With these definitions, one finds that
\begin{align}
    \epsilon =
  &
    \hat{B}_K \, {\rm Im} \lambda_t \,
    \frac{ G_F^2 f_K^2 m_K M_W^2}{12 \sqrt{2} \pi^2 \Delta M_K}
    \nonumber
  \\
  &
    \times
    \left\{ {\rm Re} \lambda_c \left[ \eta_1 S_0(x_c) 
      - \eta_3 S_0(x_c, x_t) \right] - {\rm Re} \lambda_t \,
	\eta_2 S_0(x_t) \right\} \exp(i \pi/4)
\end{align}
where $\Delta M_K$ is the mass difference between $K_L$ and $K_S$.

Thus, a determination from lattice QCD simulations of $\langle \pi \pi |
Q_i(\mu) | K \rangle$ and $\langle \overline{K}^0 | \DStwoQ(\mu) | K^0
\rangle$ matrix elements, coupled with experimental measurements of
$\epsilon^\prime$ and $\epsilon$, gives constraints on the elements of
the CKM matrix in the Standard Model.  Additionally, the lattice
calculations should also yield a value for $\omega$ which is
expected to be essentially independent of the elements of the CKM
matrix.  We now turn to some of the issues faced in the lattice
determinations of the matrix elements.

\fi


\section{Continuum Chiral Perturbation Theory and Kaon Matrix Elements}
\label{sec:cont_chiral_pert}

\ifnum\theContChiralPert=1
%
%

The calculation of decay amplitudes with multi-particle final states
presents a challenge to the Euclidean-space techniques of lattice QCD.
In a general field-theoretic context, Euclidean space and Minkowski
space are related by an analytic continuation.  Such an analytic
continuation in a numerical calculation is extremely difficult, given
that a discrete set of data points with statistical errors does not
define an analytic function.  Fortunately, there are matrix elements we
can calculate directly from lattice QCD using the usual lattice
projection technique of evaluating the large time limit of the operator
$e^{\{-H_{\rm QCD} t \}}$.  For single particle matrix elements, we
directly achieve the matrix element at the desired kinematic values.
However, for multi-particle states with non-zero relative momentum, the
state will not be the lowest energy state with a specific set of
quantum numbers and, therefore, cannot be isolated by the large time
limit of the operator $e^{\{-H_{\rm QCD} t \}}$, the Maiani-Testa
theorem \cite{Maiani:1990ca}.  As a result, $K \rightarrow \pi \pi$
transition amplitudes with physical masses cannot be directly measured
on the lattice with current techniques.  (There is a recent promising
proposal \cite{Lellouch:2000pv} to tune the finite volume of a
Euclidean-space simulation so that the physical, multi-particle final
state corresponds to a next-lowest energy, finite-volume eigenstate of
$H_{\rm QCD}$---a state that might be extracted from the time
dependence given by $e^{\{-H_{\rm QCD} t \}}$.)"

Even before the formalization of the Maiani-Testa theorem, it was
realized \cite{Bernard:1985wf} that chiral perturbation theory could be
used to relate $K \rightarrow \pi \pi $ amplitudes to $K \rightarrow
\pi $ and $K \rightarrow |0\rangle$ amplitudes (here $|0\rangle$ is the
vacuum).  In addition to circumventing the Maiani-Testa theorem, these
amplitudes should be easier to measure numerically, since they involve
fewer interpolating operators to produce the mesons.  Chiral
perturbation theory uses the effective Lagrangian representing the
pseudo-Goldstone boson degrees of freedom for QCD to determine
relations between the desired matrix elements.  It should be noted that
the chiral effective Lagrangian automatically satisfies the relevant
Ward-Takahashi identities of QCD, in the limit when these identities
are dominated by arbitrarily light pseudo-Goldstone bosons.

Our use of chiral perturbation theory in the calculation of the $K
\rightarrow \pi\pi$ weak matrix elements requires that we address a
number of issues.  We cannot currently calculate lattice matrix
elements for arbitrarily small quark mass, where the quark mass
dependence is linear, since such small masses require large volumes and
computer resources beyond those currently available.  Since our quark
masses will be as large as the strange quark mass, we must understand
the non-linear dependence expected from continuum chiral perturbation
theory for $K \rightarrow \pi$ and $K \rightarrow |0\rangle$ matrix
elements.  Such an understanding will allow us to see if our data
matches these expectations and to permit us to accurately extract the
low-energy chiral perturbation theory parameters needed to make the
connection to the desired two pion decay.  (As we will discuss in
Section \ref{sec:dwf_chiral_mod} we can also get non-linearities from a
lattice effect, domain wall fermion zero modes in quenched QCD for
finite volume.)\ \ Since our calculation is done in the quenched
approximation, we must also look for the pathologies expected from
quenched chiral perturbation theory.  Finally, our results for $K
\rightarrow \pi\pi$ weak matrix elements in the chiral limit must be
compared with the physical values measured for non-zero quark mass.
Estimates of the effects of higher order terms in chiral perturbation
theory are crucial to estimating the systematic errors in extrapolating
to the physical kaon mass.  We now turn to the results from chiral
perturbation theory relevant to our determination of weak matrix
elements.


\subsection{Lowest order Chiral Perturbation Theory}
\label{subsec:lowest_order_XPT}

Following \cite{Bernard:1985wf} and adopting their conventions for
states and normalizations (see Appendix \ref{sec:conventions} for a
summary), one must represent the various operators listed in
Eqs.\ \ref{eq:Q1} to \ref{eq:P10} in terms of the fields used in chiral
perturbation theory.  One starts with a unitary chiral matrix field,
$\Sigma$, defined by
\begin{equation}
  \Sigma \equiv \exp \left[ \frac{2 i \phi^a t^a}{f} \right],
\end{equation}
where $\phi^a$ are the real pseudo-Goldstone boson fields, $t^a$
are proportional to the Gell-Mann matrices, with ${\rm Tr}( t_a
t_b) = \delta_{ab}$, and $f$ is the pion decay constant.  In
chiral perturbation theory, the lowest order Lagrangian for QCD,
of ${\cal O}(p^2)$, is
\begin{equation}
  {\cal L}^{(2)}_{\rm QCD} = \frac{f^2}{8} {\rm Tr} \left( \partial_\mu
    \Sigma \partial^\mu \Sigma^\dagger \right) + v \, {\rm Tr}
    \left[ M \Sigma + (M \Sigma)^\dagger \right]
  \label{eq:lxpt_qcd}
\end{equation}
Here $M$ is the quark mass matrix and
\begin{equation}
  v = \frac{f^2 m_{\pi^+}^2}{4(m_u + m_d)}
\end{equation}
Thus, $v$ is the chiral condensate at zero quark mass and, as shown in
Appendix \ref{sec:conventions} $\langle \bar{u} u \rangle(m_q = 0) =
-2v$.  Note that the matrix field $\Sigma$ has $\sutlr$ quantum numbers
$(L,R) = (3,\overline{3})$.  Here $f$ is the pion decay constant in the
limit $m_q \rightarrow 0$ and we use a normalization where $f_\pi$ is
131 MeV.

Working to lowest order in chiral perturbation theory, one finds
\cite{Bernard:1985wf} that there are two possible $(8,1)$
operators with $\DSone$ and $\DDmone$, denoted by
$\tilde{\Theta}_1^{(8,1)}$ and $\tilde{\Theta}_2^{(8,1)}$ and a single
$(27,1)$ operator $\tilde{\Theta}^{(27,1)}$.  These three operators
are all that is required to represent the matrix elements of the
operators in Eq.\ \ref{eq:Q1} to \ref{eq:P10}, except $Q_7$, $Q_8$,
$P_7$ and $P_8$.  Other work \cite{Bijnens:1984ye} showed
that there is a single $(8,8)$ operator.  Thus, the correspondence
between an operator $\Theta^{(L,R)}$ given in terms of quark fields and
its representation in chiral perturbation theory is given by
\begin{alignat}{2}
    \Theta^{(8,1)}
  & \rightarrow
  & \alpha_1^{(8,1)} & \tilde{\Theta}_1^{(8,1)}
    + \alpha_2^{(8,1)} \tilde{\Theta}_2^{(8,1)}
  \label{eq:def_alpha_81}
  \\
    \Theta^{(27,1)}
  & \rightarrow
  & \; \alpha^{(27,1)} & \tilde{\Theta}^{(27,1)}
  \label{eq:def_alpha_271}
  \\
    \Theta^{(8,8)}
  & \rightarrow
  & \alpha^{(8,8)} & \tilde{\Theta}^{(8,8)}
  \label{eq:def_alpha_88}
\end{alignat}
where the $\alpha$'s are constants and
\begin{eqnarray}
    \tilde{\Theta}_1^{(8,1)}
  &
    \equiv
  &
    {\rm Tr} \left[ \Lambda \, (\partial_\mu \Sigma) \,
    ( \partial^\mu \Sigma^\dagger) \right]
    \label{eq:cpt_theta81_1}
  \\
    \tilde{\Theta}_2^{(8,1)}
  &
    \equiv
  &
    \frac{8v}{f^2} {\rm Tr} \left[ \Lambda \Sigma M
    + \Lambda (\Sigma M)^\dagger \right]
    \label{eq:cpt_theta81_2}
  \\
    \tilde{\Theta}^{(27,1)}
  &
    \equiv
  &
    T^{ij}_{kl} \, ( \Sigma \, \partial_\mu \Sigma^\dagger)^k_i \,
    ( \Sigma \, \partial^\mu \Sigma^\dagger)^l_j
    \label{eq:cpt_theta271}
  \\
    \tilde{\Theta}^{(8,8)}
  &
    \equiv
  &
    {\rm Tr} \left[ \Lambda \, \Sigma \, T_R \, \Sigma^\dagger \right]
    \label{eq:cpt_theta88}
\end{eqnarray}
Here $\Lambda_{ij} \equiv \delta_{i3} \delta_{j2}$, $T^{ij}_{kl}$ is
symmetric in $i,j$ and $k,l$ and traceless on any pair of upper and
lower indices and $T_R \equiv {\rm diag}(2, -1, -1)$.  Further detail
is given in Appendicies \ref{sec:four_quark_irreps} and
\ref{sec:cpt_decomp}, along with precise values for $T^{ij}_{kl}$ for
both the $\DIhalf$ and $\DIthalf$ components.

There is a unique set of $\alpha$'s for each four-quark operator that
is in an irreducible representation of $\sutlr$.  The operators in
Eq.\ \ref{eq:Q1} to \ref{eq:P10} are generally in reducible
representations, so we will determine the $\alpha$'s for each operator
individually.  The matrix elements of the effective operators
$\tilde{\Theta}$ given in
Eqs.\ \ref{eq:cpt_theta81_1}-\ref{eq:cpt_theta88} between states
composed of pions and kaons can be easily evaluated in chiral
perturbation theory.  For the $K \rightarrow 0$ matrix elements one
finds
\begin{eqnarray}
    \langle 0 | \Theta^{(8,1)} | K^0 \rangle
  & \; =
  & \frac{16 i v}{f^3}(m_s^\prime - m_d^\prime) \alpha_2^{(8,1)}
    \label{eq:kvacO81}
  \\
    \langle 0 | \Theta^{(27,1)} | K^0 \rangle
  & \; =
  & 0
    \label{eq:kvacO271}
  \\
    \langle 0 | \Theta^{(8,8)} | K^0 \rangle
  & \; =
  & 0
    \label{eq:kvacO88}
\end{eqnarray}
where $m_s^\prime$ and $m_d^\prime$ are the quark masses used in the
construction of the $K^0$.  Similarly
\begin{eqnarray}
    \langle \pi^+ | \Theta^{(8,1)} | K^+ \rangle
  &
    =
  &
    \frac{4 m_M^2}{f^2} \left( \alpha_1^{(8,1)}-\alpha_2^{(8,1)} \right)
    \label{eq:kpiO81}
  \\
    \langle \pi^+ | \Theta^{(27,1)} | K^+ \rangle
  &
    =
  &
    -\frac{4 m_M^2}{f^2} \alpha^{(27,1)}
    \label{eq:kpiO271}
  \\
    \langle \pi^+ | \Theta^{(8,8)} | K^+ \rangle
  &
    =
  &
    \frac{12}{f^2} \alpha^{(8,8)}
    \label{eq:kpiO88}
\end{eqnarray}
where $m_M$ is the common meson mass of the $\pi^+$ and $K^+$.
Following \cite{Bernard:1985wf}, one then finds that
the desired $K \rightarrow \pi\pi$ matrix elements are given by
\begin{eqnarray}
    \langle \pi^+ \pi^- | \Theta^{(8,1)} | K^0 \rangle
  &
    =
  &
    \frac{4 i}{f^3} \left( m_{K^0}^2 - m_{\pi^+}^2 \right)
    \alpha_1^{(8,1)}
    \label{eq:kpipiO81}
  \\
    \langle \pi^+ \pi^- | \Theta^{(27,1)} | K^0 \rangle
  &
    =
  &
    - \frac{4 i}{f^3} \left( m_{K^0}^2 - m_{\pi^+}^2 \right)
    \alpha^{(27,1)}
    \label{eq:kpipiO271}
  \\
    \langle \pi^+ \pi^- | \Theta^{(8,8)} | K^0 \rangle
  &
    =
  &
    \frac{-12i}{f^3} \alpha^{(8,8)}
    \label{eq:kpipiO88}
\end{eqnarray}
Since the (27,1) and (8,8) operators contain both $\DIhalf$ and
$\DIthalf$ parts, which we will need to measure to determine $K
\rightarrow \pi \pi$ amplitudes of definite isospin, we give the
isospin decomposition of Eqs.\ \ref{eq:kpiO271}, \ref{eq:kpiO88},
\ref{eq:kpipiO271} and \ref{eq:kpipiO88} in section
\ref{sec:cpt_decomp} of the Appendix.

These simple relations form the heart of the calculation we have
performed and a few important points are worth highlighting:
\begin{enumerate}
\item
  The current calculation is a determination of the physical parameters
  $ \alpha_1^{(8,1)}$, $\alpha^{(27,1)}$ and $\alpha^{(8,8)}$ for a
  fixed lattice spacing and volume in the quenched approximation.  As
  such, $K \rightarrow \pi \pi$ amplitudes are determined to lowest
  order in chiral perturbation theory in the quenched approximation.
\item
  The $K^+ \rightarrow \pi^+$ matrix elements of (8,1) and (27,1)
  operators vanish in the chiral limit, while for (8,8) operators
  the matrix element is non-zero.  Thus, for small enough quark
  masses, the electroweak penguin operators will dominate all
  amplitudes.  Since the electroweak penguin operators are suppressed
  by the electroweak coupling constant, the quark mass
  where they dominate is quite small.
\item
  The term $\alpha_2^{(8,1)}$ is determined by the unphysical $K^0
  \rightarrow 0$ matrix element and in general is quadratically
  divergent for regularizations which preserve chiral symmetry.  To
  determine $\alpha_1^{(8,1)}$, and hence the physical $K \rightarrow
  \pi \pi$ amplitude, requires canceling this quadratic divergence
  against the quadratic divergence in $\langle \pi^+ | \Theta^{(8,1)} |
  K^+ \rangle$.  This first-principles cancellation arises in the
  relevant Ward-Takahashi identities of QCD and is reflected in chiral
  perturbation theory, which respects these identities.  For the most
  extreme cases, the physical result is only 5\% of the size of the
  divergent terms.  This $\alpha_2^{(8,1)}$ subtraction will be
  extensively discussed in Section \ref{subsec:sub_power_div}.
\item
  The $\alpha_2^{(8,1)}$ subtraction is determined by matrix elements
  of four-quark operators in hadronic states.  As part of the
  renormalization of lattice four-quark operators, a related
  subtraction must be done for matrix elements of these operators in
  off-shell Green's functions involving quark fields.  Only the
  momentum independent divergent parts of these two subtractions are
  the same.  This issue is discussed further in Section \ref{sec:npr}.
\item
  In these lowest order chiral perturbation theory expressions, only
  $\alpha_2^{(8,1)}$ is divergent.  However, higher order terms in
  chiral perturbation theory can be multiplied by divergent
  coefficients, as happens for (8,8) operators.   Thus, Eq.\
  \ref{eq:kpiO88} is modified at next order by the addition of
  a divergent term of the form
  \begin{equation}
    \langle \pi^+ | \Theta^{(8,8)} | K^+ \rangle =
    \frac{12}{f^2} \left\{ \alpha^{(8,8)}
      + m_M^2 \alpha_{\rm div}^{(8,8)} + \cdots \right\}
    \label{eq:kpiO88div}
  \end{equation}
  where the dots represent possible non-divergent higher order terms.
  Even though the matrix element is non-zero when $m_q = 0$, the finite
  quark mass corrections enter with a power divergent coefficient.  One
  way to find the $m_q = 0$ value is to extrapolate in quark mass.  For
  domain wall fermions at finite $L_s$, the zero quark mass limit is
  not precisely known for power divergent operators.  This, coupled
  with the power divergent slope, makes the extrapolation problematic.
  One can use a subtraction to remove the divergent slope.  However, an
  even simpler approach is to use the $\Delta I = 3/2$ part of the
  (8,8) operator, which does not have divergent coefficients, to
  determine $\alpha^{(8,8)}$.

\end{enumerate}


\subsection{Full QCD 1-loop Chiral Perturbation Theory:
  $K\rightarrow \pi\pi$}
\label{subsec:one_loop_XPT}

An important early calculation in QCD revealed that in the small quark
mass limit $\mpi2$ deviates from simple linear dependence on the quark
mass, $m_q$, due to chiral logarithm terms of the form $m_q \ln m_q$
\cite{Langacker:1973hh}.  In the language of chiral perturbation theory
such logarithms arise from higher order loop effects, which for $\mpi2$
come from calculating loop corrections using ${\cal L}^{(2)}_{\rm
QCD}$.  To work to a consistent order in chiral perturbation theory
requires that if loop effects in the ${\cal O}(p^2)$ effective Lagrangian are
included, one must also include the effects of the ${\cal O}(p^4)$ terms in
the effective Lagrangian, denoted ${\cal L}^{(4)}_{\rm QCD}$.
Unfortunately, ${\cal L}^{(4)}_{\rm QCD}$ introduces new, unknown
parameters, but for on-shell particles at rest these parameters are
multiplied by $m_q^2$.  Thus, the general form for a quantity like
$\mpi2$ in full QCD is
\begin{equation}
  \mpi2 = a_1 m_q + a_{l} m_q^2 \ln m_q + a_2 m_q^2
\end{equation}

Systematic calculations of higher loop effects in chiral perturbation
theory \cite{Gasser:1984yg,Gasser:1985gg} have been done including the
up, down and strange quarks.  We will give these results in terms of
the lowest order chiral perturbation theory, or bare, meson masses,
which are given, for example, by $m_{\pi^+}^2 = 4 v (m_u+m_d)/f^2$
where $f$ and $v$ are constants.  We will set $m_u = m_d$ and denote
the subtraction point for chiral perturbation theory by $\Lcpt$.
Calculating the one loop terms in ${\cal L}_{\rm QCD}^{(2)}$ gives
\cite{Gasser:1985gg}
\begin{alignat}{4}
    (\mpi2)^{(\rm 1-loop)}
  &
    =
  &
    \; \mpi2 \Bigl\{
  &
    1
  & 
    + \; L_\chi(m_\pi)
  &
  &
    - \frac{1}{3} L_\chi(m_\eta)
  &
    + \cdots \Bigr\} \label{eq:mpi2-1loop}
  \\
    (\mK2)^{(\rm 1-loop)}
  &
    =
  &
    \; \mK2 \Bigl\{
  &
    1
  &
  &
  &
    + \frac{2}{3} L_\chi(m_\eta)
  &
    + \cdots \Bigr\} \label{eq:mK2-1loop}
  \\
    (f_\pi)^{(\rm 1-loop)}
  &
    =
  &
    \; f \Bigl\{
  &
    1
  &
    -2 L_\chi(m_\pi)
  &
    - \; L_\chi(m_K)
  &
  &
    + \cdots \Bigr\} \label{eq:fpi-1loop}
  \\
    (f_K)^{(\rm 1-loop)}
  &
    =
  &
    \; f \Bigl\{
  &
    1
  &
    -\frac{3}{4} L_\chi(m_\pi)
  &
    - \frac{3}{2} L_\chi(m_K)
  &
    - \frac{3}{4} L_\chi(m_\eta)
  &
    + \cdots \Bigr\} \label{eq:fK-1loop}
\end{alignat}
where
\begin{equation}
  L_\chi(m) \equiv \frac{m^2}{(4 \pi f )^2} \ln
    \left( \frac{ m^2}{ \Lcpt^2} \right)
\end{equation}
and the dots represent terms quadratic in the pseudoscalar masses.
The coefficients of these terms depend on parameters entering
${\cal L}_{\rm QCD}^{(4)}$.

To study matrix elements in chiral perturbation theory, one starts 
from the lowest order QCD Lagrangian in Eq.\ \ref{eq:lxpt_qcd} and adds
terms representing the effective four-fermion operators at low energies.
To ${\cal O}(p^2)$ this yields
\begin{equation}
  {\cal L}^{{\cal O}(p^2)}_{\rm eff}
    =  {\cal L}^{(2)}_{\rm QCD} + {\cal L}^{(0)}_{\DSone}
    +  {\cal L}^{(2)}_{\DSone} + {\cal L}^{(2)}_{\DStwo}
\end{equation}
for the $\DSone$ and $\DStwo$ processes of interest here.  Note that
there are terms at ${\cal O}(p^0)$ that enter the $\DSone$ part of the
chiral Lagrangian.  These are the (8,8) operators mentioned in the
previous section which represent the electroweak penguins $Q_7$ and
$Q_8$ for $\mu < m_c$, or a part of $P_7$ and $P_8$ for $\mu > m_c$.
The term ${\cal L}^{(0)}_{\DSone}$ depends on the single parameter
$\alpha^{(8,8)}$ for each operator, while ${\cal L}^{(2)}_{\DSone}$
depends on $\alpha_1^{(8,1)}$, $\alpha_2^{(8,1)}$, $\alpha^{(27,1)}$
and the coefficients for higher order (8,8) operators.  The single
operator appearing in ${\cal L}^{(2)}_{\DStwo}$ enters with a parameter
which can be shown to be related to $\alpha^{(27,1)}$.

The chiral logarithm terms in $\DSone$ and $\DStwo$ matrix elements can
be calculated using ${\cal L}^{O(p^2)}_{\rm eff}$.  Amplitudes
involving $\Theta^{(8,8)}$, which are non-zero at ${\cal O}(p^0)$ due to
${\cal L}^{(0)}_{\DSone}$, have chiral logarithms at ${\cal O}(p^2)$ due to
interaction terms in ${\cal L}^{O(p^2)}_{\rm eff}$.  These chiral logarithms
have not yet been calculated explicitly, but should modify
Eq.\ \ref{eq:kpiO88div} to the form
\begin{equation}
  \langle \pi^+ | \Theta^{(8,8)} | K^+ \rangle =
    \frac{12}{f^2} \left\{ \alpha^{(8,8)}
      \left[ 1 + \xi^{(8,8)} L_\chi(m_{M}) + \cdots \right]
      + m_M^2 \alpha_{\rm div}^{(8,8)} + \cdots \right\}
  \label{eq:kpiO88div_log}
\end{equation}
where $\xi^{(8,8)}$ is a calculable coefficient and $m_M$ is the common
mass for the $\pi^+$ and $K^+$ in this matrix element.  (For full QCD,
these terms were calculated after this manuscript was finished in
\cite{Cirigliano:2001hs}.) As previously mentioned, unless only the
$\DIthalf$ amplitude is considered, there are higher order terms in
chiral perturbation theory with power divergent coefficients, given
collectively in Eq.\ \ref{eq:kpiO88div_log} by $\alpha_{\rm
div}^{(8,8)}$.

The effective Lagrangian to the next order, ${\cal L}^{O(p^4)}_{\rm
eff}$, includes all possible ${\cal O}(p^4)$ terms and introduces many unknown
coefficients.  This Lagrangian takes the form
\begin{equation}
  {\cal L}^{{\cal O}(p^4)}_{\rm eff}
    = {\cal L}^{O(p^2)}_{\rm eff} + {\cal L}^{(4)}_{QCD} + {\cal
    L}^{(4)}_{\DSone} + {\cal L}^{(4)}_{\DStwo}
\end{equation}
For $\DSone$ processes at ${\cal O}(p^4)$, amplitudes will include loop
effects coming from ${\cal L}^{(2)}_{QCD}$ and $ {\cal
L}^{(2)}_{\DSone}$.  There are also ${\cal O}(p^4)$ contributions from two
loop corrections to ${\cal L}^{(0)}_{\DSone}$.

For $\Theta^{(8,1)}$ and $\Theta^{(27,1)}$ $\DSone$ operators,
the chiral logarithm corrections to the matrix elements of interest
in this work have been calculated \cite{Bijnens:1984ec,Bijnens:1985qt,Golterman:1998st,Golterman:2000fw,Bijnens:1998mb}.
The results for $K^+ \rightarrow \pi^+$ are:
\begin{alignat}{4}
    \langle \pi^+ | \Theta^{(8,1)} | K^+ \rangle
  &
    =
  &
    \frac{4 m_M^2}{f^2} \Bigl\{
  &
    \alpha_1^{(8,1)}
  &
    \left[ 1 + \; \, \frac{1}{3} L_\chi(m_M) \right]
  &
    -\alpha_2^{(8,1)}
  &
    \Bigr[ 1 + 2 L_\chi(m_M) \Bigl]
  &
    \Bigr\} \label{eq:kpi81}
  \\
    \langle \pi^+ | \Theta^{(27,1)} | K^+ \rangle
  &
    =
  & 
     -\frac{4 m_M^2}{f^2} \Bigl\{
  &
    \alpha^{(27,1)}
  &
    \left[ 1 - \frac{34}{3} L_\chi(m_M) \right]
  &
    \Bigr\} \label{eq:kpi271}
\end{alignat}
Similarly for $K \rightarrow |0\rangle$ one finds
\begin{alignat}{3}
    \langle 0| \Theta^{(8,1)} | K^0 \rangle
  &
    =
  &
    \frac{4 i \alpha_2^{(8,1)}}{f} (m_K^2 - m_\pi^2) \Bigl\{
  &
    1
  &
    - \frac{3}{4} L_\chi(m_\pi) - \frac{3}{2} L_\chi(m_K)
    - \frac{1}{12} L_\chi(m_\eta) \Bigr \}
    \nonumber
  \\
  &
    +
  &
    \frac{4 i \alpha_1^{(8,1)}}{f} (m_K^2 - m_\pi^2) \Bigl\{
  &
  &
    + \frac{1}{3} L_\chi(m_\eta,m_K)
    - 2 L_\chi(m_\eta, m_\pi) \Bigr \}
    \label{eq:kvac81}
  \\
    \langle 0| \Theta^{(27,1)} | K^0 \rangle
  &
    =
  &
    \; \frac{4 i \alpha^{(27,1)}}{f} (m_K^2 - m_\pi^2) \Bigl\{
  &
  &
    - 2 L_\chi(m_\eta, m_K) + 2 L_\chi(m_\eta, m_\pi)
    \Bigr\} \label{eq:kvac271}
\end{alignat}
where
\begin{equation}
  L_\chi(m_1, m_2) \equiv \frac{1}{(4 \pi f)^2}
     \; \frac{1}{m_1^2 - m_2^2} \;
     \left[ m_1^4 \ln \left( \frac{m_1^2}{\Lcpt^2} \right)
          - m_2^4 \ln \left( \frac{m_2^2}{\Lcpt^2} \right) \right]
  \label{eq:def_chiral_log}
\end{equation}

One of the most important aspects of using these formulae to determine
$K \rightarrow \pi\pi$ matrix elements is the determination of the
coefficients $\alpha_2^{(8,1)}$, which are in general quadratically
divergent in a regularization which preserves chiral symmetry.  (Since
(8,1) operators are pure $\DIhalf$, we cannot avoid $\alpha_2^{(8,1)}$
by measuring only $\DIthalf$ amplitudes, as we can avoid
$\alpha_{\rm div}^{(8,8)}$.) \ \  However, as the equations above show,
$\alpha_2^{(8,1)}$ is multiplied by chiral logarithm corrections at
subleading order.  Given the large difference possible in
$\alpha_1^{(8,1)}$ and $\alpha_2^{(8,1)}$, $\alpha_1^{(8,1)}$ can be
much smaller than $\alpha_2^{(8,1)} L_\chi(m_\pi)$.

The power divergent part of the four-quark operators can be written as
a quark bilinear times a momentum-independent coefficient.  Thus, the
chiral logarithm corrections to the power divergent parts of four-quark
operators must be the same as the chiral logarithm corrections to the
corresponding quark bilinears.  That this is indeed the case for full
QCD can be seen explicitly, since the chiral logarithms for the
bilinears are known and can be compared with Eqs.\ \ref{eq:kpi81} to
\ref{eq:kvac271}.  Following \cite{Bijnens:1985qt}, we define
\begin{equation}
  \Theta^{(3,\bar{3})} \equiv \bar{s}(1 - \gamma_5) d
  = \alpha^{(3, \bar{3})} \; {\rm Tr} \left( A \Sigma \right)
\end{equation}
to lowest order in chiral perturbation theory.  Here $A$ is a three by
three matrix with $A_{i,j} = \delta_{i,3} \delta_{j,2}$ and with our
conventions, $\alpha^{(3,\bar{3})} = -2iv$.  Then the chiral
logarithm corrections for the matrix elements of $ \langle \pi^+ |
\Theta^{(3,\bar{3})} | K^+ \rangle$ and $ \langle 0 |
\Theta^{(3,\bar{3})} | K^0 \rangle$ are given in
\cite{Bijnens:1985qt}.  We will use the value for $ \langle \pi^+ |
\Theta^{(3,\bar{3})} | K^+ \rangle$ from \cite{Bijnens:1985qt}, since
here there is a single meson mass, $m_M$.  For $ \langle 0 |
\Theta^{(3,\bar{3})} | K^0 \rangle$, where the meson masses are not
degenerate, the formula in \cite{Bijnens:1985qt} does not include
separate chiral logarithms for each of the possible meson masses,
$m_\pi$, $m_K$ and $m_\eta$.  Thus, for this matrix element, we make use
of the fact that
\begin{equation}
 \langle 0 | \Theta^{(3,\bar{3})} | K^0 \rangle \sim
   (m_K^2)^{\rm(1-loop)} (f_K)^{\rm(1-loop)} / m_q
\end{equation}
and use Eqs.\ \ref{eq:mK2-1loop} and \ref{eq:fK-1loop} to determine
the chiral logarithms.  This gives
\begin{alignat}{2}
    \langle \pi^+ | \Theta^{(3,\bar{3})} | K^+ \rangle
  &
    =
  &
    \frac{2}{f^2} \, \alpha^{(3,\bar{3})} \, \Bigl\{
  &
    1 + 2 L_\chi(m_M) \Bigr\} \label{eq:kpi33b}
  \\
    \langle 0 | \Theta^{(3,\bar{3})} | K^0 \rangle
  &
    =
  &
    \; \frac{2i}{f} \, \alpha^{(3,\bar{3})} \, \Bigl\{
  &
    1 - \frac{3}{4} L_\chi(m_\pi) - \frac{3}{2} L_\chi(m_K)
      - \frac{1}{12} L_\chi(m_\eta) \Bigr \} \label{eq:kvac33b}
\end{alignat}
Thus, we have
\begin{alignat}{2}
    \frac{ \langle 0 | \Theta^{(8,1)} | K^0 \rangle}
      {\langle 0 | \Theta^{(3,\bar{3})} | K^0 \rangle}
  &
    = 
  &
    \; 2 \, \frac{ \alpha_2^{(8,1)} }{\alpha^{(3,\bar{3})}} \,
	 \left( m_K^2 - m_\pi^2 \right)
  &
    \Bigl\{ 1 + \cdots \Bigr\} \nonumber
  \\
  &
    +
  &
    2 \, \frac{ \alpha_1^{(8,1)} }{\alpha^{(3,\bar{3})}} \,
       \left( m_K^2 - m_\pi^2 \right)
  &
    \left\{ {\rm chiral \ logs } + \cdots \right \}
    \label{eq:fix_alpha_81_2}
\end{alignat}
where the dots represent non-logarithmic higher order terms.  The
``chiral logs'' in Eq.\ \ref{eq:fix_alpha_81_2} are those given in the
second line of Eq.\ \ref{eq:kvac81} and in Eq.\ \ref{eq:kvac33b}.  As
expected, the chiral logarithms from the power divergent part of the
four-quark operator are the same as for the corresponding quark
bilinear.  The logarithms in the $\alpha_1^{(8,1)}$ term in Eq.\
\ref{eq:fix_alpha_81_2} are higher order in chiral perturbation theory
and are suppressed by the relative sizes of $\alpha_1^{(8,1)}$ and
$\alpha_2^{(8,1)}$.  For $m^2$ corrections which come from loops in the
${\cal O}(p^2)$ Lagrangian, one also expects a cancellation between the
bilinears and the four-quark operators.  This analysis leads us to
expect that the ratio in Eq.\ \ref{eq:fix_alpha_81_2} is a linear
function of $m_K^2 - m_\pi^2$ with very small corrections.  We will
investigate this numerically in Section \ref{subsec:sub_power_div}.

It is also important to note that once $\alpha_2^{(8,1)}$ has been
determined, we must numerically evaluate
\begin{equation}
    \langle \pi^+ | \Theta^{(8,1)} | K^+ \rangle
    +\alpha_2^{(8,1)} \,
    \frac{4 m_M^2}{f^2} \,
    \Bigr[ 1 + 2 L_\chi(m_M) \Bigl]
    \label{eq:use_alpha_81_2}
\end{equation}
to determine $\alpha^{(8,1)}_1$ if we are not working with arbitrarily
small quark masses.  If we could work close enough to the chiral limit,
the chiral logarithm terms in Eq.\ \ref{eq:use_alpha_81_2} would make
an arbitrarily small contribution.  This is not the case for the data
presented here, where the minimum pseudoscalar mass is 390 MeV.
Eq.\ \ref{eq:use_alpha_81_2} involves large cancellations between
divergent quantities.  Notice that the chiral logarithms are very
important in this determination, since they multiply the divergent
coefficient $\alpha_2^{(8,1)}$.  A simple way to do this, is to recall
that the power divergent part of four-quark operators should also have
the same chiral logarithms as the corresponding quark bilinear.
Eq.\ \ref{eq:kpi33b} shows this to be the case.  Thus, the combination
\begin{equation}
    \langle \pi^+ | \Theta^{(8,1)} | K^+ \rangle
    +
    2 m_M^2 \, \frac{ \alpha_2^{(8,1)} } { \alpha^{(3,\bar{3})} } \,
     \langle \pi^+ | \Theta^{(3,\bar{3})} | K^+ \rangle
    =
    \frac{4 m_M^2}{f^2} \,
    \alpha_1^{(8,1)}
    \left[ 1 + \frac{1}{3} L_\chi(m_M) \right]
    \label{eq:fix_alpha_81_1}
\end{equation}
yields a result only involving the physical coefficient
$\alpha_1^{(8,1)}$, with corrections in chiral perturbation theory that
do not involve the power divergent coefficient.  The chiral logarithms
which multiply power divergent coefficients have been removed, without
having to know their precise values.  (This subtraction technique was
originally discussed in Ref.\ \cite{Sharpe:1987xu}, although its
ability to remove power divergent terms multiplied by chiral logarithm
corrections was not discussed.) Note, this complete cancellation of the
quadratic divergence will hold as well in the quenched theory.  This is
important, since our actual calculation is done in the quenched
approximation where the coefficients of the chiral logarithms are not
known.


\subsection{Quenched 1-loop Chiral Perturbation Theory:
  $K\rightarrow \pi$ and $K \rightarrow 0$}
\label{subsec:quenched_XPT}

The discussion in the previous subsection focused on the chiral
logarithms present in various full QCD masses and matrix elements.
Similar techniques can be used to calculate the non-analytic dependence
on the quark mass for quenched simulations
\cite{Morel:1987xk,Sharpe:1990me,Bernard:1992mk}.  A surprising aspect
of these calculations is the appearance of quenched chiral logarithms,
where in addition to the $m_\pi^2 \ln m_\pi^2$ form of a conventional
QCD chiral logarithm, terms of the form $\delta \ln m_\pi^2$ also
appear.  Here $\delta$ is a constant given in terms of the parameters
which enter the low-energy effective Lagrangian for quenched QCD.
These effects are larger for small quark masses than the corresponding
conventional QCD logarithms, since they lack a factor of $m_\pi^2$.
Such effects may also appear in the matrix elements studied in this
paper and in this section we discuss the current state of analytic
results and how we will handle these effects in our simulation data.

For quenched chiral perturbation theory, a Lagrangian framework
has been developed \cite{Bernard:1992mk} and two new parameters
enter, $\alpha$ and $m_0$.  Calculating one-loop effects for the
pion mass gives
\begin{equation}
  (\mpi2)^{\rm (1-loop)} = \mpi2 \left\{ 1 + \frac{1}{8 \pi^2 f^2}
    \left[ \frac{\alpha}{3}\Lqcpt^2 - \frac{m_0^2}{3}
    - \left[ \frac{m_0^2}{3} - \frac{2 \alpha}{3} \mpi2 \right]
    \ln(\mpi2 / \Lqcpt^2) \right] \right\}
  \label{eq:qXPT_mpi}
\end{equation}
where $\Lqcpt$ is the scale used to renormalize the quenched theory.
From loops in the ${\cal O}(p^2)$ Lagrangian, one gets an ${\cal O}(m_\pi^4)$ term of
$\alpha m_\pi^4 / 24 \pi^2 f^2$, which is not shown in
Eq.\ \ref{eq:qXPT_mpi}.  It is common to define the coefficient of the
quenched chiral logarithm by $\delta$, where
\begin{equation}
  \delta \equiv \frac{m_0^2}{24 \pi^2 f^2}
\end{equation}
It is important to note that, in addition to the appearance of the
$m_0^2 \ln \mpi2$ term, the only conventional chiral logarithm appears
multiplied by $\alpha$.  In Section \ref{sec:test_dwf_chiral} we
discuss the determination of $m_0$ and $\alpha$ from our measurements
of the dependence of pion mass squared on the quark mass for
quenched domain wall fermion simulations.

For the kaon matrix elements of primary interest in this work,
quenching is also expected to modify the quark mass dependence from the
full QCD forms given in the previous subsection.  A recent calculation
of the quenched chiral logarithms for the (8,1) and (27,1) operators
has been presented in Ref.\ \cite{Golterman:2000fw}.  Calculations of
this type, including the (8,8) operators, are very useful in the
analysis of matrix elements from QCD simulations.  Unfortunately, the
currently available calculations completely remove all quark loops,
including those in the effective low energy four-quark operators.  For
the ${\cal O}(p^2)$ $\DSone$ Lagrangian of quenched chiral perturbation
theory, Eq.\ 2.2 of \cite{Golterman:2000fw} shows that the authors have
used a supertrace to represent the operators in chiral perturbation
theory.  The supertrace introduces ghost quarks to cancel loop effects
of real quarks, which is an unconventional definition of the quenched
approximation.

However, for actual numerical QCD calculations, quark loops which can
be made through self-contractions of the low-energy four-quark
operators of Eqs.\ \ref{eq:Q1} through \ref{eq:P10} are included.  Only
disconnected quark loops, generated through the quark determinant in
QCD and connected solely by gluon exchange with the four-quark
operators, are discarded.  The numerical simulations correspond to
evaluating all relevant four-quark operators, at low energies, in
background gluon fields generated without explicit vacuum polarization
quark loops.  In the quenched approximation these vacuum quark loop
effects are partially included by using an appropriately shifted value
of the bare QCD coupling $\beta$.  The existing analytic calculations
for the quenched theory correspond to evaluating all relevant
four-quark operators, including ghost quark self-contractions, in a
quenched gluon background.  Since these situations are quite distinct,
formula presented in \cite{Golterman:2000fw} are not generally
applicable to our simulation results.

There is one result from \cite{Golterman:2000fw} and earlier
calculations which is applicable to our simulations, the amplitude for
$K^+ \rightarrow \pi^+$ for the (27,1) operators.  Since there are no
self-contractions of the four-quark operators in this amplitude, it is
unaffected by the ghost-quark loops discussed in the previous
paragraphs.  Quenched chiral perturbation theory predicts that this
amplitude has the form
\begin{alignat}{4}
    \langle \pi^+ | \Theta^{(27,1)} | K^+ \rangle
  &
    =
  & 
     \; -\frac{4 m_M^2 \alpha^{(27,1)}} {f^2} \Bigl [ 1
  &
    \; - 6 L_{Q\chi}(m_M)
  \nonumber \\
  &
  &
  &
    - \frac{(m_0^2 - 2 \alpha m_M^2)}{24 \pi^2 f^2 }
    \ln(m_M^2 / \Lqcpt^2) \Bigr ] + O(m_M^4)
    \label{eq:kpi271q}
\end{alignat}
where $L_{Q\chi}(m)$ is the same as $L_{\chi}$ defined in
Eq.\ \ref{eq:def_chiral_log} except that $\Lcpt$ is replaced by
$\Lqcpt$.  Note that Eq.\ \ref{eq:kpi271q} contains both a conventional
chiral logarithm and a quenched chiral logarithm.  The conventional
chiral logarithm is quite large (its coefficient is 6) but markedly
smaller than the conventional chiral logarithm in full QCD,
Eq.\ \ref{eq:kpi271} (its coefficient is 34/3).  It is fortunate that
this quenched formula is known, since, as we will discuss in Section
\ref{sec:i_3over2_me}, the value of $\alpha^{(27,1)}$ we can determine
from our data is strongly dependent on the known analytic value for
the coefficient of the conventional chiral logarithm in quenched QCD.

For our quenched simulations, we must still perform a subtraction of
power divergent quantities to get the quenched values for
$\alpha^{(8,1)}_1$.  As we discussed in the previous subsection for full
QCD, it is vital to do the subtraction in a way which correctly removes
power divergent coefficients times both conventional and quenched
logarithms.  If the quenched formula analogous to Eqs.\ \ref{eq:kpi81}
to \ref{eq:kvac271} existed, one could in principle fit individual
amplitudes to the formulae, including logarithms, and extract the
desired coefficients.  Even with the formula, such a process could
prove difficult due to the statistical errors on the data.

However, we can make use of the fact that in chiral perturbation
theory, the power divergent parts of operators appear as lower
dimensional operators.  Thus, the logarithmic corrections, both
conventional and quenched, should be the same for the power divergent
parts of a four-quark operator and the appropriate quark bilinear.
This is the basis for the cancellation of the chiral logarithms in
Eqs.\ \ref{eq:fix_alpha_81_2} and \ref{eq:fix_alpha_81_1}.  Thus, for
the subtractions of power divergent operators, the analytic
coefficients of the logarithms are not needed.  It would, however be
useful to know the coefficients of the logarithms for the remaining
finite terms.  (As this paper was being completed, such a calculation
was reported for some of the operators of interest here
\cite{Golterman:2001qj}.)  We will have to rely on the behavior of our
data to estimate the size of these effects.

It is important to stress that this cancellation of the quadratic
divergence in the logarithmic corrections to the chiral limit discussed
above provides a concrete example of the general cancellation of
quadratic divergences implied by the form of the subtraction term which
we adopt.  As will be discussed in Section \ref{subsec:sub_power_div},
we choose to implement the subtraction required by chiral perturbation
theory in a fashion which removes the quadratic divergence completely
from the resulting $\langle \pi^+ | \Theta^{(8,1)}| K^+ \rangle$
amplitude in the limit $\mres \to 0$.  The cancellation of the
quadratic divergence is guaranteed by standard renormalization
arguments and does not rely on chiral perturbation theory.

\fi


\section{Domain Wall Fermion Modifications to Chiral Perturbation
Theory}
\label{sec:dwf_chiral_mod}

\ifnum\theDwfChiralMod=1
%
%

In the previous section, results relevant to the current calculation
from both quenched and full QCD continuum chiral perturbation theory
were discussed.  In addition to the basic lowest order relations,
chiral perturbation theory gives the logarithmic corrections for both
full and quenched QCD.  For domain wall fermions with finite extent in
the fifth dimension, exact chiral symmetry does not exist, even if only
the fermionic modes relevant for low-energy QCD physics are studied,
due to the mixing between the left- and right-handed fermion surface
states that form at the boundaries of the fifth dimension.  However,
for low energy physics this mixing appears as an additional
contribution to the fermion mass, the residual mass $\mres$, in the
low-energy effective Lagrangian describing domain wall fermion QCD at
finite values for the fifth dimension \cite{Blum:2000kn,AliKhan:2000iv}.

For the calculation at hand, we must include power divergent operators,
which are also affected by the residual chiral symmetry breaking.
However, due to their dependence on scales up to the cutoff, chiral
symmetry breaking effects here cannot be precisely described in terms
of an extra additional mass in the low-energy effective Lagrangian.
As we will see in Subsection \ref{subsec:mres_effects} below, these
effects modify the formula in Eqs.\ \ref{eq:def_alpha_81} to
\ref{eq:kpipiO88}.  These modifications will be important in the
analysis of our numerical data.

A second modification to the chiral perturbation theory formula of the
previous section comes from the presence of unsuppressed topological
near-zero modes in our quenched QCD calculation.  Without the fermionic
determinant, these modes need not occur with the distribution of full
QCD and the light-quark mass limit of quenched QCD has been seen to be
pathological \cite{Blum:2000kn}.  The effects of such modes are
suppressed for large volumes, but can be important for the volumes used
in the matrix element calculations discussed here.  Since the zero
modes can lead to nonlinear dependence on the input quark mass, just as
the chiral logarithms can, it is important to quantify their effects.
We do this through a discussion of some of the relevant Ward-Takahashi
identities in Subsection \ref{subsec:zero_mode_effects}.

The notation we use for domain wall fermions is given in
\cite{Blum:2000kn}.  In particular, we use $\Psi_i(x,s)$ to represent a
five-dimensional fermion field with four spin components and flavor
$i$.  A generic four-dimensional fermion field with four spin
components and flavor $i$ will be given by $\psi_i(x)$, while the
specific four-dimensional field defined from $\Psi_i(x,s)$ will be
given by $q_i(x)$.  For quark fields of specific flavor, $u$, $d$, $s$
and $c$ will be used to represent four-dimensional fields defined from
$\Psi_i(x,s)$.


\subsection{Residual mass effects}
\label{subsec:mres_effects}

Residual chiral symmetry breaking effects for domain wall fermions at
finite $L_s$ can be easily discussed by introducing a new term into the
action containing a special-unitary flavor matrix $\Omega$
\cite{Blum:2001sr}.  This term connects four-dimensional planes at the
mid-point of the fifth dimension and has the form
\begin{equation}
  S_{\Omega} = - \sum_x \left\{ \overline{\Psi}_{x,L_s/2-1} P_L 
    \left( \Omega^\dagger -1 \right)
    \Psi_{x,L_s/2} + \overline{\Psi}_{x,L_s/2} P_R 
    \left( \Omega -1 \right) \Psi_{x,L_s/2-1} \right\} \, .
\label{eq:omega}
\end{equation}
If we let $\Omega$ transform as
\begin{equation}
  \Omega \to U_R \, \Omega \, U_L^\dagger
\end{equation}
under $\sutlr$, then the domain wall fermion Dirac operator possesses
exact chiral symmetry.  When $L_s \to \infty$, this extra mid-point
term in the action should not matter for low-energy physics, so any
Green's function that contains a power of $\Omega$ should also contain
a factor of $\exp(-\alpha L_s)$.  (Here we assume that in the $L_s
\rightarrow \infty$ limit the residual chiral symmetry vanishes
exponentially.  For the quenched theory, the numerical data is not
conclusive on this point, but does show that the residual chiral
symmetry breaking effects can be made quite small.)  Since $\Omega$ is
a $(\overline{3},3)$ under $\sutlr$, it transforms ``like a mass
term''.

Consider a continuum effective Lagrangian description of QCD with
domain wall fermions at finite $L_s$.  The presence of the parameter
$\Omega$ implies the mass term in this Lagrangian will be
\begin{equation}
  Z_m m_f \overline{\psi}\psi +
  c \left\{ \overline{\psi} \Omega^\dagger P_R \psi
  + \overline{\psi} \Omega P_L \psi \right\}\, ,
\label{eq:mass_plus_mres}
\end{equation}
to leading order.  Here $Z_m$ is a mass renormalization constant
and $c$ is a constant with dimensions of mass that is ${\cal O}(\exp(- \alpha
L_s)/a)$ where $a$ is the lattice spacing.  With the conventional
choice $\Omega_{a,b} = \delta_{a,b}$, Eq.\ \ref{eq:mass_plus_mres}
reduces to the form
\begin{equation}
  Z_m( m_f + \mres ) \overline{\psi} \psi
\end{equation}
where $\mres \approx 10^{-3}$ for quenched lattices with $a^{-1}
\sim 2$ GeV and $L_s = 16$.

A simple case where power divergences are involved is given by the
determination of $\langle \overline{q} q \rangle$ on the lattice with
domain wall fermions.  Since this transforms as a $(\overline{3},3)$
plus $(3,\overline{3})$ in chiral perturbation theory, its dependence
on explicit chiral symmetry breaking terms is given by
\begin{equation}
  \langle \overline{q} q \rangle(m_f, L_s) \sim
   c_1(M + M^\dagger) + c_1^\prime (\Omega + \Omega^\dagger)
\end{equation}
where $c_1$ and $c_1^\prime$ are two constants.  Since $c_1$ depends on
high momentum scales and behaves as $1/a^2$, $c_1^\prime$ also depends
on high momentum and is thus not simply equal to $c_1 \mres$.  In
particular, $ c^\prime_1 \sim \exp(-\alpha L_s)/a^3$.  For the case
with $SU(3)$ flavor symmetry and the conventional choice $\Omega_{ab} =
\delta_{ab}$, the chiral condensate for domain wall fermions should
have the form
\begin{equation}
    \langle \overline{q} q \rangle(m_f, L_s)
  = \langle \overline{q} q \rangle_0 + c_1 m_f + c_1^\prime
  \label{eq:pbp_finite_ls}
\end{equation}
Notice that the value of $\langle \overline{q} q \rangle(m_f, L_s)$ for
$m_f = -\mres$ is not equal to $\langle \overline{q} q \rangle(m_f=0,
L_s = \infty)$ since there is no simple relation between $c_1$ and
$c_1^\prime$.  Thus, the residual chiral symmetry breaking effects in a
power divergent quantity are small for large $L_s$, but they cannot be
cancelled by a simple choice for the input quark mass.

The presence of the new parameter $\Omega$ for domain wall fermions
means that there is an additional operator needed to represent
$\Theta^{(8,1)}$ in chiral perturbation theory.  In particular,
replacing $M$ in Eq.\ \ref{eq:cpt_theta81_2} by $\mres \Omega$ yields
the operator
\begin{equation}
    \tilde{\Theta}_3^{(8,1)}
    \equiv
    \frac{8v}{f^2} \, \mres \, {\rm Tr} \left[ \Lambda \Sigma \Omega
    + \Lambda (\Sigma \Omega)^\dagger \right]
    \label{eq:cpt_theta81_3}
\end{equation}
and the representation of $\Theta^{(8,1)}$ in Eq.\ \ref{eq:def_alpha_81}
is modified to
\begin{equation}
  \Theta^{(8,1)} \to
  \alpha_1^{(8,1)} \tilde{\Theta}_1^{(8,1)}
+ \alpha_2^{(8,1)} \tilde{\Theta}_2^{(8,1)}
+ \alpha_3^{(8,1)} \tilde{\Theta}_3^{(8,1)}
  \label{eq:def_alpha_81_3}
\end{equation}
As mentioned in the previous section, the coefficient
$\alpha_2^{(8,1)}$ is power divergent and consequently so is
$\alpha_3^{(8,1)}$.  Because we have used an explicit factor of $\mres$
in the definition of $\tilde{\Theta}_3^{(8,1)}$, which involves power
divergences, $\alpha_3^{(8,1)} \ne \alpha_2^{(8,1)}$.  Similar to the
behavior of $ \langle \overline{q} q \rangle$ at finite $L_s$, the
chiral limit of $\Theta^{(8,1)}$ is not given by setting $m_f =
-\mres$.

The presence of this additional term in the representation of
$\Theta^{(8,1)}$ does not change Eq.\ \ref{eq:kvacO81}, since $\Omega$
is flavor symmetric and $\alpha_3^{(8,1)}$ is defined in the zero quark
mass limit.  (There can be quark mass dependence in the residual chiral
symmetry breaking effects, but this is a higher order effect.  Such
quark mass dependence has been seen in quenched simulations, but is a
small effect \cite{Blum:2000kn,AliKhan:2000iv}.) This new term does
change Eq.\ \ref{eq:kpiO81} for finite $L_s$ to
\begin{equation}
    \langle \pi^+ | \Theta^{(8,1)} | K^+ \rangle =
    \frac{4 m_M^2}{f^2} \left( \alpha_1^{(8,1)}-\alpha_2^{(8,1)} \right)
    - \frac{32v}{f^4} \, \mres \, \alpha_3^{(8,1)}
\end{equation}
where we have also taken $\Omega_{ab} = \delta_{ab}$.  Thus, we see that
$ \langle \pi^+ | \Theta^{(8,1)} | K^+ \rangle$ will not vanish at $m_f
= 0$, nor at $m_f = -\mres$, since there is no simple relation between
$\alpha_2^{(8,1)}$ and $\alpha_3^{(8,1)}$.  However, since all we
require from simulations is the value of $\alpha_1^{(8,1)}$, we see
that it can be determined from the slope of $ \langle \pi^+ |
\Theta^{(8,1)} | K^+ \rangle$ with respect to $m_f$ and the value of
$\alpha_2^{(8,1)}$ from $ \langle 0 | \Theta^{(8,1)} | K^0 \rangle$.

It is true that $\langle \pi^+ | \Theta^{(8,1)} | K^+ \rangle $ should
reach its chiral limit at $m_f = {\cal O}(-\mres)$, since the residual chiral
symmetry breaking effects still depend on the overlap between the
surface states at the ends of the fifth dimension.  We will be able to
check that our numerical results show this behavior.  In general, the
chiral limit for any divergent quantity is uncertain at finite
$L_s$.  As previously mentioned, this directly impacts the
determination of $\alpha^{(8,8)}$ from the $\DIhalf$ matrix elements of
$\langle \pi^+ | \Theta^{(8,8)} | K^+ \rangle$.  Fortunately, here
we can use the finite $\DIthalf$ matrix elements and the Wigner-Eckart
theorem to determine $\alpha^{(8,8)}$.


\subsection{Topological near-zero modes and Ward-Takahashi Identities}
\label{subsec:zero_mode_effects}

In the previous subsection we discussed how residual chiral symmetry
breaking effects from finite $L_s$ values can enter the operators of
interest in this work.  These effects make the chiral limit uncertain
for divergent operators.  A second difficulty with the chiral limit
arises for quenched domain wall simulations in finite volumes from
fermionic topological near-zero modes which are unsuppressed due to
neglecting the fermionic determinant.  The presence of these zero modes
is an important feature of domain wall fermions, but it does lead to
additional complications in the quenched simulations reported here.
Since these modes distort the chiral limit, they can produce non-linear
behavior in Green's functions that may, in a range of small quark
masses, be difficult to distinguish from the chiral logarithm effects
discussed earlier.  For the remainder of this section, we will refer to
the topological near-zero modes as zero modes, with the understanding
that their eigenvalues are not precisely zero for finite $L_s$.

The presence of zero modes in quenched simulations has been extensively
discussed in \cite{Blum:2000kn}, where their effects were seen in the
chiral condensate and hadronic masses.  In this calculation we will be
subtracting large, power divergent lattice quantities to achieve our
final physical results; it is important that the zero mode effects be
well understood for the subtraction process.  Since zero mode effects
are suppressed as the volume increases, naively down by a factor of
$1/\sqrt{V}$ relative to the fermionic modes responsible for chiral
symmetry breaking and low energy QCD physics, their effects are not
included in the infinite volume chiral perturbation theory results of
Section \ref{sec:cont_chiral_pert}.

To gain a quantitative understanding of the zero mode effects, we will
use the Ward-Takahashi identities of domain wall fermion QCD.  Since
these identities are true in the quenched theory for any quark mass and
volume, they must include the effects of zero modes.  Continuum chiral
perturbation theory is the simplest way to represent the Ward-Takahashi
identities in the infinite volume limit with arbitrarily small quark
masses.  In this limit, where zero modes do not enter, saturating
intermediate states with light pseudoscalars gives the relations of
lowest order chiral perturbation theory.  Thus, the Ward-Takahashi
identities can detail how zero mode effects alter the lowest order
chiral perturbation theory we are using to determine $K \to \pi \pi$
matrix elements.  Of course, the chiral logarithm corrections to lowest
order chiral perturbation theory are also included in the
Ward-Takahashi identities, but these are more easily handled through
chiral perturbation theory techniques.

The Ward-Takahashi identity for domain wall fermions with $SU(3)$
flavor symmetry is
 \cite{Furman:1995ky,Blum:2000kn}
\begin{equation}
  \Delta_\mu \langle {\cal A}^a_\mu(x) O(y) \rangle = 
        2m_f \langle J^a_5(x) O(y) \rangle + 2 \langle J^a_{5q}(x) O(y)
        \rangle + i \langle \delta^a O(y) \rangle.
\label{eq:ward_tak_id}
\end{equation}
Here ${\cal A}^a_\mu$ is the conserved axial current which involves all
points in the fifth dimension, $J^a_5 \equiv \bar{q} t^a \gamma_5 q$
and $J^a_{5q}$ is a similar pseudoscalar density defined at the
midpont of the fifth-dimension.  Summing over $x$ yields the
integrated form of this identity
\begin{equation}
      \sum_x  \left[ \langle ( 2m_f J^a_5(x) + 2 J^a_{5q}(x) ) \, O(y)
        \rangle + i \langle \delta^a O(y) \rangle \right] = 0
\label{eq:ward_tak_id_int}
\end{equation}
which we will use extensively.

We first consider the simple case where $O(y) = J^a_5(y)$.  Then
Eq.\ \ref{eq:ward_tak_id_int} becomes
\begin{equation}
      \sum_x  \langle ( m_f J^a_5(x) + J^a_{5q}(x) ) J^a_5(y)
        \rangle = \langle \overline{u} u(y) \rangle
	\equiv
        12 \langle \overline{u} u(y) \rangle_{\rm lat-norm}
	\qquad {(\rm no \; sum\; on \;a)}
  \label{eq:int_wi_pseudo_den}
\end{equation}
where the factor of 12 is needed since we normalize $ \langle
\overline{u} u(y) \rangle_{\rm lat-norm}$ per spin and color.  (We are
considering the case with $SU(3)$ flavor symmetry, making the chiral
condensate for $u$, $d$ and $s$ quarks is the same.) \ \ Working in
Euclidean space with correlators evaluated through the Feynman path
integral, we break the sum over $x$ into the points with $x \ne y$ and
the point with $x = y$.  For the points with $x \ne y$, the correlator
is a sum of exponentials, with the overlap between the operators
$J^a_5(x)$ and $J^a_{5q}(x)$ and the different mass states
conventionally represented as a matrix element.  For $x=y$ a
``contact'' term is generated.  Using the normalizations for the states
given in Appendix \ref{sec:conventions} gives
\begin{equation}
      \sum_{x,n}  \langle 0 | m_f J^a_5(\vec{x}, 0) + J^a_{5q}(\vec{x},
      0) | n \rangle
      \frac{\exp(-E_n(|x_0 - y_0|))}{2 V_s E_n}
      \langle n | J^a_5(\vec{y}, 0) | 0 \rangle + C(y)
      - \langle \overline{u} u(y) \rangle = 0
\end{equation}
where $V_s$ is the spatial volume and $C(y)$ is the contact term
generated when $x = y$.  The pseudo-Goldstone boson term
in the sum over $n$ gives
\begin{equation}
  \langle 0 | m_f J^a_5(0) + J^a_{5q}(0) | \pi^a \rangle
      \frac{1}{\mpi2} \langle \pi^a | J^a_5(0) | 0 \rangle
  = - \frac{m_f + \mres}{\mpi2} | \langle 0 | J^a_5(0) | \pi \rangle |^2
\end{equation}
since for the low energy physics described by the state $ | \pi^a
\rangle$ we have $J^a_{5q} = \mres J^a_5$.   This term in the sum is
not suppressed for light quark masses due to the $\mpi2$ term which
appears in the denominator.  For a general integrated Ward-Takahashi
identity, keeping only the leading terms in the $m_f \to 0$ limit,
which includes such ``pion pole saturation'' contributions, leads to
the relations of lowest order chiral perturbation theory
\cite{Gasser:1984yg}.  To apply this procedure here, we must first note
that the other states in the sum and the contact term give a
contribution of ${\cal O}(m_f)/a^2 + {\cal O}(\mres)/a^2$.  Here high
momentum modes can enter and the midpoint pseudoscalar density
$J^a_{5q}$ is not simply related to $J^a_5$.  Thus, without any effects
of zero modes, we have
\begin{equation}
  -\frac{m_f + \mres}{\mpi2} | \langle 0 | J^a_5(0) | \pi \rangle |^2
  + \frac{{\cal O}(m_f)}{a^2} + \frac{{\cal O}(\mres)}{a^2}
  =
  \langle \overline{u} u \rangle(m_f, L_s )
  \label{eq:GMOR_no_zero_modes}
\end{equation}
This relation is the same as Eq.\ \ref{eq:pbp_finite_ls} and once again
demonstrates that the chiral limit cannot be achieved at finite $L_s$
by setting $m_f = - \mres$ when divergent quantities are involved.
However, since the Ward-Takahashi identities include zero mode effects,
we can investigate their contributions to this relation.

To simplify the discussion of zero modes, we consider the $L_s \to
\infty$ limit where the contribution of the $J^a_{5q}$ term to the
Ward-Takahashi identity vanishes.  Following \cite{Blum:2000kn} we work
with generic fermion fields $\psi$, the continuum four-dimensional
Dirac operator $\slash{D}^{(4)}$ with eigenvalues and eigenvectors
given by $(\slash{D}^{(4)} + m) \psi_\lambda = (i\lambda + m )
\psi_\lambda$ and write the quark propagator as
\begin{equation}
  S^{(4)}_{x,y} = \sum_\lambda
    \frac{ \psi_\lambda(x) \psi^\dagger_\lambda(y)}
    { i \lambda + m}
\end{equation}
(Here we are considering a particular gauge field $U_\mu(x)$ and the
eigenvalues $\lambda$, eigenvectors $\psi_\lambda(x)$ and quark
propagator are functions of $U_\mu(x)$.  Green's functions result from
averaging over an appropriate distribution of gauge fields.) \ \ The
integrated Ward-Takahashi identity, Eq.\ \ref{eq:int_wi_pseudo_den},
then becomes
\begin{equation}
  -m_f \sum_{x,\lambda,\lambda^\prime}
  \, {\rm Tr} \left(
  \frac{ \psi_\lambda(y) \psi_\lambda^\dagger(x)}
	{-i \lambda + m_f} \;
  \frac{\psi_{\lambda^\prime}(x) \psi_{\lambda^\prime}^\dagger(y)}
    {i \lambda^\prime + m_f} \right)
  +
  \sum_{\lambda}
  \, {\rm Tr} \left(
  \frac{ \psi_\lambda(y) \psi_\lambda^\dagger(y)} {i \lambda + m_f}
  \right)
  =0
  \label{eq:gmor_spec_decomp}
\end{equation}
Performing the sum over $x$ in the first term gives $\delta_{\lambda,
\lambda^\prime}$ and we are left with
\begin{equation}
  -m_f \sum_{\lambda}
  \, {\rm Tr} \left(
  \frac{ \psi_\lambda(y) \psi_\lambda^\dagger(y)}{\lambda^2 + m_f^2}
  \right )
  +
  \sum_{\lambda}
  \, {\rm Tr} \left(
  \frac{ \psi_\lambda(y) \psi_\lambda^\dagger(y)} {i \lambda + m_f}
  \right )
  =0
\end{equation}
This relation is easily seen to be true, since for $\lambda \ne 0$,
there is also an eigenvalue $-\lambda$.  Also, the zero mode
contributions cancel between the two terms.  Zero modes in the left
term will alter numerical measurements of pion properties in moderate
volumes, while the right term contains the zero modes which enter in
the chiral condensate.

Consider working in moderate sized volumes where zero mode effects may
be present but enter only as small corrections to the infinite volume
results.  We decompose the sums in Eq.\ \ref{eq:gmor_spec_decomp} into
terms without zero modes and terms with zero modes.  The terms without
zero modes will give Eq.\ \ref{eq:GMOR_no_zero_modes}. Including zero
mode effects changes Eq.\ \ref{eq:GMOR_no_zero_modes} for small $m_f$
in the $L_s \to \infty$ limit to
\begin{alignat}{2}
  - m_f \left\{
    \frac{| \langle 0 | J^a_5(0) | \pi \rangle |^2}{\mpi2}
  +
  \sum_{x,\lambda = 0 \atop {\rm or} \;\lambda^\prime = 0}
  \, {\rm Tr} \left(
  \frac{ \psi_\lambda(y) \psi_\lambda^\dagger(x)}
  {-i \lambda + m_f} \;
  \frac{\psi_{\lambda^\prime}(x)\psi_{\lambda^\prime}^\dagger(y)}
    {i \lambda^\prime + m_f}  \right) \right \}
  + \frac{{\cal O}(m_f)}{a^2} \nonumber \\
  =
  \langle \overline{\psi} \psi \rangle(m_f) -
  \sum_{\lambda = 0}
  \, {\rm Tr} \left(
  \frac{ \psi_\lambda(y) \psi_\lambda^\dagger(y)} {i \lambda + m_f}
  \right )
  \label{eq:GMOR_zero_modes}
\end{alignat}
For finite $L_s$, the modifications to Eq.\ \ref{eq:GMOR_zero_modes}
come from including the mid-point term $J^a_{5q}$ and a residual chiral
symmetry breaking term for each eigenvector, referred to as $\delta
m_i$ in \cite{Blum:2000kn}.  In \cite{Blum:2000kn} it was found that a
histogram of $\delta m_i$ values for modes with eigenvalues below
$\approx \Lambda_{\rm QCD}$ was peaked very close to $\mres$.  It is
certainly possible that as $a \to 0$ the low lying eigenvalues all show
a common residual chiral symmetry breaking of $\mres$, although this has
not been demonstrated.  To proceed with our general analysis including
finite $L_s$ effects, we make this reasonable assumption and in the
sums over eigenvalues replace $m_f$ by $m_f + \mres$ for modes with
eigenvalues below $\approx \Lambda_{\rm QCD}$.  For such terms, the
factor of $m_f$ multiplying the quantity in braces on the left-hand
side is also modified to $m_f + \mres$.  For terms with eigenvalues
above $\approx \Lambda_{\rm QCD}$, such a simple modification does not
seem likely.  However these terms do not produce any effects which
diverge as $m_f \to 0 $ since the $1/m_f$ from the zero mode is
cancelled by the explicit $m_f$ multiplying the terms in braces on the
left-hand side of Eq.\ \ref{eq:GMOR_zero_modes}.  This gives us the
finite $L_s$ result
\begin{alignat}{2}
  -(m_f + \mres) & \left\{
    \frac{| \langle 0 | J^a_5(0) | \pi \rangle |^2}{\mpi2}
  +
  \sum_{x,\lambda = 0 \atop {\rm or} \;\lambda^\prime = 0}^{\lambda,
    \lambda^\prime < \Lambda_{\rm QCD}}
  \, {\rm Tr} \left(
  \frac{ \psi_\lambda(y) \psi_\lambda^\dagger(x)}
  {-i \lambda + m_f + \mres } \;
  \frac{\psi_{\lambda^\prime}(x)\psi_{\lambda^\prime}^\dagger(y)}
    {i \lambda^\prime + m_f + \mres }  \right)
    \right\} \nonumber \\
  &
  + \frac{{\cal O}(m_f)}{a^2} + \frac{{\cal O}(\mres)}{a^2}
  =
  \langle \overline{u} u \rangle(m_f,L_s) -
  \sum_{\lambda = 0}
  \, {\rm Tr} \left(
  \frac{ \psi_\lambda(y) \psi_\lambda(y)^\dagger}
    {i \lambda + m_f + \mres} \right)
  \label{eq:GMOR_zero_modes_finite_ls}
\end{alignat}

When $\langle 0 | J^a_5(0) | \pi \rangle$ and $m_\pi$ are measured from
the correlator $\langle i J^a_5(x) i J^a_5(y) \rangle$ in a numerical
simulation, some zero mode effects can be present depending on the
range of $x-y$ used.  The effects of zero modes (the second term
in braces in Eq.\ \ref{eq:GMOR_zero_modes_finite_ls}) will enter in the
measured values $\langle 0 | J^a_5(0) | \pi^\prime \rangle$ and
$m_{\pi^\prime}$, where the primes indicate quantities deviating
slightly from their infinite volume values.  We can replace the
quantity in braces in Eq.\ \ref{eq:GMOR_zero_modes_finite_ls} by
\begin{equation}
 \frac{| \langle 0 | J^a_5(0) | \pi^\prime \rangle|^2}
 {m_{\pi^\prime}^2}
 \label{eq:pi_with_zero}
\end{equation}
which is bounded by the values for $ \langle \overline{u} u \rangle
(m_f,L_s) $ measured with and without zero mode effects.

We can now do a similar analysis for the matrix element $\langle \pi^+
| \bar{s} d | K^+ \rangle$.  This is an instructive example since we
want to use measured values of $\langle \pi^+ | Q_i | K^+ \rangle$
matrix elements on the lattice to determine physical quantities and we
seek some understanding of the role of zero modes in matrix elements of
this form.  We start from Eq.\ \ref{eq:ward_tak_id_int} taking
$J^a_5(x) = [\bar{d} \gamma_5 u](x)$ and letting $O(y) \to [\bar{s}d](y)
[i\bar{u} \gamma_5 s ](z)$.  We define the pseudoscalar densities
$P_{K^-}(x) \equiv [i\bar{u} \gamma_5 s](x)$ and $P_{\pi^+}(x) \equiv
[i\bar{d} \gamma_5 u](x)$ and the scalar density $S(x) \equiv [\bar{s}
d](x)$.  (We adopt the notation $P_{K^-}(x)$ to distinguish these
pseudoscalar operators from the operators like $K^{-}(x)$ of chiral
perturbation theory, as in Eqs.\ \ref{eq:cpt_to_q_pi} and
\ref{eq:cpt_to_q_K}, which have a different normalization.) We can then
write Eq.\ \ref{eq:ward_tak_id_int} as
\begin{equation}
   \sum_x  \langle [ 2m_f P_{\pi^+}(x) + 2 P_{\pi^+}^{\rm MP}(x) ]
     \, S(y) \, P_{K^-}(z) \rangle
   -\langle P_{K^+}(y) \, P_{K^-}(z) \rangle
   +\langle S(y) \, S^\dagger(z) \rangle = 0
  \label{eq:sbard_ward_tak_id_int}
\end{equation}
where $P_{\pi^+}^{\rm MP}(x)$ is the ``mid-point'' pseudoscalar density
with the $\pi^+$ quantum numbers formed from $\Psi_i(x,s)$ for $s =
L_s/2 -1$ and $L_s/2$.  Considering the case where $L_s \to \infty$,
$y-z$ is large, there are no zero modes present and $m_f \to 0$ gives
\begin{equation}
   \frac{2 m_f}{m_\pi^2} \langle \pi^+ | \bar{s} d | K^+ \rangle
   - 1 = 0
  \label{eq:sbard_wi_no_zero}
\end{equation}
The term $\langle S(y) \, S^\dagger(z) \rangle$ plays no role in this
case, since it does not contain any contribution from the massless
pseudoscalars.

We now consider the role of zero modes for the $L_s = \infty$ case.  We
start with the complete spectral decomposition of
Eq.\ \ref{eq:sbard_ward_tak_id_int}, which is
\begin{alignat}{2}
  2 m_f \sum_{x,\lambda,\lambda^\prime, \lambda^{\prime\prime}}
  & \, {\rm Tr} \left(
    \frac{ \psi_\lambda(y) \psi_\lambda^\dagger(z)}
	{i \lambda + m_f} \;
  \frac{\psi_{\lambda^\prime}(z) \psi_{\lambda^\prime}^\dagger(x)}
    {-i \lambda^\prime + m_f} \;
  \frac{\psi_{\lambda^{\prime\prime}}(x)
  \psi_{\lambda^{\prime\prime}}^\dagger(y)}
    { i \lambda^{\prime\prime} + m_f}  \right)
  \nonumber \\
  &
  -
  \sum_{\lambda, \lambda^\prime}
  \, {\rm Tr} \left(
  \frac{ \psi_\lambda(y) \psi_\lambda^\dagger(z)} {i \lambda + m_f}
  \;
  \frac{ \psi_{\lambda^\prime}(z) \psi_{\lambda^\prime}^\dagger(y)}
  { - i \lambda^\prime + m_f} \right)
  \nonumber \\
  &
  -
  \sum_{\lambda, \lambda^\prime}
  \, {\rm Tr} \left(
  \frac{ \psi_\lambda(y) \psi_\lambda^\dagger(z)} {i \lambda + m_f}
  \;
  \frac{ \psi_{\lambda^\prime}(z) \psi_{\lambda^\prime}^\dagger(y)}
  { i \lambda^\prime + m_f} \right)
  =0
  \label{eq:sbard_spec_decomp}
\end{alignat}
The sum over $x$ allows this to be written as
\begin{alignat}{2}
  \sum_{\lambda, \lambda^\prime}
  \, {\rm Tr} \left(
  \frac{ \psi_\lambda(y) \psi_\lambda^\dagger(z)} {i \lambda + m_f}
  \;
  \psi_{\lambda^\prime}(z) \psi_{\lambda^\prime}^\dagger(y)
  \;
  \left[ \frac{ 2m_f} { (\lambda^\prime)^2 + m_f^2}
  - \frac{1}   { -i \lambda^\prime + m_f}
  - \frac{1}   {  i \lambda^\prime + m_f} \right] \right)
  =0
  \label{eq:sbard_spec_decomp_2}
\end{alignat}
The term in brackets is easily seen to be zero.  As must be the case,
the zero modes entering the spectral decomposition also satisfy the
Ward-Takahashi identity.

We now consider the modifications to Eq.\ \ref{eq:sbard_wi_no_zero}
from zero modes, when $\langle \pi^+ | S | K^+ \rangle$ is measured on
the lattice from the correlator $ \langle P_{\pi^+}(x) S(y) P_{K^-}(z)
\rangle $, with $x > y > z$.  Provided the zero modes are localized,
their effects will predominantly enter the quark propagators
$D^{-1}(x,y)$ and $D^{-1}(y,z)$, since $x-z$ can exceed the size of the
zero mode.  Thus, our measured quantities will not include the
$\lambda^{\prime} = 0$ term in the first summation of
Eq.\ \ref{eq:sbard_spec_decomp}.  Separating out this term and
again letting primes denote states where some zero mode
contamination is possible gives the following result for the
Ward-Takahashi identity when $m_f \to 0$
\begin{alignat}{2}
    \frac{2m_f}{m_{\pi^\prime}^2} \, 
    \langle 0 | P_{\pi^+} | (\pi^+)^\prime \rangle \,
    &
    \langle (\pi^+)^\prime | \, S(y) \, P_{K^-}(z) \rangle
    \nonumber \\
    & = \langle P_{K^+}(y) \, P_{K^-}(z) \rangle
     -\langle S(y) \, S^\dagger(z) \rangle
    \nonumber \\
    & \quad - 2 \sum_{\lambda^{\prime} = 0, \atop
      x,\lambda, \lambda^{\prime\prime}}
     \, {\rm Tr} \left(
     \frac{ \psi_\lambda(y) \psi_\lambda^\dagger(z)}
	{i \lambda + m_f} \;
    \; \psi_{\lambda^\prime}(z) \psi_{\lambda^\prime}^\dagger(x) \;
   \frac{\psi_{\lambda^{\prime\prime}}(x)
    \psi_{\lambda^{\prime\prime}}^\dagger(y)}
    {i \lambda^{\prime\prime} + m_f} \right)
    \nonumber \\
    & = \langle P_{K^+}(y) \, P_{K^-}(z) \rangle
     -\langle S(y) \, S^\dagger(z) \rangle
    \nonumber \\
    & \quad - 2 \sum_{\lambda^\prime = 0, \lambda}
     \, {\rm Tr} \left(
     \frac{ \psi_\lambda(y) \psi_\lambda^\dagger(z)}
	{i \lambda + m_f} \;
  \frac{\psi_{\lambda^\prime}(z) \psi_{\lambda^\prime}^\dagger(y)}
    { m_f} \right)
  \label{eq:sbard_wi_zero}
\end{alignat}
The combination $\langle P_{K^+}(y) \, P_{K^-}(z) \rangle - \langle S(y)
\, S^\dagger(z) \rangle$ has zero mode effects.  These arise from a
zero mode in either one or both quark propagators.  The $\lambda = 0$
term in the sum cancels the contribution from $\langle P_{K^+}(y) \,
P_{K^-}(z) \rangle - \langle S(y) \, S^\dagger(z) \rangle$ when both
quark propagators have a zero mode.  When $\lambda \ne 0$, the
additional term cancels half of the zero mode contribution from
$\langle P_{K^+}(y) \, P_{K^-}(z) \rangle - \langle S(y) \, S^\dagger(z)
\rangle$ due to a zero mode in only one propagator.  Since zero
mode effects enter $\langle P_{K^+}(y) \, P_{K^-}(z) \rangle$ and $
\langle S(y) \, S^\dagger(z) \rangle $ identically, the right-hand side
of Eq.\ \ref{eq:sbard_wi_zero} becomes
\begin{equation}
    \langle P_{K^+}(y) \, P_{K^-}(z) \rangle_{\rm no-zero}
  + \langle P_{K^+}(y) \, P_{K^-}(z) \rangle_{\rm one-zero}
  - \langle S(y) \, S^\dagger(z) \rangle_{\rm no-zero}
\end{equation}
Here ``no-zero'' means no zero modes included in the spectral sum and
``one-zero'' means one of the two quark propagators is a zero mode.
For small $m_f$, $\langle S(y) S^\dagger(z) \rangle_{\rm no-zero}$
plays no role leaving us with
\begin{alignat}{2}
  \frac{2m_f}{m_{\pi^\prime}^2} \, 
  \langle 0 | P_{\pi^+} | (\pi^+)^\prime \rangle \,
  &
  \langle (\pi^+)^\prime | \, S(y) \, P_{K^-}(z) \rangle
  \nonumber \\
  &
  = \langle P_{K^+}(y) \, P_{K^-}(z) \rangle_{\rm no-zero}
  + \langle P_{K^+}(y) \, P_{K^-}(z) \rangle_{\rm one-zero}
  \label{sbard_wi_zero_small_mf}
\end{alignat}
For finite $L_s$, Eq.\ \ref{sbard_wi_zero_small_mf} is modified
by replacing $2m_f$ with $2(m_f + \mres)$ since no divergent terms
appear.

For the range of $y-z$ where our matrix elements calculations are done,
we have explicit results for $\langle P_{K^+}(y) \, P_{K^-}(z) \rangle$
and $\langle S(y) \, S^\dagger(z) \rangle$.  Since
\begin{equation}
  \langle P_{K^+}(y) \, P_{K^-}(z) \rangle_{\rm no-zero} = 
  \langle P_{K^+}(y) \, P_{K^-}(z) \rangle + 
  \langle S(y) \, S^\dagger(z) \rangle
\end{equation}
we can estimate the effects of the one-zero mode term on the right side
of Eq.\ \ref{sbard_wi_zero_small_mf}.  We can compare our numerical
data to the Ward-Takahashi identity with no zero modes (Eq.\
\ref{eq:sbard_wi_no_zero}) and with zero modes
(Eq.\ \ref{sbard_wi_zero_small_mf}).  We will discuss our numerical
results for the Gell-Mann--Oakes--Renner (GMOR) relation in Section
\ref{subsec:GMOR} and for the $\overline{s} d$ Ward-Takahashi identity
in Section \ref{subsec:ward_id_sd}.


\subsection{Topological Near-zero Modes and Operator Subtraction}
\label{subsec:subtraction_zero_mode_effects}

A final part of this calculation where the features of domain wall
fermions in quenched QCD are important is the role of zero modes in
the subtraction of power divergence operators required to
determine $K \to \pi \pi $ matrix elements using chiral perturbation
theory.  As discussed in Section \ref{subsec:one_loop_XPT} and shown
in Eq.\ \ref{eq:fix_alpha_81_2}, the ratio
\begin{equation}
    \frac{ \langle 0 | \Theta^{(8,1)} | K^0 \rangle}
      {\langle 0 | \Theta^{(3,\bar{3})} | K^0 \rangle}
\end{equation}
has no chiral logarithms multiplying power divergent quantities.  This
is due to the locality of the power divergent part of the operator
$\Theta^{(8,1)}$.  The situation for zero mode effects is identical
since in the denominator they only enter the quark propagators
connecting the $K^0$ to the operator.  For the power divergent part of
the numerator, zero modes also only enter the propagators connecting
the $K^0$ to the operator and their effects cancel in the ratio.  Thus,
the linearity in $(m_K^2 -m_\pi^2)$ given in
Eq.\ \ref{eq:fix_alpha_81_2} should also be true for the
$\alpha^{(8,1)}_2$ term when zero mode effects are included.  This
linearity will make the determination of $\alpha_2^{(8,1)}$ much more
accurate and our results will not be influenced by a small zero mode
effect times a power divergent contribution.

Once $\alpha_2^{(8,1)}$ is known, we can use the combination of matrix
elements given on the left-hand side of Eq.\ \ref{eq:fix_alpha_81_1} to
determine $\alpha_1^{(8,1)}$.  Here we take a linear combination of two
$K \to \pi$ matrix elements and zero modes may enter in both.
However, once again the power divergent part of $ \langle \pi^+ |
\Theta^{(8,1)} | K^+ \rangle $ and $ \alpha_2^{(8,1)} \langle \pi^+ |
\Theta^{(3,\bar{3})} | K^+ \rangle$ are altered identically by
zero mode effects in the quark propagators between the operators
creating the pion and kaon and the $\Theta$'s.  Thus, our results will
not be altered by small zero mode effects multiplied by power divergent
terms.  There can, however, be zero mode effects left in the finite
part of the left-hand side of Eq.\ \ref{eq:fix_alpha_81_1}.  These
should be similar to the zero mode effects discussed in the preceding
section for $ \langle \pi^+ | \bar{s} d | K^+ \rangle $, whose size we
will estimate from our data in Section \ref{subsec:ward_id_sd}

\fi


\section{Basic Features of Numerical Simulations}
\label{sec:run_basics}

\ifnum\theRunBasics=1
%
%

\subsection{Simulation parameters}

The quenched gauge field ensemble used to calculate expectation values
in this study was generated at gauge coupling $\beta=6.0$ with lattice
four-volume $16^3\times 32$ (space $\times$ time).  The ensemble
comprises 400 configurations separated by 10,000 sweeps, with each
sweep consisting of a simple two-subgroup heat-bath update of each
link. The gauge coupling corresponds to a lattice cut-off of $a^{-1}=
1.922$ GeV set by the $\rho$ mass\cite{Blum:2000kn}.  The domain wall
fermion fifth dimension was $L_s=16$ sites with a domain wall height
$M_5=1.8$. These parameters yield a residual quark mass of about 3\% of
the strange quark mass~\cite{Blum:2000kn}.

The light quark masses in units of the lattice spacing were taken to be
$m_f=0.01$, 0.02, 0.03, 0.04, and 0.05.  The value of $m_f$
corresponding to a pseudo-scalar state made of degenerate quarks with
mass equal to the physical kaon at $\beta=6.0$ is
$0.018$\cite{Blum:2000kn}.  Heavier quarks were also included to allow
matrix elements to be calculated in the 4-flavor case where a charm
quark is present.  These heavy masses, with values of $m_f=0.1$, 0.2,
0.3, and 0.4 will not be discussed in this report but rather in a
subsequent publication.  Quark propagators were calculated using the
conjugate gradient method with a stopping residual $r=10^{-8}$.

Quark propagators were calculated from Coulomb gauge fixed wall sources
at time slices $t_K=5$ and $t_\pi = 27$.  The resulting propagators
were fixed to lattice Coulomb gauge (on the ``sink'' end) to reduce
fluctuations in gauge averages and to allow construction of wall-wall
correlators.  Forward and backward in time propagators were constructed
from linear combinations of propagators computed with periodic and
antiperiodic boundary conditions.  This amounts to using an unphysical
doubled lattice in the time direction with periodicity 64.  The random
wall sources used to calculate eye diagrams were spread over times
$t=14-17$, and the corresponding propagators had periodic boundary
conditions.

Before starting the production simulation, all correlation functions
were computed for a single common configuration on each of the QCDSP
machines that were to be used in the calculation. They agreed bit by
bit.  During the production simulation, we checkpointed every tenth
configuration.  All quark propagators and contractions were calculated
twice on this checkpointed configuration, in order to detect any
hardware errors. If the output from the repeated calculation did not
agree with the original, the node responsible for the failure was
tracked down and replaced. The process was repeated until bit by bit
agreement was obtained. Such hardware errors occurred very infrequently
(less than 1\% of the configurations).

\subsection{Computer code details}

We have written two completely separate production computer programs to
calculate weak matrix elements.  The first is based on the general
purpose QCD code written by the Columbia University lattice group and
runs primarily on the QCDSP supercomputers at the RIKEN-BNL Research
Center and Columbia University. The second program is based on the
general purpose QCD Code written by the MILC collaboration which was
extended by us to use domain wall fermions. We only have a single code
which calculates the propagators necessary to compute renormalization
($Z$) factors, which is part of the QCDSP version. In addition we have
three independently written analysis packages that run on workstations
which take the raw matrix elements and combine them with $Z$ factors and
Wilson coefficients to yield physical amplitudes.

We have performed several checks of these codes. Most importantly, a
completely independent check code was written to compare with the two
production versions (this does not include the $Z$ factors). Output
generated on the same configuration from each code was compared for
several test cases. In each case one code was run on a scalar
workstation and the other on a parallel machine.  The expected
agreement was obtained in each test.  We also checked the production
simulation by calculating all of the required correlation functions
with the check code on a single common gauge field configuration. All
of the production simulation parameters (volume, gauge coupling, quark
masses, sources, {\it etc.}) were used in this test. The Z factor code,
which runs on a workstation, has not been exhaustively checked by
second independent code.

As a final useful check, note that we work explicitly with the
operators defined in Eqs.\  \ref{eq:Q1} to \ref{eq:P10}.  The
$(V-A)\times(V-A)$ operators go into themselves under a Fierz
transformation. Thus, color-mixed contractions can be compared to
corresponding color-diagonal ones. We find perfect agreement in all
cases.

\fi


\section{Basic Tests of the Chiral Properties of Domain Wall Fermions}
\label{sec:test_dwf_chiral}

\ifnum\theTestDwfChiral=1
%
%

In the earlier sections we have discussed the changes in full QCD,
chiral perturbation theory relations due to quenching and using domain
wall fermions at finite $L_s$.  In this section, we will present our
numerical results for simple cases and check their consistency with the
theoretical expectations.  The cases we consider are:  1) the presence
of quenched chiral logarithms in $\mpi2$, 2) tests of the
Gell-Mann--Oakes--Renner relation for finite $L_s$ domain wall fermions
and 3) the Ward Identity satisfied by the matrix element of $\langle
\pi^+ | \bar{s}d | K^+ \rangle$.


\subsection{Quenched Chiral Logarithms in $\mpi2$}
\label{subsec:qXPT_mpi}

Numerous simulations have looked for the presence of quenched chiral
logarithms in $\mpi2$ versus $m_f$ of the form given in Eq.\
\ref{eq:qXPT_mpi}.  Recent values for $\delta$ are $\approx 0.1$
\cite{Aoki:1999yr} using Wilson fermions, the Wilson gauge action and
lattice spacings in the range 0.1 to 0.05 fm, $0.065 \pm 0.013$
\cite{Bardeen:2000cz} using clover-improved Wilson fermions, the
modified quenched approximation and a lattice spacing of 0.17 fm and
$0.07 \pm 0.04$ \cite{Blum:2000kn} using domain wall fermions, the
Wilson action and a lattice spacing of 0.2 fm.  Since $\delta$ is a
parameter of low-energy quenched QCD, the general agreement between the
results from the different lattice formulations quoted above is
encouraging and expected.

All the values for $\delta$ are below the initial estimates of $\sim
0.2$, based on the value for the $\eta^\prime$ mass in full QCD.  This
suggests that the effects of quenched chiral logarithms will only be
evident at quite small quark masses.  In this section we want to
revisit the determination of $\delta$ from $\mpi2$ versus $m_f$ for
domain wall fermions, but at a smaller lattice spacing (0.104 fm) than
our earlier determination at 0.197 fm \cite{Blum:2000kn}.  We will then
be able to assess the importance of quenched chiral logarithms in our
determination of kaon matrix elements.

In our earlier work on the chiral limit of domain wall fermions, we
found that by working on large enough volumes to suppress the effects
of topological near-zero modes, our data was consistent with the
presence of a quenched chiral logarithm and that the point where
$\mpi2$ vanished for such a fit was also in agreement with our value of
$m_{\rm res}$ determined independently.  For our current simulations,
where the volumes are not as large, we will use the previously measured
value $m_{\rm res} = 0.00124(5)$ as an input and neglect the $m_f =
0.01$ point in our analysis.  This should exclude the dominant effects
of topological near-zero modes and will also allow us to determine a
value for $\delta$.

In fitting to the general form of Eq.\ \ref{eq:qXPT_mpi} we must decide
how to handle the presence of the parameter $\alpha$ as well as
$\delta$.  We first note an important consequence of our range of pion
masses, which is that $\mpi2 \ln(\mpi2/\Lqcpt^2)$ only varies by 5\%
for $0.02 \le m_f \le 0.04$ with $\Lqcpt = 1$ GeV.  This is shown in
Figure \ref{fig:chiral_log_vs_mf} where we have used $\mpi2 =
0.0098(20) + 3.14(9) m_f$ from \cite{Blum:2000kn}.  Thus, the term
$\alpha \mpi2 \ln(\mpi2/\Lqcpt^2)$ will be approximately constant over
our range of quark masses and we cannot expect to resolve it with our
data.  The small variation in $\mpi2 \ln(\mpi2/\Lqcpt^2)$ over our pion
mass range will be an important point in fits to much of our data.

Thus, we fit our lattice data to the form
\begin{equation}
  \left( \mpi2 \right)^{\rm lat} =
      a_\pi ( m_f + m_{\rm res} )
      \left[ 1 - \delta \ln \left( \frac{ a_\pi ( m_f + m_{\rm res} )}
      {\Lqcpt^2} \right) \right] .
\end{equation}
We have used this functional form to fit $\mpi2$ from the 85
configurations used in \cite{Blum:2000kn}, where quark masses 0.015,
0.02, 0.025, 0.03, 0.035 and 0.04 were used for the fits.  These values
for $\mpi2$ come from the axial current correlator $\langle A_0^a(x)
A_0^a(0) \rangle$ to reduce the effects of topological near-zero
modes.  We have also done fits for the 400 configuration data set
generated for this matrix elements calculation, where quark masses 0.02,
0.03, 0.04 and 0.05 were used in the fits.  The pion masses for this
data set come from $ \langle \pi^a(x) A_0^a(0) \rangle$ correlators,
since we only have pseudoscalar sinks in our matrix elements programs.
We choose to quote results for $\Lqcpt = 1$ GeV and have also done fits
for $\Lqcpt = 0.77$ and 1.2 GeV.

Figure \ref{fig:mpi2_vs_mf} shows the data for both data sets and the
curve is the fit to the 85 configuration set.  For the 400
configuration data set we find $a_\pi = 3.27(2)$ and $\delta =
0.029(7)$ with $\chi^2/{\rm d.o.f} = 2.3$,  while for the earlier 85
configuration data set we find $a_\pi = 3.18(6)$ and $\delta = 0.05(2)$
with $\chi^2/{\rm d.o.f} = 0.3$.  Since these are uncorrelated fits to
correlated data, the values of $\chi^2$ are of limited validity, but,
particularly for the 85 configuration data set, show the data is
consistent with a quenched chiral logarithm form.  Varying $\Lqcpt$
only changes $a_\pi$ by $\pm2$\% and does not change $\delta$ within
errors.  The difference in the value of $\delta$ between the two data
sets is due to the $m_f = 0.015$ point only being present in the 85
configuration set.  Without this point a smaller curvature is needed,
and hence a smaller $\delta$, to make $\mpi2$ vanish at $m_f = - m_{\rm
res}$.  Also notice that the $m_f = 0.05$ value for $\mpi2$ lies
substantially above the fit line, which neglects this point.  Since we
are interested in quenched pathologies appearing at small quark masses,
we have not included ${\cal O}((m_f+m_{\rm res})^2)$ terms in our fit.
Given that $m_\pi = 790$ MeV for this heaviest quark mass, such higher
order terms are expected to be important.

Notice that we cannot determine the one-loop effects on the value of
$a_\pi$.  The combination of constants in the braces in
Eq.\ \ref{eq:qXPT_mpi}, the almost precise constancy of $\mpi2
\ln(\mpi2/\Lqcpt^2)$ and the uncertainty in $\Lqcpt$ provide too many
similar effects to be distinguished in our data.  Since $\delta$ is
small, it is reasonable to expect that ignoring these terms is a good
approximation.  Also note that a large value for $\alpha$ should
make the $\alpha m_\pi^4$ term give a noticeable non-linearity for
larger $\mpi2$.  This is not seen, implying either a small value for
$\alpha$ or a cancellation with terms from the ${\cal O}(p^4)$ Lagrangian.

Thus, we have consistency with other measurements of $\delta$ and
will use a value of $0.05$ for the remainder of this work.  The
fact that this value is small means the effects are not pronounced
for the scales of masses where we are currently simulating.


\subsection{Gell-Mann--Oakes--Renner Relation for Domain Wall Fermions}
\label{subsec:GMOR}

In Section \ref{subsec:mres_effects} and \ref{subsec:zero_mode_effects}
we discussed the role of residual mass and zero mode effects in
the Ward-Takahashi identity which is the basis for the
Gell-Mann--Oakes--Renner (GMOR) relation.  The result is given by
Eq.\ \ref{eq:GMOR_zero_modes_finite_ls}.  In this section we show
our numerical results for the quantities in this equation.

The zero mode effects in Eq.\ \ref{eq:GMOR_zero_modes_finite_ls} are
associated with $\langle \overline{u} u \rangle$ and $\langle J^a_5(x)
\, J^a_5(y) \rangle$.  For $\langle \overline{u} u \rangle$, the
effects produce a $1/m_f$ pole, as shown in \cite{Blum:2000kn}, which
can be separated out by doing an extrapolation to $m_f = 0$ from heavy
quark masses.  For $ \langle J^a_5(x) \, J^a_5(y) \rangle$, we can see
the size of the zero mode effects as a function of $x-y$ by comparing
the correlator $\langle S(y) \, S^\dagger(z) \rangle$ to $\langle
P_{K^+}(y) \, P_{K^-}(z) \rangle$.  We plot this ratio in Figure
\ref{fig:ss_over_pp}, using the the wall source, point sink propagators
from \cite{Blum:2000kn}.  In the figure one sees that this ratio is
essentially zero for $x-y > 8$ and $m_f \ge 0.02$, as it should be
since the pseudoscalar mass is much smaller than the scalar mass.
However for $m_f = 0.01$ or 0.015, the scalar correlator changes sign
and is a measurable fraction of the pseudoscalar correlator even for
$x-y > 8$.  We attribute this effect to zero modes and note
that zero mode effects are identical in the two correlators.  Thus, in
discussing the GMOR relation, we can easily remove the effects of zero
modes in $\langle \overline{u} u \rangle$, but zero modes in the
pseudoscalar correlator become $\sim 5$\% effects only for separations
greater than 12.

Since many of the terms in Eq.\ \ref{eq:GMOR_zero_modes_finite_ls} have
been measured in \cite{Blum:2000kn} for two different values of $L_s$
with the quenched Wilson gauge action at $\beta = 6.0$, we can discuss
how well the GMOR relation is satisfied.  Figure \ref{fig:gmor} shows
the terms in the GMOR relation using these measured values.  The upper
panel is for $L_s = 16$ and the lower is for $L_s = 24$.  The filled
squares are the values for $ -\langle \overline{u}u \rangle_{\rm
lat-norm}(m_f = 0, L_s)$ and the dashed line gives the $m_f/a^2$
dependence of this quantity.  The zero mode term (the sum on the
right-hand side of Eq.\ \ref{eq:GMOR_zero_modes_finite_ls}) has been
excluded by extrapolating to $m_f = 0$ from large values of $m_f$ where
zero mode effects play no role.  The open circles are

\begin{equation}
 | \langle 0 | J^a_5(0) | \pi^\prime \rangle|^2
 \; \left( \frac{m_f + \mres}
 {12 m_{\pi^\prime}^2} \right)
 \label{eq:gmor_lhs}
\end{equation}
as measured from from pseudoscalar correlators $\langle P_{\pi^+}(y) \,
P_{\pi^-}(z) \rangle$ using values of $|y-z|$ from 7 to 16.  Since this
ratio contains zero mode effects, some of the zero mode terms from
$\langle J^a_5(x) J^a_5(y) \rangle$ are included.  The solid lines are
the same quantity where a quenched chiral logarithm is included in
$m_{\pi^\prime}^2$.

For the $L_s = 16$ case, we expect the quantity in Eq.\
\ref{eq:gmor_lhs} to differ from $ -\langle \overline{u}u \rangle_{\rm
lat-norm}(m_f = 0, L_s)$ due to the presence of zero modes in this
quantity and the $\mres$ terms on the left-hand side of Eq.\
\ref{eq:GMOR_zero_modes_finite_ls}.  In Figure \ref{fig:gmor} one sees
that the $m_f \to 0 $ extrapolation of the heavier mass points lies
considerably above $-\langle \overline{u}u \rangle_{\rm lat-norm}(m_f =
0, L_s)$, revealing the size of the ${\cal O}(\mres)/a^2$ term.  Since the
slope of $-\langle \overline{u}u \rangle_{\rm lat-norm}(m_f, L_s)$ with
$m_f$ is power divergent (the dashed line), a small value for $\mres$
has a large effect.  Any zero mode effects for small $m_f$ are not
visible within our statistical errors.  Since $x-y$ in the range 7 to
16 has been used in determining the quantities in
Eq.\ \ref{eq:gmor_lhs}, Figure \ref{fig:ss_over_pp} shows that the
effects should be at the few percent level.  For $L_s = 24$ the
residual mass is much smaller, and the $m_f = 0$ extrapolation from
heavy quark masses agrees quite well with $-\langle \overline{u}u
\rangle_{\rm lat-norm}(m_f = 0, L_s)$.  Some non-linearity at small
quark masses is seen, but the errors are too large for a definite
conclusion.

Thus, we see that for $L_s = 16$, the naive GMOR relation is noticeably
modified by the presence of $\mres$, while for $L_s = 24$ the $\mres$
effects for this power divergent case appear to be smaller than 10\%.
It is important to note that $\mres$ is small for $L_s = 16$, but
$\mres/a^2$ effects are not.  We now turn to a similar comparison of
our numerical results with the Ward-Takahashi identity for $\langle
\pi^+ |\bar{s}d | K^+ \rangle $.


\subsection{Ward-Takahashi Identity for $\bar{s}d$}
\label{subsec:ward_id_sd}

In contrast to the GMOR relation discussed in the previous section, the
Ward-Takahashi identity for $\bar{s}d$ does not contain any power
divergent terms.  Thus, we can work in the large $L_s$ limit and then
replace $m_f$ with $m_f + \mres$ at the end.  We can use
Eq.\ \ref{sbard_wi_zero_small_mf} to understand the size of the zero
mode effects in $\langle \pi^+ |\bar{s}d | K^+ \rangle$.  Such zero
mode effects will appear identically in the power divergent part of
$\langle \pi^+ | Q_i | K^+ \rangle$ and will be removed in the
subtraction procedure given in Eq.\ \ref{eq:fix_alpha_81_1}.  The
remaining finite terms in the subtracted matrix element will have zero
mode effects, whose source we will understand more clearly after
investigating $\langle \pi^+ |\bar{s}d | K^+ \rangle$.

To measure $\langle \pi^+ |\bar{s}d | K^+ \rangle$, one can start
with the ratio
\begin{equation}
  R_1 \equiv
  \frac{\langle P_{\pi^+}^{\rm wall}(x_0) \, [\bar{s} d](y) \,
    P_{K^-}^{\rm wall}(z_0) \rangle }
  { \langle P_{\pi^+}^{\rm wall}(x_0) \, P_{\pi^-}(y) \rangle \,
    \langle P_{K^+}(y) \, P_{K^-}^{\rm wall}(z_0) \rangle}
  \label{eq:r1_def}
\end{equation}
where $P_{\pi^+}^{\rm wall}(x)$, etc.\  are Coulomb gauge fixed,
pseudoscalar wall sources and $x_0$ is the time coordinate at the point
$x$.  (For more details on the measurement of three-point correlators,
please see Section \ref{sec:lat_me}.) We plot this ratio in Figure
\ref{fig:sbard_PP_norm} where we take $x_0 = 5$, $z_0 = 27$ and average
over $14 \le y_0 \le 17$.  For $x \gg y \gg z$ and without zero-mode
effects, this ratio should be
\begin{equation}
  R_1 =
  \frac{ \langle \pi^+ | \overline{s} d | K^+ \rangle }
    { \langle \pi^+ | P_{\pi^-} | 0 \rangle
      \langle 0 | P_{K^+} | K^+ \rangle }
\end{equation}
which is finite and non-zero in the chiral limit.  From the
Ward-Takahashi identity, without zero modes and chiral logarithms, this
ratio is $2 m_f / (m_\pi^2 f^2)$, which is $\approx 120$ in lattice
units. (In this section, we consider the case of SU(3) flavor symmetry
so that $m_\pi = m_K = m_M$, where $m_M$ is the common meson mass first
used in Eq.\ \ref{eq:kpiO81}.) One sees from the figure that for
smaller $m_f$ the points actually are decreasing, rather than increasing
towards $\approx 120$.

Since our measurements are made with $ 9 \le x_0-y_0 \le 12$ and $ 10
\le y_0 - z_0 \le 13$, zero mode effects do enter the terms in the
denominator.  Consider a zero mode with support at $x$ and $y$.  It
produces a power of $1/m_f$ in the numerator of $R_1$ and contributions
of order $1/m_f^2$ and $1/m_f$ in the first term in the denominator of
$R_1$.  A similar argument is also true for a quark propagator
containing a zero mode at $y$ and $z$.  Thus, for very small $m_f$, the
ratio $R_1$ will go to zero due to zero modes.  We believe this to be
the source of the turnover in Figure \ref{fig:sbard_PP_norm} for small
values of $m_f$.

One can also determine $\langle \pi^+ |\bar{s}d | K^+ \rangle$ from
the ratio
\begin{equation}
  R_2 \equiv
  \frac{\langle P_{\pi^+}^{\rm wall}(x_0) \, [\bar{s} d](y) \,
    P_{K^-}^{\rm wall}(z_0) \rangle }
  { \langle P_{\pi^+}^{\rm wall}(x_0) \, P_{\pi^-}^{\rm wall}(z_0)
  \rangle}
  \label{eq:r2_def}
\end{equation}
In the denominator of $R_2$, zero modes should be negligible, since
$x_0 - z_0 = 22$ and the lattice has been doubled to make propagation
around the ends unimportant.  Thus, we are not introducing
zero mode effects into the ratio through the denominator.  Zero
modes in the numerator enter through the propagators $D^{-1}(x-y)$
and $D^{-1}(y-z)$.  Without zero mode effects, we have
\begin{equation}
  R_2 =
  \frac{ \langle \pi^+ | \overline{s} d | K^+ \rangle }
    { 2 m_\pi V_s}
  \label{eq:r2_ward_no_zero}
\end{equation}
where $V_s$ is the spatial volume.  To precisely describe our numerical
situation, we again use primes to describe states and masses which can
have zero mode effects.  For the current case, only one of the quark
propagators in the pseudoscalars in the numerator can have a zero
mode.  With this notation, we insert complete sets of states in Eq.
\ref{eq:r2_def} and find
\begin{equation}
  R_2 =
    \langle (\pi^+)^\prime | \overline{s} d | (K^+)^\prime \rangle
    \,
    \frac{ | \langle 0 | P_{\pi^+} | (\pi^+)^\prime \rangle |^2 }
         { | \langle 0 | P_{\pi^+} | \pi^+ \rangle |^2 }
    \,
    \frac{ 2 m_\pi V_s} { (2 m_{\pi^\prime} V_s)^2 }
    \,
    e^{(m_\pi -m_{\pi^\prime})( x_0 - z_0 )}
\end{equation}

The Ward-Takahashi identity result given in Eq.\
\ref{sbard_wi_zero_small_mf} can be similarly written as
\begin{alignat}{2}
  \frac{2 m_f}{m_{\pi^\prime}^2} \,
  \langle 0 | P_{\pi^-} | (\pi^+)^\prime \rangle \,
  \langle (\pi^+)^\prime | \overline{s} d | (K^+)^\prime \rangle \,
  \langle (K^+)^\prime | P_{K^-} | 0 \rangle \,
  \frac{ e^{- m_{K^\prime}(y - z_0)} } { 2 m_{K^\prime} V_s }
  \nonumber \\
  =
   \frac{|\langle 0 | P_{\pi^-} | (\pi^+)^\prime \rangle|^2}
     { 2 m_{\pi^\prime} V_s } \;
     e^{- m_{\pi^\prime}(y - z_0) }
\end{alignat}
which reduces to
\begin{equation}
  \frac{2 m_f}{m_{\pi^\prime}^2} \,
  \langle (\pi^+)^\prime | \overline{s} d | (K^+)^\prime \rangle
  = 1
  \label{eq:sbard_ward_id_one_zero}
\end{equation}
We now let $L_s$ be finite and change $m_f \to m_f + \mres$.  We are
left with
\begin{equation}
  R_2 \, \frac{ 4 V_s (m_f + \mres)}{m_\pi}
   \; \frac{|\langle 0 | P_{\pi^-} | \pi^+ \rangle|^2}
     {|\langle 0 | P_{\pi^-} | (\pi^+)^\prime \rangle|^2}
     \;
     e^{ (m_{\pi^\prime} - m_\pi)(x_0 - z_0) }
  = 1
  \label{eq:sbard_r2_ward_id}
\end{equation}
When there are no zero mode effects, $m_\pi = m_{\pi^\prime}$ and $|
\pi^+ \rangle = | (\pi^+)^\prime \rangle$ leaving $ R_2 \, 4 V_s (m_f +
\mres) /m_\pi = 1$.  Notice that a small difference in $m_\pi$ and $
m_{\pi^\prime}$ is multiplied by $x_0-z_0$, which can lead to larger
effects in $R_2$.  This is a result of the simple fact that zero modes
effect the pseudoscalar propagators in the numerator of $R_2$
differently than they effect the propagators in the denominator.

Figure \ref{fig:ward_sbard} is a plot of the value of $4 R_2 V_s ( m_f
+ \mres) /m_\pi$ versus quark mass.  We use a value for $m_\pi$ that is
not effected by zero modes.  One sees that for the smaller values of
$m_f$ this ratio deviates substantially from 1, being 16\% below 1 for
$m_f = 0.01$.  We would like to see if this is consistent with the
prediction of Eq.\ \ref{eq:sbard_r2_ward_id}.  We do not have direct
measurements of $m_{\pi^\prime}$, since this is a mass which comes from
correlators with at most one zero mode.  However, the effective mass
plots shown in Figure 21 of \cite{Blum:2000kn} give values for
$m_{\pi^{\prime \prime}}$, the mass from the pseudoscalar correlator
where any number of zero modes is allowed, for $m_f = 0.01$.  In the
range of separations 9 to 12, $m_{\pi^{\prime \prime}} = 0.211(6)$,
compared with $m_\pi = 0.199$, our best estimate for $m_\pi$ without
zero mode effects for $m_f = 0.01$.  This gives $m_{\pi^{\prime
\prime}} - m_\pi = 0.014$ and $\exp[ (m_{\pi^{\prime\prime}}-
m_{\pi})(x_0 - z_0) ] = 1.36$ for $x -z = 22$.

We do not know the relative contributions of one and two zero mode
terms to $m_{\pi^{\prime\prime}}$.  However, the zero modes have
eigenvectors where the product $\psi_\lambda(x) \psi_\lambda^\dagger(y)$
is going to zero in the range of separations we are considering.  It is
reasonable to argue that the falloff in the eigenvectors with $x-y$ is
producing $m_{\pi^{\prime \prime}} > m_\pi$, since the the two zero
mode contribution should dominate for small $m_f$, they involve
$|\psi_\lambda(x)|^2 \,| \psi_\lambda(y)|^2$ and the pseudoscalar
correlator is positive definite.  For $m_{\pi^\prime}$, only terms with
at most one zero mode contribution are included.  In this case
$\psi_\lambda(x) \psi_\lambda(y)$ enters not $|\psi_\lambda(x)|^2 \,|
\psi_\lambda(y)|^2$ and there is no positivity for the one zero mode
contribution alone.  However, naively one could expect $m_{\pi^{\prime
\prime}} > m_{\pi^\prime} > m_\pi$.  Thus it is reasonable that $1 <
\exp[ (m_{\pi^\prime}- m_{\pi})(x_0 - z_0) ] < 1.36$.  From the
determination of $f_\pi$ in \cite{Blum:2000kn} using pseudoscalar and
axial vector correlators, the zero mode effects in $ \langle 0 |
P_{\pi^-} | ( \pi^+)^\prime \rangle$ are at the few percent level.
Thus, the deviation of $4 R_2 V_s (m_f + \mres) / m_\pi$ from 1 in
Figure \ref{fig:ward_sbard} is consistent with the estimates based on
the difference in the mass of the pseudoscalar states relevant to the
numerator and denominator of $R_2$.  From
Eq.\ \ref{eq:sbard_ward_id_one_zero}, the zero mode effects in $
\langle (\pi^+)^\prime | \overline{s} d | (K^+)^\prime \rangle$ are at
most a few percent.  The small differences in the ``masses'',
$m_\pi^{\prime} - m_\pi$ indicate a substantial effect of zero modes
for time separations of the order of 10.  We believe that these effects
are responsible for the large deviation seen from the predictions of
chiral symmetry for these $\overline{s}d$ matrix elements (Figure
\ref{fig:ward_sbard}).

We now turn to the question of the extraction of matrix elements from
our lattice correlators.  As we have discussed, in the subtraction of
divergent terms zero mode effects cancel.  In the ratio $R_1$, large
zero mode effects are introduced into the denominator through the
pseudoscalar correlators acting over moderate distances.  This produces
a different effective pseudoscalar mass in the numerator and
denominator.  In the ratio $R_2$, no zero modes are introduced in the
denominator, but there is a similar mismatch in pseudoscalar masses
since the numerator can contain zero modes.  However, this mismatch is
most pronounced for the power divergent terms, which behave like the
$\overline{s}d$ matrix element above.  In the finite, subtracted
operator, a similar mass mismatch can occur for eye type diagrams, but
will not in general occur for figure eight diagrams due to the way
gamma matrices enter the traces and the fact that all zero
modes have the same chirality.  Thus, the ratio $R_2$ will not eliminate
all the effects of zero modes in the desired physical quantities, but
it minimizes them.  We will use $R_2$ for the determination of our
desired $K \to \pi$ matrix elements.

Figure \ref{fig:ktopi_sbard} shows $2 m_\pi V_s R_2$ versus $m_f$.
With no zero mode effects this equals $\langle \pi^+ | \overline{s} d |
K^+ \rangle$.  For $m_f = 0.01$, the zero modes should produce the same
relative distortions in this quantity as are shown in Figure
\ref{fig:ward_sbard}.  This matrix element is used in the operator
subtraction and as discussed previously any zero mode and chiral
logarithm effects in this matrix element will match those in the power
divergent parts of $\langle \pi^+ | Q_i | K^+ \rangle$.  Since the plot
is not obviously linear, it is important to subtract the two matrix
elements to take full advantage of the correlation of zero mode and
chiral logarithm effects between them.

\fi


\section{Wilson Coefficients}
\label{sec:wilson_coef}

\ifnum\theWilsonCoef=1

The twelve-dimensional vector of Wilson coefficients $\vec{C}(\mu)$
has been calculated at next to leading order (NLO) in QCD and
QED by the Munich~\cite{Buras:1993tc,Buras:1993zv,Buchalla:1996vs} and
Rome~\cite{Ciuchini:1994vr} groups.  In those calculations the
Callan-Symanzik equations are solved to determine the Wilson
coefficients at an energy scale $\mu \approx 1$ GeV, appropriate
for lattice calculations, starting from their values at the weak scale,
$\approx M_W$.  The solution is obtained within the approximation that
the parameters $\alpha_s$ and $\alpha$ (the fine structure constant of
electromagnetism) are small but that the products $(\alpha_s t)^n$ are
of order one, where $t = \ln{(M_W/\mu)}$.  According to this reasoning, in
leading order (LO) one sums all terms of the form $\alpha_s^n t^n$ and
$\alpha\alpha_s^n t^{n+1}$.   These terms are identified as ${\cal O}(1)$
and ${\cal O}(\alpha/\alpha_s)$ respectively.  In the next leading order
approximation (NLO) one also includes all terms of the form:
$\alpha_s^{n+1} t^n$ and $\alpha\alpha_s^n t^n$, identified as
${\cal O}(\alpha_s)$ and ${\cal O}(\alpha)$ respectively.  Terms of order
$\alpha^n$ for $n \ge 2$ are not included.

In the notation of Ref.~\cite{Buchalla:1996vs} the NLO evolution of
$\vec{C}(\mu)$ to a value of $\mu$ below the charm threshold is given by
\bea
\vec{C}(\mu) &=&
\hat{U}_3(\mu,m_c,\alpha)\,\hat{M}_4\,\hat{U}_4(m_c,m_b,\alpha)\,
\hat{M}_5\,\hat{U}_5(m_b,M_W,\alpha)\,\vec{C}(M_W),
\label{eq:evol}
\eea
where $\hat{U}_f(\mu_1,\mu_2,\alpha)$ is the renormalization group improved,
evolution matrix from the scale $\mu_2$ down to the scale $\mu_1$ in a
theory with $f$ quark flavors.  The matrix $\hat{U}_f(\mu_1,\mu_2,\alpha)$
is a $12 \times 12$ matrix for $f=4$ and 5 while it reduces to a
$10 \times 10$ matrix for $f=3$. The flavor matching matrix $\hat{M_f}$
relates the Wilson coefficients that appear in the $f$ and
$f-1$ flavor effective theories.  It is naturally written as a
$12 \times 12$ matrix for $f=4$ and 5, while for $f=3$ it is a
$10 \times 12$ array.  Here $\vec{C}(M_W)$ are the twelve
coefficients of the effective theory calculated at the scale $M_W$ by
matching to the full theory.  Evolution down to a value of $\mu$ above
the charm threshold is given by an obvious truncation of Eq.\ \ref{eq:evol}.
The matrix $\hat{U}_f(\mu,m,\alpha)$ contains terms of order ${\cal O}(1)$
and ${\cal O}(\alpha_s)$ in QCD and includes terms of ${\cal O}(\alpha/\alpha_s)$
and ${\cal O}(\alpha)$ when QED effects are included.  Following convention,
we fix $\alpha=1/128$ at $\mu=M_W$ and do not include its running in the evolution
of the Wilson coefficients.

Following Ref.~\cite{Buras:1993dy}, we express the contributions arising
from charged $W$ exchange as the sum of two terms.  The first, which evolves
with Wilson coefficients defined as $z_i(\mu)$, contains the difference of
charm and up quark fields and carries the CKM coefficients $(1-\tau)$.  The
second evolves with Wilson coefficients defined as $v_i(\mu)$, contains
the difference of the top and up quark fields and carries the CKM
coefficients $\tau$ (see Ref.~\cite{Buras:1993dy}, Eq.\ 4.4).  For the
three-flavor, ``charm-out'' case, only the ten operators
$Q_i$ appear and their Wilson coefficients are given by
\bea
C_i &=& \tau v_i + (1-\tau) z_i, \\
    &=& \tau y_i + z_i,
\eea
where $y_i = v_i - z_i$.  With this separation, the evolution of the
coefficients $z_i$ is particularly simple: The cancellation between
the charm and up quark loops (the GIM mechanism) prevents the appearance
of penguin contributions until one matches to the 3-flavor, charm-out
effective theory.  Since, with our standard choice of phase conventions,
the CP violating phase is contained in the CKM parameter $\tau$, the larger,
$\tau$-independent terms coming from $z_i$ will provide the dominant contribution to the
$CP$ conserving amplitudes Re$A_{0,2}$ while $y_i$ must appear in the
$CP$ violating amplitudes Im$A_{0,2}$ (see Eq.\ \ref{eq:DSone_ham}).

We calculate these Wilson coefficients in two steps. First
we determine $\vec{C}(\mu)$ in the NDR scheme using exactly
the formulas and procedures given in Refs.~\cite{Buras:1993dy,Buchalla:1996vs}.
In particular, when using Eq.\ \ref{eq:evol}, all ${\cal O}(\alpha_s^2,\,
\alpha_s\alpha)$, and higher order terms which are generated by
multiplication of the evolution and matching matrices are dropped so that the
final Wilson coefficients at scale $\mu$ contain all contributions up to and
including ${\cal O}(\alpha_s,\,\alpha/\alpha_s,\,\alpha)$ and no more.
An example of the breakdown of $z_i$ and $y_i$ at $\mu=1.3$ GeV in the
NDR scheme is given in Tables \ref{tab:wilson_coef_comp_z} and
\ref{tab:wilson_coef_comp_y}.
In the second step, we transform these coefficients, obtained in the NDR
scheme into the coefficients of operators defined according to the RI
scheme\footnote{This matching requires a careful definition of our
basis of operators in the NDR scheme associated with the difficulties
of defining $\gamma^5$ in dimensional regularization.  While in the RI
scheme, Fierz rearrangement of the fermion fields has no effect, this
is not true in the NDR scheme.  In fact, for the NDR calculation and
matching to RI to be described correctly, we should follow
Ref.~\cite{Buchalla:1996vs} and write our operators $Q_{1,2}$ in a
Fierz rearranged fashion.  This is the form that is used in the NDR
calculation~\cite{Buchalla:1996vs} we are following and in determining
the matching coefficients $\Delta\,r^{NDR}_{\lambda^*=0}$ in
Ref.~\cite{Ciuchini:1994vr,Ciuchini:1995cd}.  However, in our own
matrix element and NPR calculations, where the Fierz ordering is
immaterial, we find the Fierz structure shown in Eqs.\ \ref{eq:P1}-\ref{eq:P2c}
to be more convenient.} in Landau gauge using
\bea
\vec{C}_{RI}(\mu) &=&
\left(1-\frac{\alpha_s(\mu)}{4 \pi}(\Delta\,r^{NDR}_{\lambda^*=0})^{T}
\right)\vec{C}_{NDR}(\mu),
\label{eq:match}
\eea
where the matching matrix
$\Delta\,r^{NDR}_{\lambda^*=0}$ is given in Table VIII of
Ref.~\cite{Ciuchini:1995cd}.

In this paper we discuss only the three-flavor, charm-out case.  Thus,
we naturally deal with an effective theory that describes physics
at energy scales below the charm mass---the scales that dominate
the matrix elements we are computing.  However, we are concerned about
potential errors that come from using perturbation theory so close
to the non-perturbative region.  We cannot avoid the use of
perturbative matching to connect the four-flavor (charm-in) and
three-flavor (charm-out) theories since in the lattice calculations presented
in this paper we do not include a propagating charm quark.  However, the
connection between the NDR and RI Wilson coefficients, also done in
perturbation theory, can be done at a scale above the charm quark mass,
thereby reducing the perturbative uncertainties.  Note, in these
discussions the energy $\mu$ specifies the energy scale that appears
in the normalization condition that defines the operators that appear
in our effective theory.  For the case at hand, we are free to choose
this scale to be well above $m_c$ where perturbation theory may be more
reliable.  Of course, our effective theory will not describe processes
in Nature in this region of energies ($\sim m_c$), but only processes
involving lower energy scales.  Note, we are prevented from using a
very large value for $\mu$ since we do not want large lattice spacing
errors to enter our non-perturbative normalization of these operators.

In order that the product of the RI Wilson coefficients times the RI
operators be independent of the scale $\mu$ they must both be computed
in the full or the quenched theory.  Since our non-perturbative
normalization is determined in the quenched approximation, the $\mu$-dependence
of the Wilson coefficients should be determined in the quenched theory.
Therefore, we adopt the following transition to our quenched approximation.
In evolving the effective weak Hamiltonian from the W mass scale down to
a form valid in the three-quark, charm-out theory, we include all
required quark loop effects.  Making a ``quenched'' approximation here is
not necessary and would leave out physically important phenomena.  We then
interpret the resulting NDR scheme, 3-flavor effective weak Hamiltonian with
operators and coefficients defined at $\mu=m_c$ as our quenched approximation
Hamiltonian.  Thus, we use the Wilson coefficients without change but
interpret the operators as defined in the quenched approximation.
We are then free to vary the renormalization scale $\mu$, increasing it
above $m_c$ if we choose.  However, we must normalize the operators by
evaluating quenched Green's functions and evolve the Wilson coefficients
from their $\mu=m_c$ values using quenched evolution
equations.\footnote{In the results described
below, we carry out this prescription only approximately.  For the $\mu$
dependence of $\alpha_s$ we use the $f=0$ $\beta$ function and the value
of $\Lambda_{QCD}=238$MeV from the quenched calculation of
Ref.~\cite{Capitani:1998mq}.  However, we still use the 3-flavor,
2-loop anomalous dimension matrix rather than the 0-flavor matrix as
required by the above discussion.  Since the resulting evolution
only corresponds to scale changes on the order of a factor of two, there
are no large logarithms and it is appropriate to neglect such 2-loop
effects in our NLO calculation.}

Our results in the RI scheme for the three-flavor theory are given in
Tables~\ref{tab:wilson_coef_z} and \ref{tab:wilson_coef_y}.
The scales $\mu=1.51$, 2.13, 2.39, and 3.02 GeV correspond to those where the
non-perturbative, operator renormalization Z factors were calculated.
The Standard-Model parameters used to obtain these numbers are
given in Table~\ref{tab:wilson params}. Two-loop running of $\alpha_s$
is used throughout. We have performed several checks of our analysis. Our
numerical values of $\vec{C}_{NDR}(\mu)$ agree exactly with those reported
in~\cite{Buchalla:1996vs} when their values for the Standard-Model parameters
are used.  We also agree within 20\%, or much better in most cases,
with the Wilson coefficients given in~\cite{Ciuchini:1995cd} for the
NDR and RI schemes.  These differences arise because the treatment of
terms beyond NLO differs between that adopted in Ref.~\cite{Buchalla:1996vs},
which we follow, and that of \cite{Ciuchini:1995cd}.

We note that there is a potential ambiguity which arises when using the
one-loop matching given by Eq.\ \ref{eq:match}. Straight multiplication
of $\vec C_{NDR}(\mu)$ by the one loop matching matrix generates an
${\cal O}(\alpha_s\,\alpha )$ contribution which is large. After
matching we find $C_{8,RI}(\mu\approx 2 \, {\rm GeV}) \approx 0.0006$
if we drop this term, or 0.0009 if we do not.  Thus, this ${\cal
O}(\alpha_s\,\alpha )$ term increases $C_{8,RI}$ by 50\%.  The origin
of this large correction is easily understood by examining the ${\cal
O}(\alpha/\alpha_s)$ and ${\cal O}(\alpha)$ terms in $C_{8,NDR}$.  The
sub-leading ${\cal O}(\alpha)$ term is roughly 7 times the leading order
${\cal O}(\alpha/\alpha_s)$ term and they have opposite signs. The
origin of this reversal is well known; the ${\cal O}(\alpha)$ term is
dominated by the contribution proportional to $m_t^2$ which is quite
large.  This sum of leading (small) and sub-leading (large) terms is
then to be multiplied by the one loop matching for $C_8$ which is
dominated by the diagonal term which is itself anomalously large,
$(\Delta r^{NDR}_{\lambda^*=0})_{8,8}\approx 10$.

The above discussion may lead the reader to conclude that there is
a significant uncertainty in $C_8$, an important quantity in
$\epsilon^\prime$.  In fact, we believe that this is not the case.
The large corrections which arise from the matching calculation must be
included as complete factors in the Wilson coefficients to maintain the
scheme independence of the weak Hamiltonian.  Arbitrarily dropping these
higher order terms could potentially increase the scheme dependence of
our final result (we follow the general argument given
in Ref.~\cite{Buras:1999st} for the NDR and HV schemes which applies to
the RI scheme as well).  In practice, the scheme and scale dependence of
the Wilson coefficients and the renormalized operators cancels when
they are combined in the weak Hamiltonian.  Schematically,
\bea
H_W &=& \vec{Q}^T \vec{C}\\\nonumber
    &=& \vec{Q}_{RI}^T \vec{C}_{RI}\\\nonumber
    &=& \vec{Q}_{NDR}^T
\left(1+\frac{\alpha_s(\mu)}{4 \pi}(\Delta\,r^{NDR}_{\lambda^*=0})^{T}+\dots\right)
\left(1-\frac{\alpha_s(\mu)}{4 \pi}(\Delta\,r^{NDR}_{\lambda^*=0})^{T}+\dots\right)
\vec{C}_{NDR}\\\nonumber
&=& \vec{Q}_{NDR}^T \vec{C}_{NDR}.
\eea
By far the largest contribution to the ${\cal O}(\alpha)$ part
of the weak Hamiltonian is $C_8 Q_8$ which then, by itself,
must be scheme independent.  As we saw, the ${\cal O}(\alpha\alpha_s)$
contribution to the matching matrix (which by definition is scheme dependent) was
quite large. In our calculation, the renormalization of the operators
is done to all orders in QCD in the RI scheme.  Thus, the product $C_8 Q_8$
could implicitly contain a compensating large ${\cal O}(\alpha\alpha_s)$
scheme dependent contribution
coming from the ${\cal O}(\alpha)$ term in $C_8$ and the
${\cal O}(\alpha_s)$ term implicit in the non-perturbative renormalization of
$Q_8$.  Thus, it is natural to include the full matching coefficient
in the RI value of $C_8$ so that these compensating terms will both be
present in the product $C_8 Q_8$.

Recently, partial next-next-to-leading order (NNLO) calculations have
been performed\cite{Bobeth:1999mk,Buras:1999st}. We
only examine the latter case where the complete set of
${\cal O}(\alpha \alpha_s)$
and ${\cal O}(\alpha \alpha_s \sin^2{(\theta_W)} m_t^2)$ corrections to the
Wilson coefficients $C_{7-10}$ of the electroweak penguins have been
calculated. In Ref.~\cite{Buras:1999st} it is argued that these are
the dominant NNLO contributions.  We simply take the values in
Ref.~\cite{Buras:1999st} for $\mu=1.4$ GeV to estimate the change in
$C_{7-10}$ in the RI scheme, and use these values in conjunction with
the ones in Tables \ref{tab:wilson_coef_z} and \ref{tab:wilson_coef_y}
to estimate the effects of these corrections on
the $K\to\pi\pi$ amplitudes given in later sections.
We conclude that the changes in the Wilson coefficients and final
$K\to\pi\pi$ amplitudes are modest.

We explicitly tabulate the values of the Wilson coefficients at four
different scales $\mu$.  In later sections these coefficients
are combined with the non-perturbative Z-factors, computed at these
same four values of $\mu$ to determine the final physical results.
Since these final numbers should be independent of this renormalization
scale $\mu$, this comparison gives a significant indication of how well
our method is working.

\fi


\section{Operator Renormalization using NPR}
\label{sec:npr}

\ifnum\theNpr=1
%
%

As is well known, in using the lattice to calculate matrix elements,
one cannot simply transcribe the operators of the continuum theory to
the lattice.  The lattice operators and continuum operators have to be
properly renormalized and the relationship between them explicitly
known.  For this we use a two step process to take advantage of
existing continuum calculations for the Wilson coefficients.
\begin{enumerate}
\item
 We use a renormalization scheme (here the RI or regularization
 independent scheme) to define renormalized operators which is
 independent of the underlying regulator.  This ensures a common
 definition of renormalized operators on the lattice and in the
 continuum.
\item 
  We also need the relationship between operators renormalized in the RI
  scheme and those in the $\overline{\rm MS}$ scheme since the existing
  perturbative calculations of the Wilson coefficients are done in this scheme.
  The matching between RI and $\overline{\rm MS}$ with naive dimensional
  regularization (NDR) is known at one loop
  \cite{Ciuchini:1994vr,Ciuchini:1995cd}.
\end{enumerate}
An additional complication in the renormalization of the operators in
the $\Delta S=1$ Hamiltonian is the mixing between these operators and
lower dimensional operators.  This is due to the presence of quark and
antiquark fields of the same flavor in the $\Delta S=1$ operators.
Since this mixing in general involves power divergent coefficients, it
can be quite large if the lattice formulation badly breaks chiral
symmetry.  Since in our calculation with domain wall fermions chiral
symmetry breaking effects are small, this problem becomes tractable.


\subsection{Mixing for $\Delta S =1 $ Operators}

For the $\Delta S=1$ Hamiltonian, the continuum renormalized
dimension-six operators can be written in terms of bare lattice
operators as
\begin{equation}
  O_i^{\rm cont,\; ren}(\mu) =
  \sum_j Z_{ij}(\mu) \left[ O^{\rm lat}_j + \sum_k c^j_k(\mu)
  B_k^{\rm lat} \right]
  +O(a) \, .
  \label{eq:general_mixing}
\end{equation}
We have introduced the scale $\mu$ used to define the renormalized
operators.  Here $O^{\rm lat}_j$ is also a four-quark dimension six
operator and the $B_k^{\rm lat}$'s are operators that contain only two
quark fields.  Due to the $\Delta S=1$, $\Delta D=-1$ nature of the
operators we are considering, the $B_k^{\rm lat}$'s must have the $\overline{s}d$
flavor structure.  These operators can mix with coefficients $c^j_k$
that diverge as the lattice spacing tends to zero.

We will consider here the renormalization of the parity conserving part
of the $\Delta S=1$ effective Hamiltonian assuming, as in the rest of
this work, that chiral symmetry is respected.  (We have investigated
this question in detail for the renormalization of quark bilinear
operators and found no significant effects due to explicit chiral
symmetry breaking by domain wall fermions at finite $L_s$
\cite{Blum:2001sr}.)  The renormalization conditions will be imposed in
the massless limit and as such operators in different multiplets of
$SU(3)_L \otimes SU(3)_R$ or isospin do not mix under renormalization.
This imposes strong constraints on the allowed operator mixing, and in
particular on the number of quark bilinear operators that need to be
considered. The latter may be split into three classes
\cite{Dawson:1998ic,Buras:1993tc}.
\begin{enumerate}
\item Operators that vanish on shell by the equations of motion.
\item Gauge invariant operators that do not vanish by the equations
  of motion.
\item Non-gauge, but BRST, invariant operators.
\end{enumerate}
Operators of types one and three do not contribute to physical
processes and so do not have to be considered in the calculation of
hadronic matrix elements.  However, they do have to be taken into
account in operator renormalization, where amplitudes with off-shell
gauge-fixed external fields are used.

The bilinear operators $B_k$ in Eq.\ \ref{eq:general_mixing} must
contain an $\bar{s}$ and $d$ quark and conserve parity.  Thus,
their general form must be one of the following
\begin{equation}
\overline{s} X^{(1)} d
\label{npr:sdx}
\end{equation}
\begin{equation}
\overline{s}\sigma_{\mu\nu} X_{\mu\nu}^{(2)} d
\end{equation}
\begin{equation}
\overline{s}\gamma_{\mu} X_{\mu}^{(3)} d
\label{eq:sbard_gamma}
\end{equation}
where $X^{(1)}$, $X^{(2)}_{\mu \nu}$ and $X^{(3)}_\mu$ are flavor
singlet quantities which may include gluon, ghost and derivative
terms.  It is simple to see that Eq.\ \ref{npr:sdx} is in a
$(3,\overline{3}) + (\overline{3}, 3)$ representation of $\sutlr$ and
so may not mix with any of the dimension six operators we are
considering in the massless limit.

In fact, the only operator that is allowed to mix by $\sutlr$ is Eq.\
\ref{eq:sbard_gamma}, which transforms as an $(8,1)+(1,8)$. This gives
one dimension four operator,
\begin{equation}
\overline{s}
\left(
-\lvec{\slash{D}}
+\rvec{\slash{D}}
\right)
d \, ,
\label{eq:dsl}
\end{equation}
with a mixing coefficient $c^j_k$ that may behave as $1/a^2$ as $a
\rightarrow 0$, which we must consider.  BRST non-invariant operators
are allowed to mix only if they vanish by the equations of motion
\cite{Dawson:1998ic}.  This forbids the second possible dimension four
operator, $\bar{s} \slash{\partial} d $, from appearing.  This argument
allows operators of dimension five to appear.  However these operators
break chiral symmetry and are therefore forbidden.  Several dimension
six operators (for example those involving three $\slash{D}$
operations) can also occur, although their mixing coefficients diverge
at most logarithmically.

The arguments above rely on the fact the renormalization conditions
that we will be imposing are defined in the chiral limit.  The
numerical simulations that we have done to evaluate them, however, were
performed at multiple, finite values of the quark mass and the results
extrapolated to the massless limit.  As this is the case, it is also
important to study operators that may be present due to the breaking of
chiral symmetry by the quark mass and also the explicit chiral symmetry
breaking from finite $L_s$.  This allows many more operators to mix.
We will focus on the most divergent one (which diverges as $1/a^2$)
given by Eq.\ \ref{npr:sdx} with $X^{(1)} = 1$ and show that its
contributions are negligible in the chiral limit.


\subsection{Non-perturbative Renormalization}

Although, in principle, the renormalization of lattice operators can be
done by using lattice perturbation theory, in practice simple uses of
lattice perturbation theory suffer from poor convergence for currently
accessible gauge couplings ($\beta \sim6.0$).  Use of renormalized or
boosted couplings~\cite{Lepage:1993xa} improves the perturbative
behavior in many cases of interest but considerable arbitrariness
remains~\cite{Donini:1999sf}.  Furthermore, for domain wall fermions
lattice perturbation theory has the added complication that the
renormalization coefficients can depend sensitively on $M_5$, the
domain-wall
height~\cite{Blum:2001sr,Aoki:2000ee,Aoki:1999ky,Blum:1999xi}.  The
non-perturbative renormalization technique pioneered by
Rome-Southampton group \cite{Martinelli:1995ty} provides a method for
removing the uncertainties associated with perturbation theory.
(Another approach to non-perturbative renormalization has been
developed by \cite{Luscher:1997jn}.) The use of this technique here
represents one of the most complicated situations where it has been
applied.  We now give a brief overview of the method and elaborate on
its use for the $\Delta S = 1 $ case.

The NPR method starts with the computation of Green's functions of the
bare operators in question.  The Green's function is calculated using
off-shell external quark fields at large Euclidean momentum.  This
momentum defines the renormalization scale $\mu$.  The quark fields
must be in a particular gauge, and in this work we only use Landau
gauge.  We note that renormalization coefficients in the RI scheme can
be gauge dependent.  Schematically, we have
\begin{equation}
  G^{(4)}(p_1, p_1, p_2, p_2) = 
  \langle q_\alpha^i(p_1) q_\gamma^k(p_1) \, O_m \,
  \overline{q}_\beta^j(p_2) \overline{q}_\delta^l(p_2) \rangle
  \label{eq:green_schematic}
\end{equation}
and, as we will discuss in more detail later, we work with $| p_1 | = |
p_2| = | p_1 - p_2|$.  This Green's function is then amputated using
the full quark propagators calculated in the same gauge.  A
renormalization condition which fixes the $Z_{ij}$ and $c^j_k$ factors
in Eq.\ \ref{eq:general_mixing} may then be applied by requiring that
the amputated Green's function of $O^{\rm cont,\,ren}_i$ take on its
free field value for all spin and color indices on these quark
fields.  This defines the RI-scheme.  Its relationship to other
renormalization schemes requires only continuum perturbation theory,
which is better behaved than lattice perturbation theory at the low
scales ($\mu\sim 2$ GeV) used in present calculations.

The success of this method requires two important conditions to be
satisfied.
\begin{enumerate}
\item  A suitable ``window'' of momenta must exist.  The window must
  include momenta which are large enough to make non-perturbative (condensate)
  effects small.  It must also include momenta which are small enough to avoid
  artifacts due to finite lattice spacing.  Such a window was seen for quark
  bilinears in \cite{Blum:2001sr}.
\item
  Since the method of non-perturbative renormalization must eventually make a
  connection with continuum perturbation theory, our approach which uses
  Landau gauge is potentially vulnerable to the presence of Gribov gauge
  copies.  Such multiple gauge copies, present in Landau gauge lattice
  simulations, invalidate a comparison of gauge-variant quantities with
  perturbation theory, even when our calculations are performed at
  increasingly weak coupling. For the success of our method the effects of
  Gribov copies must therefore be small.
\end{enumerate}
In principle, the Gribov copy problem can be avoided by a more complete gauge
fixing procedure.  For example, we could begin with a gauge transformation to
a completely fixed axial gauge and then follow with the usual Landau gauge
fixing.  Such a procedure would guarantee that in the weak coupling, small
volume regime, a comparison with continuum perturbation theory would be accurate.
While we have not implemented this more sophisticated gauge choice for the NPR
calculations described here, we have made a non-trivial test.  We have carried
out a companion calculation for both $8^4$ and $16^4$ lattices of the
renormalization factors for both the dimension-three quark bilinear operators and
the single four-quark operator that enters the calculation of $B_K$ and found
no meaningful difference between our usual Landau gauge fixing determination
of the renormalization factors and the same determination using the more
elaborate two-step procedure described above\cite{Yuri:thesis}.  Thus, we
believe that the presence of Gribov copies is not a cause of difficulty for
the work presented here.

With this overview of NPR, we now turn to the specific issues and
conventions we use in the application of this technique to $\Delta S =
1$ operators.  We first consider the type of quark contractions that
can occur in Eq.\ \ref{eq:green_schematic} and see that there are two
types.  The first has each quark field in the operator contracted with
an external quark field, which we will call tree contractions in this
section, and the second, which we call eye contractions, have quark
propagators that begin and end on the operator.  This second class of
contractions are both theoretically and numerically challenging.  They
are theoretically challenging because it is through these diagrams that
the mixing with lower dimensional operators occurs.  They are
numerically challenging because they involve the evaluation of a
spectator quark propagator $S(p,q)$ with $ p \ne q$.  These
numerical issues will be discussed later, after the theoretical issues
are outlined.

In the RI scheme, the standard condition for determining
$c_k^j(\mu)$ is the requirement that the renormalized four-quark
operator vanish when evaluated in a Green's function with
two external quark fields.  In particular
\begin{equation}
  G^{(2)}(p,p) = \langle s_\alpha^i(p) \, O_m \, 
  \overline{d}_\beta^j(p) \rangle
  \label{eq:green_schematic_two_quark}
\end{equation}
should be zero.  As such, it is convenient 
when calculating $Z_{ij}(\mu)$ to use a
two step process where first a subtracted operator is defined by
evaluating the $c^j_k (\mu)$ in Eq.\ \ref{eq:general_mixing} through
\begin{equation}
  O^{\rm sub}_i = O_{i}^{\rm latt} + \sum_k c^i_{k} B_k \, .
  \label{eq:npr_sub}
\end{equation}
The second step consists of evaluating the four-quark Green's
function $G^{(4)}_{\rm sub}$ for the subtracted operator
using the external quark fields in Eq.\ \ref{eq:green_schematic}.
We now discuss which quark bilinears we will subtract.

A full subtraction of all the bilinear operators that could potentially
mix with the four-quark operators in question would be challenging and
prone to numerical error due to their large number. However, in the
context of the current study our accuracy is limited by the
existing one loop perturbative calculations of the matching
coefficients between the $RI$ and $\overline{MS}$ schemes and the
current Wilson coefficients, for which the finite terms are also known
only to one loop accuracy.  Therefore it is not necessary to subtract
operators that affect the renormalization factors at order
$g^4$ and above in perturbation theory, provided we have no {\em a
priori} reason to expect them to give anomalously large contributions.

Consequently, we neglect the subtraction of any bilinear operator
that is not power divergent and which mixes with the the four-quark
operators at order $g^2$ and above. The explanation for this is
straightforward.  Consider the Green's function of a generic subtracted
operator $G^{(4)}_{\rm sub}(p_1, p_1, p_2, p_2)$, evaluated in the free
case.  The bilinear operator will give no contribution to this Green's
function, due to the choice of momenta.  For interacting theories,
gluon exchange can transfer momenta, allowing a non-zero contribution
of the bilinear operator to this Green's function.
Such effects occur at order $g$.  If the lowest order contribution of
$c^j_k(\mu)$ begins at $g^2$ the total contribution will be of higher
order and may be neglected.

This counting is clearly not relevant for the bilinear operators of
dimension below six as the needed subtraction coefficients may be power
divergent as the lattice spacing tends to zero. This means we must
always consider the operator given in Eq.\ \ref{eq:dsl}, and away from
the massless limit it may be useful to subtract the operator given in
\ref{npr:sdx}, the subtraction coefficient of which has the leading
behavior $m/a^2$.  Now we have to consider various dimension six
operators.  At ${\cal O}(g^0)$ there are no dimension six bilinear operators
that mix. At one-loop, here ${\cal O}(g)$, there is a single operator 
that can mix\cite{Buras:1993tc}:
\begin{equation}
\bar{s}  \gamma_\nu d \, D_\mu F_{\mu\nu} \, .
\label{eq:sddd}
\end{equation}
To be consistent we should subtract this operator.  However, as we will
argue later, the numerical effect of neglecting this subtraction is
small.  At two loops additional gauge invariant operators which vanish
by virtue of the equations of motion and possible gauge non-invariant
operators must also be included \cite{Buras:1993tc}.  However, as
explained earlier, we can consistently ignore such order $g^4$ effects
in the present calculation

We will therefore consider the subtraction of only two bilinear
operators
\bea 
B_1 & \equiv & \overline{s} d \\ \nonumber
B_2 & \equiv & \overline{s} \left( -\lvec{\slash{D}} +\rvec{\slash{D}}
  + m_s + m_d \right) d \\ \nonumber
    &=& \overline{s} \left( -\lvec{\slash{D}} + m_s \right)
  d + \overline{s} \left( \rvec{\slash{D}} + m_d \right) d \, .  
\eea 
$B_2$ is a modification of Eq.\ \ref{eq:dsl} with additional mass
dependent terms added such that the operator vanishes on-shell both in
and out of the chiral limit, and $B_1$ is the operator in
Eq.\ \ref{npr:sdx} with $X^{(1)} = 1$. The two subtraction
coefficients, $c^i_1$ and $c^i_2$, should have leading behavior
\bea
c^{i}_1 &\propto& \frac{m_s+m_d}{a^2} + \cdots 
\label{eq:c1}
\\
c^{i}_2 &\propto& \frac{1}{a^2} + \cdots 
\label{eq:c2}
\eea
As mentioned previously, we subtract these operators
by requiring that Green's functions for the subtracted
four-quark operators $O_i^{\rm sub}$ vanish between external
quarks states with flavor structure $s \overline{d}$. 
To determine both coefficients we need to impose 
two linearly independent conditions which we choose as 
\bea
{\rm Tr} \left[
\langle s(p) \, O^{\rm sub}_i \, \overline{d}(p) \rangle_{\rm amp}
\right] &=& 0 \, ,
\label{eq:ssub}
\\
{\rm Tr} \left[ i\slash{p}
\langle s(p) \, O^{\rm sub }_i \, \overline{d}(p) \rangle_{\rm amp}
\right]
&=&0 \, ,
\label{eq:dsub}
\eea
where ``${\rm amp}$'' denotes the amputated vertex.  The momentum
$p$ where the condition is enforced is explained in detail below.

In QCD, the operators $O_i$ mix under renormalization. To account for
this mixing we define a set of suitable color, spin, and flavor
projectors which we use to implement our renormalization conditions and
thus yield the $Z_{ij}$ in the RI scheme.  First, to distinguish the
flavor structure of the operators we define a set of external quark
fields, denoted by $E^{j}_{\alpha \beta \gamma \delta}$, as
\begin{equation}
E^{j}_{\alpha \beta \gamma \delta} = 
f^{j,abcd}
q_\alpha^a (p_1) \overline{q}_\beta^b(p_2) q_\gamma^c(p_1)
\overline{q}^d_\delta(p_2)
\label{eq:external_states}
\end{equation}
where $q$ is a generic quark field; the subscripts representing spin and color
and the superscripts representing the flavor. Here $f^{j,abcd}$ is a
set of constants defining the flavor 
structure of the {\it j}th set of external quark fields. 
We then construct the  amputated Green's functions
of $O^{\mathrm{sub}}_i$ between these external quark fields
\begin{equation}
\Lambda^{i,j}_{\alpha \beta \gamma \delta}
=
\langle
\,
\ O^{\rm sub}_i \;
E^{j}_{\alpha \beta \gamma \delta}
\, \rangle_{\rm amp} 
\label{eq:lambda}
\end{equation}
and trace the result with a chosen set of projectors, $P^j$,
\begin{equation}
  P^j \left\{ \Lambda^{i,j} \right\}
  \equiv \Gamma^{j}_{\alpha \beta \gamma \delta}
\,
\Lambda^{i,j}_{ \beta \alpha \delta \gamma}
\label{eq:proj}
\end{equation}
where $\Gamma^{j}$ is a rank-four tensor in spin and color space
that defines the projector, and there is no sum over $j$ in the above
equation. The renormalization factors $Z_{ki}$ are then fixed by
requiring that, for renormalized operators with a specific
choice of the momenta appearing in Eq.\ \ref{eq:external_states},
this set of quantities be equal to its free case value,
\begin{equation}
\frac{1}{Z_q^2}Z^{ki} P^j \left\{ \Lambda^{i,j} \right\}
= F^{kj}
\end{equation}
Here $F^{ij}$ is the free case limit of 
$P^j \left\{ \Lambda^{i,j} \right\}$ and $Z_q^{1/2}$ is the
quark wave function renormalization factor from \cite{Blum:2001sr}.
This may be conveniently be written in matrix form
\begin{equation}
\frac{1}{Z_q^2} Z = F M^{-1}
\label{eq:matrixZ}
\end{equation}
with $M^{ij} \equiv P^j \left\{ \Lambda^{i,j} \right\}$.
$Z$, $M$ and $F$ are all real $N\times N$ matrices, where $N$ is
the number of operators in our basis.

As long as the external states and projectors are chosen such that a
linearly independent set of conditions is applied ($F$ is invertible),
this completely and uniquely specifies the renormalization coefficients
for any such choice of the flavor structure of the external quark
fields $f^{j,abcd}$ and projectors $\Gamma^j_{\alpha \beta \gamma
\delta}$.


\subsection{Numerical Implementation}

We now move to a discussion of the numerics of our calculation. All the
results presented were measured on quenched gauge configurations
generated using the Wilson gauge action for a lattice of size
$16^3\times32$ with $\beta=6.0$. These configurations were then fixed
into Landau gauge (see \cite{Blum:2001sr}). On these Landau gauge-fixed
configurations we then calculated the needed quark propagators using
the domain wall fermion action with $L_s=16$.

To construct the quark contractions that arise in Eqs.\ \ref{eq:ssub},
\ref{eq:dsub} and \ref{eq:lambda} three distinct quark propagators are
needed for a fixed mass.
\begin{enumerate}
\item
  The propagator from the position of the operator to a general site
  $x$ on the lattice, transformed into momentum space on $x$.
\item
  The propagator from the position of the operator back to that
  position.
\item
  A spectator propagator transformed into momentum space on both source
  and sink indices with distinct momenta, $S(p,q)$ with $p \neq q$.
\end{enumerate}
The first two of these require a single inversion of the Dirac operator
for each mass. However to calculate the last of these we inverted the
Dirac operator using a fixed momentum source, which costs an inversion
for every momenta, $q$, needed. For this reason we calculate this
propagator for only four fixed momenta and a limited range of masses.

As we are working on a finite lattice with periodic boundary conditions,
the possible values of the momenta for a given direction $i \in \left\{
x,y,z,t \right\}$ are
\begin{equation}
ap_i = \frac{2\pi n_i}{L_i} \, ,
\end{equation} 
where $L_i$ is the lattice size in direction $i$,
\begin{equation}
L_x=L_y=L_z=16 \, , \, L_t = 32
\end{equation}
and
\begin{equation}
-\frac{L_i}{2} < n_i \leq \frac{L_i}{2} 
\end{equation}


\subsubsection{Bilinear Operator Subtractions}

To evaluate the subtraction coefficients $c^i_1$ and $c^i_2$ the
spectator propagator is not needed, a single momentum space propagator
from a point being sufficient. As this is the case we have used a
separate data set from that used for the full four-quark Z-factor
calculation. We used an ensemble of 50 gauge configurations for which
we calculated the quark propagators for bare quark masses
$m_f=0.02,0.03,0.04$ and $0.05$.

From Eqs.\ \ref{eq:ssub} and
\ref{eq:dsub} we obtain
\bea
-c^i_1
- 
2 c^i_2 
\frac{{\mathrm Tr}\left[ S^{-1}(p)\right]}
{
 {\mathrm Tr} 
  \left[ \langle s(p) \left( \overline{s} d  \right) \overline{d}(p)
  \rangle_{\rm amp}
  \right]        
}
&=&
\frac{
 {\mathrm Tr}\left[
\langle s(p) \, O_{i}  \overline{d}(p) \rangle_{\rm amp}
\, \right]
}{
 {\mathrm Tr} 
\left[ \, \langle s(p) \left(\overline{s} d \right) \overline{d}(p)
  \rangle_{\rm amp} \right]
}
= \kappa^i
\label{eq:c1sub}
\eea
\bea
-c^i_1 
\frac
{
  {\mathrm Tr}
\left[
i \slash{p} 
\langle s(p) \left(\overline{s} d \right) \overline{d}(p) \rangle_{\rm amp}
\right]
}
{
2 {\mathrm Tr}
\left[ i \slash{p} S^{-1}(p) \right]
}
-c^i_2 
&=&
\frac{ 
{\mathrm Tr}
\left[i \slash{p}\,\langle s(p) \, O_{i} \overline{d}(p)  \rangle_{\rm amp}\, \right]
}{ 2 
\mathrm{Tr}
\left[ i \slash{p} S^{-1}(p)  \right]}
= \lambda^i
\label{eq:c2sub}
\eea
where we have explicitly taken the degenerate limit, $m_s=m_d=m_f$.
These two relations may be simplified by noting that
\begin{equation}
\frac{{\mathrm Tr}\left[ S^{-1}(p)\right]}
{
{\mathrm Tr} 
\left[ \langle s(p) \left(\overline{s}d\right) \overline{d}(p)  \rangle_{\rm amp}
\right]        
} = m_f
\end{equation}
to within the statistical errors given in \cite{Blum:2001sr}.  It was
also found in \cite{Blum:2001sr} that ${\mathrm Tr} \left[ i \slash{p}
\langle s(p) \left(\overline{s} d \right)\, \overline{d}(p)
\rangle_{\rm amp} \right]=0$ up to ${\cal O}(a^2)$ contributions.  As this is
the case, we extract $c_2^i$ from Eq.\ \ref{eq:c2sub} in the chiral
limit. We then substitute this value into Eq.\ \ref{eq:c1sub} to give
$c_1^i$.

It is instructive to investigate the mass dependence of $c_1^i$, since
$c_1^i$ should vanish in the chiral limit.  In addition, as shown in
Eqs.\ \ref{eq:c1} and \ref{eq:c2}, the dominant $1/a^2$ divergences in
$c_1^i$ and $c_2^i$ are momentum independent, although subleading terms
are expected to depend on $\ln(pa)$.  Thus, the dominant contributions
to $c_1^i$ and $c_2^i$ as determined through Eqs.\ \ref{eq:c1sub} and
\ref{eq:c2sub} are momentum independent as long as we are in the
required ``window'';  however, experience has shown that at the momenta
accessible for the lattice parameters we are using, discretization
errors may be important.  To check for the above features of $c_1^i$
and $c_2^i$, we first rewrite Eqs.\ \ref{eq:c1sub} and \ref{eq:c2sub}
as
\bea
\kappa^i(p) &=&  2 m_f  \, A_{\kappa^i}(p) + B_{\kappa^i}(p) \, ,
  \label{eq:npr_kappa} \\
\lambda^i(p) &=& 2 m_f \, A_{\lambda^i}(p) + B_{\lambda^i}(p) \, .
  \label{eq:npr_lambda}
\eea
where we have used notation that explicitly allows $\kappa^i$,
$\lambda^i$, $A_{\kappa^i}$, $B_{\kappa^i}$, $A_{\lambda^i}$ and
$B_{\lambda^i}$ to depend on the momentum.  (The parameters $c_1^i$ and
$c_2^i$ are given in terms of $A_{\kappa^i}$, $B_{\kappa^i}$,
$A_{\lambda^i}$ and $B_{\lambda^i}$.)  Thus,  for each momentum,
we fit our data for $\kappa^i$ and $\lambda^i$ to a linear function
of $m_f$.

Having determined $A_{\kappa^i}(p)$, $B_{\kappa^i}(p)$,
$A_{\lambda^i}(p)$ and $B_{\lambda^i}(p)$, we can now remove the
dominant effects of discretization errors.  For the momenta we are
using, these enter as ${\cal O}\left((ap)^2\right)$ effects, which we
can determine by fitting $A_{\kappa^i}(p)$, $B_{\kappa^i}(p)$,
$A_{\lambda^i}(p)$ and $B_{\lambda^i}(p)$ to the form $A_{\kappa^i}(p)
= A_{\kappa^i}^{(0)} + A_{\kappa^i}^{(2)} (ap)^2$, etc.  Momenta are
used in the fits such that $0.8 < (ap)^2 < 2.0$.  Tables \ref{tab:msx}
and \ref{tab:mix} summarize the results of the fits for $A_{\kappa^i}$
and $B_{\kappa^i}$, respectively, while Tables \ref{tab:msy} and
\ref{tab:miy} give the same information for the fits to $A_{\lambda^i}$
and $B_{\lambda^i}$.  All fits use 50 configurations, jackknifed every
one.

Tables \ref{tab:msx} and \ref{tab:mix} show that $A_{\kappa^i}^{(2)}$
and $B_{\kappa^i}^{(2)}$ are generally zero within our statistical
errors, so discretization errors for $A_{\kappa^i}(p)$ and
$B_{\kappa^i}(p)$ are not resolved.  In addition, the statistical
errors on $A_{\kappa^i}^{(2)}$ and $B_{\kappa^i}^{(2)}$ for $i = 5,6,7$
and 8 are small compared to $A_{\kappa^i}^{(0)}$ and
$B_{\kappa^i}^{(0)}$, so any discretization errors are a small effect
for $A_{\kappa^i}(p)$ and $B_{\kappa^i}(p)$.  For $A_{\lambda^i}(p)$
and $B_{\lambda^i}(p)$, the discretization errors are statistically
resolved and, for $(ap)^2 = 2$, alter $A_{\lambda^i}^{(0)}$ and
$B_{\lambda^i}^{(0)}$, by $\sim 30$\%.  Given these fits to $(ap)^2$,
we can determine $c_1^i$ and $c_2^i$ from $c_2^i =
-B_{\lambda^i}^{(0)}$ and from $c_1^i = -2 m_f ( c_2^i +
A_{\kappa^i}^{(0)}) - B_{\kappa^i}^{0}$.  Note that the combination $2
m_f A_{\kappa^i}^{(0)} + B_{\kappa^i}^{0}$ entering $c_1^i$ is
$\kappa^i(0)$, {\em i.e.} the value of Eq.\ \ref{eq:npr_kappa} for $p
= 0$.  In our final determination of $c_1^i$ this combination is found
directly from fitting $\kappa^i(p)$ to $\kappa^{i,(0)} + \kappa^{i,(2)}
(ap)^2$ and using the value of $\kappa^{i,(0)}$.  This reduces any
possible numerical imprecision from fitting both $m_f$ dependence and
$p$ dependence separately.  Our data is well fit by the various
relations given above, which shows that the data has mass dependence
predicted for $c_1^i$ and $c_2^i$ in Eqs.\ \ref{eq:c1} and \ref{eq:c2}.

Since $c_2^i$ shows no visible mass dependence, we have chosen to use
its value in the chiral limit, as just described, for the final
computation of the $Z$ factors at non-zero quark mass.  (If mass
dependence were visible in $c_2^i$, our entire approach would be
suspect.) On the other hand, since $c_1^i$ is strongly mass dependent,
we extract it at non-zero mass from Eq.\ \ref{eq:c1sub}.  The values of
$c_1^i$ used for the $m_f=0.04$ subtractions are given in Table
\ref{tab:c1}. The quoted error is only a statistical error, which
comes from the jackknife procedure.


\subsubsection{Four-Quark Operator Renormalization}

For the extraction of the four-quark renormalization factors
we have 100 configurations with two values of the quark
mass, $m_f=0.02$ and $m_f=0.04$ and a further 390 configurations for
the second mass value. The extra configurations for the heavier
mass were obtained to gain increased statistics at a reasonable cost
after the subtracted renormalization factors had been found to be
mass independent to a good degree of accuracy on the first 100
configurations.

The renormalization condition we apply is such that
all the momentum scales in the problem should be
the same, {\it i.e.},
\begin{equation}
p_1^2 = p_2^2 = (p_1-p_2)^2 = \mu^2
\end{equation} 
The values of $n_i$ corresponding to the momenta that we used
are given in Table\ \ref{npr:tab:mom}. The results are
averaged over equivalent orientations, and denoted by
the corresponding Euclidean squared momenta $(ap)^2$.

The operators below the charm threshold, $Q_{i}$ ($i=1,\cdots,10$), are
not linearly independent.  As can be seen from Eq.\ \ref{eq:matrixZ} the method
we use to calculate $Z$ requires the inverse of $M$, which is
singular in this case.  Therefore, we actually calculate $Z$ from Eq.\
\ref{eq:matrixZ} for a linearly independent subset of these operators. 

This subset was defined by eliminating $Q_4$, $Q_9$ and $Q_{10}$, 
through the identities
\bea 
\nonumber
Q_4    &=& - Q_1 + Q_2 + Q_3 \, ,\\
\nonumber 
Q_9    &=& \frac{3}{2} Q_1 - \frac{1}{2} Q_3 \, ,\\ 
Q_{10} &=& \frac{1}{2} Q_1 + Q_2 - \frac{1}{2} Q_3 \, .
\label{eq:deps}
\eea 
Since conventionally the $\Delta S=1$ Hamiltonian is given in the 
dependent basis, after calculating the $7\times7$ matrix $Z$ in the
reduced basis, we reconstructed a $10\times10$ matrix
$\hat{Z}$ in the full basis using the relations
\bea
\hat{Z}_{ij} = Z_{ij} &,&\ i,j \in \{ 1,2,3,5,6,7,8 \} \, ,\\
\hat{Z}_{ij} =  0     &,&\ i \in \{ 1,2,3,4,5,6,7,8,9,10 \} \, , \\
&& \ j \in \{ 4,9,10 \} \\
\hat{Z}_{ij} =  T^{i}_{\, \, k} Z_{kj}
&,&\
i \in \{ 4,9,10 \} \\ && \ j \in \{ 1,2,3,5,6,7,8 \} \, ,
\eea
where $T^{i}_{\, \, k}$ encodes Eq.\ \ref{eq:deps}
as $Q^{i} = T^{i}_{\, \, k} Q_{k}$ for $k=4,9$ and $10$. 

As enumerated in Appendix \ref{sec:four_quark_irreps}, the four-quark
operators we are considering are composed of elements transforming
according to the (8,1), (27,1) and (8,8) representations of $\sutlr$.
There are four distinct (8,1) representations,
two distinct (8,8) representations and a single (27,1).  Since we
renormalize in the massless limit, our $Z$ factors should not have
mixings between the (8,1), (27,1) and (8,8) representations, but there
can be mixings between the four distinct (8,1) representations and also
between the two (8,8) representations.  In particular, the calculated
values for $Z$ should be block diagonal in an operator basis where each
operator is purely an (8,1), (27,1) or (8,8).  This is already the case
for $Q_5$ and $Q_6$, which are in distinct $(8,1)$ representations, and
$Q_7$ and $Q_8$, which are in distinct $(8,8)$ representations.
However, $Q_1$, $Q_2$ and $Q_3$ are mixtures of two $(8,1)$
representations and a single $(27,1)$.  To check the chiral structure
of the renormalization factors it is convenient to consider a basis
with operators $Q'_i$ in distinct $\sutlr$ representations, given by
\bea 
Q_1' &=& 3 Q_1 + 2 Q_2 - Q_3 \, ,\\
Q_2' &=& \frac{1}{5} \left( 2 Q_1 - 2 Q_2 + Q_3 \right) \, ,\\ 
Q_3' &=& \frac{1}{5} \left( -3 Q_1 + 3 Q_2 + Q_3 \right) \, ,\\ 
Q_i' &=& Q_i \ ;\ i \in \left\{5,6,7,8 \right\} \, .
\eea 
In this new basis $Q_1'$ is in the $(27,1)$ representation and $Q_2'$ and
$Q_3'$ are in the $(8,1)$ representation. To display the chiral symmetry
properties we tabulate elements of $MF^{-1}$ in this basis in Tables \ref{mf1}
and \ref{mf2}. We tabulate $MF^{-1}$ rather than $FM^{-1}$ because the former
is linear in the quark contractions, so individual contributions are more
easily distinguished.  In terms of the elements of $MF^{-1}$ in the $Q'$ basis,
the restriction that operators in different multiplets cannot mix may be
written
\begin{equation}
\begin{array}{rcl}
\left(MF^{-1}\right)_{1i} = 
\left(MF^{-1}\right)_{i1} = 0 & ; & 
i \in \left\{ 2,3,5,6,7,8 \right\}
\\
\left(MF^{-1}\right)_{7i} = 
\left(MF^{-1}\right)_{i7} = 0
& ; &
i \in \left\{ 2,3,5,6 \right\}
\\
\left(MF^{-1}\right)_{8i} = 
\left(MF^{-1}\right)_{i8} = 0
& ; &
i \in \left\{ 2,3,5,6 \right\} \, ,
\end{array}
\end{equation}
As can be seen from Tables\ \ref{mf1} and \ref{mf2}, these relations
are satisfied to a good degree of accuracy by our data. As such, for
the calculation of the final renormalization factors we will set these
elements to be exactly zero in $MF^{-1}$, before inverting to get $Z$
to reduce the statistical error on the final result.

The final values for $\hat{Z}_{ij}/Z_q^2$ are given in Tables\
\ref{tab:zfull1} to \ref{tab:nsub2} where $(ap)^2$ is the square of the
Euclidean momenta for the external legs and $(ap_{\rm diff})^2$ is the
transferred momenta.  To display the numerical importance of the
various components of the calculation, three sets of renormalization
coefficients are given: (1) The full renormalization coefficients
(Tables \ref{tab:zfull1} and \ref{tab:zfull2}), (2) those calculated
without the eye-diagram contributions (Tables \ref{tab:nloop1} and
\ref{tab:nloop2} ), and (3) those calculated with the eye-diagrams but
without the subtraction of the lower dimensional operators (Tables
\ref{tab:nsub1} and \ref{tab:nsub2}). All values given are with
$m_f=0.04$ for $490$ configurations. The quoted error is statistical,
and was calculated by jackknifing the data in blocks of 10. To obtain
$\hat{Z}$ we use $Z_q=0.808(3)(15)$ \cite{Blum:2001sr} at $2 {\rm
GeV}$. In principle $Z_q$ should be run to the exact scale at which we
are working, however this is a very small effect \cite{Blum:2001sr}.


\subsection{Discussion}

Having completed the renormalization of our four-quark operators, we
now turn to a discussion of the size of various contributions, the
effects of discretization errors and the role of the dimension six
bilinear operators which were not included in our present work.
Turning first to the size of effects from our calculation, the
numerical results show that the eye-diagrams, even though they have a
$1/a^2$ dependence in the continuum limit, are small compared to the
other graphs.  This is in stark contrast to the matrix element case,
where as we will see in Section \ref{sec:i_1over2_me}, such divergent
graphs overshadow the physical signal by approximately two orders of
magnitude, and their subtraction is an extremely delicate operation
that must be performed with great precision.

In the matrix element study, when considering dimensionful quantities,
an order of magnitude estimate of the size of a physical signal may be
made by taking $\Lambda_{\mathrm{QCD}}$ to the relevant number of
powers. If the quantity is divergent however, the dimensions may also
be made up with inverse powers of the lattice spacing. As $a^{-1}
\approx 10 \times \Lambda_{\mathrm{QCD}}$ at the lattice spacing we are
working, the physical signal may be much smaller than the subtraction.
For the renormalization factors, however, we are studying high energy
quantities, so the relevant scale is $a^{-1} \approx \mu$.  Thus,
eye-graphs involving powers of $a^{-1}$ have a much smaller effect on
the renormalization factors than corresponding eye-graphs have on
physical hadronic matrix elements.  In addition, the eye-graphs are
suppressed as they are zero in the free case, with the non-zero signal
being due to gauge interactions.  The numerical evidence in Tables
\ref{tab:zfull1} to \ref{tab:nsub2} shows that that inclusion of the
divergent eye graphs affects the renormalization factors on the order
of a few percent.

As we are studying high energy quantities, we must also worry about the
effect of discretization errors. If the momenta, although large, still
allow lattice artifacts to be treated as small corrections, it is
possible to describe them as ${\cal O}(ap^2)$ and ${\cal O}(ap_{\rm diff}^2)$ terms.
Then, with a sufficient number of different momentum configurations,
they can be isolated and removed.  A naive estimate of the scale at
which these effects become large is $p \approx 1/a$. This is only a
rough estimate, however, and previous studies have shown that for the
lattice parameters we are using, momenta as large as $(ap)^2 = 2$
produce discretization errors of a few percent \cite{Blum:2001sr}.  As
such in this preliminary study, for which we have only a few momenta
configurations, all of which have a momenta scale of $\approx 1/a$ we
will ignore these effects.

Next we consider the effect of neglecting the subtraction of the
dimension six quark bilinear operators.  These subtractions are needed
for two reasons:
\begin{enumerate}
\item
  Discretization errors in our expressions for $B_1$ and $B_2$ are of
  ${\cal O}(a^2)$ and may be written in terms of the dimension six quark
  bilinear operators we are considering. When the Green's functions of
  these operators are multiplied by the subtraction coefficients
  $c_1^i$ and $c_2^i$, which have leading behavior $1/a^2$, this can
  lead to errors in the final results that are of ${\cal O}(1)$ in the lattice
  spacing.
\item
  The operator in Eq.\ \ref{eq:sddd} mixes at ${\cal O}(g)$ in perturbation
  theory and so should be subtracted to the order at which we are
  working.  Such a subtraction was not attempted in this first
  work, since it involves explicit external gluons.
\end{enumerate} 

Expanding on the issues raised in case one, we consider a simplified
situation involving a single dimension six operator in the continuum,
$ B^{\rm cont}_3$.  Then we can write, for example,
\begin{equation}
  B^{\rm lat}_2 = B^{\rm cont}_2  + O(a^2) B^{\rm cont}_3
  \label{eq:dim_six}
\end{equation}
When $B^{\rm lat}_2$ is multiplied by $c^j_2(\mu)$, which behaves as
$1/a^2$, then $ B^{\rm cont}_3$ is multiplied by a coefficient of
${\cal O}(1)$ in the lattice spacing.  As we have just discussed,
discretization effects are small, and so is the contribution of $
O(a^2) B^{\rm cont}_3$ to Eq.\ \ref{eq:dim_six}.  In addition, as we
have noted, the contribution of $c^j_k(\mu) B^{\rm lat}_k$ to the
four-quark renormalization factors is also small.  Hence we expect any
effects due to these discretization errors to be negligible.

A similar argument may be put forward for case two.  While this
operator should be subtracted at the order in perturbation theory in
which we are working, the subtraction coefficient associated with this
operator will be only logarithmically divergent in the lattice spacing,
rather than power divergent. Our data from the extraction of the
subtraction coefficients supports the numerical dominance of $B_1$ and
$B_2$ (Eqs.\ \ref{eq:c1} and \ref{eq:c2}) very well.  This suggests the
power divergent terms are much more important, for this set of lattice
parameters, than the logarithmically divergent terms that would
multiply the dimension six operators.  Again this indicates that we are
correctly treating the dominant part of the subtractions, which
themselves amount to only a small correction to the final
renormalization factors.

\fi


\section{Lattice calculation of $K\to\pi$ and $K\to|0\rangle$ Matrix
Elements}
\label{sec:lat_me}

\ifnum\theLatMe=1
%
%

In this section we present the lattice calculation of the $K\to\pi$ and
$K\to |0\rangle$ matrix elements. In the first two sub-sections the
lattice method and basic contractions are briefly described.  Results
for $K\to\pi$ and $K\to |0 \rangle$ matrix elements obtained by using
this methodology, which form the basis of our calculation, are given in
the last sub-section.  We continue to label pseudoscalar states with
$K$ and $\pi$ to make the discussion clear, but the matrix elements
$\langle \pi^+ | Q_i | K^+ \rangle$ are calculated with {\it
degenerate} quarks and have $m_{\pi^+} = m_{K^+}$.  Since $K \to
|0\rangle $ matrix elements vanish in this limit, we use non-degenerate
quark propagators for this case.  It is useful to keep in mind that
when the quarks are degenerate, flavor is specified by the type of
quark contraction.

\subsection{Lattice method of matrix element calculation}

In order to obtain the desired matrix elements, we work in Euclidean
space-time and calculate correlation functions. For example, a typical
$K\to\pi$ correlation function is
\begin{eqnarray}
    G_{\pi\,O\,K}(t)
  &
    \equiv
  & \sum_{z,z^\prime}\,\frac{1}{V_s}\,\sum_{y}\sum_{x,x^\prime} \langle 
     [i\bar{d}(z^\prime,t_\pi)\gamma_5 u(z,t_\pi)] O(y,t)
     [i\bar{u}(x^\prime,t_K)\gamma_5 s(x,t_K)]\rangle
  \label{eq:corr_func}
\end{eqnarray}
where $\langle \rangle$ denotes an average over gauge field
configurations, $t_\pi > t > t_K$ with $t_\pi$ and $t_K$ fixed, $V_s$
is the three-dimensional spatial volume and the factors of $i$ make the
pseudoscalar correlator positive.  We employ Coulomb gauge fixed wall
sources which have significant overlap with the pseudoscalar ground
states and the spatial average over the operator time slice enhances
the statistical average.  For fixed values of $t_K$ and $t_\pi$, a
``plateau'' in $G_{\pi\,O\,K}(t)$ emerges when $t_\pi \gg t \gg t_K$ as
then the lowest energy states ($|\pi^+\rangle$ and $|K^+\rangle$)
dominate the correlation function.  The correlation function becomes
time independent since the meson masses are equal.  Up to source matrix
elements and kinematical factors, $G_{\pi\,O\,K}(t)$ then directly
yields the desired matrix element
\begin{eqnarray}
\lim_{t_\pi \gg t \gg t_K} G_{\pi\,O\,K}(t) &\to& 
             \frac{ \langle \pi^+|O|K^+\rangle}
              {N \, (2\,m_\pi\,V_s) \, (2\,m_K\,V_s)}\,
             e^{-m_K(t-t_K)}
             e^{-m_\pi(t_\pi-t)},
  \label{eq:matrix element}
\end{eqnarray}
which is easily seen by inserting two complete sets of relativistically
normalized states between the operator and each source.  The factor $N$
represents an unknown normalization factor introduced by the the wall
sources.

One way to remove the kinematic factors and the unknown normalization
of our wall sources is to divide by the pseudoscalar two-point
correlation function from each source.  For example, with the
wall-point (spatially extended source-local sink) two-point correlation
function
\begin{eqnarray}
    G_\pi(t)
  &
    \equiv
  &
    \frac{1}{V_s}\, \sum_{x} [i\bar{u}(x,t)\gamma_5 d(x,t)]
    \sum_{z,z^\prime} [i\bar{d}(z^\prime,t_\pi)\gamma_5 u(z,t_\pi)],
\end{eqnarray}
and similarly for $G_K(t)$, we can form a ratio of the desired matrix
element to known factors.
\begin{eqnarray}
    \lim_{t_\pi \gg t \gg t_K} \frac{G_{\pi\,O\,K}(t)}{G_\pi(t) G_K(t)}
  &
    =
  &
    \frac{\langle \pi^+| O|K^+\rangle}
    {\langle \pi^+| P_{\pi^-}|0\rangle \langle 0| P_{K^+} |K^+\rangle}
  \label{eq:ratio}
\end{eqnarray}
(We use $P_{K^-}(x) \equiv [i\bar{u} \gamma_5 s](x)$ and $P_{\pi^+}(x)
\equiv [i\bar{d} \gamma_5 u](x)$ as in Section
\ref{subsec:zero_mode_effects}.) We can also normalize
Eq.\ \ref{eq:matrix element} by pseudoscalar-axial vector correlators,
which changes the denominator in Eq.\ \ref{eq:ratio} to $ \langle \pi^+
|\bar u \gamma_0 \gamma_5 d | 0 \rangle \langle 0|\bar s \gamma_0
\gamma_5 u |K^+\rangle$.  The axial current matrix elements have the
normalization given in Eq.\ \ref{eq:axial_current_me}.  These axial
current matrix elements have been calculated using point-point
correlation functions in Ref.~\cite{Blum:2000kn} and can also be
extracted from a simultaneous fit to the wall-point and wall-wall
two-point functions calculated in the present study.  As discussed in
Section \ref{subsec:ward_id_sd}, zero mode effects are introduced
through $G_\pi(t)$ and $G_K(t)$ since such effects are seen in scalar
correlators at a separation $t$.

Another possibility is to divide the three-point function by a
different three-point function.  In particular
\begin{equation}
  \lim_{t_\pi \gg t \gg t_K} \frac{G_{\pi\,O\,K}(t)}
    {G_{\pi\,\bar s d\,K}(t)}
  =
    \frac{\langle \pi^+| O|K^+\rangle}
      {\langle \pi^+| \bar s d|K^+\rangle}
  = 
    \frac{2 m_f}{m_\pi^2} \langle \pi^+| O|K^+\rangle
  \label{eq:ratio:ktopi}
\end{equation}
where we have used the Ward-Takahashi identity Eq.\
\ref{eq:sbard_wi_no_zero}, neglecting zero mode effects, in the
last step.  Since as we have seen, zero modes have a noticeable
effect on this Ward-Takahashi identity, we do not divide by this
three-point function to extract $\langle \pi^+| O|K^+\rangle$.

Our preferred approach is to divide $G_{\pi\,O\,K}(t)$
by the wall-wall two-point function computed from the
correlator from $t_\pi$ to $t_K$,
\begin{eqnarray}
    G_{\rm ww}(t_\pi,t_K)
  &
    \equiv
  &
    \frac{1}{V_s}\,\sum_{x,x^\prime} \langle
    [i\bar{u}(x,t_\pi)\gamma_5 d(x^\prime,t_\pi)] \sum_{z,z^\prime}
    [i\bar{d}(z^\prime,t_K)\gamma_5 u(z,t_K)]\rangle\, .
\end{eqnarray}
Since we work with degenerate quarks, we have
\begin{equation}
  \lim_{t_K \gg t \gg t_\pi} 
  \frac{ G_{\pi\,O\,K}(t) }{ G_{\rm ww}(t_\pi,t_K)}
  =
  \frac{ \langle \pi ^+| O | K^+ \rangle}{2 \,m_\pi}
  \label{eq:lat_me_ww}
\end{equation}
where we determine $2\,m_\pi$ from a covariant fit to the wall-point
two-point function in the range $t=12-20$ for each quark mass.
As discussed in Section \ref{subsec:ward_id_sd} this normalization
minimizes the effects of zero modes.

We have tested the various methods described above for extracting $K
\to \pi$ matrix elements from three-point correlation functions and
find the results generally consistent, within errors.  We give results
for the last method since it is the simplest, requiring only the value
for $m_\pi$, does not rely on chiral perturbation theory and minimizes
zero mode effects.  In addition, we have used two types of wall sources
to create and destroy pseudoscalar mesons: the usual pseudoscalar
source $i\,\bar{q} \gamma_5 t_a q$ and an axial-vector source $\bar{q}
\gamma_0\gamma_5 t_a q$.  They give statistically equivalent results,
but the pseudoscalar source yields somewhat smaller errors; we will
always quote the former unless otherwise specified.

As mentioned earlier, for $K \to |0 \rangle$ matrix elements we extract
the needed power divergent coefficient from the ratio
\begin{eqnarray}
       \lim_{t_\pi >> t >> t_K} \frac{G_{O\,K}(t)}{G_{K}(t)}&=&
       \frac{\langle 0| O|K^0\rangle}
       {\langle 0| \bar s\gamma_5 d|K^0\rangle},
  \label{eq:ratio:ktovac}
\end{eqnarray}
where
\begin{eqnarray}
        G_{O\,K}(t) &=& \frac{1}{V_s}\sum_{y}\sum_{x,x^\prime}
        \langle 
        O(y,t)
        [i\bar{d}(x^\prime,t_K)\gamma_5 s(x,t_K)]\rangle.
\label{eq:ktovac corr}
\end{eqnarray}
The ratio in Eq.\ \ref{eq:ratio:ktovac} is just the parity-odd analogue
of Eq.\ \ref{eq:ratio:ktopi} if we recognize the denominator of each
ratio as the parity even or odd component, respectively, of the
subtraction operator $\Theta^{(3,\bar{3})}$ discussed in
Sections~\ref{sec:cont_chiral_pert} and \ref{sec:i_1over2_me}.
However, in Eq.\ \ref{eq:ratio:ktovac} the ratio immediately gives
the needed ${\cal O}(1/a^2)$ coefficient without relying on the
Ward-Takahashi identity.

\subsection{Contractions}

To compute the $K\to \pi$ correlation function in
Eq.\ \ref{eq:corr_func}, the quark fields are Wick contracted into
propagators which are calculated by inverting the five dimensional
domain wall fermion Dirac matrix on an external source and projecting
to four dimensions in the usual way (see \cite{Blum:2000kn}).  Two
types of diagrams emerge: figure eight diagrams as shown in Figure
\ref{fig:lattice_contractions}a and eye diagrams as shown in Figure
\ref{fig:lattice_contractions}b.  The $K\to |0 \rangle$ matrix elements
are computed in an analogous fashion and require the annihilation
contraction given in Figure \ref{fig:lattice_contractions}c.  The
matrix element of $\langle \pi^+ |\bar{s}\,d |K^+ \rangle$, which is
needed to subtract the power divergent contribution, is shown in Figure
\ref{fig:lattice_contractions}d.

The figure eight diagrams are constructed from quark propagators from
the wall sources at $t_K$ and $t_\pi$ to a point $(\vec{x},t)$.
Propagators from $(\vec{x},t)$ to $t_\pi$ and $t_K$ are obtained from
the hermiticity of the quark propagators, $G(x,y)=\gamma_5
G^\dagger(y,x)\gamma_5$.   After the appropriate propagators are
combined at a point $(\vec{x}, t)$ where the weak operator is inserted,
an average over $\vec{x}$ is done.

For the eye diagrams (Figure \ref{fig:lattice_contractions}b) and $K\to
|0 \rangle$ diagrams (Figure \ref{fig:lattice_contractions}c) we also
need an additional propagator from $(\vec{x},t)$ to itself, since two
fields in the weak operator are contracted together.  To efficiently
calculate this propagator we use a common technique in lattice
simulations, calculating a propagator from a complex Gaussian random
wall source.  Since we only want the loop propagator for the weak
operator in meson states, we choose the random source to be non-zero on
time slices with $14 \le t \le 17$.  When the propagators are assembled
to form a particular contraction, we include the complex conjugate of
the random source at each sink point ($\vec{x},t$) and average over
random sources and gauge configurations to project out the desired
diagonal contribution.  This allows the spatial average of the
correlation function over the operator time slice for any number of
time slices to be done with only one (or a few) quark propagator
inversion(s) on each gauge field configuration.  We have chosen to
calculate two independent, random source quark propagators on each
configuration, corresponding to 1/3 of the computer time spent
calculating propagators.  The same random sources are used for all
quark masses on a given configuration.  The last part of the eye
diagrams is the spectator quark propagator from $t_K$ to $t_\pi$. This
is constructed using the wall source propagator from $t_K$ and using a
wall sink at $t_\pi$ where the spatial coordinates of the propagator
are summed over before inserting the propagator into the contraction.

\subsection{Lattice values for $K\to\pi$ and $K\to 0$ matrix elements}

We first demonstrate that for $t_\pi = 27$, $t = 14$ to 17 and $t_K =
5$ the ratio $2 m_\pi G_{\pi\,O\,K}(t_\pi, t, t_K) / G_{\rm
ww}(t_\pi,t_K)$ is $t$ independent.  If this is the case, then from
Eq.\ \ref{eq:lat_me_ww} this ratio is the desired matrix element.  In
Figure \ref{fig:ktopi_Q2_3half_plateau} we show $\langle \pi^+ |
Q^{3/2}_{2,\, \rm lat} |K^+ \rangle$ as a function of $t$ for $m_f =
0.01$, 0.02, 0.03, 0.04 and 0.05.  There is no visible time dependence
in the range $ 10 \le t \le 20$, demonstrating that only the lowest
energy pseudoscalar state is contributing to the matrix element and
justifying our choice of $ 14 \le t \le 17$ for the range over which
eye contractions are calculated.  The $\DIthalf$ parts of operators do
not involve any eye contractions and are easier to determine with small
statistical errors.

Having established that a plateau exists for $t$ from 14 to 17, we plot
the dependence on $t$ of the $\DIhalf$ parts of operators, where
random noise sources are used in the calculation of the eye diagrams.
Figure \ref{fig:ktopi_Q2_plateau} shows $\langle \pi^+ | Q^{1/2}_{2,\,
\rm lat} |K^+ \rangle$ as a function of $t$ for the values of $m_f$
used and Figure \ref{fig:ktopi_Q6_plateau} is the same for $\langle
\pi^+ | Q^{1/2}_{6,\, \rm lat} |K^+ \rangle$.  Note the large
difference in the vertical scale between Figures
\ref{fig:ktopi_Q2_plateau} and \ref{fig:ktopi_Q6_plateau}, which is due
to the larger divergent contribution in $Q_6$.  One sees appreciable
fluctuations between different time slices, but they agree within
errors.  This is the expectation from using a noisy estimator for the
quark loops.  Figures \ref{fig:ktovac_Q2_plateau} and
\ref{fig:ktovac_Q6_plateau} show the data for the annihilation
contractions needed for $\langle 0 | Q_i | K^0 \rangle$ matrix
elements.  These also involve random sources in the calculation of the
quark loops and we see again that the results on different time slices
agree within errors.

The results for $\langle \pi^+ | (\overline{s}d)_{\rm lat} |K^+
\rangle$, $\langle
\pi^+ | Q^{(1/2)}_{i,\, \rm lat} |K^+ \rangle$, and $\langle \pi^+ |
Q^{(3/2)}_{i,\, \rm lat} |K^+ \rangle$ are tabulated in Tables
\ref{tab:k2pi_sbard}, \ref{tab:ktopi_i0_charm_out}, and
\ref{tab:ktopi_i2_charm_out}, respectively.  Results for the ratio ${
\langle 0 | Q_i | K^0 \rangle }/ { \langle 0|\bar s \gamma_5 d | K^0
\rangle }$ are given in Tables \ref{tab:ktovac_o1_o6_charm_out} and
\ref{tab:ktovac_o7_o10_charm_out}.  In each case the matrix elements
have been averaged over time slices 14-17.  The relative statistical
error for the $\Delta I = 1/2$ matrix elements is almost 100\% for
matrix elements that are quite small (compatible with zero), {i.e.}
$\langle \pi^+|Q_1|K^+\rangle$.  For the left-left operators like $Q_2$
the statistical errors are 10-20\% and the errors fall to 0.5-3\% for
the color-mixed left-right operators. For $\Delta I=3/2$ matrix
elements the relative statistical error is 2-3\%.

\fi


\section{$\Delta I = 3/2$ Matrix Elements}
\label{sec:i_3over2_me}

\ifnum\theIThreeOverTwoMe=1
%
%

In this section we discuss the lattice $K^+ \rightarrow \pi^+$ matrix
elements for the $\DIthalf$ parts of the operators listed in
Eq.\ \ref{eq:Q1} to \ref{eq:P10}.  In lowest order chiral perturbation
theory, three constants serve to determine all of these matrix
elements.  A single value of $\alpha^{(27,1),(3/2)}_{\rm lat}$ fixes
the $\DIthalf$ parts of $Q_1$, $Q_2$, $Q_9$, $Q_{10}$, $P_1$, $P_2$,
$P_9$ and $P_{10}$.  For the electroweak penguin operators,
$\alpha^{(8,8),(3/2)}_{7,\rm lat}$ is needed for the $\DIthalf$ part of
$Q_7$ and $P_7$ and $\alpha^{(8,8),(3/2)}_{8,\rm lat}$ is needed for
$Q_8$ and $P_8$.  (The two values for the (8,8) operators arise since
Fierz transformations do not relate the electroweak operators with and
without color-mixed indices.)  The constants
$\alpha^{(27,1),(3/2)}_{\rm lat}$, $\alpha^{(8,8),(3/2)}_{7,\rm lat}$
and $\alpha^{(8,8),(3/2)}_{8,\rm lat}$ are all finite and no
subtraction is needed to determine the corresponding $K \rightarrow \pi
\pi$ matrix elements.  In addition, since there is a single (27,1)
representation for left-left operators, the value of
$\alpha^{(27,1),(3/2)}_{\rm lat}$ also provides a determination of
$\langle \Kbar^0 | Q^{(\DStwo)} | K^0 \rangle$.


\subsection{The Lattice Value of $\alpha^{(27,1),(3/2)}_{\rm lat}$}
\label{subsec:alpha_27_1}

We start with a determination of $\alpha^{(27,1),(3/2)}_{\rm lat}$.
From Eqs.\ \ref{eq:Q1_irrep}, \ref{eq:Q2_irrep}, \ref{eq:Q9_irrep},
\ref{eq:Q10_irrep} of the appendix, we see that we need the matrix
element of $Q^{\Delta s = 1, \Delta d = -1}_{LL,S,(27,1),3/2}$, defined
in \ref{eq:twenty_seven_three_half}, which is the $\DIthalf$ part of
$\Theta^{(27,1)}$.  To follow more closely the notation of subsection
\ref{subsec:lowest_order_XPT}, in Section \ref{sec:def_theta_op} of the
appendix we define $\Theta^{(27,1),(3/2)} \equiv Q^{\Delta s = 1,
\Delta d = -1}_{LL,S,(27,1),3/2}$.  Then $\alpha^{(27,1),(3/2)}_{\rm
lat}$ is defined by Eq.\ \ref{eq:def_alpha_271_32}, the generalization
of Eq.\ \ref{eq:kpiO271} for a particular isospin.  (For the (27,1)
operator the generalization is trivial and in fact
$\alpha^{(27,1),(3/2)}_{\rm lat} = \alpha^{(27,1)}_{\rm lat}$, but we
will use $\alpha^{(27,1),(3/2)}_{\rm lat}$ to make it clear that this
is determined from the $\DIthalf$ amplitude.)  The dependence of this
matrix element on the parameters of low-energy quenched QCD is given in
Eq.\ \ref{eq:kpi271q}.

Table \ref{tab:theta271_32_corr_mpi} gives our values for $\langle
\pi^+ | \Theta^{(27,1),(3/2)}_{\rm lat} | K^+ \rangle$ versus quark
mass.  The function we fit to is Eq.\ \ref{eq:kpi271q} with $\alpha =0$.
For our particular lattice spacing this takes the form
\begin{equation}
  \langle \pi^+ | \Theta^{(27,1),(3/2)}_{\rm lat} | K^+ \rangle
    = b^{(27,1)}_1 m_M^2
      \left[ 1 -
        \left( \delta + \frac{6 m_M^2}{(4 \pi f)^2} \right)
               \ln \left(3.6941 \, m_M^2
	\right)
      \right]
    + b^{(27,1)}_2 m_M^4
 \label{eq:theta271_32_corr_mpi}
\end{equation}
with $m_M^2 = 3.18 ( m_f + m_{\rm res} )$ and $1/(4 \pi f)^2 = 1.246$.
Here we have used the result for $f$ from \cite{Blum:2000kn}, which is
137(10) MeV, rather than the physical value, since we do not assume
that quenched QCD at our fixed lattice spacing agrees with the physical
world.  The factor of 3.6941 in the logarithm is $\Lqcpt^2 = 1$ GeV$^2$
in lattice units.  Figure \ref{fig:theta271_32_corr_mpi} is a plot of
the data and the solid line shows the result of a fit to
Eq.\ \ref{eq:theta271_32_corr_mpi}.  The fit uses all 5 values for the
quark mass and sets $\delta = 0.05$.  The fit is again an uncorrelated
fit to our correlated data, which results in a value of $\chi^2/{\rm
d.o.f} = 1.9$.  The other lines in the figure give the contribution to
the total of the various terms in Eq.\ \ref{eq:kpi271q}.  Of particular
importance is the chiral logarithm term (the dot-dash line) $\sim m_M^4
L_{Q\chi}(m_M)$ which is very nearly linear in $m_f$ up to $m_f =0.035$.
Numerically, this term cannot be distinguished from the simple $m_M^2$
term and, as the graph shows, the term proportional to $m_M^2$ and the
chiral logarithm term are of roughly equal size.  Thus, our value for
$b^{(27,1)}_1$ is strongly dependent on the known coefficient, $-6$,
for the chiral logarithm in Eq.\ \ref{eq:kpi271q}.  In particular,
leaving out the chiral logarithm term makes the value of $b^{(27,1)}_1$
almost a factor of two larger.

In contrast to the chiral logarithm, the quenched chiral logarithm,
shown by the short dashed line in Figure \ref{fig:theta271_32_corr_mpi}
is contributing very little to the final result.  This appears to be a
consequence of the small value for $\delta$ and the fact that we are
working with pseudoscalar masses above 390 MeV.  This particular
$\DIthalf$ amplitude has quite small statistical errors and the 1-loop
quenched chiral perturbation theory formula is known.  Since we see
very little effects of the quenched chiral logarithms here, we expect
them to have little effect on other amplitudes where the explicit
coefficient of the quenched chiral logarithm is not known.

The full range of quark masses (0.01 to 0.05) has been used in the fit
shown in Figure \ref{fig:theta271_32_corr_mpi}.  The range of
pseudoscalar masses covered by this quark mass range is 390 to 790 MeV
and from the fit it appears that 1-loop quenched chiral perturbation
theory is working reasonably well over this range.  The $\chi^2/{\rm
d.o.f}$ is somewhat large for an uncorrelated fit, with the $m_f =
0.05$ point lying somewhat above the curve from the fit.  This point
may be showing the limitations of 1-loop chiral perturbation theory.
At the other extreme, the $m_f = 0.01$ point is where chiral
perturbation theory should work the best, but this light quark mass is
the most susceptible to the effects of finite volume and topological
near-zero modes.  It is worth re-emphasizing that even for $m_f =
0.01$, the chiral logarithm contributions are about 25\% of the total
value and must be included.

To test for sensitivity to the quark mass range used in the fit, we
have done fits with different ranges and give the results in Table
\ref{tab:theta271_32_corr_mpi_fit}.  One sees essentially no difference
between the fits to $m_f = 0.02$ to 0.04 and $m_f = 0.01$ and 0.05.  On
this basis, we choose to fit to all five quark masses and find
\begin{equation}
  \alpha^{(27,1),(3/2)}_{\rm lat} = -4.13(18) \times 10^{-6}
\end{equation}


\subsection{The Lattice Value of $\alpha^{(8,8),(3/2)}_{7, \rm lat}$
  and $\alpha^{(8,8),(3/2)}_{8, \rm lat}$}
\label{subsec:alpha_8_8}

Unlike the single (27,1) operator which enters in many $Q_i$'s and
$P_i$'s, the color diagonal (8,8) enters only in $Q_7$ and $P_7$ and
the color mixed (8,8) enters only in $Q_8$ and $P_8$.  We therefore
define $\Theta^{(8,8),(3/2)}_i \equiv [Q_i]^{(3/2)}$ for $i = 7$ and 8,
as shown in more detail in Sections \ref{sec:def_theta_op} and
\ref{sec:cpt_decomp} of the appendix.
Eqs.\ \ref{eq:eight_eight_three_half}, \ref{eq:eight_eight_one_half}
and \ref{eq:Q7_isospin_decomp} give the isospin decomposition of $Q_7$
in terms of quark fields.  The results for $Q_8$ are similar, with
color mixed indices on the quark fields.  In lowest order chiral
perturbation theory, $\alpha^{(8,8),(3/2)}_{7,\rm lat}$ and
$\alpha^{(8,8),(3/2)}_{8,\rm lat}$ are determined from $ \langle \pi^+
| \Theta^{(8,8),(3/2)}_{7, \rm lat} | K^+ \rangle$ and  $ \langle \pi^+
| \Theta^{(8,8),(3/2)}_{8, \rm lat} | K^+ \rangle$ through
Eq.\ \ref{eq:def_alpha_88_32}, which is Eq.\ \ref{eq:kpiO88} decomposed
into operators of definite isospin.  Unlike the (27,1)
operator, the chiral logarithm corrections for the (8,8) operator
in quenched QCD are not currently known.

Table \ref{tab:theta88_32_7_corr_mpi} gives our values for $ \langle
\pi^+ | \Theta^{(8,8),(3/2)}_{7,\; \rm lat} | K^+ \rangle$ and Table
\ref{tab:theta88_32_8_corr_mpi} gives them for $ \langle \pi^+ |
\Theta^{(8,8),(3/2)}_{8, \; \rm lat} | K^+ \rangle$.
Since the 1-loop corrections are not known, but the general form
should be as in Eq.\ \ref{eq:kpiO88div_log}, except that the $m_M^2$
term has a finite coefficient for the $\DIthalf$ amplitudes, we
will try fitting with and without a conventional chiral logarithm term.
We will not include any quenched chiral logarithm effects, since these
were seen to be small for the (27,1), $\DIthalf$ amplitudes discussed
in the previous section.  Thus, we will fit our data to the form
\begin{equation}
  \langle \pi^+ | \Theta^{(8,8),(3/2)}_{i,\; \rm lat} | K^+ \rangle
    = b^{(8,8)}_{i,0}
      \left[ 1 +
	\left( \frac{\xi^{(8,8)}_i m_M^2}{(4 \pi f)^2} \right)
	\ln \left(3.6941 \, m_M^2 \right)
      \right]
    + b^{(8,8)}_{i,1} m_M^2
 \label{eq:theta88_32_i_corr_mpi}
\end{equation}
where $i = 7$, 8, $m_M^2 = 3.18 ( m_f + m_{\rm res} )$, $1/(4 \pi f)^2
= 1.246$ and 3.6941 is the value of $\Lqcpt^2$ = 1 GeV$^2$ in lattice
units.  Since $\xi^{(8,8)}_i$ is not known, we will do fits where it is
zero and where it is a free parameter. 

Figure \ref{fig:theta88_32_7_corr_mpi} is a plot of the values for $
\langle \pi^+ | \Theta^{(8,8),(3/2)}_{7,\; \rm lat} | K^+ \rangle$ and
Figure \ref{fig:theta88_32_8_corr_mpi} is the same for $ \langle \pi^+
| \Theta^{(8,8),(3/2)}_{8,\; \rm lat} | K^+ \rangle$.  An obvious
feature of the graphs is the nearly linear behavior of the matrix
elements.  To determine $\alpha^{(8,8),(3/2)}_{7,\rm lat}$ and
$\alpha^{(8,8),(3/2)}_{8,\rm lat}$, we must extrapolate to the chiral
limit, $m_f = -m_{\rm res}$.  Since there are no power divergences
involved in these operators, their chiral limit, up to ${\cal O}(a^2)$
corrections should be determined by $m_{\rm res}$.  Table
\ref{tab:theta88_32_corr_mpi_fit} gives the results of fits to
Eq.\ \ref{eq:theta88_32_i_corr_mpi}, where $\xi^{(8,8)}_i$ is held to
zero (simple linear fit) and allowed to be a free parameter (chiral
logarithm fit).  In Figures \ref{fig:theta88_32_7_corr_mpi} and
\ref{fig:theta88_32_8_corr_mpi} the solid lines are the linear fits and
the dashed lines include the chiral logarithm term with a free
parameter.

One sees that the value of $b^{(8,8)}_{7,0}$ changes by about 15\% with
the inclusion of a chiral logarithm term, while $b^{(8,8)}_{8, 0}$
moves by about 8\%.  Knowing $\xi^{(8,8)}_i$ analytically would
decrease the uncertainty in our extrapolation.  Without this knowledge,
we will take the chiral logarithm fits to determine the intercepts,
with the difference between the two fit choices giving an indication
of our systematic uncertainty.  Thus, we find
\begin{eqnarray}
  \alpha^{(8,8),(3/2)}_{7,\rm lat} & = & -1.61(8) \times 10^{-6} \\
  \alpha^{(8,8),(3/2)}_{8,\rm lat} & = & -4.96(27) \times 10^{-6}
\end{eqnarray}

\fi


\section{$\Delta I = 1/2$ Matrix Elements}
\label{sec:i_1over2_me}

\ifnum\theIOneOverTwoMe=1
%
%

In this section, we turn to the determination of the lattice $K^+
\rightarrow \pi^+$ matrix elements for the $\DIhalf$ parts of the
operators listed in Eq.s\ \ref{eq:Q1} to \ref{eq:P10}.  The numerical
evaluation of these matrix elements is much more involved, since the
physical quantities are found from the difference of two lattice
quantities which contain power divergences.  The basic idea behind the
subtraction of the unphysical effects was discussed in Section
\ref{subsec:lowest_order_XPT} and it is important to recall that this
subtraction is done for matrix elements in hadronic states.  A related
subtraction was discussed in Section \ref{sec:npr}, which is done in
Landau gauge fixed quark states and is used for matching operator
normalizations between the lattice and continuum perturbation theory.
An important check of our calculation is the consistency of these two
subtractions, which should receive the same contribution from the
leading momentum-independent power-divergent terms.


\subsection{Subtraction of Power Divergent Operators}
\label{subsec:sub_power_div}

All the operators in Eq.\ \ref{eq:Q1} to \ref{eq:P10} have unphysical
contributions to their $\DIhalf$, $K^+ \rightarrow \pi^+$ matrix
elements at finite quark mass, since an (8,1) or (8,8) representation
appears in each $Q_i$.  For the (8,1) parts of the operators, the
formulae in Section \ref{subsec:lowest_order_XPT} show how these
unphysical contributions are removed.  For the operators $Q_7$ and
$Q_8$, naively more options exist since they are in a single
irreducible representation of $\sutlr$.  One can 1) find the $\DIhalf$
matrix elements from the value for $\alpha^{(8,8),(3/2)}_{i, \rm lat}$
of the previous section, 2) extrapolate the divergent $\DIhalf$ matrix
elements to the chiral limit, or 3) perform a subtraction as for the
(8,1) operators at finite quark mass and then extrapolate the
remaining, non-divergent matrix element to the chiral limit.  For
domain wall fermions at finite $L_s$, only the first option is
precisely defined, since at finite $L_s$ the value of the input quark
mass yielding the chiral limit is not well defined for divergent
operators.  One only knows that the chiral limit is achieved by setting
$m_f = -O(m_{\rm res})$.  For completeness and to study the effects of
${\cal O}(m_{\rm res})$ errors, we will include the subtraction of the
$\DIhalf$ (8,8) operators in this section, but will use the values of
$\alpha^{(8,8),(3/2)}_{7, \rm lat}$ and $\alpha^{(8,8),(3/2)}_{8, \rm
lat}$ found previously to determine our final value for the
$\DIhalf$ parts of $Q_7$ and $Q_8$.

In subsection \ref{subsec:one_loop_XPT} we have argued that a
particular combination of matrix elements (Eq.\ \ref{eq:fix_alpha_81_2}
and \ref{eq:fix_alpha_81_1}) will not involve power divergent
coefficients times higher order logarithmic terms in chiral
perturbation theory.  This is extremely important for our numerical
subtraction, since such higher order logarithmic terms in chiral
perturbation theory are not small for the pseudoscalar masses we can
currently use.  In addition, there is a great benefit numerically to
dealing with quantities where such effects cancel, rather than
cancelling them through the explicit determination of extra fit
parameters.  We will also apply the same subtraction to $Q_7$ and $Q_8$
that we apply to the other operators.  This will remove the divergent
term, $m_M^2 \alpha^{(8,8)}_{\rm div}$, given in
Eq.\ \ref{eq:kpiO88div_log}, since any divergent term looks like
$\Theta^{(3, \bar{3})}$.   The finite term proportional to $m_M^2$ that
is left will not be related to the $m_M^2$ dependence of $K \rightarrow
\pi \pi$ matrix elements, since this subtraction has not properly
handled such finite corrections.  However, for these operators the
physical value we seek is the extrapolation to the chiral limit, not
the coefficient of the $m_M^2$ term, and the subtraction will only
impact our ability to extrapolate to the (approximately known for
finite $L_s$) chiral limit.

While a general approach to subtracting the power divergences is
dictated by the requirement that we obtain $\langle\pi\pi|O_i|K\rangle$
to leading order in chiral perturbation theory, the specific
subtraction procedure that we describe below is chosen so that all
quadratic divergence is removed from the subtracted amplitude if $\mres
= 0$.  This ensures that our result will not be polluted by possibly
large $1/a^2$ terms entering at higher order in chiral perturbation
theory.

In this section, we will not report our results in terms of the various
parameters $\alpha^{(8,1)}_1$ and $\alpha^{(8,1)}_2$, since there are
many different (8,1) representations present in the operators in
Eq.\ \ref{eq:Q1} to \ref{eq:P10} and each irreducible representation
has its own values for $\alpha^{(8,1)}_1$ and $\alpha^{(8,1)}_2$.
For each operator $Q_i$, we will determine a subtraction coefficient
$\eta_{1,i}$, following the form of Eq.\ \ref{eq:fix_alpha_81_2},
through
\begin{equation}
    \frac{ \langle 0 | Q_{i, \rm lat} | K^0 \rangle}
      {\langle 0 | (\bar{s} \gamma_5 d)_{\rm lat} | K^0 \rangle}
    =  \eta_{0,i} +
    \eta_{1,i} \left( m_s^\prime - m_d^\prime \right)
\label{eq:fix_eta}
\end{equation}
where $m_s^\prime$ and $m_d^\prime$ are the nondegenerate quark masses
used in the calculation of $K^{0} \rightarrow 0$ matrix elements.
Corrections to this formula from higher order effects in chiral
perturbation theory are free of power divergences.  We expect that
$\eta_{0,i}$ should be zero, but we add this free parameter to the fit
to test that expectation.  The arguments leading to
Eq.\ \ref{eq:fix_alpha_81_2} show that when, for example,
$\alpha^{(8,1)}_2$ is very large, $\langle 0 | Q_{i, \rm lat} | K^0
\rangle / \langle 0 | (\bar{s} \gamma_5 d)_{\rm lat} | K^0 \rangle$
should not show the presence of chiral logarithms, since such terms
appear only through $\alpha^{(8,1)}_1$.  Thus, for large
$\alpha^{(8,1)}_2$, where the subtraction is more delicate, the
determination of the subtraction coefficient is easier since the
linearity is better.

Starting from the values for $\langle 0 | Q_{i, \rm lat} | K^0 \rangle
/ \langle 0 | (\bar{s} \gamma_5 d)_{\rm lat} | K^0 \rangle$ given in
Tables \ref{tab:ktovac_o1_o6_charm_out} and
\ref{tab:ktovac_o7_o10_charm_out}, we have plotted this ratio versus
$m_s^\prime - m_d^\prime$ in Figures \ref{fig:ktovac_charm_out_Q2},
\ref{fig:ktovac_charm_out_Q6}, \ref{fig:ktovac_charm_out_Q1_Q3_Q4_Q5},
\ref{fig:ktovac_charm_out_Q8} and
\ref{fig:ktovac_charm_out_Q7_Q8_Q9_Q10}.  For $Q_2$, $Q_6$ and
$Q_8$, graphs are shown with better resolution.  Note that for $Q_6$
and $Q_8$ the $y$-axis is a much larger scale than for $Q_2$.  For
$Q_2$, there is some deviation for different values of $m_s^\prime$ and
$m_d^\prime$ with the same value for $m_s^\prime - m_d^\prime$, but
within our statistics no clear conclusion can be drawn.  For $Q_6$ and
$Q_8$, any such deviation is much smaller, as would be expected for
these operators with large power divergent contributions, but again
deviations are within our statistical error.

The results for uncorrelated fits to this data are given in Table
\ref{tab:eta_charm_out}.  One sees that $\eta_{1,6}$ is the largest
subtraction coefficient and has a statistical error of about 0.2\%.
The other operators with large subtraction coefficients are $Q_5$,
$Q_7$ and $Q_8$, which have comparable statistical precision.  The good
linearity of the data makes quoting such small statistical errors
sensible.  It is also vital that we know these subtraction coefficients
to this accuracy, since there are ${\cal O}(a^{-2})$ divergences to cancel
through this subtraction.  Except for $Q_7$ and $Q_8$, $\eta_{0,i}$ is
zero within statistical errors.  For $Q_7$ and $Q_8$, $\eta_{0,i}$ is
statistically non-zero, but very small in magnitude.

An important cross-check of our calculation is the comparison of the
subtraction coefficients $\eta_{1,i}$, determined from properties of
the operators in hadronic states, with the subtraction coefficients
determined by the NPR procedure of Section \ref{sec:npr}.  A similar
subtraction is performed there to remove the mixing between four quark
operators and quark bilinears.  This subtraction is done in Landau
gauge fixed quark states at momentum scales $\ge 1.5$ GeV.  Thus, the
two subtraction coefficients should not be identical.  Only the power
divergent parts should agree, since these are independent of external
momenta.  For the operators with the largest subtraction coefficients,
the agreement should be quite close, since the large subtraction comes
from the power divergent pieces dominating.

Table \ref{tab:compare_sub_charm_out} gives a comparison of the
subtraction coefficients as determined from non-perturbative
renormalization and the values from Table \ref{tab:eta_charm_out},
which were determined from chiral perturbation theory in hadronic
states.  The non-perturbative renormalization subtraction coefficients
are the values in the second column of Table \ref{tab:miy} minus the
values in the second column of Table \ref{tab:msx}.  The results in
Table \ref{tab:compare_sub_charm_out} are also plotted in Figure
\ref{fig:compare_hadronic_npr_sub.eps}.  For the $(V-A) \times (V+A)$
operators ($Q_5$, $Q_6$, $Q_7$, and $Q_8$) where the subtraction
coefficients are the largest, the agreement between the two techniques
is very good.  This gives us confidence in the subtraction procedure,
since the comparison is between quantities determined in entirely
different ways using different computer programs for data generation
and analysis.  Note that the errors from the hadronic state calculation
are considerably smaller.


\subsection{Subtracted $\Delta I = 1/2$ Matrix Elements}
\label{subsec:sub_i_1over2_me}

The combination of terms on the left-hand side of Eq.\
\ref{eq:fix_alpha_81_1} that removes chiral logarithm effects from
the divergent parts of the operators can be written as
\begin{equation}
    \langle \pi^+ | Q^{(1/2)}_{i, \rm lat} | K^+ \rangle_{\rm sub}
    \equiv
    \langle \pi^+ | Q^{(1/2)}_{i, \rm lat} | K^+ \rangle +
    \eta_{1,i} (m_s + m_d )
    \langle \pi^+ | (\bar{s}d)_{\rm lat} | K^+ \rangle
  \label{eq:ktopi_subtraction}
\end{equation}
It is easy to see that when written in this form, the subtraction
required by chiral perturbation theory removes the entire $1/a^2$
divergence present in the original $\langle \pi^+ | Q_{i,{\rm
lat}}^{(1/2)}|K^+\rangle$ matrix element if $\mres = 0$.  Usual power
counting arguments, combined with exact chiral symmetry and the CPS
symmetry of Ref. \cite{Bernard:1985wf}, dictate that all $1/a^2$
divergences which appear in the matrix elements of the operator
$Q_{i,{\rm lat}}^{(1/2)}$ can be written as a divergent coefficient
times matrix elements of the dimension-three operator
$(m_d+m_s)\overline{s}d + (m_d-m_s)\overline{s}\gamma_5 d$.
Eq.~\ref{eq:fix_eta} determines this coefficient as $\eta_{1,i}$
ensuring that the subtraction in Eq.~\ref{eq:ktopi_subtraction} removes
the entire $1/a^2$ divergent piece from the $Q_{i,{\rm lat}}^{(1/2)}$
matrix element.

In addition to chiral logarithm effects, we saw in Section
\ref{subsec:ward_id_sd} that the matrix element $ \langle \pi^+ |
(\bar{s}d)_{\rm lat} | K^+ \rangle$ is altered by zero modes for light
quark masses.  These same zero mode effects will also enter the
divergent part of $\langle \pi^+ | Q^{(1/2)}_{i, \rm lat} | K^+
\rangle$ matrix elements.  In particular, recalling Figure
\ref{fig:ktopi_sbard}, we are reminded that this matrix element is not
well represented by a simple linear dependence on $m_f$.  Again it is
simpler to let the subtraction of matrix elements in
Eq.\ \ref{eq:ktopi_subtraction} remove these non-linear terms.  Any
remaining non-linearities should be associated with the chiral
logarithms on the right-hand side of Eq.\ \ref{eq:fix_alpha_81_1} and
near-zero mode effects in the finite terms.  One once again avoids the
possibility of failing to remove a divergent term which is multiplied
by a higher order term in chiral perturbation theory.

With the values for the subtraction coefficients, $\eta_{1,i}$, from
the previous section, we have calculated the subtracted matrix
elements.  To see the extent of the subtraction, in Figure
\ref{fig:ktopi_Q6_sub_unsub_charm_out} we plot $\langle \pi^+ | Q_{6,
\rm lat} | K^+ \rangle$,  $2 m_f |\eta_{1,6}| \langle \pi^+ |
(\bar{s}d)_{\rm lat} | K^+ \rangle$ and $\langle \pi^+ | Q_{6, \rm lat}
| K^+ \rangle_{\rm sub}$.  The first two quantities show very similar
non-linearity and produce a subtracted matrix element which is much
smaller.  Given the large cancellation involved, the importance of
removing divergence terms times higher order terms in chiral
perturbation theory is clear.

The complete results for the subtracted matrix elements are given in
Table \ref{tab:ktopi_sub_12} and are plotted versus $m_f$ in Figures
\ref{fig:ktopi_Q2_sub_charm_out}, \ref{fig:ktopi_Q6_sub_charm_out},
\ref{fig:ktopi_Q1_Q3_Q4_Q5_sub_charm_out},
\ref{fig:ktopi_Q8_sub_charm_out} and
\ref{fig:ktopi_Q7_Q8_Q9_Q10_sub_charm_out}.  The subtraction is done
under a jackknife error loop, to make maximum use of  any correlations
in the values of $ \langle \pi^+ | Q^{(1/2)}_{i, \rm lat} | K^+ \rangle
$, $\eta_{1,i}$ and $ \langle \pi^+ | (\bar{s}d)_{\rm lat} | K^+
\rangle$.  The subtracted matrix elements for $Q_2$, $Q_6$ and $Q_8$
are shown on an expanded scale.  Concentrating for a moment on $Q_6$
(Figure \ref{fig:ktopi_Q6_sub_charm_out}),  the graph for the
subtracted operator reveals a number of important features:
\begin{enumerate}
\item  The presence of finite $L_s$ and power divergent operators
  means that $\langle \pi^+ | Q_{6, \rm lat} | K^+ \rangle_{\rm sub}$
  need not vanish at $m_f = 0$ or $m_f = -m_{\rm res}$.  This is
  obvious in the graph, where the matrix element vanishes around $m_f$
  of 0.02.
\item  For $Q_i$ containing an (8,1) representation, only the slope of
  the subtracted matrix element is needed, so the ambiguities of
  ${\cal O}(m_{\rm res})$ in the chiral limit are unimportant.  For (8,8)
  parts of an operator, such ambiguities prohibit a precise
  determination of the desired $\alpha$'s from the $\DIhalf$ amplitudes.
\item  The subtracted values for $Q_6$ (and also $Q_2$ and $Q_9$) show
  some non-linearity, although the effect is not conclusive given the
  statistical errors.  We have not fit to the non-linearities, since
  the coefficients of the chiral logarithms are not known for the (8,1)
  operators in quenched QCD.  For the full QCD case, where they are
  known, the coefficient is 1/3, compared to 34/3 for the (27,1)
  operators.  Thus, we use simple linear fits and expect the corrections
  in the slope we seek, due to logarithms, to be small.
\item  $Q_6$ is a pure (8,1) operator, but for $Q_1$, $Q_2$, $Q_9$
  and $Q_{10}$, which contain a (27,1) for which the chiral logarithm
  coefficient is known and large, fits could be done to incorporate
  this effect.  However, the $\DIhalf$ part of the (27,1) enters the
  total operator with a small coefficient (1/10 or 1/15).  Also, since
  $\alpha^{(27,1),(1/2)}_{\rm lat} = \alpha^{(27,1),(3/2)}_{\rm lat}$
  and $\alpha^{(27,1),(3/2)}_{\rm lat}$ is small, this particular
  chiral logarithm contribution should not be visible in our data.
\item  The lower points in the figure ($\Diamond$) are the result if
   the subtraction in Eq.\ \ref{eq:ktopi_subtraction} has $(m_s + m_d
   )$ changed to $(m_s + m_d + 2\mres)$.  This subtraction will also
   not exactly remove the ${\cal O}(\mres/a^2)$ term, but the two subtractions
   show that chiral symmetry breaking from finite $L_s$ is
   quantitatively ${\cal O}(\mres/a^2)$.
\end{enumerate}

We have fitted the subtracted operators to a linear function
parameterized by
\begin{equation}
  \langle \pi^+ | Q^{(1/2)}_{i, \rm lat} | K^+ \rangle_{\rm sub}
   = c_{0, i} + c_{1, i}  m_f
   \label{eq:fit_ktopi_sub_charm_out}
\end{equation}
with the results given in Table \ref{tab:fit_ktopi_sub_charm_out}.
These are uncorrelated linear fits to all five quark masses.  We see
that for $Q_6$, in spite of the very large subtraction involved, the
slope of $ \langle \pi^+ | Q^{(1/2)}_{i, \rm lat} | K^+ \rangle_{\rm
sub}$ is determined with a statistical error of about 10\%.

For $Q_7$ and $Q_8$, we can start from the fits given in
Table \ref{tab:fit_ktopi_sub_charm_out} and compare the value
for the $\DIhalf$ matrix elements with the value expected from
the $\DIthalf$ matrix elements.  Since $\alpha^{(8,8),(1/2)}_{i,
\rm lat} = 2 \alpha^{(8,8),(3/2)}_{i, \rm lat}$ for $i = 7$ and 8,
we can use the values for $\alpha^{(8,8),(3/2)}_{i, \rm lat}$ given
in Section \ref{subsec:alpha_8_8} to find
\begin{eqnarray}
  \alpha^{(8,8),(1/2)}_{7,\rm lat} & = & -3.22(16) \times 10^{-6} \\
  \alpha^{(8,8),(1/2)}_{8,\rm lat} & = & -9.92(54) \times 10^{-6}
\end{eqnarray}
The unsubtracted $\DIhalf$ matrix elements should have the form
\begin{equation}
  \langle \pi^+ | Q^{(1/2)}_{i, \rm lat} | K^+ \rangle
   = c_{0, i} + c_{1, i} ( m_f + m_{\rm res})
   + c^{\rm div}_{1, i}( m_f + O(m_{\rm res}) )
\end{equation}
From Table \ref{tab:ktopi_i0_charm_out} and
Eq.\ \ref{eq:ktopi_subtraction} one sees that $ c^{\rm div}_{1, 7} \sim
-1.3$ and $ c^{\rm div}_{1, 8} \sim -3.9$.  Using these values and
$m_{\rm res} = 0.00124$ gives a $m_f$ independent contribution to the
$i = 7$ and 8 matrix elements of ${\cal O}(0.0016)$ and ${\cal O}(0.0044)$ from
contact terms in the Ward-Takahashi identities.  The values for
$c_{0,i}$ for $i = 7$ and 8 are given in Table
\ref{tab:fit_ktopi_sub_charm_out} and are $-0.00720(21)$ and
$-0.0223(7)$ respectively.  Thus, the expected uncertainty due to finite
$L_s$ in determining $ \alpha^{(8,8),(1/2)}_{i,\rm lat}$ from the
subtracted $\DIhalf$ amplitudes is about 20\% in both cases.  Using
these values for $c_{0,i}$ yields
\begin{eqnarray}
    \alpha^{(8,8),(1/2)}_{7,\rm lat,sub}
  &
    =
  &
    -3.05(9) \times 10^{-6}
  \\
    \alpha^{(8,8),(1/2)}_{8,\rm lat,sub}
  &
    =
  &
    -9.44(30) \times 10^{-6}
\end{eqnarray}
The agreement with the results from the $\DIthalf$ matrix elements
is better than might be expected.  However, the $\DIthalf$ fits
include chiral logarithm corrections which change the results by
15\% for $i=7$ and 8\% for $i=8$.  The change happens to improve the
agreement with the values from the subtracted operators.  However,
this general agreement does demonstrate the reliability of the
subtraction of the power divergent operators.

Defining constants $\alpha^{(1/2)}_{i, \rm lat}$ for $i \neq 7,8$
through
\begin{equation}
  \langle \pi^+ | Q^{(1/2)}_{i, \rm lat} | K^+ \rangle_{\rm sub}
  \equiv \frac{4 m_M^2}{f^2} \alpha^{(1/2)}_{i, \rm lat}
\end{equation}
and using $m_M^2 = 3.18( m_f + m_{\rm res})$ gives the values in Table
\ref{tab:alpha_ktopi_sub_charm_out}.  We collect the $\alpha$'s
determined without requiring subtractions, in Table
\ref{tab:alpha_ktopi_nosub_charm_out}.  Finally
\ref{tab:alpha_ktopi_sub_charm_out} gives the $\DIhalf$ and $\DIthalf$
values for $\alpha_i$ for the ten operators in the basis used in the
three-quark effective theory.  These are our results for the lattice
values for the constants determining kaon matrix elements in lowest
order chiral perturbation theory from quenched QCD and domain wall
fermions.  In the next two sections we will combine these values with
the Wilson coefficients of Section \ref{sec:wilson_coef}, the $Z$
factors from Section \ref{sec:npr} and known experimental quantities to
give physical values for the real and imaginary parts of isospin zero
and two amplitudes for $K \rightarrow \pi \pi$.

\fi


\section{Physical Matrix Elements}
\label{sec:phys_me}

\ifnum\thePhysME=1
%
%

The physical values for $K \rightarrow \pi\pi$ amplitudes can now be
calculated from the effective Hamiltonian in Eq.\ \ref{eq:DSone_ham}
using the Wilson coefficients in Tables \ref{tab:wilson_coef_z} and
\ref{tab:wilson_coef_y}, the $\hat{Z}_{ij}^{\rm NPR}/Z_q^2$ values from
non-perturbative renormalization in Tables \ref{tab:zfull1} and
\ref{tab:zfull2}, the value $Z_q = 0.808(3)(15)$ from Table II of
\cite{Blum:2001sr}, the chiral perturbation theory formulae in
Eqs.\ \ref{eq:kpipiO81} and \ref{eq:kpipiO88}, the central values for
Standard-Model parameters in Table \ref{tab:sm_values} and the values
for $\alpha^{(1/2)}_{j, \rm lat}$ and $\alpha^{(3/2)}_{j, \rm lat}$
from Table \ref{tab:final_alpha_charm_out}.  The explicit formula is
\begin{align}
    \langle \pi\pi_{(I)} | \, -i \DSoneH \,
  & | K^0 \rangle
    =
    -i \sqrt{\frac{3}{4}} G_F V_{ud} V^*_{us}
    \sum_{i=1}^{10} \sum_{j=1, j \ne 4}^{8}
    \left[ z_i (\mu) + \tau y_i (\mu) \right]
    \hat{Z}^{\rm NPR}_{ij}(\mu)
  \nonumber
  \\
  &
    \times
    \left\{
    \begin{alignedat}{3}
        {\displaystyle \frac{4i}{f^3}}
      &
        \alpha^{(1/2)}_{j, \rm lat}
      &
      &
	(m_{K^0}^2 - m_{\pi^+}^2) a^{-4}
      &
	\quad I = 0,
      &
        \quad j = 1,2,3,5,6
      \\
        {\displaystyle \frac{-4\sqrt{2}i}{f^3}}
      &
        \alpha^{(3/2)}_{j, \rm lat}
      &
      &
	(m_{K^0}^2 - m_{\pi^+}^2) a^{-4}
      &
	\quad I = 2,
      &
	\quad j = 1,2,3,5,6
      \\
        {\displaystyle \frac{-12i}{f^3}}
      &
        \alpha^{(1/2)}_{j, \rm lat}
      &
      &
	a^{-6}
      &
	\quad I = 0,
      &
	\quad j = 7,8
      \\
        {\displaystyle \frac{-12\sqrt{2}i}{f^3}}
      &
        \alpha^{(3/2)}_{j, \rm lat}
      &
      &
	a^{-6} \hfill
      &
	\quad I = 2,
      &
	\quad j = 7,8
    \end{alignedat}
    \right.
  \label{eq:complete_kpipi}
\end{align}
where $a^{-1}$, the inverse lattice spacing, is 1.922 GeV
\cite{Blum:2000kn}.  Before discussing the numerical values produced
from our data, we will outline our strategy for making the transition
from the quenched QCD matrix elements we have calculated to the full
QCD matrix elements needed for comparison with the physical world.  We
can then assess the impact of the known chiral logarithms in full QCD
on our results and also discuss how sensitive our results are to the
values of the Standard-Model parameters given in Table
\ref{tab:sm_values}.

For our lattice calculation we have used a quenched value for $f$,
which is defined in the chiral limit, of 137 MeV \cite{Blum:2000kn}.
There is no reason why this value must agree with the full QCD value of
$f_{\rm QCD} \approx 120$ MeV.  In quenched chiral perturbation theory,
$f^{(\rm 1-loop)}_{\pi}$ and $f^{(\rm 1-loop)}_{K}$ do not
contain any conventional chiral logarithms, only quenched chiral
logarithms which we have argued are small.  This is consistent with the
linear quark mass behavior seen in \cite{Blum:2000kn} in the
determination of $f$.  In relating lattice $K \rightarrow \pi$ matrix
elements to lattice $K \rightarrow \pi \pi$ matrix elements, one should
use this $f$.  For small quark masses, the resulting lattice $K
\rightarrow \pi \pi$ matrix elements should be equal to those
explicitly calculated via a technique such as has been proposed by
Lellouch and Luscher \cite{Lellouch:2000pv}, provided the quenched
theory does not corrupt the full QCD relations between $K \rightarrow
\pi$ and $K \rightarrow \pi \pi$.

We will make the transition from the quenched theory to full QCD at the
level of the matrix elements $\langle \pi \pi | Q_i | K^0 \rangle$ and
not at the level of the lattice constants $\alpha_{i, \rm lat}$.  Since
the $\alpha_{i, \rm lat}$ factors in Eq.\ \ref{eq:complete_kpipi} are
multiplied by $f^{-3}$, changing from $f$ to $f_{\rm QCD}$ would be a
large effect and a factor of $f^2$ has already entered in the
calculation of the $\alpha_{i, \rm lat}$ from our lattice data.  For
the $\langle \pi \pi | Q_i | K^0 \rangle$ matrix elements which vanish
in the chiral limit, we have actually only determined the slope of the
matrix element.  The matrix element itself involves using chiral
perturbation theory to extrapolate to the kaon mass.  This
extrapolation introduces an additional choice in relating quenched
matrix elements to those in full QCD.

With this strategy of using the values for quenched $K \rightarrow \pi
\pi$ matrix elements as estimates for full QCD, we consider two choices
for the extrapolation to the kaon scale.  The first choice involves
extrapolating to the kaon mass for (8,1) and (27,1) operators using
lowest order chiral perturbation theory in the quenched theory.  The
second extrapolates to the kaon scale in the full theory and
incorporates the known and estimated chiral logarithms for the $K
\rightarrow \pi \pi$ matrix elements in full QCD.  We now discuss these
choices in detail.
\begin{enumerate}
\item
  Physical values for $m_{K^0}^2$ and $m_{\pi^+}^2$ are used in
  Eq.\ \ref{eq:complete_kpipi}.  For (8,1) and (27,1) operators, this
  can be thought of as an extrapolation to the physical kaon mass in
  quenched QCD using lowest order chiral perturbation theory, since we
  have found the quenched chiral logarithms to be small and there are
  no conventional chiral logarithms in these masses in the quenched
  theory.  These quenched $K \rightarrow \pi \pi$ matrix elements with
  $m_{K^0}^2$ and $m_{\pi^+}^2$ taking their physical values are taken
  as the matrix elements for full QCD.  The same results would be
  achieved by a lowest order extrapolation in full QCD, except that
  the the use of the physical kaon and pion masses is somewhat
  ambiguous, since physical masses include chiral logarithm
  corrections if the quark masses are taken as known input parameters.
  This ambiguity would change the matrix elements at the 10\% level.
\item
  We extrapolate to the physical kaon mass in full QCD, including the
  chiral logarithm corrections.  For the (8,1) and $\DIthalf$ part of
  the (27,1) operators the quenched slope is taken for the full QCD
  value and the known chiral logarithms in full QCD
  \cite{Bijnens:1985qt,Golterman:1997wb} are used in the
  extrapolation.  For (8,8) operators, the non-zero value in the
  quenched chiral limit is taken directly to full QCD.  Recent work on
  the electroweak penguins \cite{Cirigliano:1999pv} allows us to
  estimate the coefficients of the chiral logarithm term.  These
  authors write the matrix elements for the electroweak penguins at
  ${\cal O}(p^2)$ as $M_I = M_I^{(0)}(1 + \Delta_I)$ where $M_I^{(0)}$ is the
  lowest order value as given in Eq.\ \ref{eq:kpipiO88}.  They find
  $\Delta_0 = 0.98 \pm 0.55$ and $\Delta_2 = 0.27 \pm 0.27$ and state
  that $\Delta_I$ only includes the contributions from chiral
  logarithms.  The errors they quote come from varying $\Lqcpt$.  If we
  assume the correction is all from a chiral logarithm term
  $L_\chi(m_K)$, then the coefficient of this term would be $\sim -8.4$
  for $I = 0$ and $\sim -2.3$ for $I=2$.
\end{enumerate}

Thus, for our second extrapolation choice, where chiral logarithms are
included, we modify the second line of Eq.\ \ref{eq:complete_kpipi} to
\begin{equation}
    \times
    \left\{
     \begin{alignedat}{4}
        {\displaystyle \frac{4i}{f^3}}
      &
        \alpha^{(1/2)}_{j, \rm lat}
      &
      &
	(m_{K^0}^2 - m_{\pi^+}^2) a^{-4}
      &
      &
        \left[
          {\displaystyle 1 - \frac{97}{27} L_\chi(m_K) }
	\right]
      &
	\quad I = 0,
      &
        \quad j = 1,2,3,5,6
      \\
        {\displaystyle \frac{-4\sqrt{2}i}{f^3}}
      &
        \alpha^{(3/2)}_{j, \rm lat}
      &
      &
	(m_{K^0}^2 - m_{\pi^+}^2) a^{-4}
      &
      &
        \left[
          {\displaystyle 1 - \frac{3}{2} L_\chi(m_K) }
	\right]
      &
	\quad I = 2,
      &
	\quad j = 1,2,3,5,6
      \\
        {\displaystyle \frac{-12i}{f^3}}
      &
        \alpha^{(1/2)}_{j, \rm lat}
      &
      &
	a^{-6}
      &
      &
        \left[ 1 - 8.4 L_\chi(m_K) \right]
      &
	\quad I = 0,
      &
	\quad j = 7,8
      \\
        {\displaystyle \frac{-12\sqrt{2}i}{f^3}}
      &
        \alpha^{(3/2)}_{j, \rm lat}
      &
      &
	a^{-6} \hfill
      &
      &
        \left[ 1 - 2.3 L_\chi(m_K) \right]
      &
	\quad I = 2,
      &
	\quad j = 7,8
    \end{alignedat}
    \right.
  \label{eq:complete_kpipi_log}
\end{equation}
In these equations, the physical values for $m_{K^0}^2$ and
$m_{\pi^+}^2$ should be used.  We use our quenched value for $f$ in the
$1/(4 \pi f )^2$ factor in the chiral logarithms.  In addition to
estimating the coefficient of the chiral logarithm term for the (8,8)
operators, we have also used the (8,1) chiral logarithm for all of the
non-electroweak $\DIhalf$ matrix elements.  This is a very good
approximation, since the $\DIhalf$ part of the (27,1) operator
contributes very little here as can be seen from the size of
$\alpha^{(27,1),(1/2)}_{\rm lat}$.

\fi


\section{Real $A_0$, $A_2$ and $B_K$}
\label{sec:real_a0_a2}

\ifnum\theRealAZeroATwo=1
%
%

Following the procedure of the previous section, we now proceed to our
results for Re($A_0$) and Re($A_2$) and the $\DIhalf$ rule.  These
amplitudes are expected to come predominantly from the current-current
operators $Q_1$ and $Q_2$, as seen in the relative sizes of the Wilson
coefficients $z_i(\mu)$ and $y_i(\mu)$ given in Tables
\ref{tab:wilson_coef_z} and \ref{tab:wilson_coef_y}.  (Such a statement
depends on the scale $\mu$ under consideration, since the operators mix
under renormalization.) As such, they are quite independent of $V_{td}$
and CP violation effects in the Standard Model and provide an
independent forum for comparison between our quenched lattice QCD
calculations and experimental results.  We conclude with our results
for $B_K$, since it is determined by the matrix elements of the same
$(27,1)$ operator that determines Re($A_2$).

Using our data and Eqs.\ \ref{eq:complete_kpipi} and
\ref{eq:complete_kpipi_log} produces the values for Re($A_0$),
Re($A_2$), Im($A_0$) and Im($A_2$) in Tables
\ref{tab:ReA0_ReA2_mom1.tab} to \ref{tab:ImA0_ImA2_mom4.tab}.  Here the
contribution to $ \langle \pi\pi_{(I)} | \, -i \DSoneH \, | K^0
\rangle$ is decomposed into contributions for each value of the index
$i$ in Eqs.\ \ref{eq:complete_kpipi} and \ref{eq:complete_kpipi_log}.
We will refer to this as the full contribution to $ \langle
\pi\pi_{(I)} | \, -i \DSoneH \, | K^0 \rangle$ from the continuum
operator $Q_{i, \rm cont}$.  These tables use the central values for
Standard-Model parameters given in Table \ref{tab:sm_values}.  The
matching scale $\mu$ is 1.51 GeV for Tables
\ref{tab:ReA0_ReA2_mom1.tab} and \ref{tab:ImA0_ImA2_mom1.tab}, 2.13 GeV
for Tables \ref{tab:ReA0_ReA2_mom2.tab} and
\ref{tab:ImA0_ImA2_mom2.tab}, 2.39 GeV for Tables
\ref{tab:ReA0_ReA2_mom3.tab} and \ref{tab:ImA0_ImA2_mom3.tab} and 3.02
GeV for Tables \ref{tab:ReA0_ReA2_mom4.tab} and
\ref{tab:ImA0_ImA2_mom4.tab}.  It should be noted that the continuum
operators mix when this scale is changed, so the decomposition of the
physical amplitudes into particular $Q_{i, \rm cont}$ contributions
will change.  Only the complete amplitude should be insensitive to
scale and this will only occur if the Wilson coefficients and
non-perturbative renormalization factors are known to all orders in
$\alpha_S$.  In addition, we always use $Z_q(\mu)$ for $\mu = 2.0$ GeV
in the matching, since in the determination of $Z_q(\mu)$ the running
effects were found to be quite small \cite{Blum:2001sr}.   (The
one-loop anomalous dimension for $Z_q$ vanishes in Landau gauge.) The
scale dependence of our results will be an important test of our
calculation.

Results for the two choices for extrapolation discussed in Section
\ref{sec:phys_me} are given in Tables \ref{tab:ReA0_ReA2_mom1.tab} to
\ref{tab:ImA0_ImA2_mom4.tab}.  The first choice, a 0-loop extrapolation
in quenched QCD, and the second, a 1-loop extrapolation in full QCD,
differ by no more than $\sim 40$\%, except for the contributions to
$A_0$ coming from $Q_{7, \rm cont}$ and $Q_{8, \rm cont}$.  These
contributions change by almost a factor of two, due to the large
coefficient of the chiral logarithm term.  As we will see, these play
no role in our final results, due to the small size of $\DIhalf$
effects from electroweak penguin operators compared to the $\DIhalf$
effects from exchange and gluon penguin operators.  Table
\ref{tab:phys_parm_charm_out_vs_choice_mu_2_13} shows the values for
Re($A_0$), Re($A_2$) and Re($A_0$)/Re($A_2$) = $1/\omega$ for the two
extrapolation choices for $\mu = 2.13$ GeV.  In addition, we plot
Re($A_0$), Re($A_2$) and Re($A_0$)/Re($A_2$) = $1/\omega$ for $\mu =
2.13$ GeV in Figures \ref{fig:Re_A0}, \ref{fig:Re_A2} and
\ref{fig:ReA0_over_ReA2} as a function of a parameter $\xi$, which we
introduce into Eqs.\ \ref{eq:complete_kpipi} and
\ref{eq:complete_kpipi_log} by replacing all the squared pseudoscalar
masses $m_{PS}^2$ by $ \xi m_{PS}^2$.  The chiral limit is
given by $\xi = 0$ and the physical point corresponds to $\xi = 1$.
The experimental values are given by the filled triangles. The
difference between the two extrapolations gives an indication of the
contribution expected from including all the ${\cal O}(p^4)$ terms, rather
than just the logarithms.  We comment that the dependence of the chiral
logarithms on the scale $\Lcpt$ must be canceled by a similar
dependence in the ${\cal O}(p^4)$ coefficients.

Starting with Re($A_0$) and its dependence as a function of $\xi$ shown
in Figure \ref{fig:Re_A0}, we see that the chiral logarithms are
producing a 42\% change in the value at the physical point.  Given this
large correction, the close agreement between our choice 2 value of
$2.96(17) \times 10^{-7}$ GeV and the experimental value of $3.33
\times 10^{-7}$ GeV must be viewed as coincidental, but it is
encouraging that the chiral logarithms move the quenched theoretical
prediction closer to the experimental value.  Similar consideration of
Re($A_2$) and Figure \ref{fig:Re_A2} shows that inclusion of the chiral
logarithms only changes the extrapolated value by 18\%, also in the
direction of the experimental value.  Our choice 2 extrapolation value
of $1.172(53) \times 10^{-8}$ GeV is 22\% below the experimental value
of $1.50 \times 10^{-8}$ GeV.

For Re($A_0$)/Re($A_2$), the differences in the extrapolations are
smaller.  The chiral logarithms for the (8,1) and (27,1) operators
which dominate Re($A_0$) and Re($A_2$), respectively, have the same
sign but different amplitudes.  From Figure \ref{fig:ReA0_over_ReA2},
it is readily apparent that the logarithms have little effect on the
answer and it is in good agreement with the experimental value of
22.2.

We choose to quote as our best estimates for Re($A_0$), Re($A_2$) and
Re($A_0$)/Re($A_2$) the values using the choice 2 extrapolation (1-loop
full QCD).  This extrapolation includes the most information currently
available for corrections to lowest order chiral perturbation theory,
but is not a complete higher order calculation.  The value of $\mu$ to
use for our final answer should, in principle, not matter.  However,
for $\mu = 1.51$ GeV, non-perturbative low-energy QCD effects could be
causing a systematic shift in the values for $Z^{\rm NPR}_{ij}$.  For
$\mu =3.02$ GeV, finite lattice spacing effects could begin to play a
role.  In Table \ref{tab:choice2_phys_parm_charm_out} we give the $\mu$
dependence of our results.  For Re($A_0$) and Re($A_2$), the $\mu$
dependence is plotted in Figure \ref{fig:ReA0_ReA2_mu_dep}, while for
Re($A_0$)/Re($A_2$) the $\mu$ dependence is plotted in Figure
\ref{fig:omega_epe_mu_dep}.  No statistically significant $\mu$
dependence is seen, so choosing to quote results at $\mu = 2.13$ GeV,
where systematic effects should be smallest, does not alter the quoted
values

Our final results for Re($A_0$), Re($A_2$) and Re($A_0$)/Re($A_2$) for
the choice 2 extrapolation (1-loop full QCD chiral perturbation theory)
with $\mu = 2.13$ GeV are given in Table
\ref{tab:choice2_mu_2_13_charm_out_vs_exp}.  Figure
\ref{fig:ReA0_ReA2_break.eps} shows a breakdown of the contribution of
$Q_{i, \rm cont}$ to Re($A_0$) (upper panel) and Re($A_2$) (lower
panel).  The solid filled bars in the graph denote positive quantities
and the hashed represent negative quantities.  One clearly sees that
the dominant contributions are from $Q_{i, \rm cont}$ for $i = 1,2$.
The good agreement with experiment is very encouraging, although better
than might be expected given the approximations inherent in the current
calculation.

We end this section with our results for the kaon $B$ parameter, $B_K$,
discussed in Section \ref{subsec:theory_to_exp} and defined in Eq.\
\ref{eq:bk_def}.  In the $SU(3)$ flavor limit, one has
\begin{eqnarray}
    \langle \Kbar^0 | \DStwoQ_{\rm lat} | K^0 \rangle
  &
    =
  &
    3\, \langle\pi^+|[Q_1+Q_2]^{(3/2)}_{\rm lat} |K^+\rangle
  \\ \nonumber
  &
    =
  &
    2\, \langle\pi^+|\Theta^{(27,1),(3/2)}_{\rm lat} |K^+\rangle.
\end{eqnarray}
For the determination of $B_K$, we need $\langle \pi^+ |
\Theta^{(27,1),(3/2)}_{\rm lat} | K^+ \rangle$ at $m_f=0.018$, a quark
mass which gives a kaon made from degenerate quarks its physical mass.
This matrix element has been fit to the form given in
Eq.\ \ref{eq:theta271_32_corr_mpi} with the fit parameters given in the
second line of Table~\ref{tab:theta271_32_corr_mpi_fit}.  To convert
from the lattice matrix element to one with a continuum $\overline{\rm
MS}$ normalization we use $Z_{Q^{\Delta S=2}}(2\,{\rm
GeV})/Z_A^2=0.928$ \cite{Npr:2001zz}, $Z_A=0.7555$~\cite{Blum:2000kn}
and the one loop matching between the $RI$ and $\overline{MS}$ schemes
from \cite{Crisafulli:1996ad}.  This one-loop matching has a value of
1.02 in this case.

To complete the determination of $B_K$, values for $m_K$ and $f_K$ are
needed (Eq.\ \ref{eq:bk_def}).  Although $f_K$ and $m_K$ are given in
Tables XIX and XXXI in \cite{Blum:2000kn}, the current calculation
contains 400 configurations compared to the 85 of \cite{Blum:2000kn},
producing a reduced statistical error.  To extract $f_K$ and $m_K$, we
simultaneously fit wall-wall pseudoscalar correlators and wall-point
pseudoscalar axial-current correlators to determine the pseudoscalar
mass, $\langle 0 | P_{K^+, {\rm wall}} | K^+ \rangle$ and $\langle 0 |
A_{0, {\rm pt}} | K^+ \rangle$.  The fits use correlators a distance $t
= 12$ to 19 from the wall source and, as can be seen in Figure
\ref{fig:ss_over_pp}, in this range zero mode effects should be small.

From the 400 configuration data set of this work, the values we find
for $m_{PS}$ and $f_{PS}$ are given in Table \ref{tab:B_PS} for $m_f =
0.01$ to 0.05.  (Here the subscript $PS$ added to the mass, decay
constant and $B$ parameter is a label for a generic pseudoscalar meson
which could be the $\pi$, $K$, etc.) Including the determinations of
$m_{PS}$, $f_{PS}$ and the fit to
$\langle\pi^+|\Theta^{(27,1),(3/2)}_{\rm lat} |K^+\rangle$ under a
jackknife loop produces the values for $B_{PS}^{\rm wall}$ in the fifth
column of Table \ref{tab:B_PS}.  Adding an interpolation to $m_f =
0.018$ in the jackknife loop, we find $B_{K,\overline{MS}}^{\rm
wall}(2\,{\rm GeV})=0.532(11)$ where the error is statistical only.  We
can also calculate the value in the chiral limit, $m_f = -\mres$, and
this gives $B_{PS,\overline{MS}}^{\rm wall}(2\,{\rm GeV}) (m_f =
-\mres) = 0.267(14)$.  This method of extracting $B_K$ using wall-wall
correlators for the matrix element avoids introducing zero mode effects
through normalization factors in a similar fashion to the techniques we
have used in the analysis of $K \to \pi$ matrix elements.

This result agrees within errors with the value 0.538(8) that we
obtained on a subset of 200 configurations from the present
ensemble~\cite{Blum:2000tt}.  There the traditional method of
calculating $B_K$ was used, where a $\Delta S = 2$ Green's function is
normalized with axial current-pseudoscalar correlators.  This ratio of
Green's functions should be free of quenched chiral logarithms, but the
axial current correlator can introduce zero modes for small quark
masses.  We revisit this earlier determination here with the full 400
configuration data set and use $B_{PS}^{\rm AA}$ to denote the result
from this method.  To get the value in the $\overline{MS}$ scheme at
$\mu = 2$ GeV, the lattice $\Delta S = 2$ matrix element must be
multiplied by the value of $Z_{Q^{\Delta S=2}}/Z_A^2$ given above and a
factor of 1.02, which is the value for the one-loop matching between
the $RI$ and $\overline{MS}$ schemes.  In the fourth column of Table
\ref{tab:B_PS}, our values for $B_{PS}^{\rm AA}$ in the $\overline{MS}$
scheme at $\mu = 2$ GeV are given for $m_f = 0.01$ to 0.05.  Fitting
this data to the form given by 1-loop quenched chiral perturbation
theory for degenerate mesons \cite{Sharpe:1992ft,Golterman:2000fw},
\begin{eqnarray}
  B_{PS}^{\rm AA} &=&
  b_0^{\rm AA}\left(1.0 - \frac{1}{(4\pi f)^2}
    \left( 6\, m_{PS}^2\,\log{\frac{m_{PS}^2}{\Lqcpt^2}}
    \right)\right) + b_1^{\rm AA} m_{PS}^2,
  \label{eq:B_PS}
\end{eqnarray}
we find
\begin{equation}
  B_{K,\overline{MS}}^{\rm AA}(2\,{\rm GeV}) = 0.536(6)
\end{equation}
which is in very good agreement with our earlier result on 200
configurations.  The fit gives $b_0^{\rm AA} = 0.285(4)$ and $b_1^{\rm
AA} = 1.44(6)$ and we note that $b_0^{\rm AA}$ is the chiral limit
value for the unrenormalized $B$ parameter using this method.
Including renormalization factors yields $B_{PS,\overline{MS}}^{\rm
AA}(2\,{\rm GeV}) (m_f = -\mres) = 0.270(4)$.  The details of the fit
are the same as those following Eq.\ \ref{eq:theta271_32_corr_mpi} in
Section~\ref{sec:i_3over2_me} where the extraction of
$\alpha^{(27,1),(3/2)}_{\rm lat}$ was discussed.  The one difference is
that no quenched chiral logarithm appears in Eq.\ \ref{eq:B_PS}, since
they cancel in the ratio of the matrix element and its vacuum
saturation approximation.

The two methods described above have produced quite similar results for
$B_K$; 0.536(6) using the axial current-pseudoscalar normalization and
0.532(11) using the wall-wall correlator normalization.  The results
given in Table \ref{tab:B_PS} and plotted in Figure \ref{fig:B_PS} show
very good agreement in $B_{PS}^{\rm AA}$ and $B_{PS}^{\rm wall}$ for
$m_f = 0.03$, 0.04 and 0.05.  For smaller quark masses, differences at
the one standard deviation level occur.  Since both analysis methods
use the same raw data, the difference may be correlated and have
statistical significance, but we have not pursued this question.  The
solid line in Figure \ref{fig:B_PS} is a fit of $B_{PS}^{\rm AA}$ to
the form given in Eq.\ \ref{eq:B_PS}.  This fit goes below the $m_f =
0.01$ data point and agrees well in the chiral limit with the value
determined from the wall normalization.  Since the non-linearity due to
the chiral logarithm is quite pronounced near the chiral limit, the
extrapolation to this limit likely requires further study to understand
all the systematic effects.

Because the wall-wall normalization does not introduce additional zero
mode effects and it is the technique we have used for all the $K \to
\pi$ matrix elements, we will use the results from this approach for
our final values.  Therefore, for our single lattice spacing ($a^{-1} =
1.922(40)$ GeV) and volume ($16^3 \times 32$) we find
$B_{K,\overline{MS}}(2\,{\rm GeV})=0.532(11)$ and the value for
$B_{PS}$ in the chiral limit is 0.267(14).  Our value for $B_K$ is
smaller than that found in Ref.~\cite{AliKhan:2001wr} using the
RG-improved gauge action of Iwasaki at a similar lattice spacing and
volume.  In the $\overline{MS}$ scheme at 2 GeV they find $B_K(q^{*} =
1/a) = 0.564(14)$, where perturbation theory has been used to determine
the renormalization factors.  (The dependence of their result on $q^*$
is smaller than the error quoted above.)  The two central values differ
by about 6\%, while the quoted errors are about 2-3\%.  The difference
between the calculations is small and could be merely a statistical
fluctuation.  However, there are also systematic differences between
the two calculations that are not reflected in the statistical errors;
the gauge actions are different and Ref.~\cite{AliKhan:2001wr} uses
perturbation theory to calculate $Z$ factors, whereas this work has
used NPR.  The smaller value for $\mres$ for the Iwasaki action seems
unlikely to effect $B_K$, given that no power divergent operators are
involved and the large mass of the kaon.  For $Z$ factors for this
four-quark operator, a direct comparison has not been done, due to the
difference in gauge actions.  A direct comparison of perturbative and
non-perturbative $Z$ factors for quark bilinears was given in Ref.
\cite{Blum:2001sr} where agreement at the 5\% level was found between
mean field perturbative results and NPR.  This could be responsible for
much of the difference in the central values for $B_K$ with domain wall
fermions.

Our value for $B_{PS}$ in the chiral limit, 0.267(14), is markedly
lower than the value of 0.412 given in Ref.~\cite{AliKhan:2001wr} for
the chiral and continuum limit.  The continuum limit extrapolation in
Ref.~\cite{AliKhan:2001wr} is likely not responsible for this
difference; rather, it is the form of the extrapolation to the chiral
limit.  Our data uses the analytically calculated coefficient of the
chiral logarithm term as a fixed input parameter, whereas
Ref.~\cite{AliKhan:2001wr} fits for this coefficient, using data with
the AA normalization.  Their fit results in a coefficient $\sim 3$
times smaller than the analytic result, which will have a pronounced
effect on the chiral limit and explains much of the difference in the
two results.  We find our data is well fit using the analytically known
coefficient, provided zero mode effects are minimized, and have used
this coefficient consistently in both the determination of the (27,1)
$\DIthalf$ $K \to \pi$ matrix elements and the extrapolation of $B_{PS}$
to the chiral limit.

Our domain wall fermion result is more than one standard deviation
lower than the continuum limit quenched value of
0.628(42)~\cite{Aoki:1998nr} computed with Kogut-Susskind fermions.
For Kogut-Susskind fermions, large ${\cal O}(a^2)$ effects are seen.
Our result does not include an extrapolation to the continuum, but in
Ref.~\cite{AliKhan:2001wr} this extrapolation is done for domain wall
fermions, yielding 0.5746(61)(191).  Thus it appears that domain wall
fermions are giving a quenched value of $B_K$ about 10\% smaller than
the quenched value computed with Kogut-Susskind fermions and slightly
more than one standard deviation away.  Given the relatively small
statistical errors currently possible, these systematic differences
need to be reconciled.

\fi


\section{Imaginary $A_0$ and $A_2$}
\label{sec:imag_a0_a2}

\ifnum\theImagAZeroATwo=1
%
%


In the previous section, we saw that the results for the real $K
\rightarrow \pi \pi$ amplitudes from this single lattice
spacing, quenched calculation were quite consistent with the known
experimental values.  We now present our results for the imaginary $K
\rightarrow \pi \pi$ amplitudes and $\repe$.  These are all directly
proportional to the parameter $\eta$ in the CKM matrix and we will use
the central value for $\eta$ from Table \ref{tab:sm_values}.

Values for Im($A_0$) and Im($A_2$) are given in Tables
\ref{tab:ImA0_ImA2_mom1.tab}, \ref{tab:ImA0_ImA2_mom2.tab},
\ref{tab:ImA0_ImA2_mom3.tab} and \ref{tab:ImA0_ImA2_mom4.tab} for $\mu
= 1.51$, 2.13, 2.39 and 3.02 GeV, respectively.  The tables include
both extrapolation choices.  The values in the tables reflect the
long-standing expectation that the dominant part of Im($A_0$) is
produced by $Q_{6, \rm cont}$, although $Q_{4, \rm cont}$ is $\sim
35$\% of the size of $Q_{6, \rm cont}$ and of the opposite sign and
$Q_{8, \rm cont}$ is $\sim 10$\% of $Q_{6, \rm cont}$ and of the same
sign.  Since we choose to work in a a basis where $Q_{4, \rm cont}$ is linearly
dependent, most of its value is coming from $Z^{\rm NPR}_{41} \langle
\pi^+ | Q^{(1/2)}_{1, \rm lat} | K^+ \rangle_{\rm sub}$ and $Z^{\rm
NPR}_{42} \langle \pi^+ | Q^{(1/2)}_{2, \rm lat} | K^+ \rangle_{\rm
sub}$.  Since the values for $\alpha^{(1/2)}_{1, \rm lat}$ and
$\alpha^{(1/2)}_{2, \rm lat}$ in Table \ref{tab:final_alpha_charm_out}
have opposite sign and $Z^{\rm NPR}_{41}$ and $Z^{\rm NPR}_{42}$ also
have opposite sign, these contributions add in $Q_{4, \rm cont}$.
Finally we note that $y_4(\mu)$ and $y_6(\mu)$ are of similar size,
resulting in the sizable contribution of $Q_{4, \rm cont}$ to
Im($A_0$).  Im($A_2$) is dominated by $Q_{8, \rm cont}$ and receives
only $\sim 10$\% contributions from the next largest source, $Q_{9, \rm
cont}$.

The values for Im($A_0$) and Im($A_2$) and their dependence on the
choice of extrapolation to the physical kaon mass is given in Table
\ref{tab:phys_parm_charm_out_vs_choice_mu_2_13} for $\mu = 2.13$ GeV.
Figures \ref{fig:Im_A0} and \ref{fig:Im_A2} show Im($A_0$) and
Im($A_2$), respectively as a function of $\xi$ for $\mu = 2.13$ GeV.
We note that Im($A_0$) does not vanish as $\xi \rightarrow 0$, due to
the contribution from the electroweak penguins.  The chiral logarithms
change the extrapolated value of Im($A_0$) by 47\% and Im($A_2$) by
28\%.  The $\mu$ dependence of Im($A_0$) and Im($A_2$), using
extrapolation choice 2, is given in Table
\ref{tab:choice2_phys_parm_charm_out} and plotted in Figure
\ref{fig:ImA0_ImA2_mu_dep}.  The results for Im($A_0$) show no
statistically significant $\mu$ dependence, while Im($A_2$) varies by
about 25\% over this range of $\mu$.

We can now discuss our results for $\repe$.  Considering only the
contribution from the dominant operators $Q_2$, $Q_6$ and $Q_8$
(represented by $Q_2 \sim \alpha_2 m_{K^0}^2 \xi$ $Q_6 \sim \alpha_6
m_{K^0}^2 \xi$ and $Q_8 \sim \alpha_8$) and assuming $Z^{\rm
NPR}_{i,j}$ has small off-diagonal elements yields a schematic formula
for $\repe$ giving the rough size and mass dependence of the various
contributions.
\begin{equation}
    \repe
    =
    \left( \frac{\omega}{ \sqrt{2} | \epsilon | } \right)_{\rm exp}
    \left\{
    \left[ \frac{ \alpha_{\rm W} \, \alpha_8} 
      { \alpha_{\rm W} \, \alpha_8 + \alpha_2
	\, m_{K^0}^2 \,\xi }
    \right]^{(3/2)}
  - \left[ \frac{ \alpha_{\rm W} \, \alpha_8
      + \alpha_{\rm S} \, \alpha_6 \, m_{K^0}^2 \, \xi }
      { \alpha_{\rm W} \, \alpha_8
      + \alpha_2 \, m_{K^0}^2 \,  \xi }
    \right]^{(1/2)}
    \right\}
  \label{eq:epe_schematic}
\end{equation}
where $\alpha_W$ is the electroweak fine structure constant and
$\alpha_{\rm S}$ is for QCD.  Here we take $\omega$ and $|\epsilon|$
from experiment, since we will concentrate on the mass dependence of
$P_2 - P_0$.  Recalling the $\DIhalf$ rule gives
\begin{equation}
  \left[ \alpha_2 \, m_{K^0}^2 \,\xi  \right]^{(3/2)}
     = \omega
  \left[ \alpha_2 \, m_{K^0}^2 \,\xi  \right]^{(1/2)},
\end{equation}
which makes the $I = 3/2$ contribution at the physical point ($\xi =
1$) ${\cal O}(\alpha_{\rm W}/ \omega)$ rather than ${\cal
O}(\alpha_{\rm W})$.  Eq.\ \ref{eq:epe_schematic} shows that in the
chiral limit ($\xi = 0$), the electroweak penguins dominate Re($A_0$),
Re($A_2$), Im($A_0$) and Im($A_2$) and produce $\repe = 0$.  Since in
this limit, both the $I = 1/2$ and $I = 3/2$ amplitudes come from the
same source, there is no phase difference between them.  This limit is
quite different from the case with physical quark masses, where the
source of Im($A_0$) is primarily the gluonic penguins, Im($A_2$) the
electroweak penguins and Re($A_0$) and Re($A_2$) the exchange
operators.

To examine the $\xi$ dependence of $P_2 - P_0$ when the operators
important to the physically relevant case are noticeable,
we plot in Figure \ref{fig:epe_all} the quantity
\begin{equation}
  \left( \frac{\omega}{ \sqrt{2} | \epsilon | } \right)_{\rm exp}
    ( P_2(\xi) - P_0(\xi))
  \label{eq:epe_xi}
\end{equation}
starting at $\xi = 0.2$.  The data is for $\mu = 2.13$ GeV and we
remark that for $\xi = 1$, the quantity in Eq.\ \ref{eq:epe_xi} is
$\repe$.  One sees that for both extrapolation choices,
Eq.\ \ref{eq:epe_xi} starts out large and negative and becomes very
small for $\xi = 1$.  The large negative value arises when Re($A_2$) is
receiving very little contribution from the exchange operators and this
diminishes as Re($A_2$) grows with $\xi$.  For the 1-loop full QCD
extrapolation, we show the individual contributions $\repetwo$ and
$-\repezero$ in Figure \ref{fig:epe_choice2} for $\mu = 2.13$ GeV.
The contribution proportional to $P_2$ is going to zero with increasing
$\xi$ due to the increase in Re($A_2$).  The term proportional to
$-P_0$ is constant in lowest order chiral perturbation theory, once
$\xi$ is large enough that the electroweak penguins play no role, and
has no chiral logarithm corrections.  At the physical point $\xi = 1$,
the two terms are largely cancelling.  The $\mu$ dependence of $\repe$
is given in Table \ref{tab:choice2_phys_parm_charm_out} and plotted in
Figure \ref{fig:omega_epe_mu_dep}.  The $\mu$ dependence is coming
largely from the $\mu$ dependence of Im($A_2$).   We will take the
value for $\repe$ at $\mu =2.13$ GeV for our final result.

We can also study the contribution of the individual continuum
operators to the imaginary amplitudes entering $\repe$.  To do this, we
define $P_I^i$ by $P_I^i \equiv {\rm Im} \left( \langle (\pi \pi)_I |
-i \DSoneH | K^0 \rangle \right)_i/{\rm Re}(A_I)$, where the subscript
$i$ on the matrix element in the numerator means that only the
contribution from the renormalized continuum operator $Q_{i, \rm cont}$
is included.  Figure \ref{fig:epe_mom2_break.eps} shows a breakdown of
the contributions of $-\repezeroi$ (upper panel) and $\repetwoi$
(lower panel) to $\repe$ and Table \ref{tab:epe_mom2_break.tab} gives
the numerical values.  The solid filled bars in the graph denote
positive quantities and the hashed bars represent negative quantities.
This figure shows the importance of $Q_{4, \rm cont}$ and $Q_{6, \rm
cont}$ to $-\repezero$ and that $\repetwo$ comes primarily from
$Q_{8, \rm cont}$.

In spite of the near cancellation in $P_2 - P_0$ visible in Figure
\ref{fig:epe_choice2}, the statistical error on the final answer, $\pm
2.3 \times 10^{-4}$ is quite encouraging.  The figure also shows that
the magnitude of the contribution to $\repe$ from the term proportional
to $P_2$ is about the magnitude of the experimental value, as is also
true for $P_0$.  In Table \ref{tab:choice2_mu_2_13_charm_out_vs_exp} we
give our final values for the main physical quantities calculated in
this work.  Whether $\omega$ is taken from experiment or from this
calculation is not very significant in $\repe$, as can be seen from
Table \ref{tab:choice2_phys_parm_charm_out}.  Given the general
agreement with the experimental values for real $K \rightarrow \pi \pi$
amplitudes and the relatively small statistical error on $\repe$, the
difference between the current calculation for $\repe$ and experiment
is surprising.

\fi


\section{Conclusions}
\label{sec:conclusions}

\ifnum\theConclusions=1
%
%

We have reported the details and results of our calculation of the $K
\to \pi \pi$ matrix elements relevant for the $\DIhalf$ rule and $\epe$
in quenched lattice QCD using domain wall fermions.  In addition, we
have also reported a value for $B_K$, which is needed to determine
$\epsilon$ from the Standard Model.  Our value for $B_K$ is slightly
smaller than with other approaches, but the differences are at the 10\%
percent level.  Our results for Re($A_0$) and Re($A_2$) are $10-20$\%
smaller than experimental values, but our value for their ratio
is within 10\% of the experimental value.  This is a very encouraging
result, since a large enhancement of the $I=0$ amplitude is being seen
from the hadronic matrix elements, calculated using a technique where
the current approximations can be reduced in the future.  The
perturbative enhancement through the QCD running of the $I=0$ and $I=2$
Wilson coefficients is almost an order of magnitude smaller than the
experimentally observed enhancement.  Improvements of these
calculations will provide reliable systematic errors and fewer
approximations, leading to a more precise test of this initial
agreement between theory and experiment.

For $\epe$, the situation is more complex and more interesting.  Our
results quantitatively support the long standing expectation from
simple estimates that the two isospin contributions to $\epe$ are of
the same order and opposite sign.  Of course, such a large cancellation
may be dramatically altered by removing the approximations in the
current calculation.  While a subtraction of power divergences is
needed for Re($A_0$), it is quantitatively much smaller than the
subtraction for $Q_6$, which is the major contribution to Im($A_0$).
(No subtraction is required for the contributions to Im($A_2$).) As we
have shown, the dominant term in the subtraction procedure is not
affected by chiral logarithm and zero mode effects, making the
subtraction seem quite robust given our current understanding.
Thus, it appears that domain wall fermions, with their small chiral
symmetry breaking for finite lattice spacing, have removed the
problems found in earlier attempts where chiral symmetry breaking
effects were large.

The many approximations in this calculation could affect the real and
imaginary amplitudes in different ways, although at present we have no
insight into how this might occur.  We can estimate the size of the
effects introduced by the approximations acting singly.  The quenched
approximation has been generally found to agree with experimental
results at the $10-20$\% level, except for QCD near the finite
temperature phase transition where light quarks play a large role.  The
lowest order chiral perturbation theory results for the $K \to \pi \pi$
matrix elements are altered at the $\sim 30$\% level when the
extrapolation to the physical kaon mass includes the known chiral
logarithms.  We see a $\sim 25$\% variation in Im($A_i$) with the scale
$\mu$, which indicates the reliability of the combination of: using
continuum perturbation theory below 1.3 GeV, one-loop matching from the
NDR to RI schemes and our implementation of non-perturbative
renormalization where some operators, of order $g_S^2$ which are argued
to be small, are neglected.  We have used linear fits to our lattice
data in many cases, since analytic results for the chiral logarithm
terms are not known, and this could easily contribute errors on the 10\%
scale.  We have not included any effects of isospin breaking in our
results.  Finally, we have only worked at one lattice spacing, but our
experience with hadron masses calculated with domain wall fermions
makes it likely that changes of no more than 10\% will be encountered
in taking the continuum limit.

Each of these approximations could individually produce a $\sim 25$\%
change in Re($A_0$), Re($A_2$), Im($A_0$) or Im($A_2$).  Cumulatively,
these approximations could markedly alter our result for $\epe$, but
there is currently no identified single approximation that could easily
explain the discrepancy between our results and the experimental
value.  Lacking a single ``worst'' approximation to focus on we do not
have enough information at present to even estimate how these effects
act in concert for a quantity like $\epe$, which is the difference of
the ratio of amplitudes.  With further work, improved calculations
involving fewer approximations and reliable systematic errors will be
possible.

Removing the uncontrolled effects introduced by the quenched
approximation will simplify the calculation in addition to deleting a
significant possible systematic error.  The simplification comes from
the removal of the effects of unsuppressed zero modes present in
quenched QCD and the change from quenched chiral perturbation theory,
where new free parameters appear in the Lagrangian, to full or
partially quenched chiral perturbation theory.  A recent calculation in
quenched chiral perturbation theory \cite{Golterman:2001qj} has shown
that a quenched chiral logarithm appears in the determination of the
subtraction coefficient $\alpha_2^{(8,1)}$, multiplied by a new free
parameter.  From the linearity of our data with $m_s - m_d$, we
conclude that this parameter is small, but the presence of such terms
makes fitting to numerical results less precise and offers new ways in
which the quenched approximation can exhibit pathologies.

We have also calculated all the lattice matrix elements and
renormalization coefficients necessary to repeat the current
calculation in the context of the four-flavor effective low-energy
theory, where the charm quark is not integrated out.  For the
four-flavor theory, continuum perturbation theory need only be used to
a scale of $\sim 2$ GeV to match to our lattice.  This should decrease
the errors coming from the Wilson coefficients.  However, the quenched
lattice calculation is now required to well approximate full QCD
running between the scales of 2 GeV and $\sim 500$ MeV, the scale of
the kaon physics we are studying.  This will clearly be a worse
approximation than in the current calculation where the quenched
running must approximate full QCD only between 1.3 GeV and $\sim 500$
MeV.  Finally, in the four-flavor theory, operators with dimension
greater than six in the effective Lagrangian are suppressed by powers
of $\sim (0.5 \; {\rm GeV}/ 5.0 \; {\rm GeV})$ compared to powers of
$\sim (0.5 \; {\rm GeV}/ 1.3 \; {\rm GeV})$ in the current calculation.
The different systematic errors inherent in the use of the four-flavor
theory will provide insight into the stability of our current results
from the three-flavor theory.

We conclude by noting that attempts to use lattice QCD to calculate $K
\to \pi \pi $ matrix elements have been ongoing for almost 20 years.
The entire framework for successful calculations is in place and all
the current approximations can be steadily improved.  These
calculations rely on the continuum calculations of the Wilson
coefficients, which represents a very substantial effort.  The current
calculation demonstrates that: statistical errors are not a limiting
factor; the domain wall fermion formulation, in addition to being a
major theoretical advance, can be used in practical simulations and
that the complicated matching of continuum and lattice $\DSone$
operators can be done with non-perturbative renormalization and domain
wall fermions.  This presents a very exciting future for precise
calculations of experimentally important quantities using analytic
techniques and lattice QCD.

During the completion of this work, results from a similar study
were also reported \cite{Noaki:2001un}.

\fi


\ifnum\theAcknowledgments=1
\section*{Acknowledgments}

The authors would like to thank Andrzej Buras, Mike Creutz, Maarten
Golterman and Yigal Shamir for useful discussions.  We also acknowledge
use of the MILC collaboration software ({\tt
http://physics.indiana.edu/$\tilde{\,}$sg/milc.html}) for some of the
tests we performed on our computer programs.  We would like to thank
T.~D.~Lee for valuable scientific discussions and his support in all
phases of this work.

The calculations reported here were done on the 400 Gflops QCDSP
computer \cite{Chen:1998cg} at Columbia University and the 600 Gflops
QCDSP computer \cite{Mawhinney:2000fx} at the RIKEN-BNL Research
Center.  We thank the Information Technology Division at BNL for their
support, particularly the technical support staff led by Ed McFadden.
We thank RIKEN, Brookhaven National Laboratory and the U.S.\
Department of Energy for providing the facilities essential for the
completion of this work.

This research was supported in part by the DOE under grant \#
DE-FG02-92ER40699 (Columbia), in part by the NSF under grant \#
NSF-PHY96-05199 (Vranas), in part by the DOE under grant
DE-AC02-98CH10886 (Soni-Dawson), in part by the RIKEN-BNL Research
Center (Blum-Wingate-Ohta) and in part by the Max-Kade Foundation
(Siegert).
\fi


\appendix

\ifnum\theAppendix=1

\section{Conventions for States and Operators}
\label{sec:conventions}

Comparing the Lagrangian of chiral perturbation theory described in
\ref{subsec:lowest_order_XPT} with the Lagrangian of QCD defines the
relationship between quantities expressed in terms of the pseudoscalar
fields of chiral perturbation theory and the quark fields used in our
simulations.  Our conventions follow \cite{Bernard:1989nb}, where more
details can be found.  We start with the Lagrangian given in Eq.\
\ref{eq:lxpt_qcd} and the Minkowski space QCD Lagrangian
\begin{equation}
  {\cal L}_{\rm QCD} = -\frac{1}{4} (F_{\mu \nu}^a)^2
    + \overline{\psi} \left( i\slash{D} - m \right) \psi
\end{equation}
We use the conventional assignment of pseudoscalars to the chiral
perturbation theory fields
\begin{equation}
  \overline{\pi} \equiv \phi^a t^a = \left(
  \begin{array}{ccc}
      \pi^0/\sqrt{2}+\eta/\sqrt{6}
    &
      \pi^+
    &
      K^+
    \\
      \pi^-
    &
      -\pi^0/\sqrt{2}+\eta/\sqrt{6}
    &
      K\,^0
    \\
      K^-
    &
      \overline{K}\,^0
    &
      -2\eta/\sqrt{6}.		
  \end{array}
  \right)
\end{equation}
We work with relativistically normalized states
\begin{equation}
  \langle \pi^a (\vec{p}) | \pi^b ( \vec{p}^{ \,\prime} \rangle
  = \delta^{ab} \, (2 E_{\vec{p}}) \, ( 2 \pi)^3
    \, \delta^3( \vec{p} - \vec{p}^{\, \prime})
  @>>{\rm lattice}>
  \delta^{ab} \, (2 E_{\vec{p}}) \, V_s
  \, \delta_{\vec{p}, \vec{p}^\prime }
\end{equation}

By considering global axial transformations with $U_L = \exp(-ia_a t_a)$
and $U_R = \exp(i a_a t_a)$, we find for the axial currents $A_\mu^a$
\begin{alignat}{2}
    A^\mu_a = \;
  &
    \frac{i f^2}{4} \left[
    {\rm Tr} \left( \Sigma \, t_a \, \partial^\mu \Sigma^\dagger \right)
  - {\rm Tr} \left( \Sigma^\dagger \, t_a \, \partial^\mu \Sigma \right)
    \right] \qquad
  &
    \cpt 
  \\
    A^\mu_a = \;
  &
    \overline{\psi} \gamma^u \gamma_5 t_a \psi
  &
    \text{QCD}
\end{alignat}
The divergence of the axial currents is
\begin{alignat}{2}
    \partial_\mu A^\mu_a = \;
  &
    iv \; {\rm Tr}  \left[ t_a \right( \{ M, \Sigma \}
      - \{ M^\dagger, \Sigma^\dagger \} \left) \right] \qquad
  &
    \cpt \label{eq:cpt_axial_div}
  \\
    \partial_\mu A^\mu_a = \;
  &
    2m \left[ i \overline{\psi} \gamma_5 t_a \psi \right]
  &
    \text{QCD}
\end{alignat}
For degenerate quark masses, Eq.\ \ref{eq:cpt_axial_div} becomes
\begin{equation}
    \partial_\mu A^\mu_a =
    2imv \; {\rm Tr}  \left[ t_a ( \Sigma - \Sigma^\dagger ) \right]
    \qquad
    \cpt
\end{equation}
Thus, in lowest order in chiral perturbation theory, we can  
make the associations
\begin{alignat}{2}
    i \bar{d}(x) \gamma_5 u(x) = 
  &
    \; i \overline{\psi}(x) \gamma_5 \frac{t_1 + i t_2}
   {\sqrt{2}} \psi(x)
  &
    \quad \Longleftrightarrow \quad
  &
    \frac{-4v}{f} \pi^+(x)
  \label{eq:cpt_to_q_pi}
  \\
    i \bar{s}(x) \gamma_5 u(x) = 
  &
    \; i \overline{\psi}(x) \gamma_5 \frac{t_4 + i t_5}
   {\sqrt{2}} \psi(x)
  &
    \quad \Longleftrightarrow \quad
  & 
    \frac{-4v}{f} K^+(x)
  \label{eq:cpt_to_q_K}
\end{alignat}
States $|K^+ \rangle$ created by the operator $K^-(x)$ therefore
have
\begin{equation}
  \langle 0| \, i \bar{s} \gamma_5 u \, | K^+ \rangle = -\frac{4v}{f}
\end{equation}
and to lowest order in chiral perturbation theory
\begin{equation}
  \langle 0| \, \bar{d}(x) \gamma^u \gamma_5 u(x) \, | \pi^+ \rangle
  = -i f p^\mu e^{-ip \cdot x}
  \label{eq:axial_current_me}
\end{equation}
where $f > 0 $.

We define a pseudoscalar density in chiral perturbation theory by
\begin{equation}
  P^{\cpt}_a \equiv - \frac{4v}{f} \phi_a
\end{equation}
and a corresponding QCD pseudoscalar density as
\begin{equation}
  P^{\rm QCD}_a \equiv i \overline{\psi} \gamma_5 t_a \psi
\end{equation}
Then for degenerate quark masses, the Minkowski space Ward-Takahashi
identify governing the pseudoscalar masses is
\begin{alignat}{3}
    i \partial_\mu^x \langle A^\mu_a(x) \; P_b(y) \rangle
  &
    \; =
  &
    2mi \langle P_a(x) P_b(y) \rangle
  &
    \; - 4 v \delta_{a,b} \delta^4(x-y)
  &
     \qquad\cpt
  \\
    i \partial_\mu^x \langle A^\mu_a(x) \; P_b(y) \rangle
  &
    \; =
  &
    2mi \langle P_a(x) P_b(y) \rangle
  &
    \; + 2 \langle \bar{u} u(x) \rangle \delta_{a,b} \delta^4(x-y)
  &
    \qquad \text{QCD}
\end{alignat}
where the chiral perturbation theory result is valid in lowest order.
Here we see the relation $\langle \bar{u} u \rangle = -2v$ between the
chiral condensate in QCD and in chiral perturbation theory.


\section{Flavor and Isospin Decomposition of Four-quark Operators}
\label{sec:four_quark_irreps}

As discussed in \cite{Bernard:1989nb}, one can apply the tensor method
for finding irreducible representations of groups to the operators in
Eqs.\ \ref{eq:Q1} to \ref{eq:P10}.  We start first with the left-left
operators and note the general term $\bar{q}_{L,i} \bar{q}_{L,j}
q_{L,k} q_{L,l}$, where $i,j,k$ and $l$ are flavor indices, is a member
of a representation of $SU(3)_L$ with dimension 81.  Denoting this term
by $(T_L)^{i,j}_{k,l}$, the irreducible representations are found by
appropriately symmetrizing $T_L$.
\begin{center}
\begin{tabular}{c|cccc}
  Symmetry of $T_L$
  & $ (T_L)^{\{i,j\}}_{\{k,l\}} $
  & $ (T_L)^{\{i,j\}}_{ [k,l] } $
  & $ (T_L)^{ [i,j] }_{\{k,l\}} $
  & $ (T_L)^{ [i,j] }_{ [k,l] } $
  \\ \hline
  Dimension
  & 36
  & 18
  & 18
  & 9
  \\
  Irrep. Dimension
  & 27,8,1
  & 8,8,1,1
  & 8,8,1,1
  & 8,1
\end{tabular}
\end{center}
The irreducible representations in the last line are found by tracing
on pairs of upper and lower indices.  For example, the 27
representation is completely symmetric in all indices and traceless on
any pair of upper and lower indices, while the completely symmetric
representation, which has a non-zero trace, is dimension 8.

We can now determine the number of irreducible representations that
$Q_1 = (\bar{s} d)_{V-A} (\bar{u} u)_{V-A}$ enters.  Here we will
suppress the color indices and only consider the color unmixed case,
so the terms in parentheses will have their color indices contracted
together.  Since $(\bar{s} d)_{V-A}$ and $(\bar{u} u)_{V-A}$ commute
with each other, left-left four quark current operators are symmetric
under simultaneous exchange of quark and anti-quark indices.  Thus,
left-left operators must belong to $ (T_L)^{\{i,j\}}_{\{k,l\}} $ or $
(T_L)^{ [i,j] }_{ [k,l] } $ and they have either $(L,R) = (8,1)$ or
$(L,R) = (27,1)$.  We will also want to simultaneously separate the
operators into representations of definite isospin.

The operator
\begin{equation}
  (\bar{s} d)_{V-A} (\bar{u} u)_{V-A} +
  (\bar{s} u)_{V-A} (\bar{u} d)_{V-A}
  \label{eq:Q1start}
\end{equation}
is completely symmetric on all indices.  To get a $(27,1)$ with
$I=3/2$, we must add terms so it is simultaneously traceless in
$SU(3)_L$ and isospin.  Eq.\ \ref{eq:Q1start} has $(T_L)^{3,1}_{2,1} =
(T_L)^{3,1}_{1,2} = (T_L)^{1,3}_{2,1} = (T_L)^{1,3}_{1,2} = 1/2$, so if
we add $(T_L)^{3,2}_{2,2} = (T_L)^{2,3}_{2,2} = -1/2$ with all other
elements zero, we have tracelessness in $SU(3)_L$ and isospin.  Thus,
we have for left-left operators, symmetric in all indices, a (27,1)
representation with $I = 3/2$ given by
\begin{equation}
  Q^{\Delta s = 1, \Delta d = -1}_{LL,S,(27,1),3/2}
  =
  (\bar{s} d)_{V-A} (\bar{u} u)_{V-A} +
  (\bar{s} u)_{V-A} (\bar{u} d)_{V-A} -
  (\bar{s} d)_{V-A} (\bar{d} d)_{V-A}
  \label{eq:twenty_seven_three_half}
\end{equation}
Returning again to \ref{eq:Q1start} we can find the $I=1/2$ operator
by making \ref{eq:Q1start} symmetric under $u \leftrightarrow d$
and then making the results traceless on pairs of upper and lower
indices.  This gives
\begin{eqnarray}
  Q^{\Delta s = 1, \Delta d = -1}_{LL,S,(27,1),1/2}
  & = &
    (\bar{s} d)_{V-A} (\bar{u} u)_{V-A} +
    (\bar{s} u)_{V-A} (\bar{u} d)_{V-A} \nonumber \\
  & + &
  2 (\bar{s} d)_{V-A} (\bar{d} d)_{V-A} -
  3 (\bar{s} d)_{V-A} (\bar{s} s)_{V-A}
  \label{eq:twenty_seven_one_half}
\end{eqnarray}
corresponding to $(T_L)^{3,1}_{2,1} = (T_L)^{3,1}_{1,2} =
(T_L)^{1,3}_{2,1} = (T_L)^{1,3}_{1,2} = 1/2$, $(T_L)^{3,2}_{2,2} =
(T_L)^{2,3}_{2,2} =  1$ and $(T_L)^{3,3}_{2,3} = (T_L)^{3,3}_{3,2} =
-3/2$, with other elements zero.

For the $(8,1)$ from $(T_L)^{\{i,j\}}_{\{k,l\}}$  we start again from
\ref{eq:Q1start}, again symmetrizing \ref{eq:Q1start} under $u
\leftrightarrow d$ to get $I=1/2$.  However, demanding that
the operator not be traceless on contraction of upper and lower
indices while still being symmetric on exchange of upper or lower
indices gives
\begin{eqnarray}
  Q^{\Delta s = 1, \Delta d = -1}_{LL,S,(8,1),1/2}
  & = &
    (\bar{s} d)_{V-A} (\bar{u} u)_{V-A} +
    (\bar{s} u)_{V-A} (\bar{u} d)_{V-A} \nonumber \\
  & + &
  2 (\bar{s} d)_{V-A} (\bar{d} d)_{V-A} +
  2 (\bar{s} d)_{V-A} (\bar{s} s)_{V-A}
\end{eqnarray}
The final (8,1) comes from $ (T_L)^{[i,j]}_{[k,l]} $, which
is antisymmetric on pairs of upper and lower indices, and
is easily seen to be
\begin{equation}
  Q^{\Delta s = 1, \Delta d = -1}_{LL,A,(8,1),1/2} =
  (\bar{s} d)_{V-A} (\bar{u} u)_{V-A} -
  (\bar{s} u)_{V-A} (\bar{u} d)_{V-A}
\end{equation}

Thus, we have found that there are three irreducible representations of
left-left, $\Delta s = 1, \Delta d = -1$ four-quark operators under
$SU(3)_L \otimes SU(3)_R$; a (27,1) and two (8,1) representations.  The
(27,1) contains both $I=1/2$ and $I=3/2$ parts.  We can write $Q_1$,
$Q_2$, $Q_3$, $Q_4$, $Q_9$ and $Q_{10}$ in terms of these
representations, yielding
\begin{alignat}{5}
    Q_1
  & = \quad
      \frac{1}{10} & Q^{\Delta s = 1, \Delta d = -1}_{LL,S,(8,1),1/2}
  & + \frac{1}{2}  & Q^{\Delta s = 1, \Delta d = -1}_{LL,A,(8,1),1/2}
  & + \frac{1}{15} & Q^{\Delta s = 1, \Delta d = -1}_{LL,S,(27,1),1/2}
  & + \frac{1}{3}  & Q^{\Delta s = 1, \Delta d = -1}_{LL,S,(27,1),3/2}
  \label{eq:Q1_irrep}
  \\ 
    Q_2
  & = \quad
      \frac{1}{10} & Q^{\Delta s = 1, \Delta d = -1}_{LL,S,(8,1),1/2}
  & - \frac{1}{2}  & Q^{\Delta s = 1, \Delta d = -1}_{LL,A,(8,1),1/2}
  & + \frac{1}{15} & Q^{\Delta s = 1, \Delta d = -1}_{LL,S,(27,1),1/2}
  & + \frac{1}{3}  & Q^{\Delta s = 1, \Delta d = -1}_{LL,S,(27,1),3/2}
  \label{eq:Q2_irrep}
  \\ 
    Q_3
  & = \quad
      \frac{1}{2}  & Q^{\Delta s = 1, \Delta d = -1}_{LL,S,(8,1),1/2}
  & + \frac{1}{2}  & Q^{\Delta s = 1, \Delta d = -1}_{LL,A,(8,1),1/2}
  &  &
  &  &
  \\ 
    Q_4
  & = \quad
      \frac{1}{2}  & Q^{\Delta s = 1, \Delta d = -1}_{LL,S,(8,1),1/2}
  & - \frac{1}{2}  & Q^{\Delta s = 1, \Delta d = -1}_{LL,A,(8,1),1/2}
  &  &
  &  &
  \\ 
    Q_9
  & = - \frac{1}{10} & Q^{\Delta s = 1, \Delta d = -1}_{LL,S,(8,1),1/2}
  & + \frac{1}{2}  & Q^{\Delta s = 1, \Delta d = -1}_{LL,A,(8,1),1/2}
  & + \frac{1}{10} & Q^{\Delta s = 1, \Delta d = -1}_{LL,S,(27,1),1/2}
  & + \frac{1}{2}  & Q^{\Delta s = 1, \Delta d = -1}_{LL,S,(27,1),3/2}
  \label{eq:Q9_irrep}
  \\ 
    Q_{10}
  & = - \frac{1}{10} & Q^{\Delta s = 1, \Delta d = -1}_{LL,S,(8,1),1/2}
  & - \frac{1}{2}  & Q^{\Delta s = 1, \Delta d = -1}_{LL,A,(8,1),1/2}
  & + \frac{1}{10} & Q^{\Delta s = 1, \Delta d = -1}_{LL,S,(27,1),1/2}
  & + \frac{1}{2}  & Q^{\Delta s = 1, \Delta d = -1}_{LL,S,(27,1),3/2}
  \label{eq:Q10_irrep}
\end{alignat}

For left-right operators, we can perform a similar construction.
For the gluonic penguins, the right-handed currents are singlets
under $SU(3)_R$ due to the sum over $u,d$ and $s$ quarks, with
equal weight for each quark.  Including the charm quark still produces
an $(8,1)$ since the charm quark is also an $SU(3)_R$ singlet.

For the left-right electroweak penguins, a bit more work is required.
Now we have three representation matrices for each operator,
$(T_L)^i_j$, $(T_R)^k_l$, and $(T_I)^{k}_{j,l}$, for $SU(3)_L$,
$SU(3)_R$ and isospin, respectively.  For the isospin case,
we restrict $j,k$ and $l$ to be 1 or 2. Notice that both left- and
right-handed quarks appear in the $T$ for isospin and to get the
desired isospin decomposition, we will have to symmetrize,
anti-symmetrize and trace on these indices.  To get (8,8)
representations, we must have $(T_L)^i_i = 0$ and $(T_R)^k_k = 0$.

We start with a part of $Q_7$ and see how many irreducible
representations it enters by appropriate symmetrizations, etc.\ on
the quarks.  The first term in $Q_7$ is
\begin{equation}
  (\bar{s} d)_{V-A} (\bar{u} u)_{V+A}
  \label{eq:Q7start}
\end{equation}
To make an $I=3/2$ operator $(T_I)^{k}_{j,l}$ must be symmetric
on $j$ and $l$ and traceless on $k$ and either $j$ or $l$.  Symmetrizing
gives
\begin{equation}
  (\bar{s} d)_{V-A} (\bar{u} u)_{V+A} +
  (\bar{s} u)_{V-A} (\bar{u} d)_{V+A}
\end{equation}
and tracelessness in both isospin and $SU(3)_R$ gives
\begin{equation}
  Q^{\Delta s = 1, \Delta d = -1}_{LR,(8,8),S,3/2}
  =
  (\bar{s} d)_{V-A} (\bar{u} u)_{V+A} +
  (\bar{s} u)_{V-A} (\bar{u} d)_{V+A} -
  (\bar{s} d)_{V-A} (\bar{d} d)_{V+A}
  \label{eq:eight_eight_three_half}
\end{equation}
From Eq.\ \ref{eq:Q7start} we can make an $I=1/2$ operator by putting
the quarks in an $I=1$ state and then adding the antiquark such that
the total isospin is 1/2.  We symmetrize $(T_I)^{k}_{j,l}$ on $j$ and
$l$ and require that $(T_I)^{1}_{j,1} = (T_I)^{2}_{j,2}$ to get
isospin 1/2.  This yields
\begin{equation}
  (\bar{s} d)_{V-A} (\bar{u} u)_{V+A} +
  (\bar{s} u)_{V-A} (\bar{u} d)_{V+A} +
  2 (\bar{s} d)_{V-A} (\bar{d} d)_{V+A}
\end{equation}
The last step requires tracelessness on only the $SU(3)_R$ index,
to give an $8_R$.  Thus, we get
\begin{eqnarray}
  Q^{\Delta s = 1, \Delta d = -1}_{LR,(8,8),S,1/2}
  & = &
    (\bar{s} d)_{V-A} (\bar{u} u)_{V+A} +
    (\bar{s} u)_{V-A} (\bar{u} d)_{V+A}
    \nonumber \\
  & + &
    2 (\bar{s} d)_{V-A} (\bar{d} d)_{V+A} -
    3 (\bar{s} d)_{V-A} (\bar{s} s)_{V+A}
\end{eqnarray}

From Eq.\ \ref{eq:Q7start} we can make a second $I=1/2$ operator by
putting the quarks in an $I=0$ state and then adding the antiquark.  We
anti-symmetrize $(T_I)^{k}_{j,l}$ on $j$ and $l$ and require that
$(T_R)^{k}_{l} = 0$ to produce an $8_R$.  This yields
\begin{eqnarray}
  Q^{\Delta s = 1, \Delta d = -1}_{LR,(8,8),A,1/2}
  & = &
    (\bar{s} d)_{V-A} (\bar{u} u)_{V+A} -
    (\bar{s} u)_{V-A} (\bar{u} d)_{V+A}
    \nonumber \\
  & - &
    (\bar{s} d)_{V-A} (\bar{s} s)_{V+A}
  \label{eq:eight_eight_one_half}
\end{eqnarray}

With these isospin representations of an (8,8) color unmixed
operator, we can write
\begin{equation}
  Q_7 =
  \frac{1}{2} Q^{\Delta s = 1, \Delta d = -1}_{LR,(8,8),S,3/2} +
  \frac{1}{2} Q^{\Delta s = 1, \Delta d = -1}_{LR,(8,8),A,1/2}
  \label{eq:Q7_isospin_decomp}
\end{equation}
The result for $Q_8$ is identical, except the color indices are
mixed.


\section{Definitions of $\Theta$ Operators}
\label{sec:def_theta_op}

In this section we give the relations between the $\Theta$
operators of chiral perturbation theory and the four-quark
operators defined in section \ref{sec:four_quark_irreps} of the
Appendix.  We define
\begin{alignat}{2}
    \Theta^{(27,1),(3/2)}
  &
    \; \equiv
  &
    Q^{\Delta s = 1, \Delta d = -1}_{LL,S,(27,1),3/2}
  &
    \quad \quad \quad {\rm (Eq. \; \ref{eq:twenty_seven_three_half})}
  \\
    \Theta^{(27,1),(1/2)}
  &
    \; \equiv
  &
    Q^{\Delta s = 1, \Delta d = -1}_{LL,S,(27,1),1/2}
  &
    \quad \quad \quad {\rm (Eq. \; \ref{eq:twenty_seven_one_half})}
  \\
    \Theta^{(8,8),(3/2)}_7
  &
    \; \equiv \; \frac{1}{2}
  &
    Q^{\Delta s = 1, \Delta d = -1}_{LR,(8,8),S,3/2}
  &
    \quad \quad \quad {\rm (Eq. \; \ref{eq:eight_eight_three_half})}
    \label{eq:theta_eight_eight_three_half}
  \\
    \Theta^{(8,8),(1/2)}_7
  &
    \; \equiv \; \frac{1}{2}
  &
    Q^{\Delta s = 1, \Delta d = -1}_{LR,(8,8),A,1/2}
  &
    \quad \quad \quad {\rm (Eq. \; \ref{eq:eight_eight_one_half})}
    \label{eq:theta_eight_eight_one_half}
\end{alignat}
The definitions for $\Theta^{(8,8),(3/2)}_8$ and
$\Theta^{(8,8),(1/2)}_8$ are the same as in Eqs.\
\ref{eq:theta_eight_eight_three_half} and
\ref{eq:theta_eight_eight_one_half}, except that the four-quark
operator has color mixed indices.  For $i = 7,8$ this gives
\begin{equation}
  Q_i = \Theta^{(8,8)}_i =
    \Theta^{(8,8),(3/2)}_i + \Theta^{(8,8),(1/2)}_i
\end{equation}
In terms of the parameters $\alpha^{(27,1)}$ and $\alpha^{(8,8)}$
defined in Eqs.\ \ref{eq:def_alpha_81}, \ref{eq:def_alpha_271} and
\ref{eq:def_alpha_88} we have
\begin{eqnarray}
    \Theta^{(27,1),(1/2)}
  &
    =
  &
    \alpha^{(27,1)} \tilde{\Theta}^{(27,1),(1/2)}
  \\
    \Theta^{(27,1),(3/2)}
  &
    =
  &
    \alpha^{(27,1)} \tilde{\Theta}^{(27,1),(3/2)}
  \\
    \Theta^{(8,8),(1/2)}
  &
    =
  &
    \alpha^{(8,8)} \tilde{\Theta}^{(8,8),(1/2)}
  \\
    \Theta^{(8,8),(3/2)}
  &
    =
  &
    \alpha^{(8,8)} \tilde{\Theta}^{(8,8),(3/2)}
\end{eqnarray}


\section{Isospin Decomposition of Operators in Chiral Perturbation
  Theory}
\label{sec:cpt_decomp}

In Appendix \ref{sec:four_quark_irreps} we have given the decomposition
of our $\DSone$, $\DDmone$ four-quark operators into irreducible
representations of $\sutlr$ with well-defined isospin.  In this
section, we give the explicit decomposition of the chiral perturbation
theory operators $\tilde{\Theta}^{(27,1)}$ and $\tilde{\Theta}^{(8,8)}$
into definite isospin components.  From this one can easily work out
the relations between the $\DIhalf$ and $\DIthalf$ parts of matrix
elements.

For $\tilde{\Theta}^{(8,8)}$, we use the definition in 
\cite{Bijnens:1984ye} and write
\begin{equation}
  \tilde{\Theta}^{(8,8)} = {\rm Tr}
  \left[
    \left(
      \begin{array}{ccc}
        0  &  0  &  0  \\
        0  &  0  &  0  \\
        0  &  1  &  0  \\
      \end{array}
    \right)
    \Sigma
    \left(
      \begin{array}{ccc}
        2  &  0  &  0  \\
        0  & -1  &  0  \\
        0  &  0  & -1  \\
      \end{array}
    \right)
    \Sigma^\dagger
  \right]
\end{equation}
The non-zero element of the first matrix in the equation above
reproduces the $\bar{s}d$ factor in equation \ref{eq:Q7} while
the diagonal terms in the second matrix represent the terms
in the sum over quarks in \ref{eq:Q7}.  The isospin decomposition
can be immediately read off from Eqs.\ \ref{eq:eight_eight_three_half}
and \ref{eq:eight_eight_one_half} giving
\begin{alignat}{2}
  \tilde{\Theta}^{(8,8),(3/2)} & \; = {\rm Tr}
  \left[
    \left(
      \begin{array}{ccc}
        0  &  0  &  0  \\
        0  &  0  &  0  \\
        0  &  1  &  0  \\
      \end{array}
    \right)
    \Sigma
    \left(
      \begin{array}{ccc}
        1  &  0  &  0  \\
        0  & -1  &  0  \\
        0  &  0  &  0  \\
      \end{array}
    \right)
    \Sigma^\dagger
  \right ] 
  + {\rm Tr}
  \left[
    \left(
      \begin{array}{ccc}
        0  &  0  &  0  \\
        0  &  0  &  0  \\
        1  &  0  &  0  \\
      \end{array}
    \right)
    \Sigma
    \left(
      \begin{array}{ccc}
        \; 0  &  \; 1  &  0  \\
        \; 0  &  \; 0  &  0  \\
        \; 0  &  \; 0  &  0  \\
      \end{array}
    \right)
    \Sigma^\dagger
  \right] \\ 
  \tilde{\Theta}^{(8,8),(1/2)} & \; = {\rm Tr}
  \left[
    \left(
      \begin{array}{ccc}
        0  &  0  &  0  \\
        0  &  0  &  0  \\
        0  &  1  &  0  \\
      \end{array}
    \right)
    \Sigma
    \left(
      \begin{array}{ccc}
        1  &  0  &  0  \\
        0  &  0  &  0  \\
        0  &  0  & -1  \\
      \end{array}
    \right)
    \Sigma^\dagger
  \right ]
   + {\rm Tr}
  \left[
    \left(
      \begin{array}{ccc}
        0  &  0  &  0  \\
        0  &  0  &  0  \\
        1  &  0  &  0  \\
      \end{array}
    \right)
    \Sigma
    \left(
      \begin{array}{ccc}
         0  & -1  &  0  \\
         0  &  0  &  0  \\
         0  &  0  &  0  \\
      \end{array}
    \right)
    \Sigma^\dagger
  \right] 
\end{alignat}
where $\tilde{\Theta}^{(8,8)} = \tilde{\Theta}^{(8,8),(1/2)}
+ \tilde{\Theta}^{(8,8),(3/2)}$.

With this explicit isospin decomposition, one finds
\begin{alignat}{3}
    \langle \pi^+ | \Theta^{(8,8),(3/2)} | K^+ \rangle
  &
    \; \equiv
  &
    \; \frac{12}{f^2} \alpha^{(8,8),(3/2)}
  &
    \; =
  &
    \; \frac{4}{f^2} \alpha^{(8,8)}
    \label{eq:def_alpha_88_32}
  \\
    \langle \pi^+ | \Theta^{(8,8),(1/2)} | K^+ \rangle
  &
    \; \equiv
  &
    \; \frac{12}{f^2} \alpha^{(8,8),(1/2)}
  &
    \; =
  &
    \; \frac{8}{f^2} \alpha^{(8,8)}
    \label{eq:def_alpha_88_12}
\end{alignat}
which yields $ \alpha^{(8,8),(1/2)} = 2 \alpha^{(8,8),(3/2)}$
where $ \alpha^{(8,8)} = \alpha^{(8,8),(1/2)} + \alpha^{(8,8),(3/2)}$.
Similarly one finds
\begin{alignat}{3}
    \langle \pi^+ \pi^- | \Theta^{(8,8),(3/2)} | K^0 \rangle
  &
    \; =
  &
    \; \frac{-12i}{f^3} \alpha^{(8,8),(3/2)}
  &
    \; =
  &
    \; \frac{-4i}{f^3} \alpha^{(8,8)}
  \\
    \langle \pi^+ \pi^- | \Theta^{(8,8),(1/2)} | K^0 \rangle
  &
    \; =
  &
    \; \frac{-12i}{f^3} \alpha^{(8,8),(1/2)}
  &
    \; =
  &
    \; \frac{-8i}{f^3} \alpha^{(8,8)}
\end{alignat}

For $\tilde{\Theta}^{(27,1)}$, we use the definition in 
\cite{Bernard:1985wf} and write
\begin{eqnarray}
   \tilde{\Theta}^{(27,1),(3/2)}
 &
   =
 &
   {\rm Tr}
   \left[
     \left(
       \begin{array}{ccc}
         0  &  0  &  0  \\
         0  &  0  &  0  \\
         0  &  1  &  0  \\
       \end{array}
     \right)
     \Sigma \; \partial_\mu \Sigma^\dagger
   \right]
   {\rm Tr}
   \left[
     \left(
       \begin{array}{ccc}
         1  &  0  &  0  \\
         0  & -1  &  0  \\
         0  &  0  &  0  \\
       \end{array}
     \right)
     \Sigma \; \partial^\mu \Sigma^\dagger
   \right ] 
  \nonumber \\
 &
   +
 &
   {\rm Tr}
   \left[
     \left(
       \begin{array}{ccc}
         0  &  0  &  0  \\
         0  &  0  &  0  \\
         1  &  0  &  0  \\
       \end{array}
     \right)
     \Sigma \; \partial_\mu \Sigma^\dagger
   \right]
   {\rm Tr}
   \left[
     \left(
       \begin{array}{ccc}
         0  &  1  &  0  \\
         0  &  0  &  0  \\
         0  &  0  &  0  \\
       \end{array}
     \right)
     \Sigma \; \partial^\mu \Sigma^\dagger
   \right ] 
  \\
   \tilde{\Theta}^{(27,1),(1/2)}
 &
   =
 &
   {\rm Tr}
   \left[
     \left(
       \begin{array}{ccc}
         0  &  0  &  0  \\
         0  &  0  &  0  \\
         0  &  1  &  0  \\
       \end{array}
     \right)
     \Sigma \; \partial_\mu \Sigma^\dagger
   \right]
   {\rm Tr}
   \left[
     \left(
       \begin{array}{ccc}
         1  &  0  &  0  \\
         0  &  2  &  0  \\
         0  &  0  & -3  \\
       \end{array}
     \right)
     \Sigma \; \partial^\mu \Sigma^\dagger
   \right ] 
  \nonumber \\
 &
   +
 &
   {\rm Tr}
   \left[
     \left(
       \begin{array}{ccc}
         0  &  0  &  0  \\
         0  &  0  &  0  \\
         1  &  0  &  0  \\
       \end{array}
     \right)
     \Sigma \; \partial_\mu \Sigma^\dagger
   \right]
   {\rm Tr}
   \left[
     \left(
       \begin{array}{ccc}
         0  &  1  &  0  \\
         0  &  0  &  0  \\
         0  &  0  &  0  \\
       \end{array}
     \right)
     \Sigma \; \partial^\mu \Sigma^\dagger
   \right ] 
  \\
\end{eqnarray}
Working in lowest order chiral perturbation theory then gives
\begin{alignat}{3}
    \langle \pi^+ | \Theta^{(27,1),(3/2)} | K^+ \rangle
  &
    \; \equiv
  &
    \; - \frac{4 m_M^2}{f^2} \alpha^{(27,1),(3/2)}
  &
    \; =
  &
    \; - \frac{4 m_M^2}{f^2} \alpha^{(27,1)}
    \label{eq:def_alpha_271_32}
  \\
    \langle \pi^+ | \Theta^{(27,1),(1/2)} | K^+ \rangle
  &
    \; \equiv
  &
    \; - \frac{4 m_M^2}{f^2} \alpha^{(27,1),(1/2)}
  &
    \; =
  &
    \; -\frac{4 m_M^2}{f^2} \alpha^{(27,1)}
    \label{eq:def_alpha_271_12}
\end{alignat}
and
\begin{eqnarray}
    \langle \pi^+ \pi^- | \Theta^{(27,1),(3/2)} | K^0 \rangle
  &
     =
  &
    - \frac{4 i}{f^3} \left( m_{K^0}^2 - m_{\pi^+}^2 \right)
    \alpha^{(27,1),(3/2)}
  \nonumber \\
  &
     =
  &
    - \frac{4 i}{f^3} \left( m_{K^0}^2 - m_{\pi^+}^2 \right)
    \alpha^{(27,1)}
  \\
    \langle \pi^+ \pi^- | \Theta^{(27,1),(1/2)} | K^0 \rangle
  &
     =
  &
    - \frac{4 i}{f^3} \left( m_{K^0}^2 - m_{\pi^+}^2 \right)
    \alpha^{(27,1),(1/2)}
  \nonumber \\
  &
    =
  &
    - \frac{4 i}{f^3} \left( m_{K^0}^2 - m_{\pi^+}^2 \right)
    \alpha^{(27,1)}
\end{eqnarray}


\section{Definitions for Standard Model Parameters}

We follow \cite{PDBook} and define the Cabibbo-Kobayashi-Maskawa
matrix as
\begin{eqnarray}
    V 
  &
    \equiv
  &
    \left(
      \begin{array}{ccc}
        V_{ud}	& V_{us}	& V_{ub}	\\
        V_{cd}	& V_{cs}	& V_{cb}	\\
        V_{td}	& V_{ts}	& V_{tb}	\\
      \end{array}
    \right)
  \nonumber
  \\
  &
    \approx
  &
    \left(
      \begin{array}{ccc}
           1 - \lambda^2 / 2
	&
	  \lambda
	&
	  A \lambda^3( \rho - i \eta)
	\\
          - \lambda
	&
	  1 - \lambda^2 / 2
	&
	  A \lambda^2
	\\
	  A \lambda^3 ( 1 - \rho - i \eta )
	&
	  - A \lambda^2
	&
	  1
      \end{array}
    \right)
\end{eqnarray}
Outside of this section, we use $\lambda_{\rm CKM} = \lambda$, $A_{\rm
CKM} = A$ and $\rho_{\rm CKM} = \rho$ to avoid confusion.  Recent
reviews have quoted values for
\begin{eqnarray}
   \bar{\rho}
  &
    \equiv
  &
    \rho \left( 1 - \frac{\lambda^2}{2} \right)
  \\
   \bar{\eta}
  &
    \equiv
  &
    \eta \left( 1 - \frac{\lambda^2}{2} \right)
\end{eqnarray}
Our values for $V_{td}$ are determined from
\begin{equation}
  V_{td} =  A \lambda^3 ( 1 - \rho - i \eta )
\end{equation}

\fi


\bibliography{paper}


\ifnum\theTables=1

%
   
\begin{table}
\begin{tabular}{cccccc}
$i$ &   ${\cal O}(1) $  &  ${\cal O}(\alpha_s)$ &
 ${\cal O}(\alpha)$ & ${\cal O}(\alpha/\alpha_s)$ & total \cr
\hline
1 & -0.517171	& 0.119497	& 0.00160768	& -0.00393867
  & -0.400005	\\
2 &  1.26603	&  -0.0670242	& -0.00253	& 0.00964183
  & 1.20612	\\
3 &  0.0	& 0.00421037	& 0.0000320653	& 0.0
  &  0.00424243 \\
4 &  0.0	& -0.0126311	& -0.0000961959	& 0.0
  & -0.0127273	\\
5 & 0.0		& 0.00421037	& 0.0000320653	& 0.0
  & 0.00424243	\\
6 & 0.0		& -0.0126311	& -0.0000961959 & 0.0
  & -0.0127273	\\
7 &  0.0	& 0.0		& 0.0000525882	& 0.0
  & 0.0000525882\\
8 & 0.0		& 0.0		& 0.0		& 0.0
  & 0.0		\\
9 & 0.0		& 0.0		& 0.0000525882	& 0.0
  & 0.0000525882 \\
10& 0.0		& 0.0		& 0.0		& 0.0
  & 0.0
\end{tabular}

\caption{Decomposition of the next-to-leading order (NLO) Wilson
coefficients into contributions of a given order.  The coefficients
$z_i$ at $\mu = 1.3$ GeV (the charm quark mass) are given in the NDR
scheme for the 3-flavor case where the charm quark has been integrated
out.}

\label{tab:wilson_coef_comp_z}
\end{table}

%
   
\begin{table}
\begin{tabular}{cccccc}
$i$ &   ${\cal O}(1) $  &  ${\cal O}(\alpha_s)$ &
 ${\cal O}(\alpha)$ & ${\cal O}(\alpha/\alpha_s)$ & total \cr
\hline
1 & 0.0		& 0.0		& 0.0		& 0.0
  & 0.0		\\
2 & 0.0		& 0.0		& 0.0		& 0.0
  & 0.0		\\
3 & 0.0266933	& -0.000750255	& 0.00143301	& 0.000130383
  & 0.0275065	\\
4 & -0.051399	& -0.00254918	& -0.0010719	& -0.000277595
  & -0.0552976	\\
5 & 0.0132739	& -0.00788698	& 0.000117102	& 0.0000774746 
  & 0.00558151	\\
6 & -0.0775222	& -0.00534437	& -0.000868366	& -0.000372801
  & -0.0841077	\\
7 & 0.0		& 0.0		& 0.000700858	& -0.000878706
  & -0.000177847 \\
8 & 0.0		& 0.0		& 0.0012366	& -0.000180252
  & 0.00105634	\\
9 & 0.0		& 0.0		& -0.0107664	& -0.000999603
  & -0.011766	\\
10& 0.0		& 0.0		& 0.00406102	& 0.000173261
  & 0.00423429
\end{tabular}

\caption{Decomposition of the next-to-leading order (NLO) Wilson
coefficients into contributions of a given order.  The coefficients
$y_i$ at $\mu = 1.3$ GeV (the charm quark mass) are given in the NDR
scheme for the 3-flavor case where the charm quark has been integrated
out.}

\label{tab:wilson_coef_comp_y}
\end{table}

   
\begin{table}
\begin{tabular}{ccccc}
$i$ &   1.51  &   2.13 & 2.39 &  3.02 (GeV)\cr
\hline
1 & -0.346301	& -0.304999	& -0.292757	& -0.269806 \\
2 & 1.17384	& 1.14951	& 1.14247	& 1.12947 \\
3 & 0.00404856	& 0.00181346	& 0.00121441	& 0.000164314 \\
4 & -0.0129397	& -0.00573613	& -0.00368611	& 0.0000666811 \\
5 & 0.00476383	& 0.00281554	& 0.00222381	& 0.00109864 \\
6 & -0.0146471	& -0.00656106	& -0.00440476	& -0.00061269 \\
7 & 0.0000530348 & 0.0000666811	& 0.0000739361	& 0.0000922512 \\
8 & -0.0000223135 & -0.0000625724 & -0.0000721542 & -0.0000875988 \\
9 & 0.0000415803 &  0.0000340103 &  0.0000356813 &  0.0000443653 \\
10& 0.0000159289 &  0.0000422636 &  0.0000493559 &  0.0000617442

\end{tabular}

\caption{The Wilson coefficients $z_i(\mu)$ in the RI scheme for the
3-flavor case.  Starting from the 3-flavor, NDR scheme Wilson
coefficients in full QCD at the charm mass, the Wilson coefficients are
evolved to the $\mu$ values in this table using the quenched 3-loop
value for $\Lambda_{\overline{MS}}$ and the 2-loop quenched
$\alpha_s$.  At this $\mu$ they are converted to the RI scheme.}

\label{tab:wilson_coef_z}
\end{table}

%
   
\begin{table}
\begin{tabular}{ccccc}
$i$ &   1.51  &   2.13 & 2.39 &  3.02 (GeV)\cr
\hline
1 & 0.0		& 0.0		& 0.0		& 0.0	\\
2 & 0.0		& 0.0		& 0.0		& 0.0	\\
3 & 0.0238943	& 0.0224644	& 0.0220211	& 0.0211685 \\
4 & -0.0505155	& -0.0511484	& -0.0513014	& -0.0515536 \\
5 & 0.00583245	& 0.00719003	& 0.00756092	& 0.008223 \\
6 & -0.0912935	& -0.0817901	& -0.0792629	& -0.0748307 \\
7 & -0.000176754& -0.000155239	& -0.000148013	& -0.000133228 \\
8 & 0.00115608	& 0.000971975	& 0.00092186	& 0.000832504 \\
9 & -0.0114196	& -0.0111436	& -0.0110649	& -0.010921 \\
10& 0.00368473	& 0.00325191	& 0.00312729	& 0.00289785

\end{tabular}

\caption{The Wilson coefficients $y_i(\mu)$ in the RI scheme for the
3-flavor case.  Starting from the 3-flavor, NDR scheme Wilson
coefficients in full QCD at the charm mass, the Wilson coefficients are
evolved to the $\mu$ values in this table using the quenched 3-loop
value for $\Lambda_{\overline{MS}}$ and the 2-loop quenched
$\alpha_s$.  At this $\mu$ they are converted to the RI scheme.}

\label{tab:wilson_coef_y}
\end{table}

   
\begin{table}

\begin{tabular}{cc}
Parameter & Value \\ \hline
$M_W$ & 80.419\cr
$m_t(M_W)$ & 175.5\cr
$m_b(m_b)$ & 4.4\cr
$m_c(m_c)$ & 1.3 \cr
$\alpha$ & 1/128 \cr
$\Lambda^{(f=5)}_{\overline{MS}}$ & $0.208\pm0.025$\cr
$\sin^2{\theta_{W}}$ & 0.23117 \cr
\end{tabular}

\caption{Standard Model parameters used to generate the Wilson
coefficients. Dimensionful parameters are in GeV.}

\label{tab:wilson params}

\end{table}


\begin{table}
\begin{center}
\begin{tabular}{rrr}
$i$ & $A_{\kappa^i}^{(0)}$ & $A_{\kappa^i}^{(2)}$  \\ 
\hline
$1$ & $2.2(16) \times 10^{-3}$ & $-7.2(54) \times 10^{-4}$ \\
$2$ & $8.3(89) \times 10^{-3}$ & $-1.3(30) \times 10^{-3}$ \\
$3$ & $2.3(19) \times 10^{-2}$ & $-4.6(65) \times 10^{-3}$ \\
$4$ & $2.9(27) \times 10^{-2}$ & $-5.2(92) \times 10^{-3}$ \\
$5$ & $6.73(12) \times 10^{-1}$ & $1.9(31) \times 10^{-3}$ \\
$6$ & $2.037(44)$ & $1.3(97) \times 10^{-3}$ \\
$7$ & $-3.330(73) \times 10^{-1}$ & $-1.7(13) \times 10^{-3}$ \\
$8$ & $-9.95(20) \times 10^{-1}$ & $-5.6(33) \times 10^{-3}$ \\
$9$ & $-8.3(89) \times 10^{-3}$ & $1.3(30) \times 10^{-3}$ \\
\end{tabular}
\end{center}

\caption{The ${\cal O}(a)^2$ errors in the lower dimensional operator
subtractions are eliminated by fitting the coefficient $A_{\kappa^i}(p)$
in Eq.\ \ref{eq:npr_kappa} to the form $A_{\kappa^i}(p)  =
A_{\kappa^i}^{(0)} + A_{\kappa^i}^{(2)}(ap)^2$.  This table gives
results for $A_{\kappa^i}^{(0)}$ and $A_{\kappa^i}^{(2)}$.}

\label{tab:msx}

\end{table}


\begin{table}
\begin{center}
\begin{tabular}{rrr}
$i$ & $B_{\kappa^i}^{(0)}$ & $B_{\kappa^i}^{(2)}$  \\ 
\hline
$1$ & $-1.6(21) \times 10^{-4}$ & $1.02(72) \times 10^{-4}$ \\
$2$ & $-8(13) \times 10^{-4}$ & $3.5(43) \times 10^{-4}$ \\
$3$ & $-2.0(29) \times 10^{-3}$ & $9.8(93) \times 10^{-4}$ \\
$4$ & $-2.7(41) \times 10^{-3}$ & $1.2(13) \times 10^{-3}$ \\
$5$ & $1.05(19) \times 10^{-2}$ & $-7.3(47) \times 10^{-4}$ \\
$6$ & $2.49(62) \times 10^{-2}$ & $1(13) \times 10^{-4}$ \\
$7$ & $-5.4(11) \times 10^{-3}$ & $4.1(21) \times 10^{-4}$ \\
$8$ & $-1.52(30) \times 10^{-2}$ & $9.4(46) \times 10^{-4}$ \\
$9$ & $8(13) \times 10^{-4}$ & $-3.5(43) \times 10^{-4}$ \\
\end{tabular}
\end{center}

\caption{The ${\cal O}(a)^2$ errors in the lower dimensional operator
subtractions are eliminated by fitting the coefficient $B_{\kappa^i}(p)$
in Eq.\ \ref{eq:npr_kappa} to the form $B_{\kappa^i}(p)  =
B_{\kappa^i}^{(0)} + B_{\kappa^i}^{(2)}(ap)^2$.  This table gives
results for $B_{\kappa^i}^{(0)}$ and $B_{\kappa^i}^{(2)}$.}

\label{tab:mix}

\end{table}


\begin{table}
\begin{center}
\begin{tabular}{rrr}
$i$ & $A_{\lambda^i}^{(0)}$ & $A_{\lambda^i}^{(2)}$  \\ 
\hline
$1$ & $7(64) \times 10^{-5}$ & $1(21) \times 10^{-5}$ \\
$2$ & $-1.03(29) \times 10^{-2}$ & $2.93(98) \times 10^{-3}$ \\
$3$ & $-1.97(60) \times 10^{-2}$ & $5.7(20) \times 10^{-3}$ \\
$4$ & $-3.02(87) \times 10^{-2}$ & $8.6(30) \times 10^{-3}$ \\
$5$ & $-3.62(34) \times 10^{-2}$ & $8.85(96) \times 10^{-3}$ \\
$6$ & $-1.32(12) \times 10^{-1}$ & $3.36(34) \times 10^{-2}$ \\
$7$ & $1.60(14) \times 10^{-2}$ & $-3.85(39) \times 10^{-3}$ \\
$8$ & $4.59(41) \times 10^{-2}$ & $-1.10(11) \times 10^{-2}$ \\
$9$ & $1.03(29) \times 10^{-2}$ & $-2.93(98) \times 10^{-3}$ \\
\end{tabular}
\end{center}

\caption{The ${\cal O}(a)^2$ errors in the lower dimensional operator
subtractions are eliminated by fitting the coefficient
$A_{\lambda^i}(p)$ in Eq.\ \ref{eq:npr_lambda} to the form
$A_{\lambda^i}(p)  = A_{\lambda^i}^{(0)} + A_{\lambda^i}^{(2)}(ap)^2$.
This table gives results for $A_{\lambda^i}^{(0)}$ and
$A_{\lambda^i}^{(2)}$.}

\label{tab:msy}

\end{table}


\begin{table}
\begin{center}
\begin{tabular}{rrr}
$i$ & $B_{\lambda^i}^{(0)}$ & $B_{\lambda^i}^{(2)}$  \\ 
\hline
$1$ & $-2.02(48) \times 10^{-3}$ & $2.06(69) \times 10^{-4}$ \\
$2$ & $1.14(14) \times 10^{-2}$ & $-2.31(36) \times 10^{-3}$ \\
$3$ & $1.67(32) \times 10^{-2}$ & $-4.02(79) \times 10^{-3}$ \\
$4$ & $3.01(44) \times 10^{-2}$ & $-6.5(11) \times 10^{-3}$ \\
$5$ & $4(14) \times 10^{-4}$ & $-5.3(23) \times 10^{-4}$ \\
$6$ & $4.26(45) \times 10^{-2}$ & $-8.8(12) \times 10^{-3}$ \\
$7$ & $-5.95(95) \times 10^{-4}$ & $1.32(26) \times 10^{-4}$ \\
$8$ & $-1.82(28) \times 10^{-3}$ & $4.13(78) \times 10^{-4}$ \\
$9$ & $-1.14(14) \times 10^{-2}$ & $2.31(36) \times 10^{-3}$ \\
\end{tabular}
\end{center}

\caption{The ${\cal O}(a)^2$ errors in the lower dimensional operator
subtractions are eliminated by fitting the coefficient
$B_{\lambda^i}(p)$ in Eq.\ \ref{eq:npr_lambda} to the form
$B_{\lambda^i}(p)  = B_{\lambda^i}^{(0)} + B_{\lambda^i}^{(2)}(ap)^2$.
This table gives results for $B_{\lambda^i}^{(0)}$ and
$B_{\lambda^i}^{(2)}$.}
\label{tab:miy}

\end{table}


\begin{table}
\begin{center}
\begin{tabular}{rr}
$i$ & $c_1^i$  \\ \hline
$1$ & $1.2(12) \times 10^{-4}$ \\
$2$ & $-8.0(77) \times 10^{-4}$\\
$3$ & $-1.2(17) \times 10^{-3}$\\
$4$ & $-2.2(24) \times 10^{-3}$\\
$5$ & $-6.42(13) \times 10^{-2}$\\
$6$ & $-1.916(34) \times 10^{-1}$\\
$7$ & $3.204(65) \times 10^{-2}$\\
$8$ & $9.49(17) \times 10^{-2}$\\
$9$ & $8.0(77) \times 10^{-4}$\\
\end{tabular}
\end{center}

\caption{The lower dimensional operator subtraction coefficients
$c_1^i$ used in the $m_f=0.04$ subtraction.}
\label{tab:c1}

\end{table}


\begin{table}
\begin{center}
\begin{tabular}{ccc}
 $(ap)^2$ & $n_1$ & $n_2$ \\
\hline
1.23 & & \\\hline
 &[0, 2, 2, 0] & [2, 2, 0, 0] \\
 &[0, 2, 2, 0] & [-2, 2, 0, 0] \\
 &[0, 2, 2, 0] & [2, 0, 2, 0] \\
 &[0, 2, 2, 0] & [-2, 0, 2, 0] \\
 &[0, 2, 2, 0] & [0, 2, 0, 4] \\
 &[0, 2, 2, 0] & [0, 2, 0, -4] \\
 &[0, 2, 2, 0] & [0, 0, 2, 4] \\
 &[0, 2, 2, 0] & [0, 0, 2, -4] \\
\hline
1.54 & & \\\hline
 &[1, 1, 2, 4] & [1, -2, 1, 4] \\
 &[1, 1, 2, 4] & [1, 2, -1, 4] \\
 &[1, 1, 2, 4] & [-2, 1, 1, 4] \\
 &[1, 1, 2, 4] & [2, 1, -1, 4] \\
 &[1, 1, 2, 4] & [-2, 1, 2, 2] \\
 &[1, 1, 2, 4] & [2, 1, 2, -2] \\
 &[1, 1, 2, 4] & [1, -2, 2, 2] \\
 &[1, 1, 2, 4] & [1, 2, 2, -2] \\
\end{tabular}
\end{center}

\caption{The discrete Euclidean four-momenta used in the four-quark
operator renormalization calculation. Values are given in the order [x,
y, z, t].}

\label{npr:tab:mom}

\end{table}


\begin{table}[!htb]
\begin{center}
\begin{tabular}{rrrr}
 & 1 & 2 & 3 \\ \hline
1 & $1.1380(35)$ & $3(11) \times 10^{-5}$ & $2.1(70) \times 10^{-5}$\\
2 & $6(245) \times 10^{-8}$ & $1.052(12)$ & $7.03(98) \times 10^{-2}$\\
3 & $-8(368) \times 10^{-8}$ & $8.0(19) \times 10^{-2}$ & $1.086(22)$\\
5 & $-6(45) \times 10^{-20}$ & $4.8(32) \times 10^{-2}$ & $1.8(24) \times 10^{-2}$\\
6 & $-1(112) \times 10^{-20}$ & $-2.1(60) \times 10^{-2}$ & $1.3(73) \times 10^{-2}$\\
7 & $1.11(37) \times 10^{-4}$ & $5.1(41) \times 10^{-3}$ & $9.9(50) \times 10^{-3}$\\
8 & $-1.5(20) \times 10^{-5}$ & $1.6(12) \times 10^{-2}$ & $3.0(15) \times 10^{-2}$\\
\hline
\\ 
 & 5 & 6 & 7 \\\hline
1 & $1.52(80) \times 10^{-5}$ & $-2.87(33) \times 10^{-5}$ & $1.71(36) \times 10^{-3}$\\
2 & $9.7(38) \times 10^{-3}$ & $-8(21) \times 10^{-4}$ & $-2.3(18) \times 10^{-4}$\\
3 & $-2.2(61) \times 10^{-3}$ & $2.1(11) \times 10^{-2}$ & $-1.08(77) \times 10^{-3}$\\
5 & $1.039(12)$ & $9.00(77) \times 10^{-2}$ & $1.1(16) \times 10^{-3}$\\
6 & $3.2(23) \times 10^{-2}$ & $1.218(35)$ & $-2.2(50) \times 10^{-3}$\\
7 & $-4(15) \times 10^{-4}$ & $-1.8(22) \times 10^{-3}$ & $1.0562(29)$\\
8 & $-1.3(45) \times 10^{-3}$ & $-5.1(64) \times 10^{-3}$ & $6.10(25) \times 10^{-2}$\\
\hline
\\
 & 8 &  &  \\\hline
1 & $-5.4(15) \times 10^{-4}$ &  &\\
2 & $8(16) \times 10^{-5}$ &  &\\
3 & $7.3(65) \times 10^{-4}$ &  &\\
5 & $1.2(18) \times 10^{-3}$ &  &\\
6 & $8.5(55) \times 10^{-3}$ &  &\\
7 & $8.31(17) \times 10^{-2}$ &  &\\
8 & $1.1354(43)$ &  &\\
\end{tabular}
\end{center}

\caption{The inverse of the four-quark renormalization matrix,
$MF^{-1}$, in the block diagonal basis of irreducible representations
of $SU(3)_L\times SU(3)_R$. $Q_1^\prime$ is in the (27,1)
representation, $Q_2^\prime$ $Q_3^\prime$ $Q_5^\prime$ and $Q_6^\prime$
are in (8,1) representations, and $Q_{7,8}^\prime$ belong to (8,8)
representations.  Note that entries connecting the various
representations are either zero within statistical errors or very
small.  The renormalization point is $(ap)^2 = (ap)_{\rm diff}^2 =
1.23$.}

\label{mf1}

\end{table}


\begin{table}
\begin{center}
\begin{tabular}{rrrr}
 & 1 & 2 & 3 \\\hline
1 & $1.1516(36)$ & $-5(99) \times 10^{-6}$ & $5(59) \times 10^{-6}$\\
2 & $2(235) \times 10^{-8}$ & $1.0665(95)$ & $8.95(76) \times 10^{-2}$\\
3 & $-4(353) \times 10^{-8}$ & $7.3(15) \times 10^{-2}$ & $1.066(18)$\\
5 & $1(13) \times 10^{-20}$ & $-8(23) \times 10^{-3}$ & $-9(21) \times 10^{-3}$\\
6 & $5(68) \times 10^{-20}$ & $-4.8(53) \times 10^{-2}$ & $-1.5(61) \times 10^{-2}$\\
7 & $7.5(31) \times 10^{-5}$ & $-1.9(25) \times 10^{-3}$ & $-4(33) \times 10^{-4}$\\
8 & $-1.0(15) \times 10^{-5}$ & $-6.1(76) \times 10^{-3}$ & $-2.1(97) \times 10^{-3}$\\
\hline
\\
 & 5 & 6 & 7 \\\hline
1 & $1.9(78) \times 10^{-6}$ & $-2.25(23) \times 10^{-5}$ & $1.02(32) \times 10^{-3}$\\
2 & $-3(28) \times 10^{-4}$ & $-1.8(13) \times 10^{-3}$ & $1.6(11) \times 10^{-4}$\\
3 & $-9.6(63) \times 10^{-3}$ & $3.29(77) \times 10^{-2}$ & $-2.4(41) \times 10^{-4}$\\
5 & $1.0684(82)$ & $8.65(67) \times 10^{-2}$ & $-3.1(10) \times 10^{-3}$\\
6 & $4.7(21) \times 10^{-2}$ & $1.246(26)$ & $-8.8(31) \times 10^{-3}$\\
7 & $-6(11) \times 10^{-4}$ & $-2.0(18) \times 10^{-3}$ & $1.0626(26)$\\
8 & $-1.8(33) \times 10^{-3}$ & $-5.8(52) \times 10^{-3}$ & $7.57(18) \times 10^{-2}$\\
\hline
\\
 & 8 &  &  \\\hline
1 & $-6(12) \times 10^{-5}$ &  &\\
2 & $7(11) \times 10^{-5}$ &  &\\
3 & $2.1(62) \times 10^{-4}$ &  &\\
5 & $8(10) \times 10^{-4}$ &  &\\
6 & $2.6(40) \times 10^{-3}$ &  &\\
7 & $8.80(18) \times 10^{-2}$ &  &\\
8 & $1.1234(41)$ &  &\\
\end{tabular}
\end{center}

\caption{The same as Table~\ref{mf1} except the renormalization point
is $(ap)^2 = (ap)_{\rm diff}^2 = 1.54$.}

\label{mf2}

\end{table}


\begin{table}
\begin{center}
\begin{tabular}{rrrr}
 & 1 & 2 & 3 \\\hline
1 & $9.466(27) \times 10^{-1}$ & $-6.79(26) \times 10^{-2}$ & $3.1(35) \times 10^{-3}$\\
2 & $-5.65(72) \times 10^{-2}$ & $9.353(70) \times 10^{-1}$ & $-4.7(59) \times 10^{-3}$\\
3 & $9.1(14) \times 10^{-2}$ & $-9.1(14) \times 10^{-2}$ & $8.79(16) \times 10^{-1}$\\
4 & $-9.13(20) \times 10^{-1}$ & $9.13(20) \times 10^{-1}$ & $8.71(19) \times 10^{-1}$\\
5 & $-1.03(51) \times 10^{-2}$ & $1.03(51) \times 10^{-2}$ & $-1.13(92) \times 10^{-2}$\\
6 & $1.4(21) \times 10^{-2}$ & $-1.4(21) \times 10^{-2}$ & $2(18) \times 10^{-3}$\\
7 & $0.0(0)$ & $0.0(0)$ & $0.0(0)$\\
8 & $0.0(0)$ & $0.0(0)$ & $0.0(0)$\\
9 & $1.3746(73)$ & $-5.65(72) \times 10^{-2}$ & $-4.347(61) \times 10^{-1}$\\
10 & $3.715(35) \times 10^{-1}$ & $9.466(27) \times 10^{-1}$ & $-4.424(40) \times 10^{-1}$\\
\hline
\\
 & 5 & 6 & 7 \\\hline
1 & $-9.2(37) \times 10^{-3}$ & $2.4(19) \times 10^{-3}$ & $0.0(0)$\\
2 & $3.1(53) \times 10^{-3}$ & $-1.61(85) \times 10^{-2}$ & $0.0(0)$\\
3 & $-2.1(12) \times 10^{-2}$ & $-2.5(15) \times 10^{-2}$ & $0.0(0)$\\
4 & $-9(15) \times 10^{-3}$ & $-4.3(24) \times 10^{-2}$ & $0.0(0)$\\
5 & $9.65(11) \times 10^{-1}$ & $-7.11(64) \times 10^{-2}$ & $0.0(0)$\\
6 & $-2.6(18) \times 10^{-2}$ & $8.23(24) \times 10^{-1}$ & $0.0(0)$\\
7 & $0.0(0)$ & $0.0(0)$ & $9.508(25) \times 10^{-1}$\\
8 & $0.0(0)$ & $0.0(0)$ & $-5.11(20) \times 10^{-2}$\\
9 & $-3.1(53) \times 10^{-3}$ & $1.61(85) \times 10^{-2}$ & $0.0(0)$\\
10 & $9.2(37) \times 10^{-3}$ & $-2.4(19) \times 10^{-3}$ & $0.0(0)$\\
\hline
\\
 & 8 &  &  \\\hline
1 & $0.0(0)$ &  &\\
2 & $0.0(0)$ &  &\\
3 & $0.0(0)$ &  &\\
4 & $0.0(0)$ &  &\\
5 & $0.0(0)$ &  &\\
6 & $0.0(0)$ &  &\\
7 & $-6.96(12) \times 10^{-2}$ &  &\\
8 & $8.845(34) \times 10^{-1}$ &  &\\
9 & $0.0(0)$ &  &\\
10 & $0.0(0)$ &  &\\
\end{tabular}
\end{center}

\caption{The four-quark operator renormalization factors
$\hat{Z}_{ij}/Z_q^2$ at the renormalization point $(ap)^2=1.23$ ($\mu =
2.13$ GeV) for the 3-flavor case.  Values are given in the full
over-complete basis of operators as explained in the text.}

\label{tab:zfull1}

\end{table}


\begin{table}
\begin{center}
\begin{tabular}{rrrr}
 & 1 & 2 & 3 \\\hline
1 & $9.458(23) \times 10^{-1}$ & $-7.74(22) \times 10^{-2}$ & $-9(26) \times 10^{-4}$\\
2 & $-7.14(57) \times 10^{-2}$ & $9.397(60) \times 10^{-1}$ & $1.9(48) \times 10^{-3}$\\
3 & $9.0(10) \times 10^{-2}$ & $-9.0(10) \times 10^{-2}$ & $8.70(12) \times 10^{-1}$\\
4 & $-9.28(16) \times 10^{-1}$ & $9.28(16) \times 10^{-1}$ & $8.72(15) \times 10^{-1}$\\
5 & $-2.4(33) \times 10^{-3}$ & $2.4(33) \times 10^{-3}$ & $1.9(70) \times 10^{-3}$\\
6 & $9(17) \times 10^{-3}$ & $-9(17) \times 10^{-3}$ & $9(15) \times 10^{-3}$\\
7 & $0.0(0)$ & $0.0(0)$ & $0.0(0)$\\
8 & $0.0(0)$ & $0.0(0)$ & $0.0(0)$\\
9 & $1.3739(65)$ & $-7.14(57) \times 10^{-2}$ & $-4.361(49) \times 10^{-1}$\\
10 & $3.567(31) \times 10^{-1}$ & $9.458(23) \times 10^{-1}$ & $-4.333(31) \times 10^{-1}$\\
\hline
\\
 & 5 & 6 & 7 \\\hline
1 & $-6(25) \times 10^{-4}$ & $3.5(11) \times 10^{-3}$ & $0.0(0)$\\
2 & $9.6(53) \times 10^{-3}$ & $-2.57(54) \times 10^{-2}$ & $0.0(0)$\\
3 & $1.7(12) \times 10^{-2}$ & $-4.1(10) \times 10^{-2}$ & $0.0(0)$\\
4 & $2.8(16) \times 10^{-2}$ & $-7.0(15) \times 10^{-2}$ & $0.0(0)$\\
5 & $9.389(70) \times 10^{-1}$ & $-6.54(47) \times 10^{-2}$ & $0.0(0)$\\
6 & $-3.5(16) \times 10^{-2}$ & $8.05(17) \times 10^{-1}$ & $0.0(0)$\\
7 & $0.0(0)$ & $0.0(0)$ & $9.464(22) \times 10^{-1}$\\
8 & $0.0(0)$ & $0.0(0)$ & $-6.38(14) \times 10^{-2}$\\
9 & $-9.6(53) \times 10^{-3}$ & $2.57(54) \times 10^{-2}$ & $0.0(0)$\\
10 & $6(25) \times 10^{-4}$ & $-3.5(11) \times 10^{-3}$ & $0.0(0)$\\
\hline
\\
 & 8 &  &  \\\hline
1 & $0.0(0)$ &  &\\
2 & $0.0(0)$ &  &\\
3 & $0.0(0)$ &  &\\
4 & $0.0(0)$ &  &\\
5 & $0.0(0)$ &  &\\
6 & $0.0(0)$ &  &\\
7 & $-7.42(12) \times 10^{-2}$ &  &\\
8 & $8.951(32) \times 10^{-1}$ &  &\\
9 & $0.0(0)$ &  &\\
10 & $0.0(0)$ &  &\\
\end{tabular}
\end{center}

\caption{The same as Table~\ref{tab:zfull1} except the
renormalization point is $(ap)^2=1.54$ ($\mu = 2.39$ GeV).}

\label{tab:zfull2}

\end{table}


\begin{table}
\begin{center}
\begin{tabular}{rrrr}
 & 1 & 2 & 3 \\\hline
1 & $9.484(26) \times 10^{-1}$ & $-6.96(16) \times 10^{-2}$ & $9(380) \times 10^{-8}$\\
2 & $-6.96(16) \times 10^{-2}$ & $9.484(26) \times 10^{-1}$ & $9(380) \times 10^{-8}$\\
3 & $6.96(16) \times 10^{-2}$ & $-6.96(16) \times 10^{-2}$ & $8.787(27) \times 10^{-1}$\\
4 & $-9.484(26) \times 10^{-1}$ & $9.484(26) \times 10^{-1}$ & $8.787(27) \times 10^{-1}$\\
5 & $3.3(54) \times 10^{-5}$ & $-3.3(54) \times 10^{-5}$ & $-3.1(10) \times 10^{-4}$\\
6 & $-4.72(71) \times 10^{-4}$ & $4.72(71) \times 10^{-4}$ & $5.6(48) \times 10^{-5}$\\
7 & $0.0(0)$ & $0.0(0)$ & $0.0(0)$\\
8 & $0.0(0)$ & $0.0(0)$ & $0.0(0)$\\
9 & $1.3877(38)$ & $-6.96(16) \times 10^{-2}$ & $-4.394(13) \times 10^{-1}$\\
10 & $3.698(24) \times 10^{-1}$ & $9.484(26) \times 10^{-1}$ & $-4.394(13) \times 10^{-1}$\\
\hline
\\
 & 5 & 6 & 7 \\\hline
1 & $-1.29(38) \times 10^{-4}$ & $-1.7(13) \times 10^{-5}$ & $0.0(0)$\\
2 & $-1.14(18) \times 10^{-4}$ & $1.05(16) \times 10^{-4}$ & $0.0(0)$\\
3 & $-6.1(14) \times 10^{-4}$ & $1.58(50) \times 10^{-4}$ & $0.0(0)$\\
4 & $-6.0(12) \times 10^{-4}$ & $2.80(55) \times 10^{-4}$ & $0.0(0)$\\
5 & $9.510(23) \times 10^{-1}$ & $-7.03(11) \times 10^{-2}$ & $0.0(0)$\\
6 & $-5.108(98) \times 10^{-2}$ & $8.823(31) \times 10^{-1}$ & $0.0(0)$\\
7 & $0.0(0)$ & $0.0(0)$ & $9.509(23) \times 10^{-1}$\\
8 & $0.0(0)$ & $0.0(0)$ & $-5.103(98) \times 10^{-2}$\\
9 & $1.14(18) \times 10^{-4}$ & $-1.05(16) \times 10^{-4}$ & $0.0(0)$\\
10 & $1.29(38) \times 10^{-4}$ & $1.7(13) \times 10^{-5}$ & $0.0(0)$\\
\hline
\\
 & 8 &  &  \\\hline
1 & $0.0(0)$ &  &\\
2 & $0.0(0)$ &  &\\
3 & $0.0(0)$ &  &\\
4 & $0.0(0)$ &  &\\
5 & $0.0(0)$ &  &\\
6 & $0.0(0)$ &  &\\
7 & $-7.02(11) \times 10^{-2}$ &  &\\
8 & $8.823(31) \times 10^{-1}$ &  &\\
9 & $0.0(0)$ &  &\\
10 & $0.0(0)$ &  &\\
\end{tabular}
\end{center}

\caption{The four-quark operator renormalization factors
$\hat{Z}_{ij}/Z_q^2$ at the renormalization point $(ap)^2=1.23$ ($\mu =
2.13$ GeV) for the 3-flavor case except that the eye diagrams (and
consequently the lower dimensional operator subtractions) have been
omitted in the calculation of $\hat{Z}_{ij}/Z_q^2$.}

\label{tab:nloop1}

\end{table}


\begin{table}
\begin{center}
\begin{tabular}{rrrr}
 & 1 & 2 & 3 \\\hline
1 & $9.465(23) \times 10^{-1}$ & $-7.81(16) \times 10^{-2}$ & $1(47) \times 10^{-7}$\\
2 & $-7.81(16) \times 10^{-2}$ & $9.465(23) \times 10^{-1}$ & $1(47) \times 10^{-7}$\\
3 & $7.81(16) \times 10^{-2}$ & $-7.81(16) \times 10^{-2}$ & $8.684(27) \times 10^{-1}$\\
4 & $-9.465(23) \times 10^{-1}$ & $9.465(23) \times 10^{-1}$ & $8.684(27) \times 10^{-1}$\\
5 & $3.3(36) \times 10^{-5}$ & $-3.3(36) \times 10^{-5}$ & $-2.07(84) \times 10^{-4}$\\
6 & $-1.47(52) \times 10^{-4}$ & $1.47(52) \times 10^{-4}$ & $3.8(35) \times 10^{-5}$\\
7 & $0.0(0)$ & $0.0(0)$ & $0.0(0)$\\
8 & $0.0(0)$ & $0.0(0)$ & $0.0(0)$\\
9 & $1.3807(34)$ & $-7.81(16) \times 10^{-2}$ & $-4.342(13) \times 10^{-1}$\\
10 & $3.560(26) \times 10^{-1}$ & $9.465(23) \times 10^{-1}$ & $-4.342(13) \times 10^{-1}$\\
\hline
\\
 & 5 & 6 & 7 \\\hline
1 & $-8.4(33) \times 10^{-5}$ & $-1.1(10) \times 10^{-5}$ & $0.0(0)$\\
2 & $-5.7(14) \times 10^{-5}$ & $3.0(12) \times 10^{-5}$ & $0.0(0)$\\
3 & $-3.6(12) \times 10^{-4}$ & $2.6(45) \times 10^{-5}$ & $0.0(0)$\\
4 & $-3.4(10) \times 10^{-4}$ & $6.7(47) \times 10^{-5}$ & $0.0(0)$\\
5 & $9.474(22) \times 10^{-1}$ & $-7.45(11) \times 10^{-2}$ & $0.0(0)$\\
6 & $-6.07(11) \times 10^{-2}$ & $8.943(29) \times 10^{-1}$ & $0.0(0)$\\
7 & $0.0(0)$ & $0.0(0)$ & $9.474(22) \times 10^{-1}$\\
8 & $0.0(0)$ & $0.0(0)$ & $-6.07(11) \times 10^{-2}$\\
9 & $5.7(14) \times 10^{-5}$ & $-3.0(12) \times 10^{-5}$ & $0.0(0)$\\
10 & $8.4(33) \times 10^{-5}$ & $1.1(10) \times 10^{-5}$ & $0.0(0)$\\
\hline
\\
 & 8 &  &  \\\hline
1 & $0.0(0)$ &  &\\
2 & $0.0(0)$ &  &\\
3 & $0.0(0)$ &  &\\
4 & $0.0(0)$ &  &\\
5 & $0.0(0)$ &  &\\
6 & $0.0(0)$ &  &\\
7 & $-7.45(11) \times 10^{-2}$ &  &\\
8 & $8.942(29) \times 10^{-1}$ &  &\\
9 & $0.0(0)$ &  &\\
10 & $0.0(0)$ &  &\\
\end{tabular}
\end{center}

\caption{The same as Table~\ref{tab:nloop1} except the renormalization
point is $(ap)^2=1.54$ ($\mu = 2.39$ GeV).}

\label{tab:nloop2}

\end{table}


\begin{table}
\begin{center}
\begin{tabular}{rrrr}
 & 1 & 2 & 3 \\\hline
1 & $9.463(28) \times 10^{-1}$ & $-6.75(26) \times 10^{-2}$ & $4.2(35) \times 10^{-3}$\\
2 & $-5.52(73) \times 10^{-2}$ & $9.340(71) \times 10^{-1}$ & $-9.0(66) \times 10^{-3}$\\
3 & $9.2(14) \times 10^{-2}$ & $-9.2(14) \times 10^{-2}$ & $8.73(17) \times 10^{-1}$\\
4 & $-9.09(21) \times 10^{-1}$ & $9.09(21) \times 10^{-1}$ & $8.60(21) \times 10^{-1}$\\
5 & $-9.0(49) \times 10^{-3}$ & $9.0(49) \times 10^{-3}$ & $-1.06(93) \times 10^{-2}$\\
6 & $2.4(21) \times 10^{-2}$ & $-2.4(21) \times 10^{-2}$ & $-1.6(19) \times 10^{-2}$\\
7 & $0.0(0)$ & $0.0(0)$ & $0.0(0)$\\
8 & $0.0(0)$ & $0.0(0)$ & $0.0(0)$\\
9 & $1.3733(75)$ & $-5.52(73) \times 10^{-2}$ & $-4.303(67) \times 10^{-1}$\\
10 & $3.719(34) \times 10^{-1}$ & $9.463(28) \times 10^{-1}$ & $-4.436(41) \times 10^{-1}$\\
\hline
\\
 & 5 & 6 & 7 \\\hline
1 & $-7.5(36) \times 10^{-3}$ & $-3(19) \times 10^{-4}$ & $0.0(0)$\\
2 & $-3.8(60) \times 10^{-3}$ & $-5.1(88) \times 10^{-3}$ & $0.0(0)$\\
3 & $-3.0(13) \times 10^{-2}$ & $-1.1(16) \times 10^{-2}$ & $0.0(0)$\\
4 & $-2.6(17) \times 10^{-2}$ & $-1.6(25) \times 10^{-2}$ & $0.0(0)$\\
5 & $9.70(11) \times 10^{-1}$ & $-7.99(75) \times 10^{-2}$ & $0.0(0)$\\
6 & $-4.1(17) \times 10^{-2}$ & $8.42(23) \times 10^{-1}$ & $0.0(0)$\\
7 & $0.0(0)$ & $0.0(0)$ & $9.521(24) \times 10^{-1}$\\
8 & $0.0(0)$ & $0.0(0)$ & $-4.67(17) \times 10^{-2}$\\
9 & $3.8(60) \times 10^{-3}$ & $5.1(88) \times 10^{-3}$ & $0.0(0)$\\
10 & $7.5(36) \times 10^{-3}$ & $3(19) \times 10^{-4}$ & $0.0(0)$\\
\hline
\\
 & 8 &  &  \\\hline
1 & $0.0(0)$ &  &\\
2 & $0.0(0)$ &  &\\
3 & $0.0(0)$ &  &\\
4 & $0.0(0)$ &  &\\
5 & $0.0(0)$ &  &\\
6 & $0.0(0)$ &  &\\
7 & $-7.32(25) \times 10^{-2}$ &  &\\
8 & $8.720(73) \times 10^{-1}$ &  &\\
9 & $0.0(0)$ &  &\\
10 & $0.0(0)$ &  &\\
\end{tabular}
\end{center}

\caption{The four-quark operator renormalization factors
$\hat{Z}_{ij}/Z_q^2$ at the renormalization point $(ap)^2=1.23$ ($\mu =
2.13$ GeV) for the 3-flavor case except that lower dimensional operator
subtractions have been omitted in the calculation of
$\hat{Z}_{ij}/Z_q^2$.}

\label{tab:nsub1}
\end{table}


\begin{table}[!htb]
\begin{center}
\begin{tabular}{rrrr}
 & 1 & 2 & 3 \\\hline
1 & $9.451(24) \times 10^{-1}$ & $-7.67(22) \times 10^{-2}$ & $-8(24) \times 10^{-4}$\\
2 & $-6.86(59) \times 10^{-2}$ & $9.370(62) \times 10^{-1}$ & $1.7(48) \times 10^{-3}$\\
3 & $9.3(11) \times 10^{-2}$ & $-9.3(11) \times 10^{-2}$ & $8.69(13) \times 10^{-1}$\\
4 & $-9.21(16) \times 10^{-1}$ & $9.21(16) \times 10^{-1}$ & $8.72(16) \times 10^{-1}$\\
5 & $-3.6(29) \times 10^{-3}$ & $3.6(29) \times 10^{-3}$ & $3.0(67) \times 10^{-3}$\\
6 & $1.7(18) \times 10^{-2}$ & $-1.7(18) \times 10^{-2}$ & $1.2(15) \times 10^{-2}$\\
7 & $0.0(0)$ & $0.0(0)$ & $0.0(0)$\\
8 & $0.0(0)$ & $0.0(0)$ & $0.0(0)$\\
9 & $1.3712(68)$ & $-6.86(59) \times 10^{-2}$ & $-4.359(49) \times 10^{-1}$\\
10 & $3.575(31) \times 10^{-1}$ & $9.451(24) \times 10^{-1}$ & $-4.333(30) \times 10^{-1}$\\
\hline
\\
 & 5 & 6 & 7 \\\hline
1 & $-4(24) \times 10^{-4}$ & $1.1(11) \times 10^{-3}$ & $0.0(0)$\\
2 & $8.7(51) \times 10^{-3}$ & $-1.68(57) \times 10^{-2}$ & $0.0(0)$\\
3 & $1.6(12) \times 10^{-2}$ & $-3.0(11) \times 10^{-2}$ & $0.0(0)$\\
4 & $2.5(15) \times 10^{-2}$ & $-4.8(16) \times 10^{-2}$ & $0.0(0)$\\
5 & $9.407(67) \times 10^{-1}$ & $-6.79(53) \times 10^{-2}$ & $0.0(0)$\\
6 & $-3.3(15) \times 10^{-2}$ & $8.34(18) \times 10^{-1}$ & $0.0(0)$\\
7 & $0.0(0)$ & $0.0(0)$ & $9.475(23) \times 10^{-1}$\\
8 & $0.0(0)$ & $0.0(0)$ & $-5.98(17) \times 10^{-2}$\\
9 & $-8.7(51) \times 10^{-3}$ & $1.68(57) \times 10^{-2}$ & $0.0(0)$\\
10 & $4(24) \times 10^{-4}$ & $-1.1(11) \times 10^{-3}$ & $0.0(0)$\\
\hline
\\
 & 8 &  &  \\\hline
1 & $0.0(0)$ &  &\\
2 & $0.0(0)$ &  &\\
3 & $0.0(0)$ &  &\\
4 & $0.0(0)$ &  &\\
5 & $0.0(0)$ &  &\\
6 & $0.0(0)$ &  &\\
7 & $-7.48(18) \times 10^{-2}$ &  &\\
8 & $8.927(58) \times 10^{-1}$ &  &\\
9 & $0.0(0)$ &  &\\
10 & $0.0(0)$ &  &\\
\end{tabular}
\end{center}

\caption{The same as Table~\ref{tab:nsub1} except the renormalization
point is $(ap)^2=1.54$ ($\mu = 2.39$ GeV).}

\label{tab:nsub2}
\end{table}


\begin{table}
\begin{tabular}{ccc}
 $m_f$  & $m_\pi$ (85 conf.) & $m_\pi$ (400 conf.) \\ \hline
  0.01  &  0.203(3)  &  0.2052(17) \\
  0.02  &  0.270(3)  &  0.2699(14) \\
  0.03  &  0.324(2)  &  0.3231(12) \\
  0.04  &  0.371(2)  &  0.3700(12) \\
  0.05  &            &  0.4129(11)
\end{tabular}
\caption{Values for $m_\pi$ versus $m_f$ from 85 configurations using
$\langle A^a(x) A_0^a(0) \rangle$ and from the 400
configurations of this work using $\langle \pi^a(x) A_0^a(0) \rangle$.}

\label{tab:mpi}
\end{table}


\begin{table}
\begin{tabular}{cc}
$m_f$ & $ \langle \pi^+ | (\bar{s} d)_{\rm lat} | K^+ \rangle $ \\
\hline

0.01 & 1.510(25) \\
0.02 & 1.548(16) \\
0.03 & 1.599(12) \\
0.04 & 1.660(10) \\
0.05 & 1.722(9)

\end{tabular}

\caption{The values for $\langle \pi^+ | \bar{s} d_{\rm lat} | K^+
\rangle $ for each light quark mass studied.  These matrix elements are
used in the subtraction needed in the determination of $K \to \pi \pi$
matrix elements from $K \to \pi$ and $K \to |0 \rangle$ matrix
elements.}

\label{tab:k2pi_sbard}
\end{table}


\begin{table}
\begin{tabular}{clllll}

$i$ & $ m_f = 0.01$ & $ m_f = 0.02$ & $ m_f = 0.03$ &
   $ m_f = 0.04$ & $ m_f = 0.05$  \\
\hline

1 & \ \ 0.030(24) & \ \ 0.024(26)  & \ \ 0.007(27)  & \ -0.012(28)
  & \ -0.032(29) \\
2 & \ -0.058(12) & \ -0.117(13)  & \ -0.176(14)  & \ -0.233(14)
  & \ -0.290(15) \\
3 & \ -0.03(8)  & \ -0.18(9) & \ -0.37(10)  & \ -0.56(10)
  & \ -0.75(10) \\
4 & \ -0.12(7)  & \ -0.32(8)  & \ -0.55(8)  & \ -0.78(8)
  & \ -1.01(9) \\
5 & \ \ 2.10(8)  & \ \  4.12(9)  & \ \ 6.28(10)  & \ \ 8.61(10)
  & \ 11.09(11) \\
6 &  \ \ 5.92(12)  & \ 11.79(14)  & \ 18.07(16)  & \ 24.84(18)
  & \ 32.03(19) \\
7 & \ -1.805(34)  & \ -2.989(34)  & \ -4.227(36)  & \ -5.553(38)
  & \ -6.955(40) \\
8 & \ -5.56(10)  & \ -9.16(11)  & -12.93(11)  & -16.98(12)  &
  -21.26(12) \\
9 & \ \ 0.063(12)  & \ \ 0.127(13)  & \ \ 0.194(14)  & \ \ 0.261(14)
  & \ \ 0.329(15) \\
10 & \ -0.026(24)  & \ -0.013(26)  & \ \ 0.011(27) & \ \ 0.040(28)
  & \ \ 0.071(29)

\end{tabular}

\caption{The values for $\langle \pi^+ | Q^{1/2}_{i,\rm lat} | K^+
\rangle \times 10^2$ for each light quark mass studied.}

\label{tab:ktopi_i0_charm_out}

\end{table}


\begin{table}
\begin{tabular}{clllll}

$i$ & $ m_f = 0.01$ & $ m_f = 0.02$ & $ m_f = 0.03$ &
   $ m_f = 0.04$ & $ m_f = 0.05$  \\
\hline

1 & \ \ \ \ 0.914(30)  & \ \ \ \ 2.106(45)  & \ \ \ \ 3.64(7)
  & \ \ \ \ 5.55(9)  & \ \ \ \ 7.85(12) \\
2 & \ \ \ \ 0.914(30)  & \ \ \ \ 2.106(45)  & \ \ \ \ 3.64(7)
  & \ \ \ \ 5.55(9)  & \ \ \ \ 7.85(12) \\
7 & \ -44.7(12)  & \ -54.3(11)  & \ -64.0(12)  & \ -74.8(13)
  & \ -86.8(14) \\
8 & -137.5(38) & -162.1(35) & -185.8(35) & -211.9(37) & -240.1(40)\\
9 & \ \ \ \ 1.370(44)  & \ \ \ \ 3.16(7)  & \ \ \ \ 5.46(10)
  & \ \ \ \ 8.33(14)  & \ \ \ \ 11.78(18) \\
10& \ \ \ \ 1.370(44)  & \ \ \ \ 3.16(7)  & \ \ \ \ 5.46(10)
  & \ \ \ \ 8.33(14)  & \ \ \ \ 11.78(18)

\end{tabular}

\caption{The values for $\langle \pi^+ | Q^{3/2}_{i,\rm lat} | K^+
\rangle \times 10^4$ for each light quark mass studied.}

\label{tab:ktopi_i2_charm_out}

\end{table}


\begin{table}
\begin{tabular}{cccccc}
$i$ & $m_s$ & $m_d=0.01 $ &  $m_d=0.02$  &  $m_d=0.03$ &
  $m_d=0.04$ \\
\hline        

1  & 0.02 & -0.009(43) $\times 10^{-3}$  & 		
	 & 				& 				\\ 
  & 0.03 & -0.056(50) $\times 10^{-3}$  & -0.013(40) $\times 10^{-3}$ 
	 &				&				\\ 
  & 0.04 & -0.098(56) $\times 10^{-3}$  & -0.053(43) $\times 10^{-3}$   
	 & -0.019(39) $\times 10^{-3}$  &				\\ 
  & 0.05 & -0.138(62) $\times 10^{-3}$  &  -0.090(47) $\times 10^{-3}$  
         & -0.054(40) $\times 10^{-3}$  &  -0.026(38) $\times 10^{-3}$  \\
2 & 0.02 & 0.338(22) $\times 10^{-3}$   &	
	 &				&				\\ 
  & 0.03 & 0.663(26) $\times 10^{-3}$   &   0.323(20) $\times 10^{-3}$
	 &                              &				\\ 
  & 0.04 & 0.979(29) $\times 10^{-3}$   & 0.634(22) $\times 10^{-3}$   
	 & 0.311(20) $\times 10^{-3}$ &			      \\ 
  & 0.05 & 1.287(32) $\times 10^{-3}$  & 0.938(24) $\times 10^{-3}$  
	 & 0.612(21) $\times 10^{-3}$  & 0.301(19) $\times 10^{-3}$  \\ 
3 & 0.02 & 0.065(15) $\times 10^{-2}$   &	
	 &				&				\\ 
  & 0.03 & 0.116(18) $\times 10^{-2}$   & 0.061(14) $\times 10^{-2}$
	 &				&				\\ 
  & 0.04 & 0.166(21) $\times 10^{-2}$   & 0.111(16) $\times 10^{-2}$   
	 & 0.056(14) $\times 10^{-2}$   &				\\ 
  & 0.05 & 0.215(23) $\times 10^{-2}$   & 0.160(17) $\times 10^{-2}$ 
	 & 0.106(15) $\times 10^{-2}$   & 0.052(14) $\times 10^{-2}$    \\ 
4 & 0.02 & 0.100(13) $\times 10^{-2}$   &
	 &				&				\\ 
  & 0.03 & 0.187(15) $\times 10^{-2}$   & 0.094(12) $\times 10^{-2}$    
	 &				&				\\ 
  & 0.04 & 0.273(17) $\times 10^{-2}$   & 0.179(13) $\times 10^{-2}$   
	 & 0.089(12) $\times 10^{-2}$   &				\\ 
  & 0.05 & 0.357(19) $\times 10^{-2}$   &  0.263(15) $\times 10^{-2}$ 
	 & 0.172(13) $\times 10^{-2}$   & 0.085(12) $\times 10^{-2}$   \\ 
5 & 0.02 & -0.635(13) $\times 10^{-2}$  &	
	 &				&				\\ 
  & 0.03 & -1.293(15) $\times 10^{-2}$  &   -0.644(12) $\times 10^{-2}$ 
	 &				&				\\ 
  & 0.04 & -1.950(18) $\times 10^{-2}$  & -1.302(13) $\times 10^{-2}$   
	 & -0.647(12) $\times 10^{-2}$  &				\\ 
  & 0.05 & -2.605(19) $\times 10^{-2}$  & -1.958(15) $\times 10^{-2}$ 
	 & -1.303(12) $\times 10^{-2}$  & -0.648(11) $\times 10^{-2}$   \\ 
6 & 0.02 & -1.870(11) $\times 10^{-2}$  &
	 &				&				\\ 
  & 0.03 & -3.775(13) $\times 10^{-2}$  & -1.8956(88) $\times 10^{-2}$ 
	 &				&				\\ 
  & 0.04 & -5.680(15) $\times 10^{-2}$  & -3.803(10) $\times 10^{-2}$   
	 & -1.8970(80) $\times 10^{-2}$ &				\\ 
  & 0.05 & -7.576(17) $\times 10^{-2}$  & -5.700(12) $\times 10^{-2}$   
	 & -3.7962(90) $\times 10^{-2}$ & -1.8900(74) $\times 10^{-2}$

\end{tabular}

\caption{The values for the ratio $ \langle 0 | Q_{i, \rm lat} | K^0
\rangle / \langle 0 | (\bar{s} \gamma_5 d)_{\rm lat} | K^0 \rangle $
for $i = 1$ to 6 for each non-degenerate pair of light quark masses.
These ratios are used in the determination of the subtraction
coefficient required to relate $K \to \pi$ matrix elements to $K \to
\pi \pi$ matrix elements.}

\label{tab:ktovac_o1_o6_charm_out}
\end{table}


\begin{table}
\begin{tabular}{cccccc}
$i$ & $m_s$ & $m_d=0.01 $ &  $m_d=0.02$  &  $m_d=0.03$ &
  $m_d=0.04$ \\
\hline        

7 & 0.02 & 3.4616(68) $\times 10^{-3}$  &
	 &				&				\\ 
  & 0.03 &  6.911(11) $\times 10^{-3}$  &  3.4359(47) $\times 10^{-3}$
	 &				&				\\ 
  & 0.04 & 10.333(15) $\times 10^{-3}$  & 6.8498(86) $\times 10^{-3}$   
	 & 3.4074(40) $\times 10^{-3}$  &				\\ 
  & 0.05 & 13.723(19) $\times 10^{-3}$  & 10.235(12) $\times 10^{-3}$   
	 & 6.7876(77) $\times 10^{-3}$  & 3.3762(37) $\times 10^{-3}$   \\ 
8 & 0.02 & 10.402(20) $\times 10^{-3}$  &
	 &				&				\\ 
  & 0.03 & 20.759(34) $\times 10^{-3}$  & 10.316(14) $\times 10^{-3}$  
	 &				&				\\ 
  & 0.04 & 31.031(45) $\times 10^{-3}$  & 20.563(26) $\times 10^{-3}$   
	 & 10.226(12) $\times 10^{-3}$  &				\\ 
  & 0.05 & 41.207(55) $\times 10^{-3}$  & 30.722(37) $\times 10^{-3}$  
	 & 20.370(23) $\times 10^{-3}$  & 10.131(11) $\times 10^{-3}$  \\ 
9 & 0.02 & -0.338(22) $\times 10^{-3}$  & 
	 &				&				\\ 
  & 0.03 & -0.662(26) $\times 10^{-3}$  &  -0.323(20) $\times 10^{-3}$ 
	 &				&				\\ 
  & 0.04 & -0.976(29) $\times 10^{-3}$  & -0.632(22) $\times 10^{-3}$   
	 & -0.311(20) $\times 10^{-3}$  &				\\ 
  & 0.05 & -1.281(32) $\times 10^{-3}$  & -0.934(24) $\times 10^{-3}$  
	 & -0.610(21) $\times 10^{-3}$  & -0.301(19) $\times 10^{-3}$   \\ 
10 & 0.02 & 0.010(43) $\times 10^{-3}$   &					 &				&				\\ 
   & 0.03& 0.057(50) $\times 10^{-3}$   &  0.014(40) $\times 10^{-3}$ 
	 &				&				\\ 
   & 0.04&  0.101(57) $\times 10^{-3}$  & 0.055(43) $\times 10^{-3}$   
	 &  0.020(39) $\times 10^{-3}$  &				\\ 
   & 0.05& 0.144(62) $\times 10^{-3}$   & 0.094(47) $\times 10^{-3}$  
	 & 0.056(40) $\times 10^{-3}$   & 0.026(38) $\times 10^{-3}$

\end{tabular}
\caption{The values for the ratio $ \langle 0 | Q_{i, \rm lat} | K^0
\rangle / \langle 0 | (\bar{s} \gamma_5 d)_{\rm lat} | K^0 \rangle $
matrix elements for $i = 7$ to 10 for each non-degenerate pair of light
quark masses.}

\label{tab:ktovac_o7_o10_charm_out}
\end{table}


\begin{table}
\begin{tabular}{cc}
 $m_f$  & $\langle \pi^+ | \Theta^{(27,1),(3/2)}_{\rm lat} | K^+
   \rangle$ \\ \hline
  0.01  &  0.000274(9) \\
  0.02  &  0.000632(14) \\
  0.03  &  0.001092(20) \\
  0.04  &  0.001665(27) \\
  0.05  &  0.002356(36) \\
\end{tabular}
\caption{Values for $\langle \pi^+ | \Theta^{(27,1),(3/2)}_{\rm lat} |
K^+ \rangle $ versus $m_f$.}
\label{tab:theta271_32_corr_mpi}
\end{table}


\begin{table}
\begin{tabular}{cccc}
 $m_f$ range  & $b^{(27,1)}_1$  & $b^{(27,1)}_2$ & $\chi^2/{\rm d.o.f}$
   \\ \hline
  0.01-0.04  &  0.00345(16)  &  0.0497(22)  &  1.1(4) \\
  0.01-0.05  &  0.00325(14)  &  0.0542(18)  &  1.9(6) \\
  0.02-0.04  &  0.00320(14)  &  0.0537(18)  &  0.4(1) \\
  0.02-0.05  &  0.00301(13)  &  0.0575(15)  &  0.7(2) \\
\end{tabular}
\caption{The dependence of the fit parameters in Eq.\
\ref{eq:theta271_32_corr_mpi} on the range of quark masses used.}
\label{tab:theta271_32_corr_mpi_fit}
\end{table}


\begin{table}
\begin{tabular}{cc}
 $m_f$  & $\langle \pi^+ | \Theta^{(8,8),(3/2)}_{7,\; \rm lat} | K^+
   \rangle$ \\ \hline
  0.01  &  -0.00447(12) \\
  0.02  &  -0.00543(11) \\
  0.03  &  -0.00640(12) \\
  0.04  &  -0.00748(13) \\
  0.05  &  -0.00868(13) \\
\end{tabular}
\caption{Values for $\langle \pi^+ | \Theta^{(8,8),(3/2)}_{7,\; \rm lat}
| K^+ \rangle $ versus $m_f$.}
\label{tab:theta88_32_7_corr_mpi}
\end{table}


\begin{table}
\begin{tabular}{cc}
 $m_f$  & $\langle \pi^+ | \Theta^{(8,8),(3/2)}_{8,\; \rm lat} | K^+
   \rangle$ \\ \hline
  0.01  &  -0.0137(4) \\
  0.02  &  -0.0162(3) \\
  0.03  &  -0.0186(4) \\
  0.04  &  -0.0212(4) \\
  0.05  &  -0.0240(4) \\
\end{tabular}
\caption{Values for $\langle \pi^+ | \Theta^{(8,8),(3/2)}_{8,\; \rm lat}
| K^+ \rangle $ versus $m_f$.}
\label{tab:theta88_32_8_corr_mpi}
\end{table}


\begin{table}
\begin{tabular}{ccccc}
 $i$ & $b^{(8,8)}_{i,0}$  & $b^{(8,8)}_{i,1}$ & $\xi^{(8,8)}_i$ &
   $\chi^2/{\rm d.o.f}$ \\ \hline
  7 &  -0.00323(13)  &  -0.0328(9)  & set to 0 & 0.6(2) \\ 
  7 &  -0.00380(20)  &  -0.0334(9)  & 1.5(2)   & 0.1(3) \\
  8 &  -0.0108(4)    &  -0.0801(27) & set to 0 & 0.2(1) \\ 
  8 &  -0.0117(6)    &  -0.0809(25) & 0.8(3)   & 0.1(2) \\
\end{tabular}
\caption{The results for fits to $\langle \pi^+ |
\Theta^{(8,8),(3/2)}_{i,\; \rm lat} | K^+ \rangle$ using the
parameterization of Eq.\ \ref{eq:theta88_32_i_corr_mpi}.  The data
gives $O(1)$ coefficients for the chiral logarithm term, which are not
currently known analytically.}
\label{tab:theta88_32_corr_mpi_fit}
\end{table}    


\begin{table}
\begin{tabular}{ccc}

$i$ & $\eta_{0,i}$ & $\eta_{1,i}$ \\ \hline        

1 & 0.024(35) $\times 10^{-3}$  & -0.040(12) $\times 10^{-1}$ \\
2 & -0.005(18) $\times 10^{-3}$   & 3.220(59) $\times 10^{-2}$ \\
3 & 0.006(13) $\times 10^{-2}$   & 0.521(42) $\times 10^{-1}$  \\
4 & 0.004(11) $\times 10^{-2}$  & 0.883(36) $\times 10^{-1}$ \\
5 & 0.010(10) $\times 10^{-2}$   & -6.543(37) $\times 10^{-1}$ \\
6 & 0.077(71) $\times 10^{-3}$   & -18.978(36) $\times 10^{-1}$ \\
7 & -1.285(74) $\times 10^{-5}$   & 34.326(46) $\times 10^{-2}$ \\
8 & -0.401(19) $\times 10^{-4}$   & 10.307(14) $\times 10^{-1}$ \\
9 & 0.004(18) $\times 10^{-3}$   & -3.203(59) $\times 10^{-2}$ \\
10 & -0.025(35) $\times 10^{-3}$  & 0.042(12) $\times 10^{-1}$

\end{tabular}

\caption{Results for uncorrelated fits of $ \langle 0 | Q_{i, \rm lat} |
K^0 \rangle / \langle 0 | (\bar{s} \gamma_5 d)_{\rm lat} | K^0 \rangle$
to the form  $\eta_{0,i} + \eta_{1,i} \left( m_s^\prime - m_d^\prime
\right) $.  For $Q_7$ and $Q_8$ the value for $\eta_{0,i}$ is very
small, but statistically non-zero.}

\label{tab:eta_charm_out}
\end{table}    


\begin{table}
\begin{tabular}{ccc}

$i$ & $\eta_{1,i}$ &  NPR \\ \hline        
  $
  \begin{aligned}
     &1 \\
     &2 \\
     &3 \\
     &4 \\
     &5 \\
     &6 \\
     &7 \\
     &8 \\
     &9 \\
     &10 \\
  \end{aligned}
  $
&
  $
  \begin{aligned}
    -0.&0040(12) \\
     0.&03220(59) \\
     0.&0521(42) \\
     0.&0883(36) \\
    -0.&6543(37) \\
    -1.&8978(36) \\
     0.&34326(46) \\
     1.&0307(14) \\
    -0.&03203(59) \\
     0.&0042(12) \\
  \end{aligned}
  $
&
  $
  \begin{aligned}
   -0.&0042(17) \\
    0.&0031(91) \\
   -0.&006(20) \\
    0.&001(28) \\
   -0.&672(12) \\
   -1.&995(45) \\
    0.&332(7) \\
    0.&993(20) \\
   -0.&0031(91) \\
      &       \\
  \end{aligned}
  $
\end{tabular}

\caption{A comparison of the subtraction coefficients in hadronic
states, $\eta_{1,i}$, with those found from Landau gauge-fixed quark
states.  Divergent contributions, which are independent of external
momenta, should give the same contribution to the two coefficients.
For operators with large power divergent subtractions, like $Q_6$ and
$Q_8$, the two coefficients are very similar.}

\label{tab:compare_sub_charm_out}
\end{table}


\begin{table}

\begin{tabular}{cccccc}

$i$ & $m_f=0.01 $ &  $m_f=0.02$  &  $m_f=0.03$ &  $m_f=0.04$ &
  $m_f=0.05$   \\
\hline

1  & 0.018(24) & -0.001(27)     & -0.031(29)     & -0.065(32)     & -0.101(36)    \\ 
2  & 0.039(11)     & 0.082(13)     & 0.133(14)     & 0.194(16)     & 0.265(17)    \\ 
3  & 0.123(84)     & 0.139(96)     & 0.13(10)     & 0.13(11)     & 0.14(13)    \\ 
4  & 0.144(70)     & 0.222(80)     & 0.298(87)     & 0.393(96)     & 0.51(11)    \\ 
5  & 0.127(74)     & 0.067(85)     & -0.001(93)     & -0.08(10)     & -0.18(11)    \\ 
6  & 0.193(75)     & 0.037(81)     & -0.141(89)     & -0.362(97)     & -0.65(11)    \\ 
7  & -0.768(20)     & -0.864(16)     & -0.934(15)     & -0.995(15)     & -1.046(16)    \\ 
8  & -2.450(63)     & -2.784(53)     & -3.045(50)     & -3.292(51)     & -3.520(53)    \\ 
9  & -0.034(11)     & -0.071(13)     & -0.113(14)     & -0.164(16)     & -0.223(17)    \\ 
10  & -0.013(24)     & 0.013(27)     & 0.051(29)     & 0.095(32)     & 0.143(36)

\end{tabular}
\caption{Values for the $\DIhalf$ matrix elements of the subtracted
operators, $ \langle \pi^+ | Q^{(1/2)}_{i, \rm lat} | K^+ \rangle_{\rm
sub} \times 10^2$.  This subtraction is done in hadronic states and
removes the unphysical contribution to this matrix element for
$i \ne 7$ and 8.  For $Q_7$ and $Q_8$, the subtraction leaves a finite
matrix element, whose value in the chiral limit is related to physical
quantities.}

\label{tab:ktopi_sub_12}
\end{table}


\begin{table}

\begin{tabular}{clll}

$i$ &  \multicolumn{1}{c}{$c_{0, i}$}  & \multicolumn{1}{c}{$c_{1, i}$}
   & $\chi^2/{\rm d.o.f}$ \\ \hline

1  &  \ 0.00053(27)  &   -0.0297(78) & 0.05(8) \\
2  &   -0.00024(13)  &  \ 0.0555(40) & 0.6(3)  \\
3  &  \ 0.00123(97)  &  \ 0.0036(284)& 0.004(12) \\
4  &  \ 0.00047(80)  &  \ 0.089(24)  & 0.04(6) \\
5  &  \ 0.00210(84)  &   -0.074(25)  & 0.02(4) \\
6  &  \ 0.00426(84)  &   -0.203(25)  & 0.3(2) \\
7  &   -0.00720(21)  &   -0.0675(37) & 0.8(3) \\
8  &   -0.0223(7)    &   -0.262(12)  & 0.4(2) \\
9  &  \ 0.00018(13)  &   -0.0464(40) & 0.4(2) \\
10 &   -0.00059(27)  &  \ 0.0389(79) & 0.09(10)

\end{tabular}
\caption{Results for linear fits of $ \langle \pi^+ | Q^{(1/2)}_{i, \rm
lat} | K^+ \rangle_{\rm sub}$ to the form of
Eq.\ \ref{eq:fit_ktopi_sub_charm_out}.  The slope of the fit, given by
$c_{1,i}$, is related to the low energy constant needed to determine $K
\rightarrow \pi \pi$ matrix elements for $i \ne 7$ and 8.  For $i = 7$
and 8, the matrix element in the chiral limit is the physical quantity
we seek , but the chiral limit is uncertain for these power divergent
operators at finite $L_s$.  For these operators, we use the $\DIthalf$
part of the operator to determine the $\DIhalf$ part.}

\label{tab:fit_ktopi_sub_charm_out}
\end{table}


\begin{table}

\begin{tabular}{cr}

$i$ &  \multicolumn{1}{c}{$\alpha^{(1/2)}_{i, \rm lat}$} \\ \hline

1  &  $ -1.19(31) \; \times 10^{-5} $ \\
2  &  $  2.22(16) \; \times 10^{-5} $ \\
3  &  $  0.15(113)   \times 10^{-5}$  \\
4  &  $  3.55(96) \; \times 10^{-5} $ \\
5  &  $ -2.97(100)   \times 10^{-5}$  \\
6  &  $ -8.12(98)   \times 10^{-5}$  \\
9  &  $ -1.85(16) \; \times 10^{-5} $ \\
10 &  $  1.55(31) \; \times 10^{-5} $

\end{tabular}
\caption{The lattice values for the low energy, chiral perturbation
theory constants for $\DIhalf$ amplitudes for $i \ne 7$ and 8.  These
were determined from subtracted $K^+ \rightarrow \pi^+$ matrix
elements.}

\label{tab:alpha_ktopi_sub_charm_out}
\end{table}


\begin{table}

\begin{tabular}{lr}

Parameter &  Value \\ \hline

$\alpha^{(27,1),(1/2)}_{\rm lat}$   &  $ -4.13(18) \times 10^{-6}$ \\
$\alpha^{(8,8),(1/2)}_{7,\rm lat}$  &  $ -3.22(16) \times 10^{-6}$ \\
$\alpha^{(8,8),(1/2)}_{8,\rm lat}$  &  $ -9.92(54) \times 10^{-6}$ \\

$\alpha^{(27,1),(3/2)}_{\rm lat}$   &  $ -4.13(18) \times 10^{-6}$ \\
$\alpha^{(8,8),(3/2)}_{7,\rm lat}$  &  $ -1.61(8) \times 10^{-6}$ \\
$\alpha^{(8,8),(3/2)}_{8,\rm lat}$  &  $ -4.96(27) \times 10^{-6}$

\end{tabular}
\caption{The lattice values for the $\DIhalf$ and $\DIthalf$ low
energy, chiral perturbation theory constants determined from $K^+
\rightarrow \pi^+$ matrix elements not requiring subtraction.}

\label{tab:alpha_ktopi_nosub_charm_out}
\end{table}


\begin{table}

\begin{tabular}{crr}

$i$ &  \multicolumn{1}{c}{$\alpha^{(1/2)}_{i, \rm lat}$}
    &  \multicolumn{1}{c}{$\alpha^{(3/2)}_{i, \rm lat}$} \\ \hline

1  &  $ -1.19(31) \; \times 10^{-5} $	& $-1.38(6) \; \times 10^{-6}$\\
2  &  $  2.22(16) \; \times 10^{-5} $	& $-1.38(6) \; \times 10^{-6}$\\
3  &  $  0.15(113)   \times 10^{-5}$	&   0.0			      \\
4  &  $  3.55(96) \; \times 10^{-5} $	&   0.0			      \\
5  &  $ -2.97(100)   \times 10^{-5}$	&   0.0			      \\
6  &  $ -8.12(98) \; \times 10^{-5}$	&   0.0			      \\
7  &  $ -3.22(16) \; \times 10^{-6} $   & $-1.61(8) \; \times 10^{-6}$\\
8  &  $ -9.92(54) \; \times 10^{-6} $   & $-4.96(27)   \times 10^{-6}$\\
9  &  $ -1.85(16) \; \times 10^{-5} $	& $-2.07(9) \; \times 10^{-6}$\\
10 &  $  1.55(31) \; \times 10^{-5} $	& $-2.07(9) \; \times 10^{-6}$

\end{tabular}
\caption{The lattice values for the low energy, chiral perturbation
theory constants decomposed by isospin for $Q_1$ to $Q_{10}$.}

\label{tab:final_alpha_charm_out}
\end{table}


\begin{table}

\begin{tabular}{lcl}
  Quantity &  Central Value &  Comments and References \\ \hline
  $m_{\pi^+}$		& 139.57 MeV	& \\
  $m_{\pi^0}$		& 134.98 MeV	&  \\
  $f_{\pi^+}$		& 130.7 MeV	&  \\
  $m_{K^+}$		& 493.68 MeV	&  \\
  $m_{K^0}$		& 497.67 MeV	&  \\
  $f_{K^+}$		& 159.8 MeV	&  \\
  $G_F$			& $1.166 \times 10^{-5}$ GeV$^{-2}$ \\
  $\lambda_{\rm CKM}$	& 0.2237	& \cite{Ciuchini:2000de} \\
  $A_{\rm CKM}$		& 0.819		& \cite{Ciuchini:2000de} \\
  $\bar{\rho}_{\rm CKM}$ & 0.222	& \cite{Ciuchini:2000de} \\
  $\bar{\eta}_{\rm CKM}$ & 0.316	& \cite{Ciuchini:2000de} \\
  $\rho_{\rm CKM}$	& 0.228		&
    From $\lambda_{\rm CKM}$, $\bar{\rho}_{\rm CKM}$ and
    $\bar{\eta}_{\rm CKM}$ \\
  $\eta_{\rm CKM}$	& 0.324		&
    From $\lambda_{\rm CKM}$, $\bar{\rho}_{\rm CKM}$ and
    $\bar{\eta}_{\rm CKM}$ \\
  $|V_{us}|$		& 0.2237	& $ \equiv \lambda_{\rm CKM}$ \\
  $|V_{ud}|$		& 0.9747	& \\
  $|V_{cb}|$		& 0.0410	&
    $ = A_{\rm CKM} \lambda_{\rm CKM}^2 $ \\ 
  $V_{td}$		& $0.00708 -0.00297 i$		& \\
  $\tau$		& $0.00133 -0.000559 i$		& \\
  $\epsilon$		& $2.271 \times 10^{-3}$	& \\
  Re $A_0$		& $3.33\times10^{-7}$ GeV  & \\
  $ \omega$		& 0.045		& \\
  $\repe$		& $(20.7 \pm 2.8) \times 10^{-4}$ &
    KTEV \cite{Alavi-Harati:web} \\
  			& $(15.3 \pm 2.6) \times 10^{-4}$ &
    NA48 \cite{Fanti:web} \\
\end{tabular}

\caption{Central values for standard model parameters and experimental
results relevant to the calculations presented in this paper.  All
values are from the 2000 Particle Data Book unless otherwise noted.
The central values for $\lambda_{\rm CKM}$, $A_{\rm CKM}$,
$\bar{\rho}_{\rm CKM}$ and $\bar{\eta}_{\rm CKM}$ are taken, without
errors.  Current errors on all quantities in the table which enter
as inputs in our calculation have virtually no effect on our results.}

\label{tab:sm_values}

\end{table}


\begin{table}
\begin{tabular}{rrrrr}
	& \multicolumn{2}{c}{Real $A_0$} 
	& \multicolumn{2}{c}{Real $A_2$} \\
	$i$ 
	& \multicolumn{1}{c}{choice 1} 
	& \multicolumn{1}{c}{choice 2} 
	& \multicolumn{1}{c}{choice 1} 
	& \multicolumn{1}{c}{choice 2} \\ \hline

$
  \begin{alignedat}
   1 & 1 \\
     & 2 \\
     & 3 \\
     & 4 \\
     & 5 \\
     & 6 \\
     & 7 \\
     & 8 \\
     & 9 \\
     & 10 

  \end{alignedat}
  $
& 
 $ 
 \begin{alignedat}{2} 
3.&02(68) & \times 10&^{-8} \\ 
2.&00(18) & \times 10&^{-7} \\ 
1.&4(29) & \times 10&^{-10} \\ 
-3.&80(84) & \times 10&^{-9} \\ 
-6.&9(29) & \times 10&^{-10} \\ 
4.&99(77) & \times 10&^{-9} \\ 
4.&04(21) & \times 10&^{-11} \\ 
-5.&74(32) & \times 10&^{-11} \\ 
-3.&91(39) & \times 10&^{-12} \\ 
2.&27(41) & \times 10&^{-12} \\ 
\end{alignedat} 
 $ 
& 
 $ 
 \begin{alignedat}{2} 
4.&28(97) & \times 10&^{-8} \\ 
2.&83(25) & \times 10&^{-7} \\ 
2.&0(41) & \times 10&^{-10} \\ 
-5.&4(12) & \times 10&^{-9} \\ 
-9.&8(41) & \times 10&^{-10} \\ 
7.&1(11) & \times 10&^{-9} \\ 
8.&00(42) & \times 10&^{-11} \\ 
-1.&137(63) & \times 10&^{-10} \\ 
-5.&54(56) & \times 10&^{-12} \\ 
3.&23(59) & \times 10&^{-12} \\ 
\end{alignedat} 
 $ 
& 
 $ 
 \begin{alignedat}{2} 
-4.&11(18) & \times 10&^{-9} \\ 
1.&392(62) & \times 10&^{-8} \\ 
0.&0 & & \\ 
0.&0 & & \\ 
0.&0 & & \\ 
0.&0 & & \\ 
2.&86(15) & \times 10&^{-11} \\ 
-4.&06(22) & \times 10&^{-11} \\ 
4.&69(21) & \times 10&^{-13} \\ 
3.&70(17) & \times 10&^{-13} \\ 
\end{alignedat} 
 $ 
& 
 $ 
 \begin{alignedat}{2} 
-4.&82(22) & \times 10&^{-9} \\ 
1.&635(73) & \times 10&^{-8} \\ 
0.&0 & & \\ 
0.&0 & & \\ 
0.&0 & & \\ 
0.&0 & & \\ 
3.&63(19) & \times 10&^{-11} \\ 
-5.&15(28) & \times 10&^{-11} \\ 
5.&51(25) & \times 10&^{-13} \\ 
4.&35(20) & \times 10&^{-13} \\ 
\end{alignedat} 
 $

\end{tabular}

\caption{The contribution in GeV from the renormalized continuum
operator $Q_{i, \rm cont}$ to the real parts of $\langle (\pi \pi)_I |
-i \DSoneH | K^0 \rangle$ for $\mu = 1.51$ GeV. The central values for
the standard model parameters given in Table \ref{tab:sm_values} have
been used.}

\label{tab:ReA0_ReA2_mom1.tab}

\end{table}


\begin{table}
\begin{tabular}{rrrrr}
	& \multicolumn{2}{c}{Imaginary $A_0$} 
	& \multicolumn{2}{c}{Imaginary $A_2$} \\
	$i$ 
	& \multicolumn{1}{c}{choice 1} 
	& \multicolumn{1}{c}{choice 2} 
	& \multicolumn{1}{c}{choice 1} 
	& \multicolumn{1}{c}{choice 2} \\ \hline

$
  \begin{alignedat}
   1 & 1 \\
     & 2 \\
     & 3 \\
     & 4 \\
     & 5 \\
     & 6 \\
     & 7 \\
     & 8 \\
     & 9 \\
     & 10 

  \end{alignedat}
  $
& 
 $ 
 \begin{alignedat}{2} 
0.&0 & & \\ 
0.&0 & & \\ 
-4.&7(94) & \times 10&^{-13} \\ 
8.&2(18) & \times 10&^{-12} \\ 
4.&7(20) & \times 10&^{-13} \\ 
-1.&72(27) & \times 10&^{-11} \\ 
7.&57(39) & \times 10&^{-14} \\ 
-1.&787(98) & \times 10&^{-12} \\ 
-9.&45(95) & \times 10&^{-13} \\ 
-2.&25(41) & \times 10&^{-13} \\ 
\end{alignedat} 
 $ 
& 
 $ 
 \begin{alignedat}{2} 
0.&0 & & \\ 
0.&0 & & \\ 
-7.&(13) & \times 10&^{-13} \\ 
1.&17(26) & \times 10&^{-11} \\ 
6.&7(28) & \times 10&^{-13} \\ 
-2.&45(38) & \times 10&^{-11} \\ 
1.&498(78) & \times 10&^{-13} \\ 
-3.&54(19) & \times 10&^{-12} \\ 
-1.&34(14) & \times 10&^{-12} \\ 
-3.&19(58) & \times 10&^{-13} \\ 
\end{alignedat} 
 $ 
& 
 $ 
 \begin{alignedat}{2} 
0.&0 & & \\ 
0.&0 & & \\ 
0.&0 & & \\ 
0.&0 & & \\ 
0.&0 & & \\ 
0.&0 & & \\ 
5.&35(28) & \times 10&^{-14} \\ 
-1.&263(70) & \times 10&^{-12} \\ 
1.&135(51) & \times 10&^{-13} \\ 
-3.&66(16) & \times 10&^{-14} \\ 
\end{alignedat} 
 $ 
& 
 $ 
 \begin{alignedat}{2} 
0.&0 & & \\ 
0.&0 & & \\ 
0.&0 & & \\ 
0.&0 & & \\ 
0.&0 & & \\ 
0.&0 & & \\ 
6.&78(35) & \times 10&^{-14} \\ 
-1.&602(88) & \times 10&^{-12} \\ 
1.&334(60) & \times 10&^{-13} \\ 
-4.&30(19) & \times 10&^{-14} \\ 
\end{alignedat} 
 $

\end{tabular}

\caption{The contribution in GeV from the renormalized continuum
operator $Q_{i, \rm cont}$ to the imaginary parts of $\langle (\pi
\pi)_I | -i \DSoneH | K^0 \rangle$ for $\mu = 1.51$ GeV.  The central
values for the standard model parameters given in Table
\ref{tab:sm_values} have been used.}

\label{tab:ImA0_ImA2_mom1.tab}
\end{table}


\begin{table}
\begin{tabular}{rrrrr}
	& \multicolumn{2}{c}{Real $A_0$} 
	& \multicolumn{2}{c}{Real $A_2$} \\
	$i$ 
	& \multicolumn{1}{c}{choice 1} 
	& \multicolumn{1}{c}{choice 2} 
	& \multicolumn{1}{c}{choice 1} 
	& \multicolumn{1}{c}{choice 2} \\ \hline

$
  \begin{alignedat}
   1 & 1 \\
     & 2 \\
     & 3 \\
     & 4 \\
     & 5 \\
     & 6 \\
     & 7 \\
     & 8 \\
     & 9 \\
     & 10 

  \end{alignedat}
  $
& 
 $ 
 \begin{alignedat}{2} 
2.&69(61) & \times 10&^{-8} \\ 
1.&81(12) & \times 10&^{-7} \\ 
1.&(13) & \times 10&^{-11} \\ 
-1.&46(33) & \times 10&^{-9} \\ 
-4.&4(18) & \times 10&^{-10} \\ 
3.&09(38) & \times 10&^{-9} \\ 
5.&32(27) & \times 10&^{-11} \\ 
-1.&785(97) & \times 10&^{-10} \\ 
-2.&59(20) & \times 10&^{-12} \\ 
5.&14(94) & \times 10&^{-12} \\ 
\end{alignedat} 
 $ 
& 
 $ 
 \begin{alignedat}{2} 
3.&82(87) & \times 10&^{-8} \\ 
2.&57(17) & \times 10&^{-7} \\ 
2.&(18) & \times 10&^{-11} \\ 
-2.&07(47) & \times 10&^{-9} \\ 
-6.&3(26) & \times 10&^{-10} \\ 
4.&38(54) & \times 10&^{-9} \\ 
1.&054(54) & \times 10&^{-10} \\ 
-3.&53(19) & \times 10&^{-10} \\ 
-3.&68(28) & \times 10&^{-12} \\ 
7.&3(13) & \times 10&^{-12} \\ 
\end{alignedat} 
 $ 
& 
 $ 
 \begin{alignedat}{2} 
-3.&64(16) & \times 10&^{-9} \\ 
1.&371(61) & \times 10&^{-8} \\ 
0.&0 & & \\ 
0.&0 & & \\ 
0.&0 & & \\ 
0.&0 & & \\ 
3.&76(19) & \times 10&^{-11} \\ 
-1.&262(68) & \times 10&^{-10} \\ 
3.&43(15) & \times 10&^{-13} \\ 
8.&33(37) & \times 10&^{-13} \\ 
\end{alignedat} 
 $ 
& 
 $ 
 \begin{alignedat}{2} 
-4.&27(19) & \times 10&^{-9} \\ 
1.&610(72) & \times 10&^{-8} \\ 
0.&0 & & \\ 
0.&0 & & \\ 
0.&0 & & \\ 
0.&0 & & \\ 
4.&77(24) & \times 10&^{-11} \\ 
-1.&601(87) & \times 10&^{-10} \\ 
4.&03(18) & \times 10&^{-13} \\ 
9.&79(44) & \times 10&^{-13} \\ 
\end{alignedat} 
 $

\end{tabular}

\caption{The contribution in GeV from the renormalized continuum
operator $Q_{i, \rm cont}$ to the real parts of $\langle (\pi \pi)_I |
-i \DSoneH | K^0 \rangle$ for $\mu = 2.13$ GeV.  The central values for
the standard model parameters given in Table \ref{tab:sm_values} have
been used.}

\label{tab:ReA0_ReA2_mom2.tab}
\end{table}


\begin{table}
\begin{tabular}{rrrrr}
	& \multicolumn{2}{c}{Imaginary $A_0$} 
	& \multicolumn{2}{c}{Imaginary $A_2$} \\
	$i$ 
	& \multicolumn{1}{c}{choice 1} 
	& \multicolumn{1}{c}{choice 2} 
	& \multicolumn{1}{c}{choice 1} 
	& \multicolumn{1}{c}{choice 2} \\ \hline

$
  \begin{alignedat}
   1 & 1 \\
     & 2 \\
     & 3 \\
     & 4 \\
     & 5 \\
     & 6 \\
     & 7 \\
     & 8 \\
     & 9 \\
     & 10 

  \end{alignedat}
  $
& 
 $ 
 \begin{alignedat}{2} 
0.&0 & & \\ 
0.&0 & & \\ 
-7.&(87) & \times 10&^{-14} \\ 
7.&2(16) & \times 10&^{-12} \\ 
6.&3(26) & \times 10&^{-13} \\ 
-2.&12(26) & \times 10&^{-11} \\ 
6.&95(36) & \times 10&^{-14} \\ 
-1.&583(86) & \times 10&^{-12} \\ 
-8.&43(64) & \times 10&^{-13} \\ 
-2.&01(37) & \times 10&^{-13} \\ 
\end{alignedat} 
 $ 
& 
 $ 
 \begin{alignedat}{2} 
0.&0 & & \\ 
0.&0 & & \\ 
-1.&(12) & \times 10&^{-13} \\ 
1.&02(23) & \times 10&^{-11} \\ 
9.&0(36) & \times 10&^{-13} \\ 
-3.&00(37) & \times 10&^{-11} \\ 
1.&376(70) & \times 10&^{-13} \\ 
-3.&13(17) & \times 10&^{-12} \\ 
-1.&196(91) & \times 10&^{-12} \\ 
-2.&85(52) & \times 10&^{-13} \\ 
\end{alignedat} 
 $ 
& 
 $ 
 \begin{alignedat}{2} 
0.&0 & & \\ 
0.&0 & & \\ 
0.&0 & & \\ 
0.&0 & & \\ 
0.&0 & & \\ 
0.&0 & & \\ 
4.&91(25) & \times 10&^{-14} \\ 
-1.&119(61) & \times 10&^{-12} \\ 
1.&114(50) & \times 10&^{-13} \\ 
-3.&25(15) & \times 10&^{-14} \\ 
\end{alignedat} 
 $ 
& 
 $ 
 \begin{alignedat}{2} 
0.&0 & & \\ 
0.&0 & & \\ 
0.&0 & & \\ 
0.&0 & & \\ 
0.&0 & & \\ 
0.&0 & & \\ 
6.&23(32) & \times 10&^{-14} \\ 
-1.&419(77) & \times 10&^{-12} \\ 
1.&309(59) & \times 10&^{-13} \\ 
-3.&82(17) & \times 10&^{-14} \\ 
\end{alignedat} 
 $

\end{tabular}

\caption{The contribution in GeV from the renormalized continuum
operator $Q_{i, \rm cont}$ to the imaginary parts of $\langle (\pi
\pi)_I | -i \DSoneH | K^0 \rangle$ for $\mu = 2.13$ GeV.  The central
values for the standard model parameters given in Table
\ref{tab:sm_values} have been used.}

\label{tab:ImA0_ImA2_mom2.tab}
\end{table}


\begin{table}
\begin{tabular}{rrrrr}
	& \multicolumn{2}{c}{Real $A_0$} 
	& \multicolumn{2}{c}{Real $A_2$} \\
	$i$ 
	& \multicolumn{1}{c}{choice 1} 
	& \multicolumn{1}{c}{choice 2} 
	& \multicolumn{1}{c}{choice 1} 
	& \multicolumn{1}{c}{choice 2} \\ \hline

$
  \begin{alignedat}
   1 & 1 \\
     & 2 \\
     & 3 \\
     & 4 \\
     & 5 \\
     & 6 \\
     & 7 \\
     & 8 \\
     & 9 \\
     & 10 

  \end{alignedat}
  $
& 
 $ 
 \begin{alignedat}{2} 
2.&69(59) & \times 10&^{-8} \\ 
1.&87(11) & \times 10&^{-7} \\ 
9.&(87) & \times 10&^{-12} \\ 
-9.&9(22) & \times 10&^{-10} \\ 
-3.&5(14) & \times 10&^{-10} \\ 
2.&03(25) & \times 10&^{-9} \\ 
5.&76(29) & \times 10&^{-11} \\ 
-2.&08(11) & \times 10&^{-10} \\ 
-2.&97(21) & \times 10&^{-12} \\ 
6.&1(11) & \times 10&^{-12} \\ 
\end{alignedat} 
 $ 
& 
 $ 
 \begin{alignedat}{2} 
3.&82(84) & \times 10&^{-8} \\ 
2.&65(16) & \times 10&^{-7} \\ 
1.&(12) & \times 10&^{-11} \\ 
-1.&40(31) & \times 10&^{-9} \\ 
-5.&0(20) & \times 10&^{-10} \\ 
2.&88(35) & \times 10&^{-9} \\ 
1.&140(58) & \times 10&^{-10} \\ 
-4.&12(22) & \times 10&^{-10} \\ 
-4.&21(30) & \times 10&^{-12} \\ 
8.&7(15) & \times 10&^{-12} \\ 
\end{alignedat} 
 $ 
& 
 $ 
 \begin{alignedat}{2} 
-3.&45(15) & \times 10&^{-9} \\ 
1.&346(60) & \times 10&^{-8} \\ 
0.&0 & & \\ 
0.&0 & & \\ 
0.&0 & & \\ 
0.&0 & & \\ 
4.&07(21) & \times 10&^{-11} \\ 
-1.&472(80) & \times 10&^{-10} \\ 
3.&70(17) & \times 10&^{-13} \\ 
9.&46(42) & \times 10&^{-13} \\ 
\end{alignedat} 
 $ 
& 
 $ 
 \begin{alignedat}{2} 
-4.&05(18) & \times 10&^{-9} \\ 
1.&582(71) & \times 10&^{-8} \\ 
0.&0 & & \\ 
0.&0 & & \\ 
0.&0 & & \\ 
0.&0 & & \\ 
5.&16(26) & \times 10&^{-11} \\ 
-1.&87(10) & \times 10&^{-10} \\ 
4.&35(20) & \times 10&^{-13} \\ 
1.&111(50) & \times 10&^{-12} \\ 
\end{alignedat} 
 $

\end{tabular}

\caption{The contribution in GeV from the renormalized continuum
operator $Q_{i, \rm cont}$ to the real parts of $\langle (\pi \pi)_I |
-i \DSoneH | K^0 \rangle$ for $\mu = 2.39$ GeV. The central values for
the standard model parameters given in Table \ref{tab:sm_values} have
been used.}

\label{tab:ReA0_ReA2_mom3.tab}
\end{table}


\begin{table}
\begin{tabular}{rrrrr}
	& \multicolumn{2}{c}{Imaginary $A_0$} 
	& \multicolumn{2}{c}{Imaginary $A_2$} \\
	$i$ 
	& \multicolumn{1}{c}{choice 1} 
	& \multicolumn{1}{c}{choice 2} 
	& \multicolumn{1}{c}{choice 1} 
	& \multicolumn{1}{c}{choice 2} \\ \hline

$
  \begin{alignedat}
   1 & 1 \\
     & 2 \\
     & 3 \\
     & 4 \\
     & 5 \\
     & 6 \\
     & 7 \\
     & 8 \\
     & 9 \\
     & 10 

  \end{alignedat}
  $
& 
 $ 
 \begin{alignedat}{2} 
0.&0 & & \\ 
0.&0 & & \\ 
-9.&(86) & \times 10&^{-14} \\ 
7.&5(17) & \times 10&^{-12} \\ 
6.&6(27) & \times 10&^{-13} \\ 
-1.&99(24) & \times 10&^{-11} \\ 
6.&46(33) & \times 10&^{-14} \\ 
-1.&512(82) & \times 10&^{-12} \\ 
-8.&76(61) & \times 10&^{-13} \\ 
-1.&99(35) & \times 10&^{-13} \\ 
\end{alignedat} 
 $ 
& 
 $ 
 \begin{alignedat}{2} 
0.&0 & & \\ 
0.&0 & & \\ 
-1.&(12) & \times 10&^{-13} \\ 
1.&07(23) & \times 10&^{-11} \\ 
9.&4(38) & \times 10&^{-13} \\ 
-2.&83(34) & \times 10&^{-11} \\ 
1.&279(65) & \times 10&^{-13} \\ 
-2.&99(16) & \times 10&^{-12} \\ 
-1.&243(87) & \times 10&^{-12} \\ 
-2.&83(50) & \times 10&^{-13} \\ 
\end{alignedat} 
 $ 
& 
 $ 
 \begin{alignedat}{2} 
0.&0 & & \\ 
0.&0 & & \\ 
0.&0 & & \\ 
0.&0 & & \\ 
0.&0 & & \\ 
0.&0 & & \\ 
4.&57(23) & \times 10&^{-14} \\ 
-1.&069(58) & \times 10&^{-12} \\ 
1.&093(49) & \times 10&^{-13} \\ 
-3.&09(14) & \times 10&^{-14} \\ 
\end{alignedat} 
 $ 
& 
 $ 
 \begin{alignedat}{2} 
0.&0 & & \\ 
0.&0 & & \\ 
0.&0 & & \\ 
0.&0 & & \\ 
0.&0 & & \\ 
0.&0 & & \\ 
5.&79(30) & \times 10&^{-14} \\ 
-1.&356(74) & \times 10&^{-12} \\ 
1.&285(58) & \times 10&^{-13} \\ 
-3.&63(16) & \times 10&^{-14} \\ 
\end{alignedat} 
 $

\end{tabular}

\caption{The contribution in GeV from the renormalized continuum
operator $Q_{i, \rm cont}$ to the imaginary parts of $\langle (\pi
\pi)_I | -i \DSoneH | K^0 \rangle$ for $\mu = 2.39$ GeV. The central
values for the standard model parameters given in Table
\ref{tab:sm_values} have been used.}

\label{tab:ImA0_ImA2_mom3.tab}
\end{table}


\begin{table}
\begin{tabular}{rrrrr}
	& \multicolumn{2}{c}{Real $A_0$} 
	& \multicolumn{2}{c}{Real $A_2$} \\
	$i$ 
	& \multicolumn{1}{c}{choice 1} 
	& \multicolumn{1}{c}{choice 2} 
	& \multicolumn{1}{c}{choice 1} 
	& \multicolumn{1}{c}{choice 2} \\ \hline

$
  \begin{alignedat}
   1 & 1 \\
     & 2 \\
     & 3 \\
     & 4 \\
     & 5 \\
     & 6 \\
     & 7 \\
     & 8 \\
     & 9 \\
     & 10 

  \end{alignedat}
  $
& 
 $ 
 \begin{alignedat}{2} 
2.&46(54) & \times 10&^{-8} \\ 
1.&72(12) & \times 10&^{-7} \\ 
-2.&(13) & \times 10&^{-12} \\ 
-4.&5(11) & \times 10&^{-13} \\ 
-1.&64(69) & \times 10&^{-10} \\ 
3.&51(42) & \times 10&^{-10} \\ 
6.&79(35) & \times 10&^{-11} \\ 
-2.&58(14) & \times 10&^{-10} \\ 
-3.&92(30) & \times 10&^{-12} \\ 
7.&4(13) & \times 10&^{-12} \\ 
\end{alignedat} 
 $ 
& 
 $ 
 \begin{alignedat}{2} 
3.&48(77) & \times 10&^{-8} \\ 
2.&45(16) & \times 10&^{-7} \\ 
-3.&(18) & \times 10&^{-12} \\ 
-6.&4(15) & \times 10&^{-13} \\ 
-2.&33(98) & \times 10&^{-10} \\ 
4.&98(60) & \times 10&^{-10} \\ 
1.&344(69) & \times 10&^{-10} \\ 
-5.&10(28) & \times 10&^{-10} \\ 
-5.&56(43) & \times 10&^{-12} \\ 
1.&05(19) & \times 10&^{-11} \\ 
\end{alignedat} 
 $ 
& 
 $ 
 \begin{alignedat}{2} 
-3.&09(14) & \times 10&^{-9} \\ 
1.&294(58) & \times 10&^{-8} \\ 
0.&0 & & \\ 
0.&0 & & \\ 
0.&0 & & \\ 
0.&0 & & \\ 
4.&80(25) & \times 10&^{-11} \\ 
-1.&821(99) & \times 10&^{-10} \\ 
5.&12(23) & \times 10&^{-13} \\ 
1.&127(51) & \times 10&^{-12} \\ 
\end{alignedat} 
 $ 
& 
 $ 
 \begin{alignedat}{2} 
-3.&63(16) & \times 10&^{-9} \\ 
1.&520(68) & \times 10&^{-8} \\ 
0.&0 & & \\ 
0.&0 & & \\ 
0.&0 & & \\ 
0.&0 & & \\ 
6.&09(31) & \times 10&^{-11} \\ 
-2.&31(13) & \times 10&^{-10} \\ 
6.&02(27) & \times 10&^{-13} \\ 
1.&324(59) & \times 10&^{-12} \\ 
\end{alignedat} 
 $

\end{tabular}

\caption{The contribution in GeV from the renormalized continuum
operator $Q_{i, \rm cont}$ to the real parts of $\langle (\pi \pi)_I |
-i \DSoneH | K^0 \rangle$ for $\mu = 3.02$ GeV. The central values for
the standard model parameters given in Table \ref{tab:sm_values} have
been used.}

\label{tab:ReA0_ReA2_mom4.tab}
\end{table}


\begin{table}
\begin{tabular}{rrrrr}
	& \multicolumn{2}{c}{Imaginary $A_0$} 
	& \multicolumn{2}{c}{Imaginary $A_2$} \\
	$i$ 
	& \multicolumn{1}{c}{choice 1} 
	& \multicolumn{1}{c}{choice 2} 
	& \multicolumn{1}{c}{choice 1} 
	& \multicolumn{1}{c}{choice 2} \\ \hline

$
  \begin{alignedat}
   1 & 1 \\
     & 2 \\
     & 3 \\
     & 4 \\
     & 5 \\
     & 6 \\
     & 7 \\
     & 8 \\
     & 9 \\
     & 10 

  \end{alignedat}
  $
& 
 $ 
 \begin{alignedat}{2} 
0.&0 & & \\ 
0.&0 & & \\ 
1.&2(80) & \times 10&^{-13} \\ 
6.&7(16) & \times 10&^{-12} \\ 
6.&8(29) & \times 10&^{-13} \\ 
-2.&06(25) & \times 10&^{-11} \\ 
5.&49(28) & \times 10&^{-14} \\ 
-1.&386(75) & \times 10&^{-12} \\ 
-8.&01(62) & \times 10&^{-13} \\ 
-1.&82(33) & \times 10&^{-13} \\ 
\end{alignedat} 
 $ 
& 
 $ 
 \begin{alignedat}{2} 
0.&0 & & \\ 
0.&0 & & \\ 
2.&(11) & \times 10&^{-13} \\ 
9.&5(23) & \times 10&^{-12} \\ 
9.&6(41) & \times 10&^{-13} \\ 
-2.&93(35) & \times 10&^{-11} \\ 
1.&087(56) & \times 10&^{-13} \\ 
-2.&74(15) & \times 10&^{-12} \\ 
-1.&137(88) & \times 10&^{-12} \\ 
-2.&58(46) & \times 10&^{-13} \\ 
\end{alignedat} 
 $ 
& 
 $ 
 \begin{alignedat}{2} 
0.&0 & & \\ 
0.&0 & & \\ 
0.&0 & & \\ 
0.&0 & & \\ 
0.&0 & & \\ 
0.&0 & & \\ 
3.&88(20) & \times 10&^{-14} \\ 
-9.&80(53) & \times 10&^{-13} \\ 
1.&049(47) & \times 10&^{-13} \\ 
-2.&78(12) & \times 10&^{-14} \\ 
\end{alignedat} 
 $ 
& 
 $ 
 \begin{alignedat}{2} 
0.&0 & & \\ 
0.&0 & & \\ 
0.&0 & & \\ 
0.&0 & & \\ 
0.&0 & & \\ 
0.&0 & & \\ 
4.&92(25) & \times 10&^{-14} \\ 
-1.&243(67) & \times 10&^{-12} \\ 
1.&232(55) & \times 10&^{-13} \\ 
-3.&27(15) & \times 10&^{-14} \\ 
\end{alignedat} 
 $

\end{tabular}

\caption{The contribution in GeV from the renormalized continuum
operator $Q_{i, \rm cont}$ to the imaginary parts of $\langle (\pi
\pi)_I | -i \DSoneH | K^0 \rangle$ for $\mu = 3.02$ GeV. The central
values for the standard model parameters given in Table
\ref{tab:sm_values} have been used.}

\label{tab:ImA0_ImA2_mom4.tab}
\end{table}


\begin{table}
\begin{tabular}{lllll}
  $m_f$ & $m_{PS}$ & $f_{PS}$ & $B_{PS}^{\rm AA} $ & $B_{PS}^{\rm wall}$
  \\ \hline
  0.01 &  0.2073(19) & 0.0769(7) & 0.478(10) & 0.466(14) \\
  0.02 &  0.2713(16) & 0.0797(6) & 0.554(6)  & 0.547(11) \\
  0.03 &  0.3245(15) & 0.0837(6) & 0.602(5)  & 0.600(10) \\
  0.04 &  0.3716(14) & 0.0876(6) & 0.636(4)  & 0.635(9)  \\
  0.05 &  0.4147(13) & 0.0915(7) & 0.662(3)  & 0.662(9)  \\
\end{tabular}

\caption{Values for the pseudoscalar mass $m_{PS}$ and decay constant
$f_{PS}$ (both in lattice units) versus the quark mass $m_f$, along with
the pseudoscalar $B$ parameter, $B_{PS}$, determined from two different
normalizations.  $B_{PS}^{\rm AA}$ is found by normalizing the
desired $\Delta S = 2$ Green's function by axial current-pseudoscalar
Green's functions, which may introduce zero mode effects.  $B_{PS}^{\rm
wall}$ is determined by normalizing with the wall-wall correlators used
for the $K \to \pi$ matrix elements, which we have argued should not
introduce zero modes through the normalization.  Values for $B_{PS}$
are given in the $\overline{MS}$ scheme at $\mu = 2$ GeV.  The results
for each value of $m_f$ are averaged over the time-slice range $14\,\le
t\,\le\,17$.  The physical value $B_K = 0.532(11)$ is found by choosing
$m_f = 0.018$ so that a kaon made with degenerate quarks has its
physical mass.}

\label{tab:B_PS}

\end{table}


\begin{table}
\begin{tabular}{rrrr}
	 & \multicolumn{1}{c}{Choice 1}
	 & \multicolumn{1}{c}{Choice 2} \\
  Quantity
	 & \multicolumn{1}{c}{(0-loop quenched)}
	 & \multicolumn{1}{c}{(1-loop full)} \\ \hline

$
  \begin{alignedat}
   1 & {\rm Re A_0}\\
     & {\rm Re A_2}\\
     & {\rm Im A_0}\\
     & {\rm Im A_2}\\
     & {\rm Re A_0/Re A_2}\\
     & (\epsilon^{'}/\epsilon)_{\omega \; {\rm exp}} \\
     & (\epsilon^{'}/\epsilon)_{\omega \; {\rm th}} 
		
  \end{alignedat}
  $
& 
 $ 
 \begin{alignedat}{2} 
2.&09(12) & \times 10&^{-7} \\ 
9.&98(45) & \times 10&^{-9} \\ 
-1.&60(28) & \times 10&^{-11} \\ 
-9.&91(56) & \times 10&^{-13} \\ 
2.&09(15) & \times 10&^{1} \\ 
-3.&2(22) & \times 10&^{-4} \\ 
-3.&4(23) & \times 10&^{-4} \\ 
\end{alignedat} 
 $ 
& 
 $ 
 \begin{alignedat}{2} 
2.&96(17) & \times 10&^{-7} \\ 
1.&172(53) & \times 10&^{-8} \\ 
-2.&35(40) & \times 10&^{-11} \\ 
-1.&264(72) & \times 10&^{-12} \\ 
2.&53(18) & \times 10&^{1} \\ 
-4.&0(23) & \times 10&^{-4} \\ 
-3.&5(19) & \times 10&^{-4} \\ 
\end{alignedat} 
 $

\end{tabular}

\caption{The dependence of physical quantities on the extrapolation
choice for $\mu =2.13$ GeV.}

\label{tab:phys_parm_charm_out_vs_choice_mu_2_13}
\end{table}


\begin{table}
\begin{tabular}{rrrrr}
Quantity & $\mu = 1.51$ GeV & $\mu = 2.13$ GeV & $\mu = 2.39$ GeV &
  $\mu = 3.02$ GeV \\ \hline

$
  \begin{alignedat}
   1 & {\rm Re A_0}\\
     & {\rm Re A_2}\\
     & {\rm Im A_0}\\
     & {\rm Im A_2}\\
     & {\rm Re A_0/Re A_2}\\
     & (\epsilon^{'}/\epsilon)_{\omega \; {\rm exp}} \\
     & (\epsilon^{'}/\epsilon)_{\omega \; {\rm th}} 
		
  \end{alignedat}
  $
& 
 $ 
 \begin{alignedat}{2} 
3.&27(25) & \times 10&^{-7} \\ 
1.&151(52) & \times 10&^{-8} \\ 
-1.&78(44) & \times 10&^{-11} \\ 
-1.&444(83) & \times 10&^{-12} \\ 
2.&84(24) & \times 10&^{1} \\ 
-9.&9(23) & \times 10&^{-4} \\ 
-7.&8(16) & \times 10&^{-4} \\ 
\end{alignedat} 
 $ 
& 
 $ 
 \begin{alignedat}{2} 
2.&96(17) & \times 10&^{-7} \\ 
1.&172(53) & \times 10&^{-8} \\ 
-2.&35(40) & \times 10&^{-11} \\ 
-1.&264(72) & \times 10&^{-12} \\ 
2.&53(18) & \times 10&^{1} \\ 
-4.&0(23) & \times 10&^{-4} \\ 
-3.&5(19) & \times 10&^{-4} \\ 
\end{alignedat} 
 $ 
& 
 $ 
 \begin{alignedat}{2} 
3.&04(16) & \times 10&^{-7} \\ 
1.&163(52) & \times 10&^{-8} \\ 
-2.&12(37) & \times 10&^{-11} \\ 
-1.&206(68) & \times 10&^{-12} \\ 
2.&61(17) & \times 10&^{1} \\ 
-4.&8(20) & \times 10&^{-4} \\ 
-4.&1(16) & \times 10&^{-4} \\ 
\end{alignedat} 
 $ 
& 
 $ 
 \begin{alignedat}{2} 
2.&79(17) & \times 10&^{-7} \\ 
1.&140(51) & \times 10&^{-8} \\ 
-2.&26(39) & \times 10&^{-11} \\ 
-1.&103(63) & \times 10&^{-12} \\ 
2.&45(18) & \times 10&^{1} \\ 
-2.&2(24) & \times 10&^{-4} \\ 
-2.&0(21) & \times 10&^{-4} \\ 
\end{alignedat} 
 $

\end{tabular}

\caption{The dependence of the physical quantities we have calculated
on the scale used to match from continuum perturbation theory to the
lattice calculation for extrapolation choice 2.  The dependence on
$\mu$ indicates the reliability of the combination of: using continuum
perturbation theory below 1.3 GeV (needed to define the three-quark
effective theory), one-loop matching from the NDR to RI schemes and our
implementation of non-perturbative renormalization.  }

\label{tab:choice2_phys_parm_charm_out}
\end{table}


\begin{table}
\begin{tabular}{rrrrr}
  & \multicolumn{2}{c}{$-\repezeroi$}
  & \multicolumn{2}{c}{$-\repetwoi$}
	\\
        $i$ 
        & \multicolumn{1}{c}{choice 1} 
        & \multicolumn{1}{c}{choice 2} 
        & \multicolumn{1}{c}{choice 1} 
        & \multicolumn{1}{c}{choice 2} \\ \hline

$
  \begin{alignedat}
   1 & 1 \\
     & 2 \\
     & 3 \\
     & 4 \\
     & 5 \\
     & 6 \\
     & 7 \\
     & 8 \\
     & 9 \\
     & 10 

  \end{alignedat}
  $
& 
 $ 
 \begin{alignedat}{2} 
0.&0 & & \\ 
0.&0 & & \\ 
5.&(58) & \times 10&^{-6} \\ 
-4.&8(11) & \times 10&^{-4} \\ 
-4.&2(17) & \times 10&^{-5} \\ 
1.&42(19) & \times 10&^{-3} \\ 
-4.&66(36) & \times 10&^{-6} \\ 
1.&061(84) & \times 10&^{-4} \\ 
5.&65(23) & \times 10&^{-5} \\ 
1.&35(24) & \times 10&^{-5} \\ 
\end{alignedat} 
 $ 
& 
 $ 
 \begin{alignedat}{2} 
0.&0 & & \\ 
0.&0 & & \\ 
5.&(58) & \times 10&^{-6} \\ 
-4.&8(11) & \times 10&^{-4} \\ 
-4.&2(17) & \times 10&^{-5} \\ 
1.&42(19) & \times 10&^{-3} \\ 
-6.&50(50) & \times 10&^{-6} \\ 
1.&48(12) & \times 10&^{-4} \\ 
5.&65(23) & \times 10&^{-5} \\ 
1.&35(24) & \times 10&^{-5} \\ 
\end{alignedat} 
 $ 
& 
 $ 
 \begin{alignedat}{2} 
0.&0 & & \\ 
0.&0 & & \\ 
0.&0 & & \\ 
0.&0 & & \\ 
0.&0 & & \\ 
0.&0 & & \\ 
6.&90(33) & \times 10&^{-5} \\ 
-1.&571(77) & \times 10&^{-3} \\ 
1.&56383(69) & \times 10&^{-4} \\ 
-4.&5635(20) & \times 10&^{-5} \\ 
\end{alignedat} 
 $ 
& 
 $ 
 \begin{alignedat}{2} 
0.&0 & & \\ 
0.&0 & & \\ 
0.&0 & & \\ 
0.&0 & & \\ 
0.&0 & & \\ 
0.&0 & & \\ 
7.&45(35) & \times 10&^{-5} \\ 
-1.&697(84) & \times 10&^{-3} \\ 
1.&56493(75) & \times 10&^{-4} \\ 
-4.&5668(22) & \times 10&^{-5} \\ 
\end{alignedat} 
 $

\end{tabular}

\caption{The contribution from the renormalized continuum operator
$Q_{i, \rm cont}$ to the imaginary parts of $\langle (\pi \pi)_I | -i
\DSoneH | K^0 \rangle$ is used to calculate $P_I^i \equiv {\rm Im}
\left( \langle (\pi \pi)_I | -i \DSoneH | K^0 \rangle \right)_i/{\rm
Re}(A_I)$.  Here we tabulate $-\repezeroi$ and $\repetwoi$ for our two
extrapolation choices for $\mu = 2.13$ GeV.  One sees that the largest
contribution to the $I = 0$ channel is from $Q_{6,\rm cont}$ and the
largest contribution to for the $I=2$ channel is from $Q_{8, \rm
cont}$.  The very small errors for the contribution of $Q_{9, \rm
cont}$ and $Q_{10, \rm cont}$ to $\repetwoi$ is due to the fact that
the (27,1) operator is dominating the numerator and denominator.
Since the errors in the $Q_{i, \rm cont}$ are correlated, the error for
$\repe$ is not simply related to the errors from the individual
contributions in this table.  The experimental values for $\omega$ and
$|\epsilon|$ are used here.}

\label{tab:epe_mom2_break.tab}
\end{table}


\begin{table}
\begin{tabular}{l|c|c}
  & & This calculation \\
Quantity	& Experiment	& (statistical errors only) \\ \hline
  $
  \begin{alignedat}{1}
    & {\rm Re} \; A_0 {\rm (GeV) } \\
    & {\rm Re} \; A_2 {\rm (GeV) } \\
    & \omega^{-1} 		\\
    & {\rm Re} \; (\epsilon^\prime/\epsilon) \\
    &
  \end{alignedat}
  $
&
  $
  \begin{alignedat}{2}
    3.&33    & \times 10&^{-7} \\
    1.&50    & \times 10&^{-8} \\
   22.&2     & & \\
  (15.&3 \pm 2.6) & \times 10&^{-4}
     \; {\rm (NA48)} \\
   (20.&7 \pm 2.8) & \times 10&^{-4}
     \; {\rm (KTEV)} \\
  \end{alignedat}
  $
&
  $
  \begin{alignedat}{2}
    (2.&96 \pm 0.17)  & \times 10&^{-7} \\
    (1.&172 \pm 0.053) & \times 10&^{-8} \\
   (25.&3 \pm 1.8)   & & \\
   (-4.&0 \pm 2.3)   & \times 10&^{-4} \\
   & & &
  \end{alignedat}
  $
\end{tabular}

\caption{Our final values for physical quantities using 1-loop full QCD
extrapolations to the physical kaon mass (choice 2) and a value of $\mu
= 2.13$ GeV for the matching between the lattice and continuum.  The
errors for our calculation are statistical only.}

\label{tab:choice2_mu_2_13_charm_out_vs_exp}

\end{table}

\fi


\ifnum\theFigures=1
%
%

%
%
%

\begin{figure}
\epsfxsize=\hsize
\begin{center}
\epsfbox{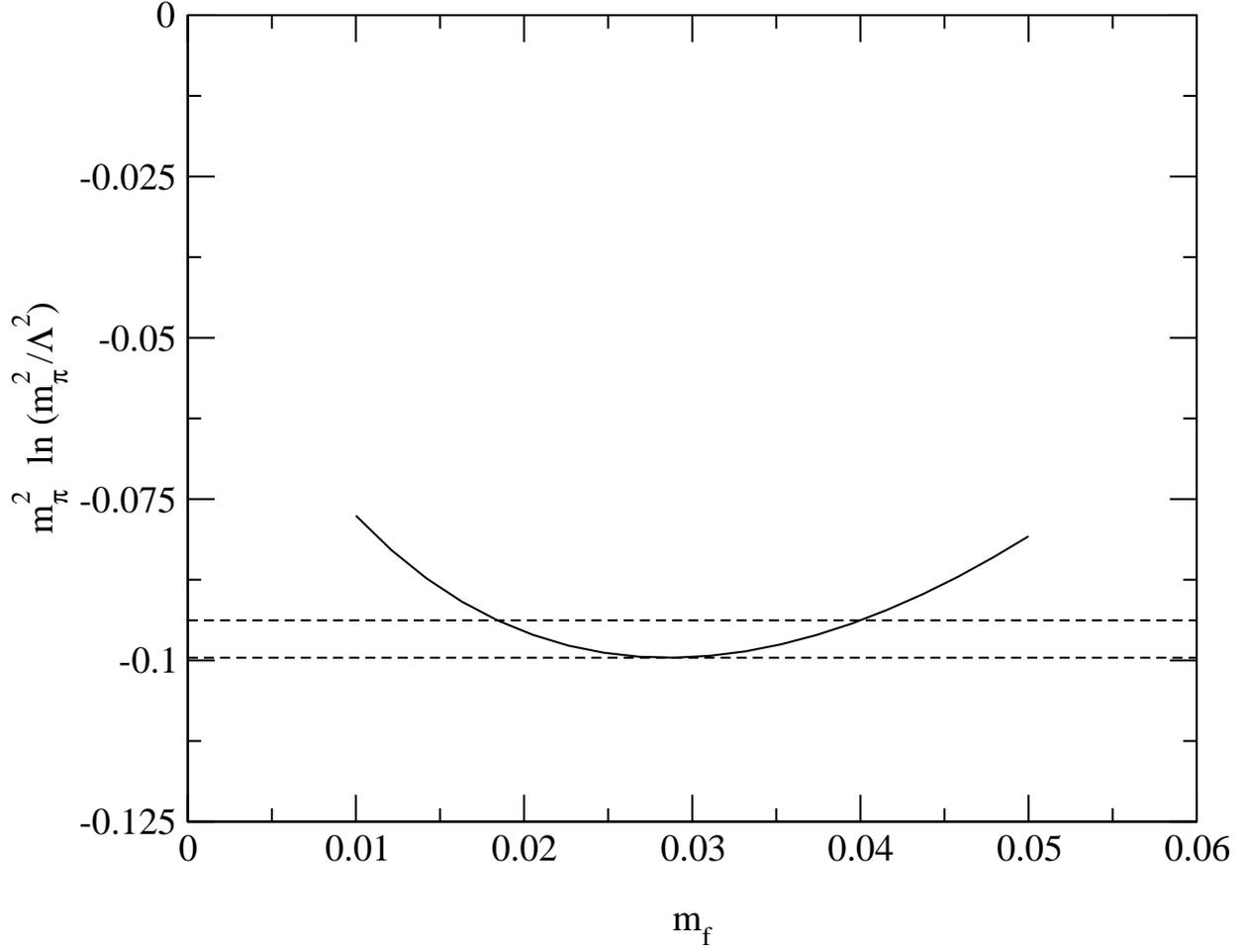}
\end{center}
\caption{The value of $\mpi2 \ln( \mpi2/\Lqcpt^2 )$ versus $m_f$ for
the range of quark masses used in our simulations.  The dashed lines
have $\mpi2 \ln( \mpi2/\Lqcpt^2 ) = -0.0938$ and -0.0996.  For $0.02
\le m_f \le 0.04$ the variation in $\mpi2 \ln( \mpi2/\Lqcpt^2 )$ is
about 5\%.}
\label{fig:chiral_log_vs_mf}
\end{figure}

%
%
%

\begin{figure}
\epsfxsize=\hsize
\begin{center}
\epsfbox{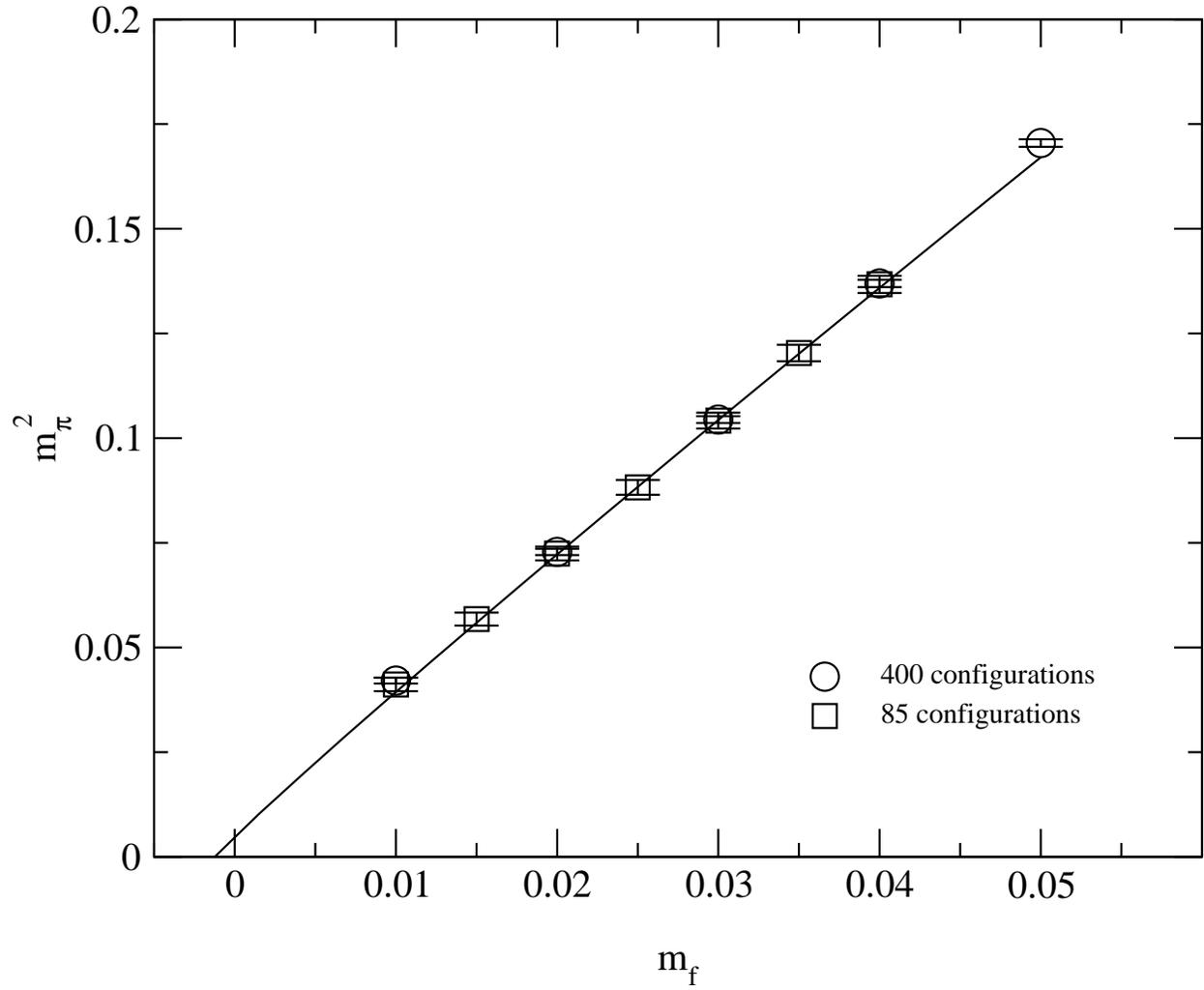}
\end{center}
\caption{The data for $\mpi2$ from 85 configurations and 400
configurations.  The line is a fit to the 85 configuration data,
excluding the $m_f = 0.01$ point, and gives $\delta = 0.05(2)$.}
\label{fig:mpi2_vs_mf}
\end{figure}

%
%
%

\begin{figure}
\epsfxsize=\hsize
\begin{center}
\epsfbox{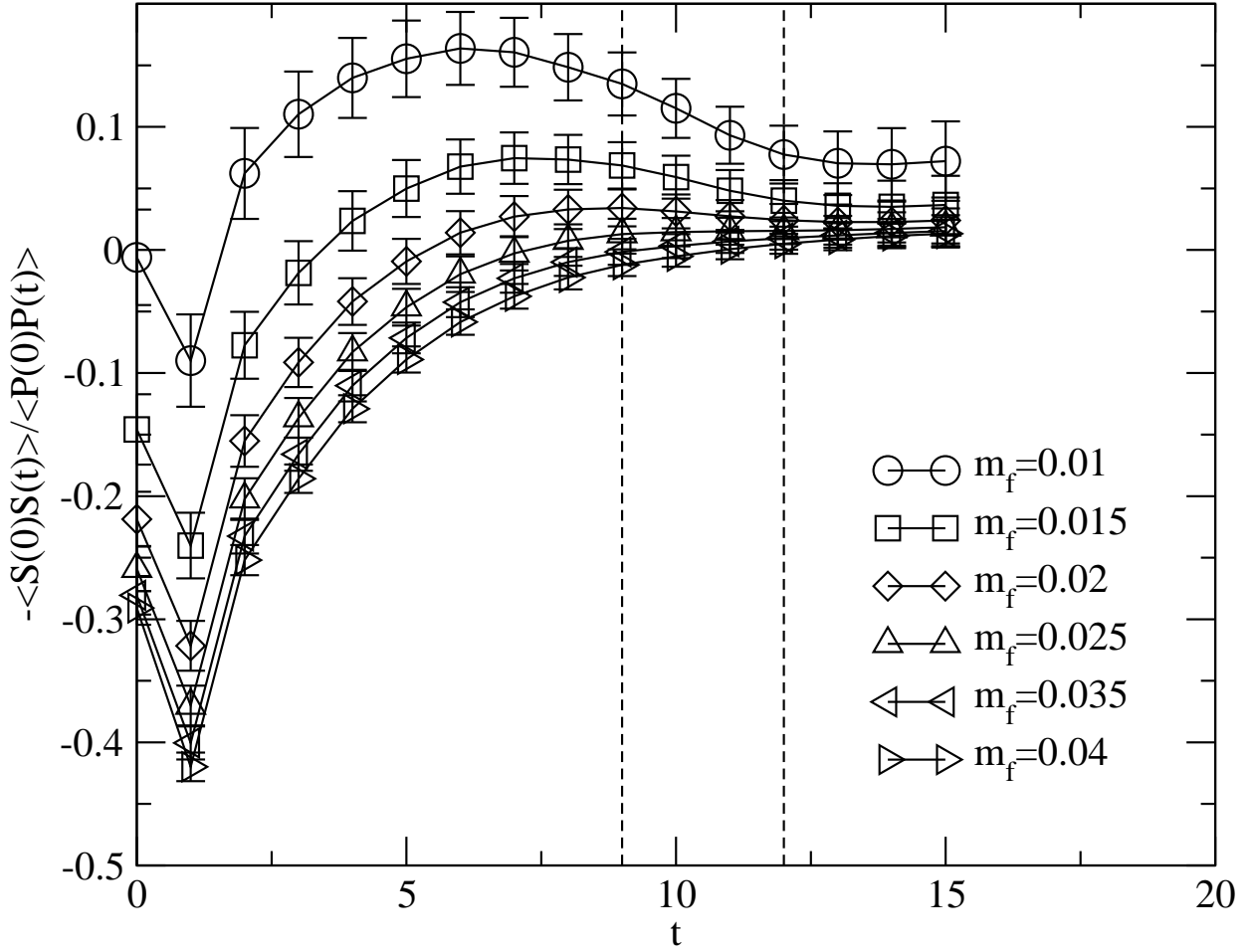}
\end{center}

\caption{The ratio $ - \langle S^{\rm wall}(0) \, S^\dagger(t) \rangle
/ \langle P_{K^+}^{\rm wall}(0) \, P_{K^-}(t) \rangle$ of the scalar
and pseudoscalar correlators, as a function of temporal separation.
Without zero mode effects the ratio should be zero for $m_f$ small,
since the pseudoscalar mass is vanishing.  Zero mode effects are
present at the $\sim 10$\% level for $t = 9$ to 12.  This is the
separation used in our evaluation of lattice matrix elements.}

\label{fig:ss_over_pp}
\end{figure}

%
%
%

\begin{figure}
\epsfxsize=\hsize
\begin{center}
\epsfbox{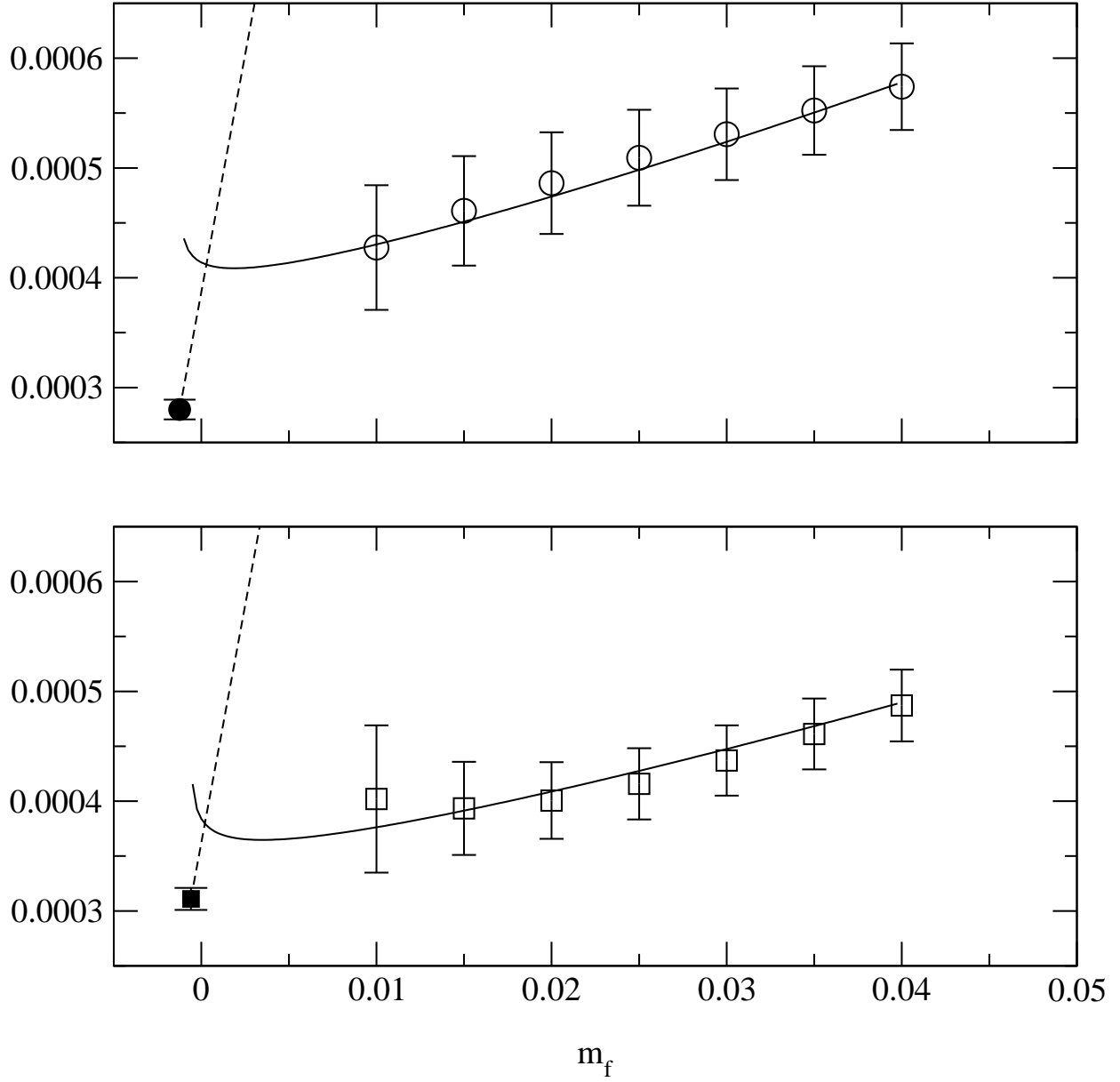}
\end{center}
\caption{The GMOR relation for $L_s = 16$ (upper panel) and $L_s =24$
(lower panel).  The open symbols are $ (m_f + \mres) | \langle 0 |
J^a_5(0) | \pi^\prime \rangle|^2 / ( 12 m_{\pi^\prime}^2)$
 and the filled symbols are $- \langle \overline{u}u \rangle_{\rm
lat-norm}(m_f = 0, L_s)$.  The prime on the states and masses indicates
that zero mode effects may be present.  The dashed line gives the
$m_f/a^2$ dependence of $ -\langle \overline{u}u \rangle_{\rm
lat-norm}(m_f, L_s)$ as determined from large quark masses where
zero mode effects are absent.  The solid line includes the effects of
quenched chiral logarithms in $m_\pi^2$.}
\label{fig:gmor}
\end{figure}

%
%
%

\begin{figure}
\epsfxsize=\hsize
\begin{center}
\epsfbox{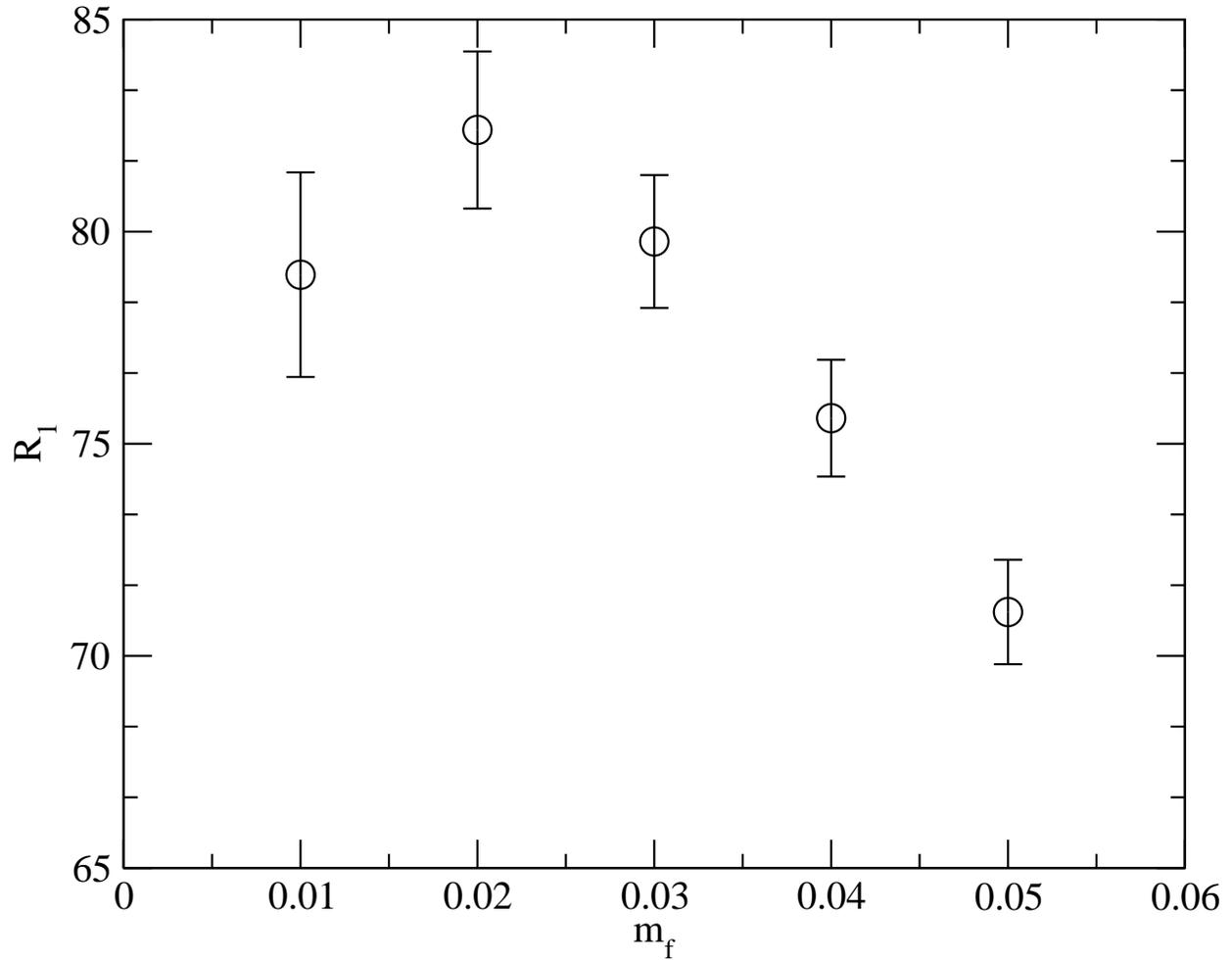}
\end{center}
\caption{A plot of $R_1$, the ratio of a three point correlator
to two, two-point correlators, defined in Eq.\ \ref{eq:r1_def}.
The larger zero mode effects in the two-point correlators should make
this quantity vanish in the $m_f \to 0 $ limit, a marked change
from the chiral limit value of $\approx 120$ expected without these
chiral pathologies.}
\label{fig:sbard_PP_norm}
\end{figure}

%
%
%

\begin{figure}
\epsfxsize=\hsize
\begin{center}
\epsfbox{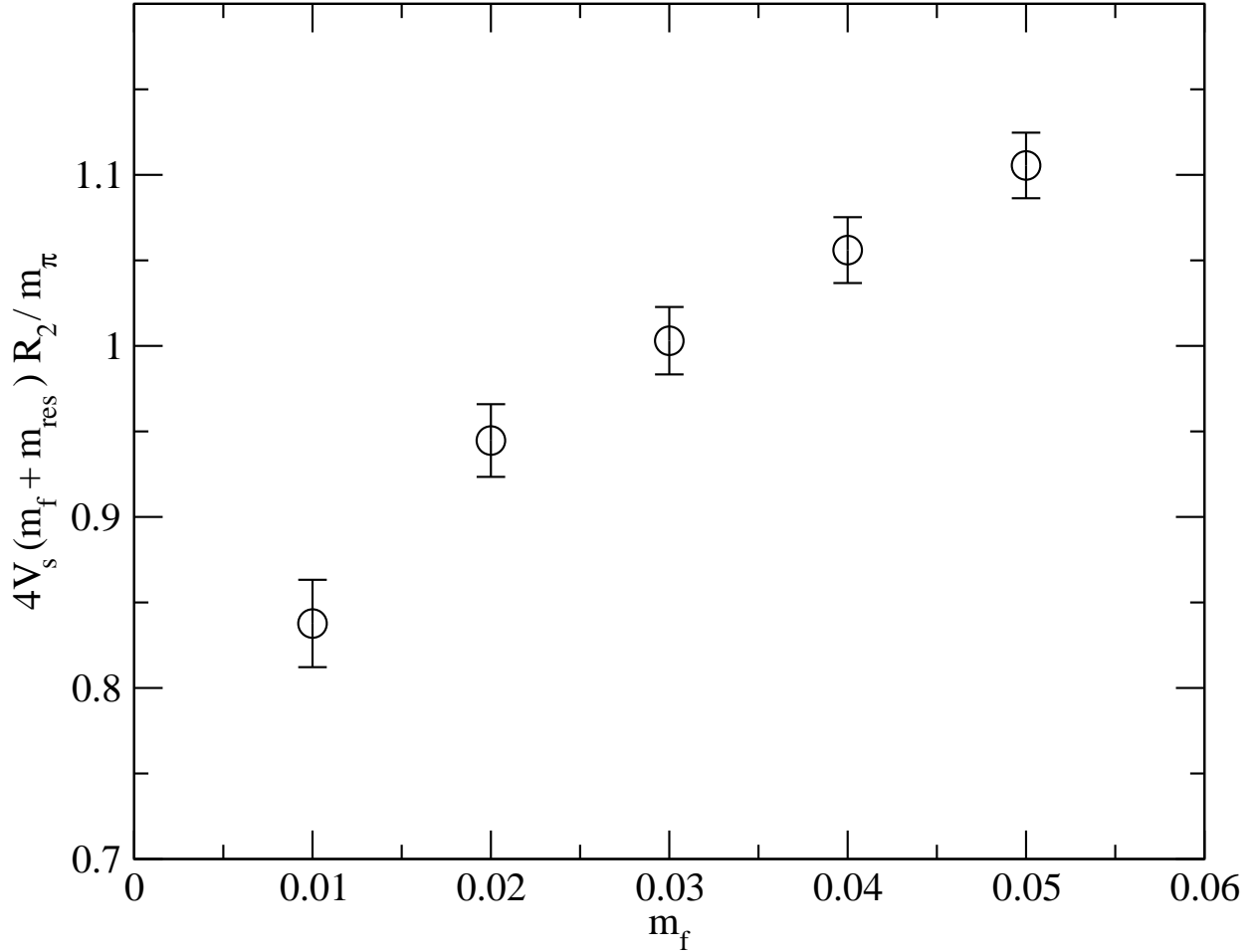}
\end{center}
\caption{ A plot of $4 V_s (m_f + \mres) R_2/m_\pi$ versus $m_f$, where
$R_2$ is defined in Eq.\ \ref{eq:r2_def}.  The Ward-Takahashi identity
determines that this value should be 1 for $m_f \to 0$ if there are no
zero mode effects present.  The deviation from 1 for small $m_f$ is
consistent with estimates of the different effective pseudoscalar
masses entering in the Green's functions in the numerator and
denominator of $R_2$.  The different effective pseudoscalar masses
arise through zero mode effects, as discussed in Sections
\ref{subsec:zero_mode_effects} and \ref{subsec:ward_id_sd}.}
\label{fig:ward_sbard} \end{figure}

%
%
%

\begin{figure}
\epsfxsize=\hsize
\begin{center}
\epsfbox{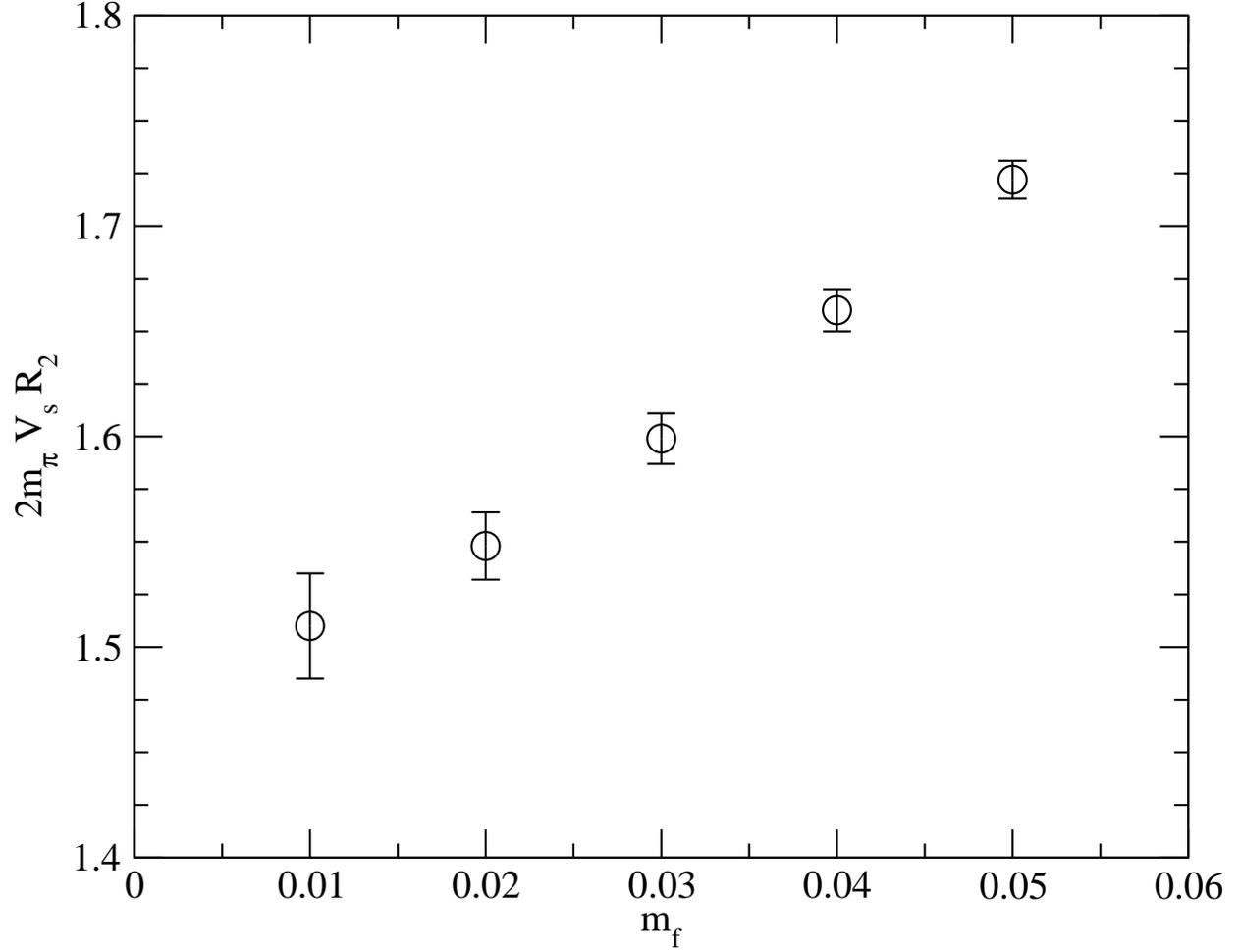}
\end{center}
\caption{ A graph of $2 m_\pi V_s R_2$ versus $m_f$.  Without zero mode
effects, this quantity is $\langle \pi^+ | \overline{s} d | K^+
\rangle$.  In the operator subtraction, any non-linearities 
in the power divergent parts of $\langle \pi^+ | Q_i | K^+ \rangle$
will exactly match the non-linearities in this plot.  The resulting
subtracted operator will not have chiral logarithm and zero mode
effects multiplied by power divergent terms.}
\label{fig:ktopi_sbard}
\end{figure}

%
%

\begin{figure}
\epsfxsize=\hsize
\begin{center}
\epsfbox{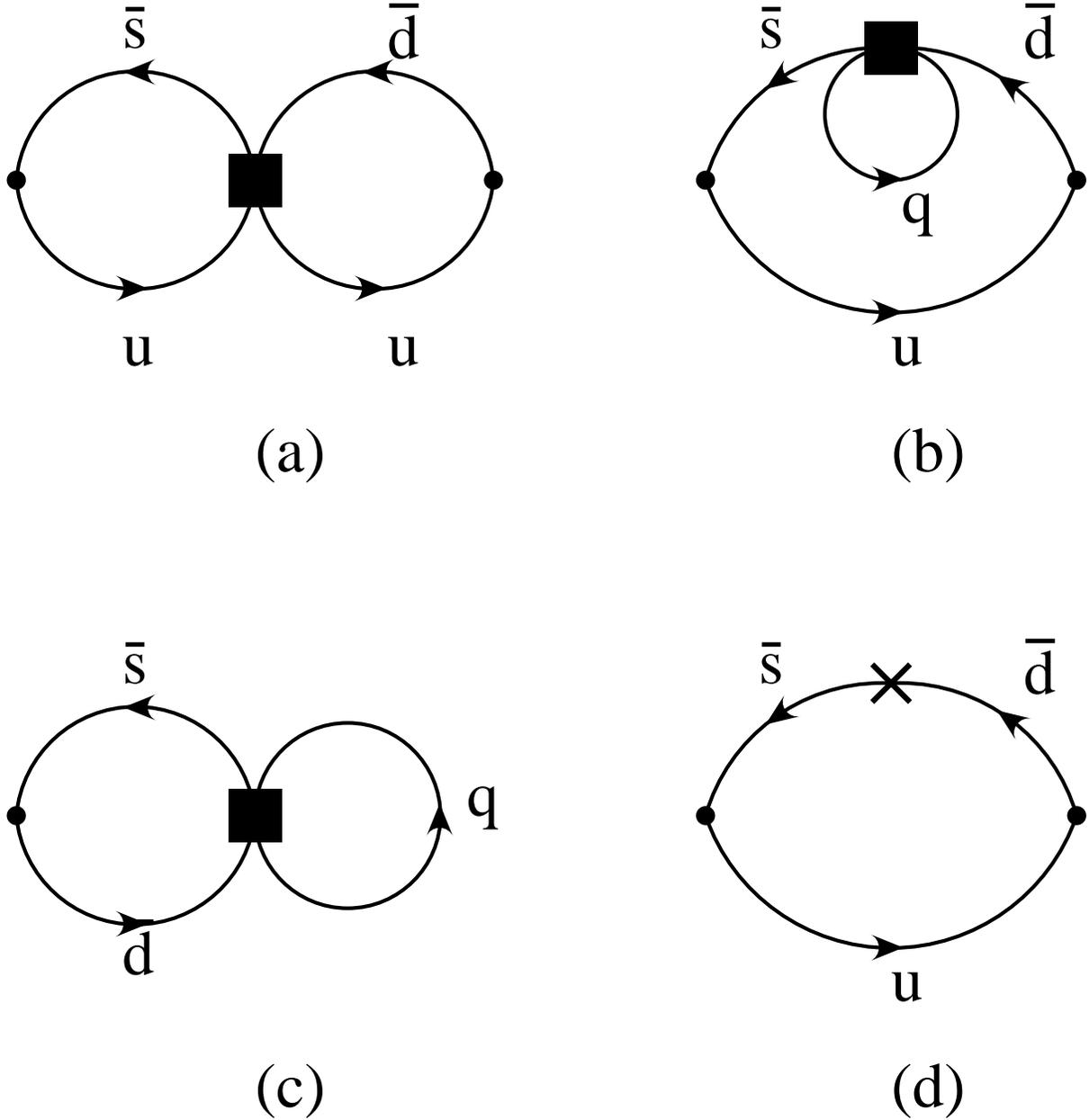}
\end{center}
\caption{The quark contractions needed for $\langle \pi^+ | Q_i | K^+
\rangle$ matrix elements are the figure eight (a) and eye (b)
contractions.  If the quark loop in (b) contains a $d$ or $s$ quark,
there are two different eye contractions possible.  This is the case
for $Q_3$ through $Q_{10}$.  For $\langle 0 | Q_i |K^0 \rangle$ matrix
elements, the annihilation contraction (c) is needed.  For the
determination of $\langle \pi^+ | Q_i | K^+ \rangle_{\rm sub}$ the
contraction shown in (d) is needed, where the cross is an insertion of
the quark bilinear $\overline{s}d$.  The filled boxes represent
insertions of a generic four-fermion operator, and the filled dots the
creation and annhilation of the pseudoscalar states.  Depending on the
particular weak operator, the quark loops in (b) and (c) may contain
$q=u,d,s$ quarks (and $c$ if charm is an active flavor).}
\label{fig:lattice_contractions}
\end{figure}

%
%
%

\begin{figure}
\epsfxsize=\hsize
\begin{center}
\epsfbox{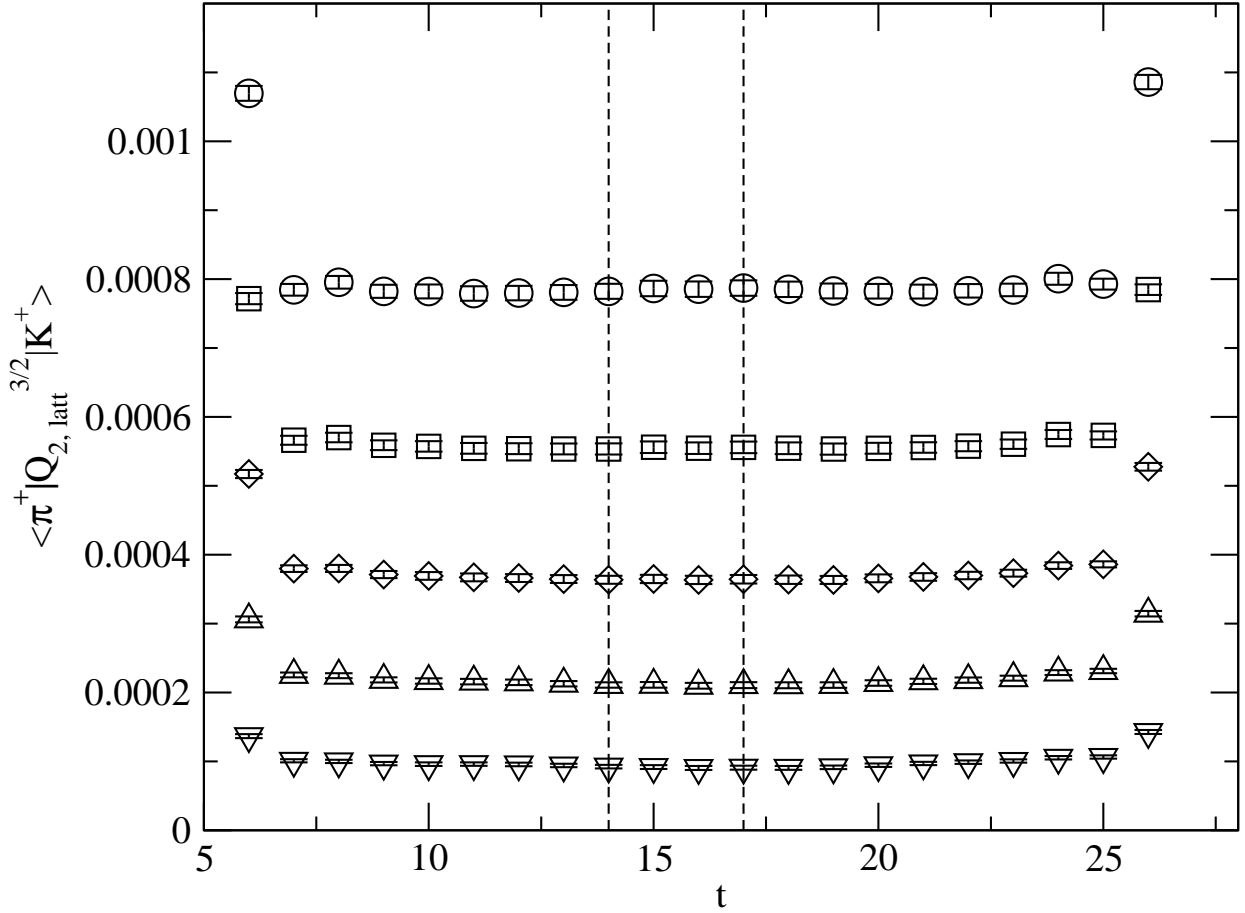}
\end{center}

\caption{$\langle \pi^+ | Q^{(3/2)}_{2, \rm lat} |K^+ \rangle$ for each
Euclidean time slice $t$ where the four quark operator was inserted.
The different $m_f$ values shown are: 0.01 ($\bigtriangledown$), 0.02
($\bigtriangleup$), 0.03 ($\Diamond$), 0.04 ($\Box$) and
0.05($\bigcirc$).  The matrix element is time-independent for the range
$14 \le t \le 17$ for each mass (vertical dashed lines), showing that
only the lowest energy pseudoscalar states are contributing.}

\label{fig:ktopi_Q2_3half_plateau}
\end{figure}

%
%
%

\begin{figure}
\epsfxsize=\hsize
\begin{center}
\epsfbox{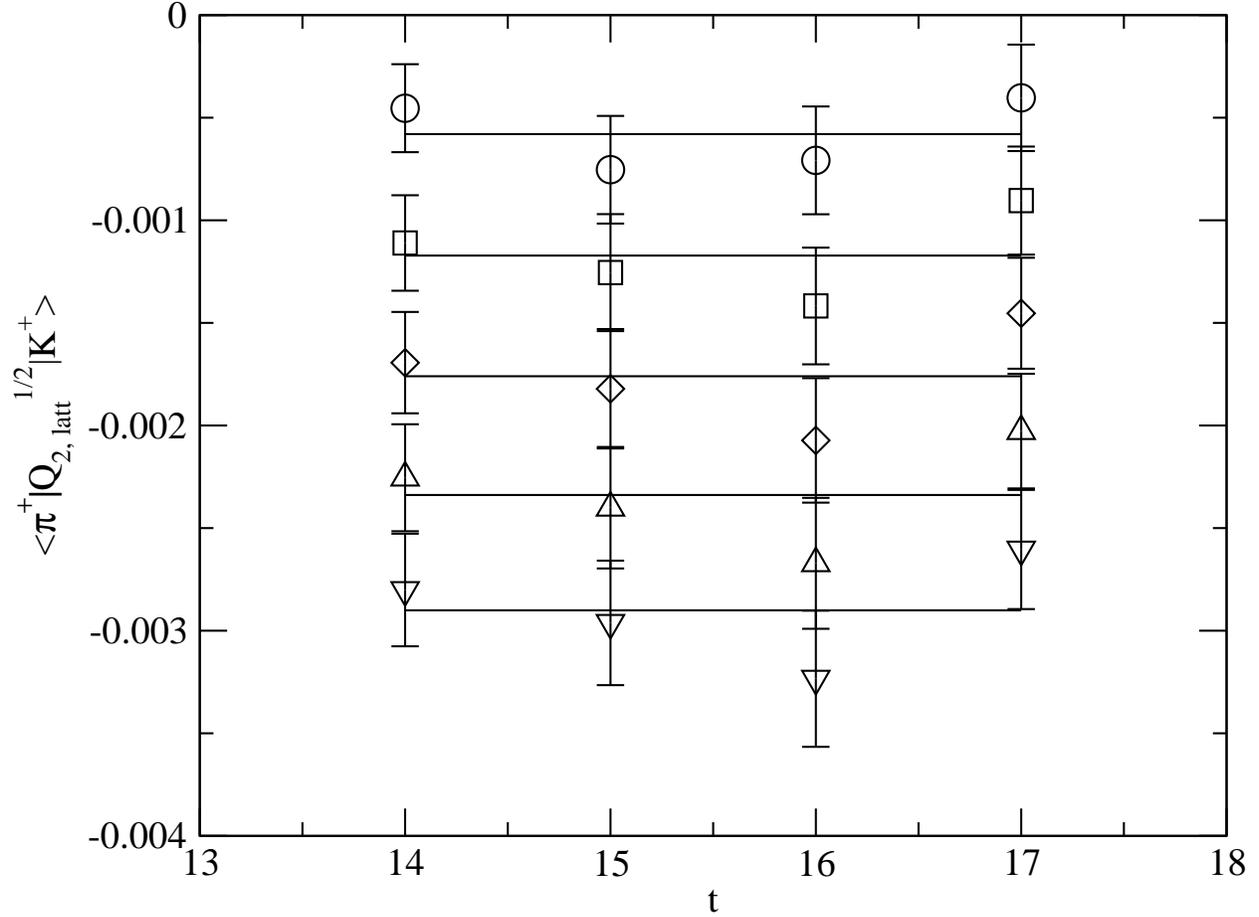}
\end{center}

\caption{ $\langle \pi^+ | Q^{(1/2)}_{2, \rm lat} |K^+ \rangle$ for the
time slices $ 14 \le t \le 17$.  This matrix element involves a noisy
estimator for the fermion loop in the eye contractions.  The symbols
denote different values for $m_f$, as in Figure
\ref{fig:ktopi_Q2_3half_plateau}, and the lines are the average over
time slices for a single $m_f$.  The values on different time slices
agree within the quoted statistical errors.}

\label{fig:ktopi_Q2_plateau}
\end{figure}

%
%
%

\begin{figure}
\epsfxsize=\hsize
\begin{center}
\epsfbox{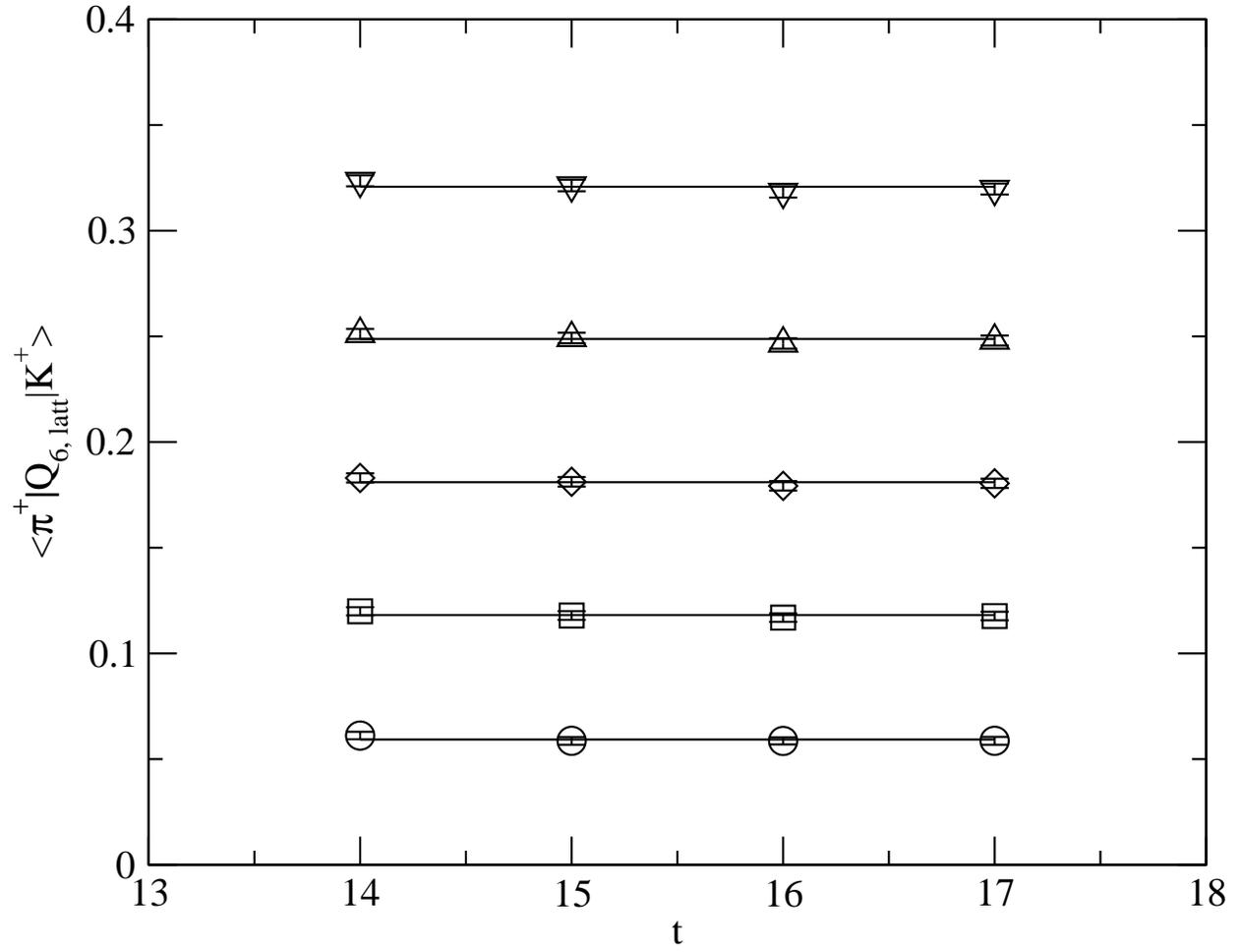}
\end{center}

\caption{ The same as in Figure \ref{fig:ktopi_Q2_plateau}, except that
$\langle \pi^+ | Q_{6, \rm lat} |K^+ \rangle$ is shown.  Again values
agree on different time slices within errors.}

\label{fig:ktopi_Q6_plateau}
\end{figure}

%
%
%

\begin{figure}
\epsfxsize=\hsize
\begin{center}
\epsfbox{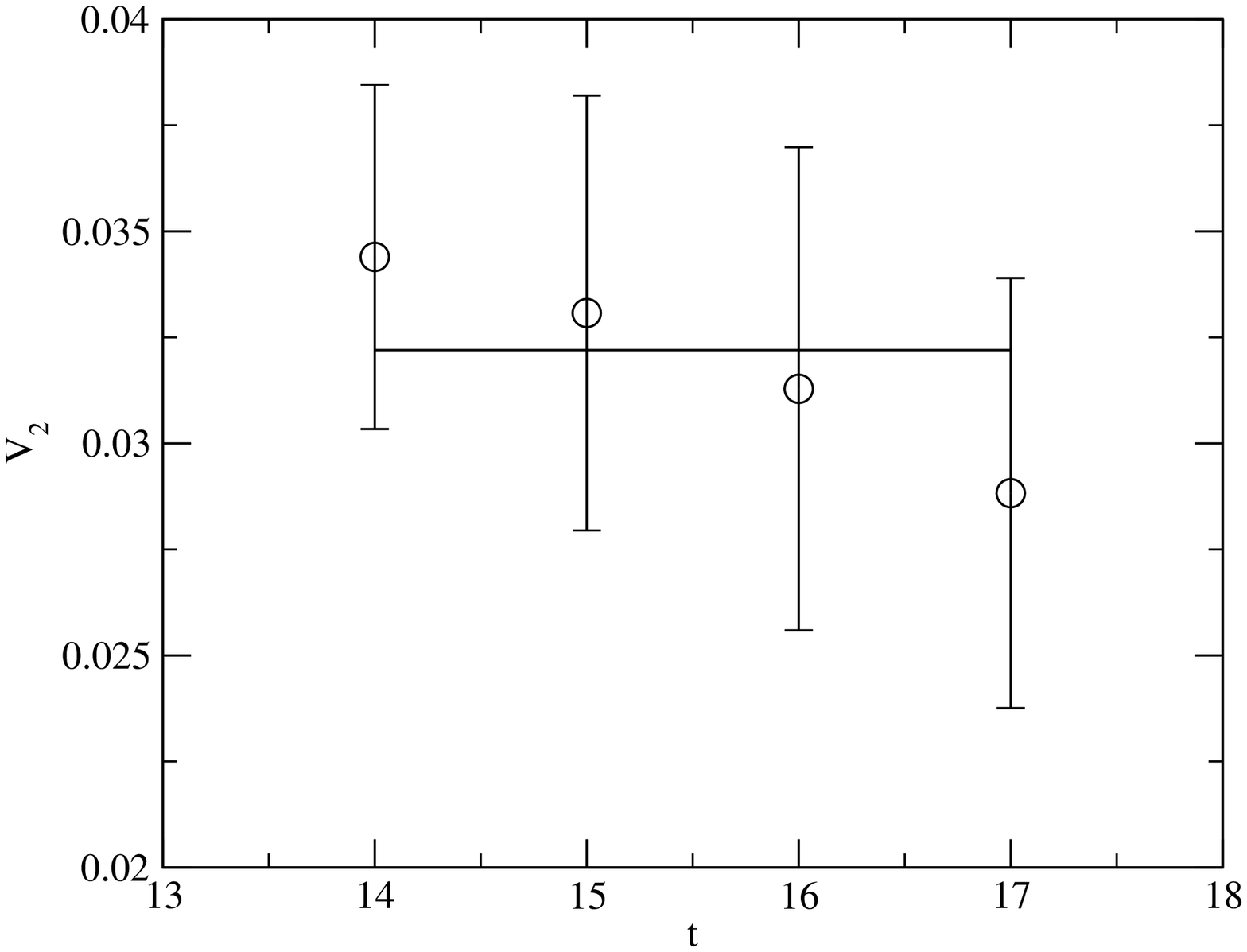}
\end{center}

\caption{A graph of $V_2 \equiv \langle 0|Q_{2, \rm lat} | K^0 \rangle
/ ((m_s-m_d) \langle 0 | \bar{s} \gamma_5 d | K \rangle)$ for each
Euclidean time slice where the operator was inserted.  The data is for
$m_d=0.01$ and $m_s=0.02$.  A noisy estimator is used for the closed
fermion loop and the values on each time slice agree within errors.
The line is the average over $t$.  }

\label{fig:ktovac_Q2_plateau}
\end{figure}

%
%
%

\begin{figure}
\epsfxsize=\hsize
\begin{center}
\epsfbox{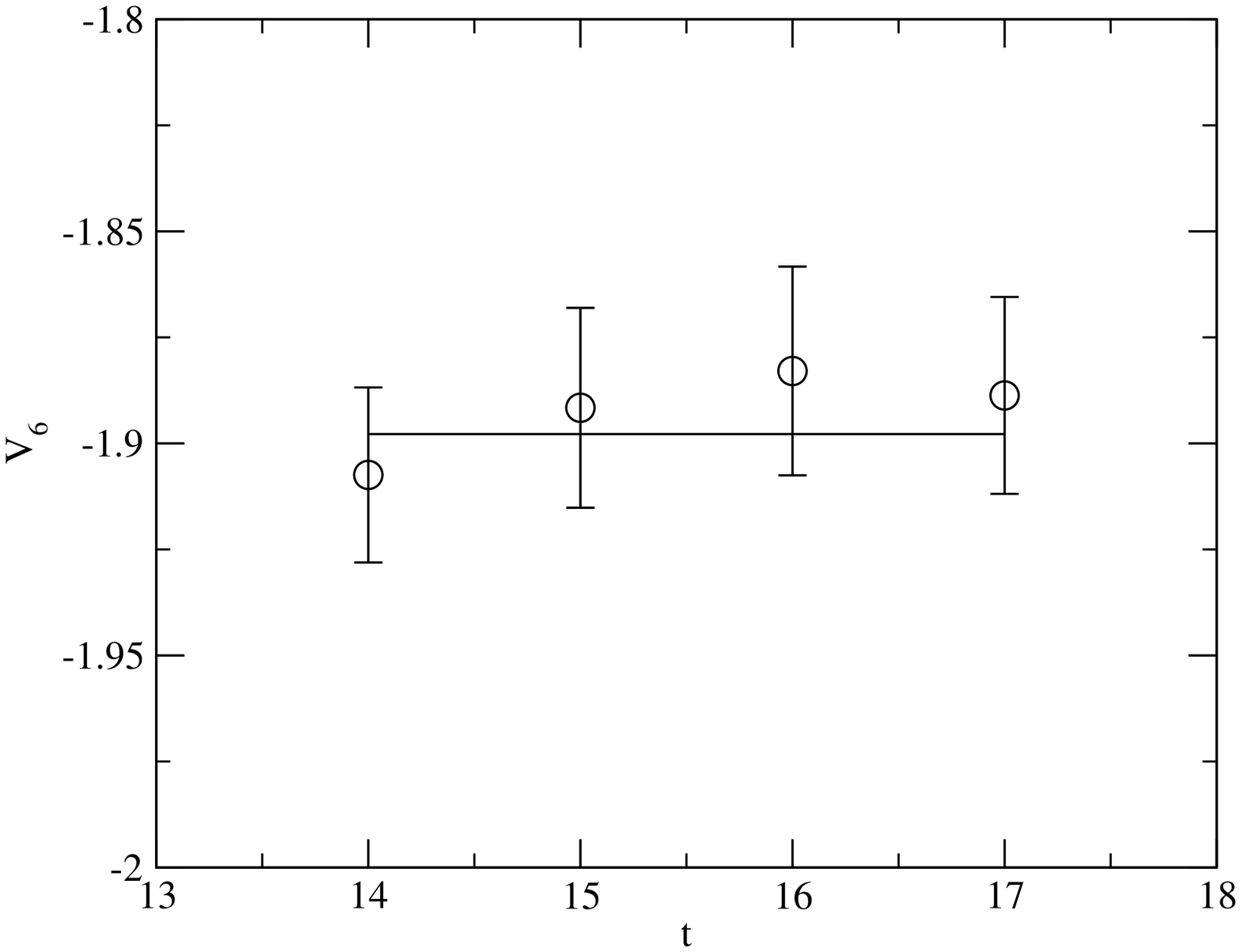}
\end{center}

\caption{A graph of $V_6 \equiv \langle 0|Q_{6, \rm lat} | K^0 \rangle
/ ((m_s-m_d) \langle 0 | \bar{s} \gamma_5 d | K \rangle)$ for each
Euclidean time slice where the operator was inserted.  The data is for
$m_d=0.01$ and $m_s=0.02$.  A noisy estimator is used for the closed
fermion loop and the values on each time slice agree within errors.
The line is the average over $t$.  }

\label{fig:ktovac_Q6_plateau}
\end{figure}

%
%
%

\begin{figure}
\epsfxsize=\hsize
\begin{center}
\epsfbox{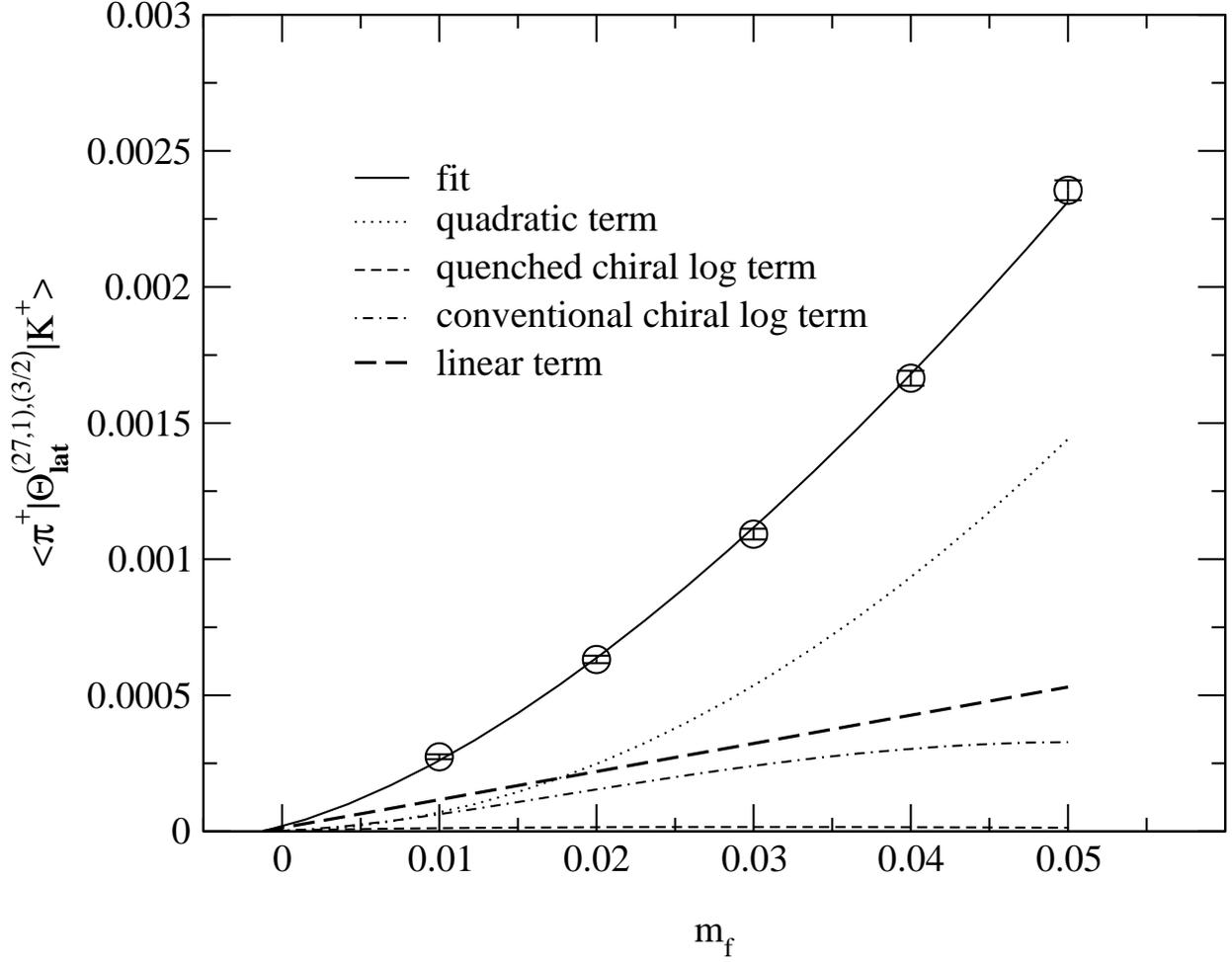}
\end{center}
\caption{The matrix element for $\Theta^{(27,1),(3/2)}_{\rm lat}$, which
shows noticeable non-linearity as a function of quark mass.  The solid
line is a fit to Eq.\ \ref{eq:theta271_32_corr_mpi}, using all five
quark masses.  The contributions from the various terms in Eq.\
\ref{eq:theta271_32_corr_mpi} are shown, with the conventional chiral
logarithm term (the dot-dashed line) of particular importance due to
its essential linearity over most of our quark mass range.  To extract
a value of $\alpha^{(27,1),(3/2)}_{\rm lat}$ from this data, we rely on
the known analytic value for the conventional chiral logarithm.}
\label{fig:theta271_32_corr_mpi}
\end{figure}

%
%
%

\begin{figure}
\epsfxsize=\hsize
\begin{center}
\epsfbox{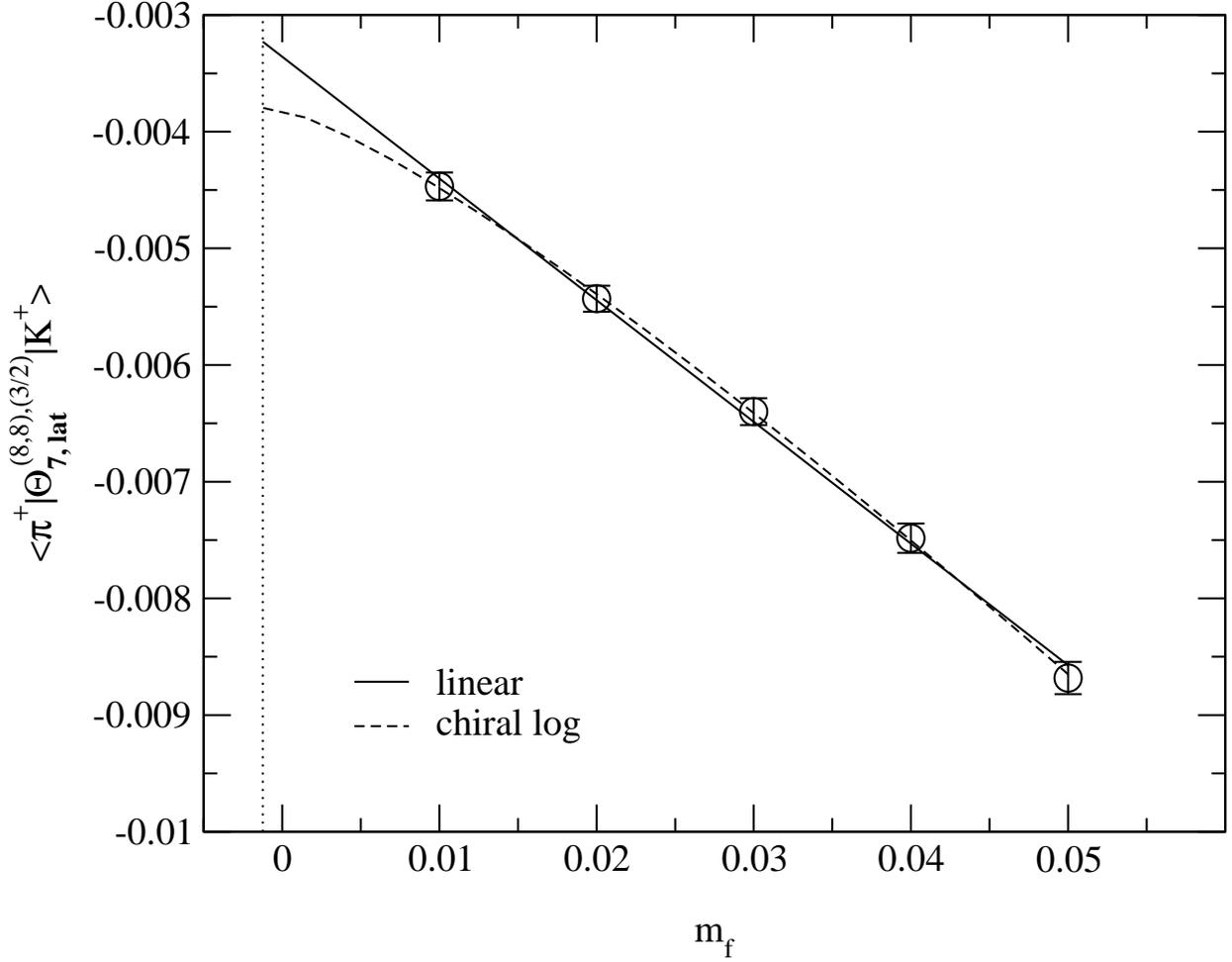}
\end{center}
\caption{The lattice matrix element for $\Theta^{(8,8),(3/2)}_{7,\; \rm
lat}$, fit to Eq.\ \ref{eq:theta88_32_i_corr_mpi}.  All five quark
masses are used in the fit and any non-linearity in the data is small.
The vertical dashed line is drawn at $m_f = -m_{\rm res}$.  There is no
analytic result for the coefficient of the conventional chiral
logarithm in the quenched theory for this matrix element, so we have
done both simple linear fits ($\xi^{(8,8)}_7 = 0$) and fits where the
chiral logarithm is included with a free coefficient.  The linearity of
the data shows the chiral logarithm is not nearly as important here as
for the fits to $\Theta^{(27,1),(3/2)}_{\rm lat}$.}

\label{fig:theta88_32_7_corr_mpi}
\end{figure}

%
%
%

\begin{figure}
\epsfxsize=\hsize
\begin{center}
\epsfbox{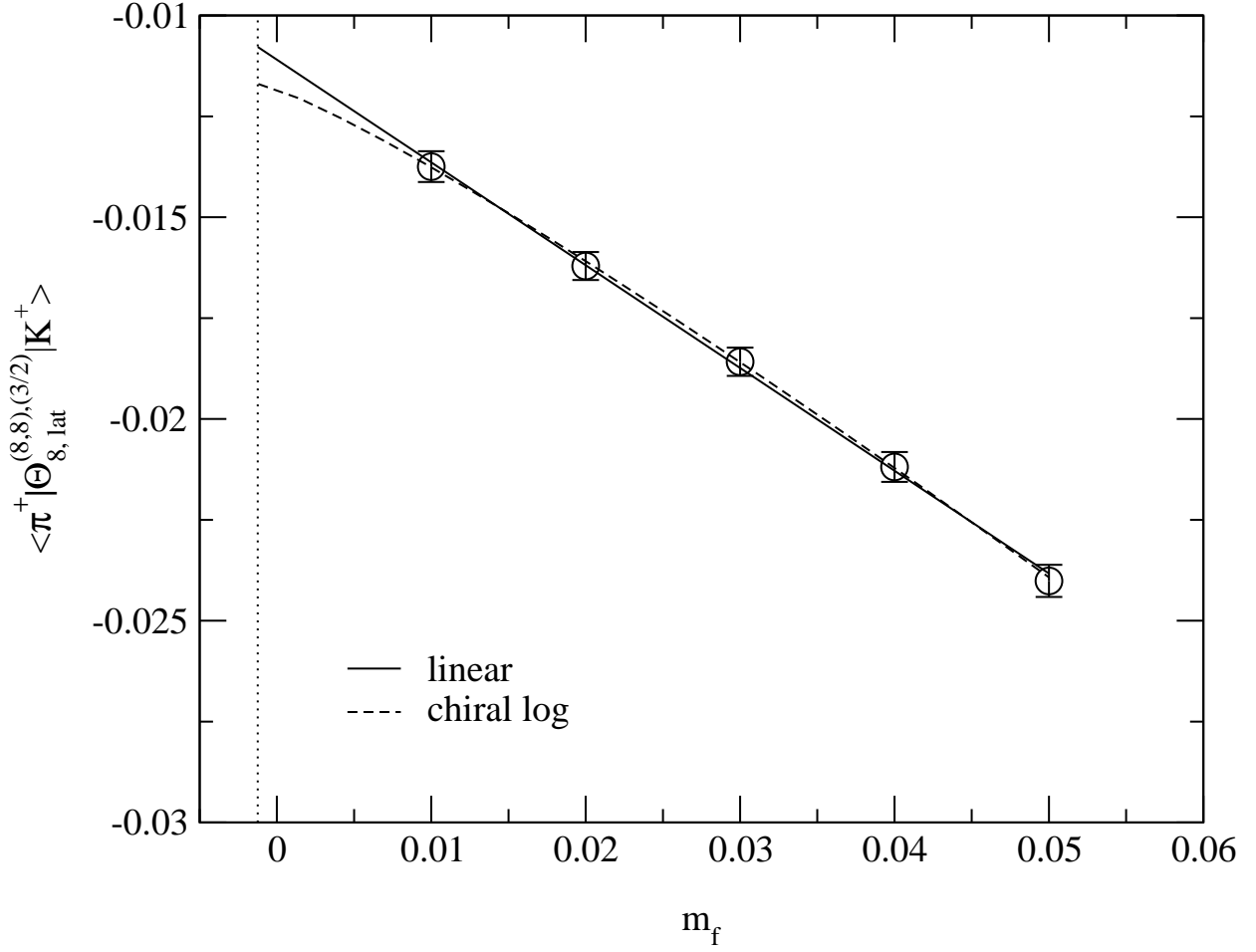}
\end{center}
\caption{The lattice matrix element for $\Theta^{(8,8),(3/2)}_{8,\; \rm
lat}$, fit to Eq.\ \ref{eq:theta88_32_i_corr_mpi}.  All five quark
masses are used in the fit and there is little non-linearity in the
data.  The vertical dashed line is drawn at $m_f = -m_{\rm res}$.  Both
linear fits ($\xi^{(8,8)}_8 = 0$) and fits where the chiral logarithm
is included with a free coefficient are shown.  The linearity of the
data shows the chiral logarithm is not nearly as important here as for
the fits to $\Theta^{(27,1),(3/2)}_{\rm lat}$.}

\label{fig:theta88_32_8_corr_mpi}
\end{figure}


\begin{figure}
\begin{center}
\epsfxsize=\hsize
\epsfbox{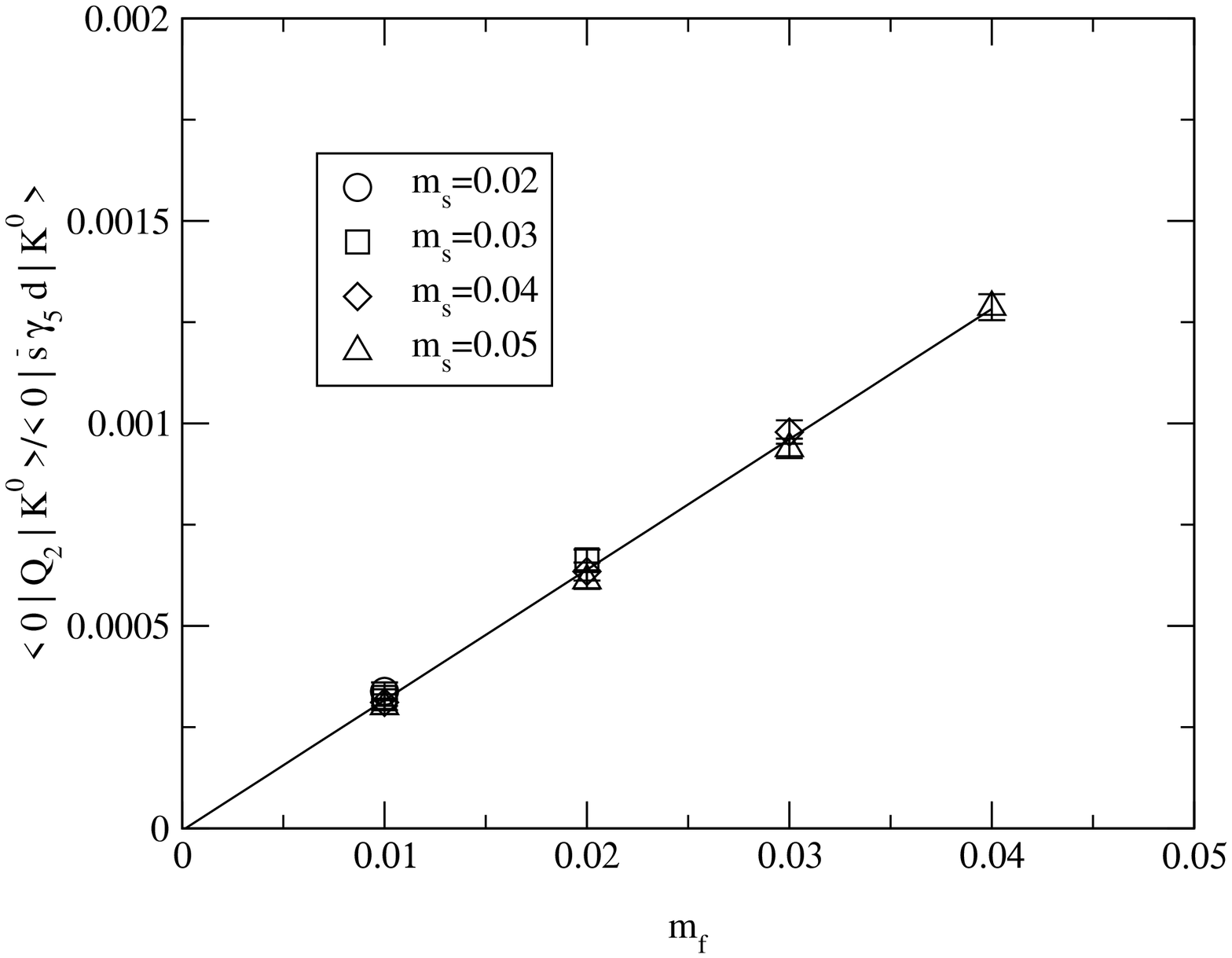}

\caption{The ratio $\langle 0 | Q_2 | K^0 \rangle/ \langle 0 | \bar{s}
\gamma_5 d| K^0 \rangle$ versus $m_s^\prime - m_d^\prime$. The line is
a linear fit of the form given in Eq.\ \ref{eq:fix_eta}.}

\label{fig:ktovac_charm_out_Q2}
\end{center}
\end{figure}


\begin{figure}
\begin{center}
\epsfxsize=\hsize
\epsfbox{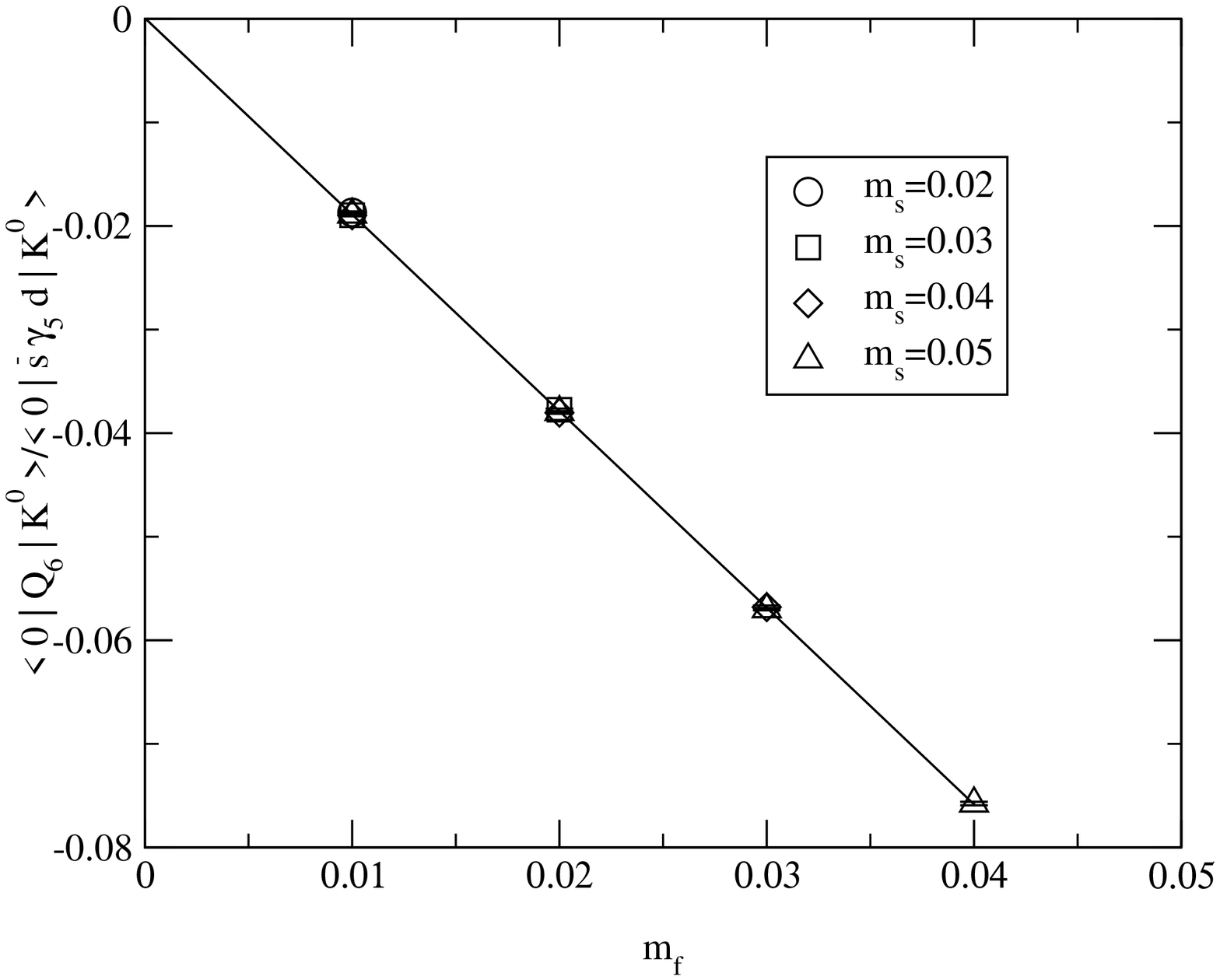}

\caption{The ratio $\langle 0 | Q_6 | K^0 \rangle/ \langle 0 | \bar{s}
\gamma_5 d| K^0 \rangle$ versus $m_s^\prime - m_d^\prime$. The line is
a linear fit of the form given in Eq.\ \ref{eq:fix_eta}.}

\label{fig:ktovac_charm_out_Q6}
\end{center}
\end{figure}


\begin{figure}
\begin{center}
\epsfxsize=\hsize
\epsfbox{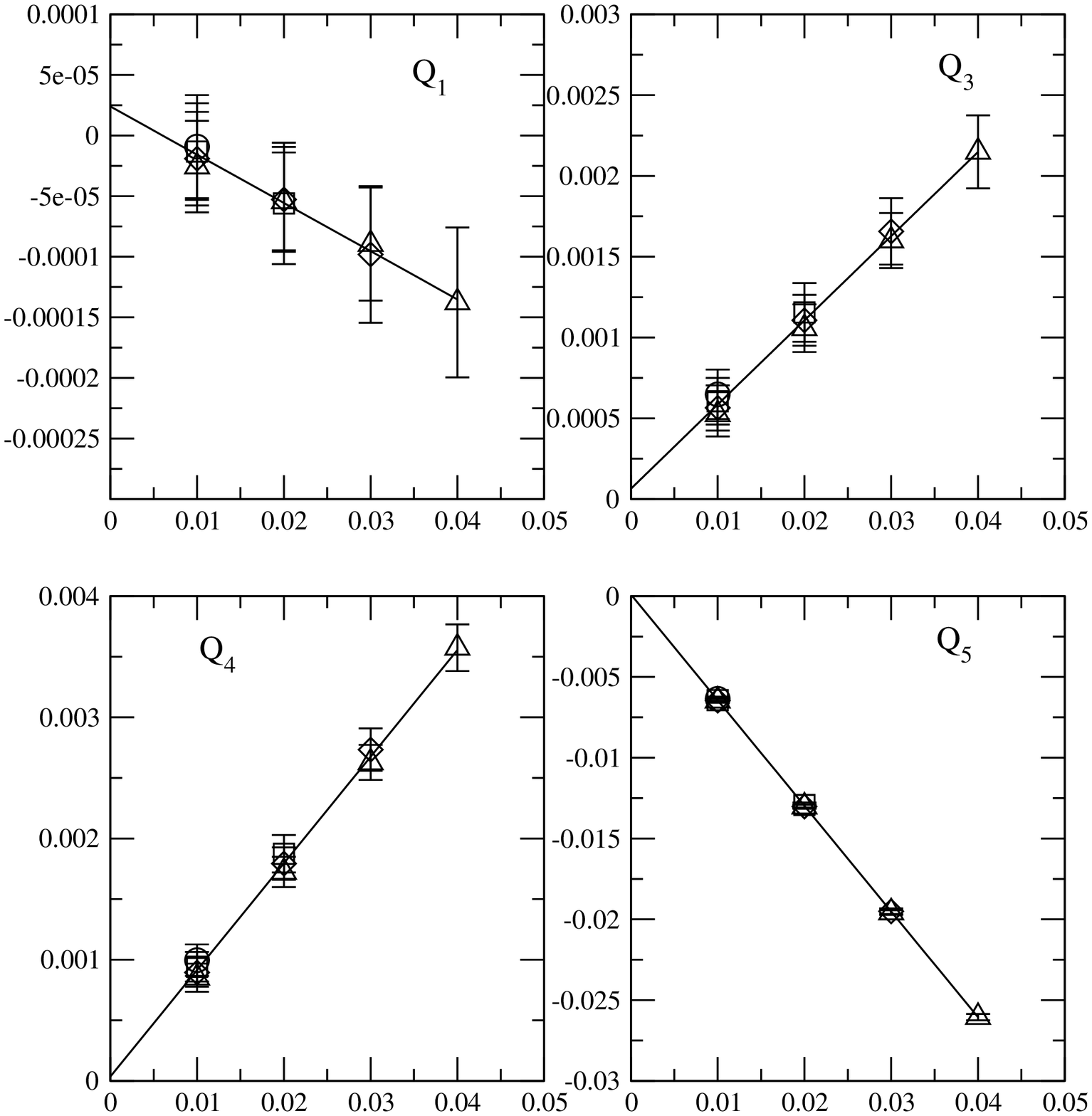}
\vspace{0.2in}

\caption{The ratio $\langle 0 | Q_i | K^0 \rangle/ \langle 0 | \bar{s}
\gamma_5 d| K^0 \rangle$ versus $m_s^\prime - m_d^\prime$ for $i =
1$, 3, 4 and 5.  The line is a linear fit of the form given in Eq.\
\ref{eq:fix_eta}.  The symbols have the same meaning as in Figure
\ref{fig:ktovac_charm_out_Q6}. }

\label{fig:ktovac_charm_out_Q1_Q3_Q4_Q5}
\end{center}
\end{figure}


\begin{figure}
\begin{center}
\epsfxsize=\hsize
\epsfbox{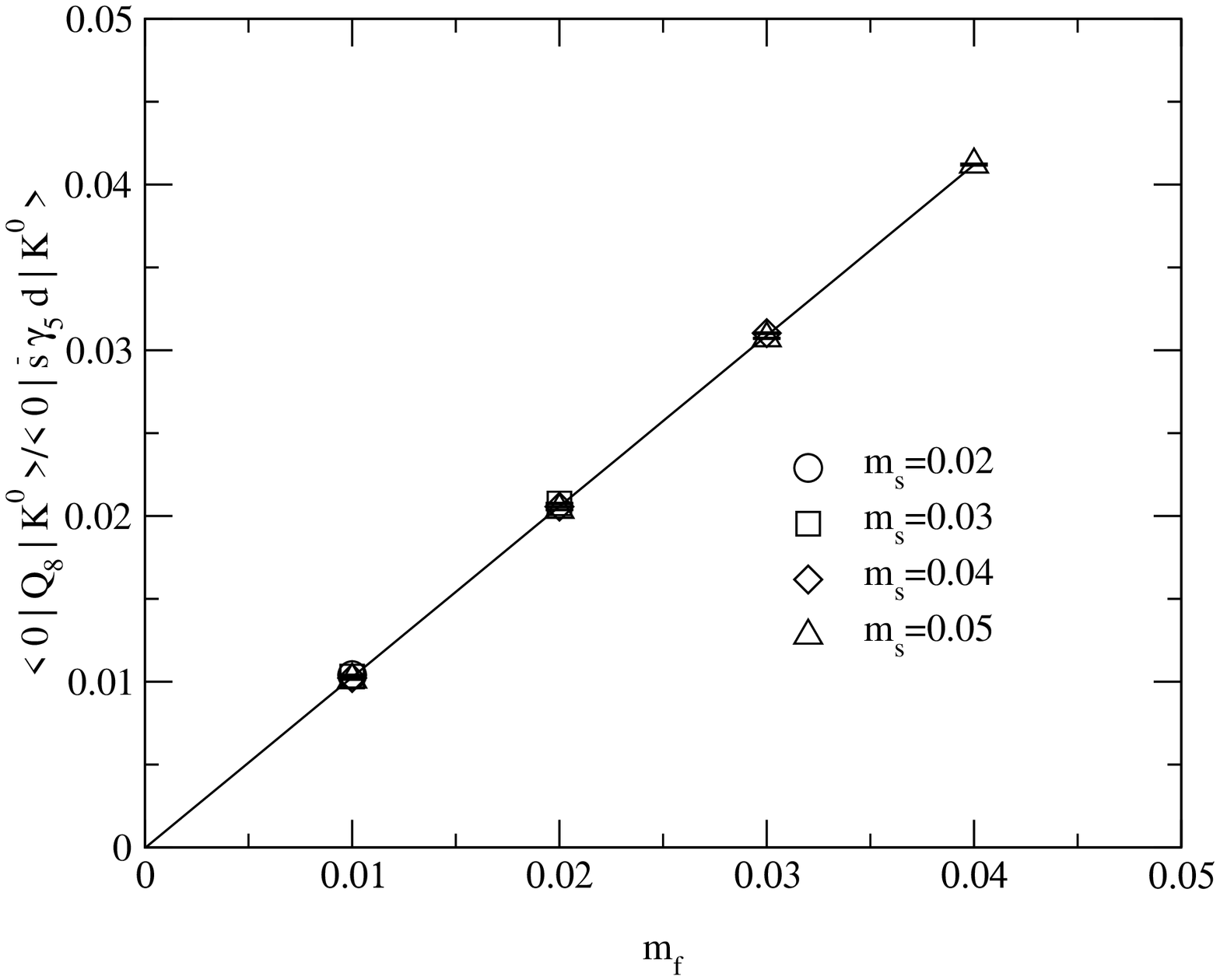}

\caption{The ratio $\langle 0 | Q_8 | K^0 \rangle/ \langle 0 | \bar{s}
\gamma_5 d| K^0 \rangle$ versus $m_s^\prime - m_d^\prime$.  The line is
a linear fit of the form given in Eq.\ \ref{eq:fix_eta}.}

\label{fig:ktovac_charm_out_Q8}
\end{center}
\end{figure}


\begin{figure}
\begin{center}
\epsfxsize=\hsize
\epsfbox{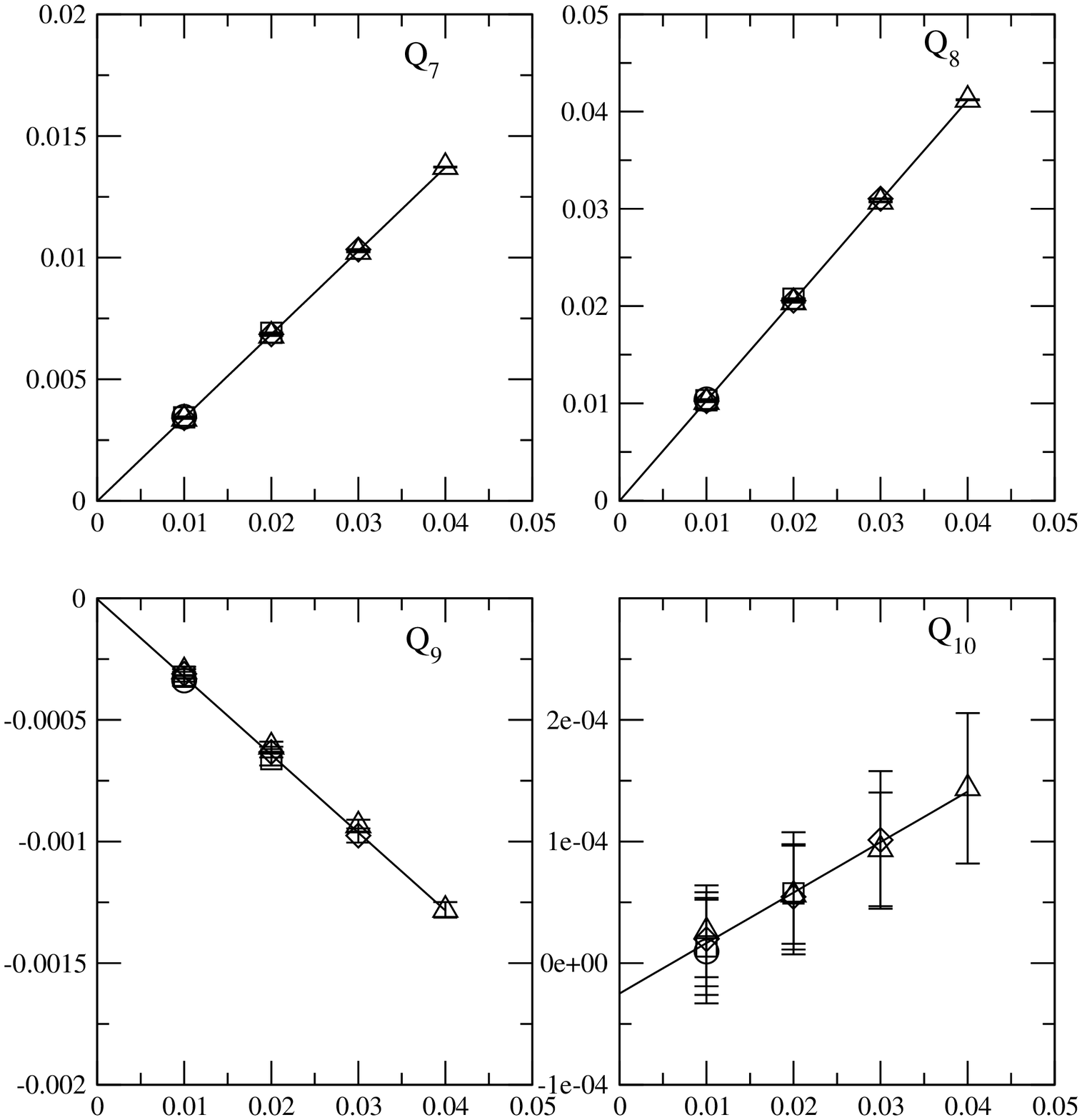}
\vspace{0.2in}

\caption{The ratio $\langle 0 | Q_i | K^0 \rangle/ \langle 0 | \bar{s}
\gamma_5 d| K^0 \rangle$ versus $m_s^\prime - m_d^\prime$ for $i = 7$,
8, 9 and 10.  The line is a linear fit of the form given in Eq.\
\ref{eq:fix_eta}.  The symbols have the same meaning as in Figure
\ref{fig:ktovac_charm_out_Q6}. }

\label{fig:ktovac_charm_out_Q7_Q8_Q9_Q10}
\end{center}
\end{figure}


\begin{figure}
\begin{center}
\epsfxsize=\hsize
\epsfbox{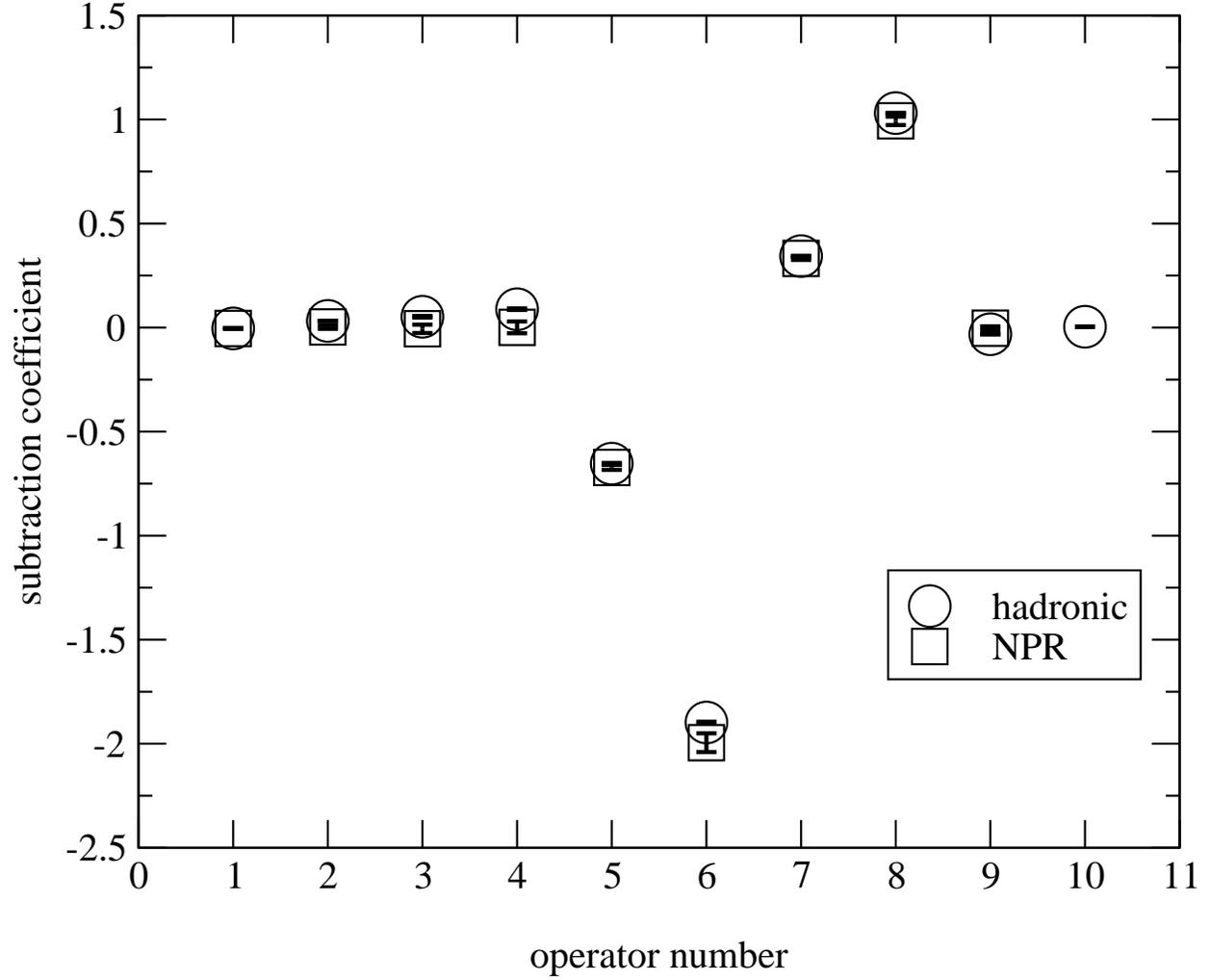}
\caption{The subtraction coefficients determined in hadronic states
($\bigcirc$) compared with those determined in Landau gauge fixed
quark states at $\mu = 2.13$ GeV ($\Box$).  For the operators
with large power divergences, the subtraction coefficients agree
well since the external momentum does not enter the power divergent
coefficient.}
\label{fig:compare_hadronic_npr_sub.eps}
\end{center}
\end{figure}


\begin{figure}
\begin{center}
\epsfxsize=\hsize
\epsfbox{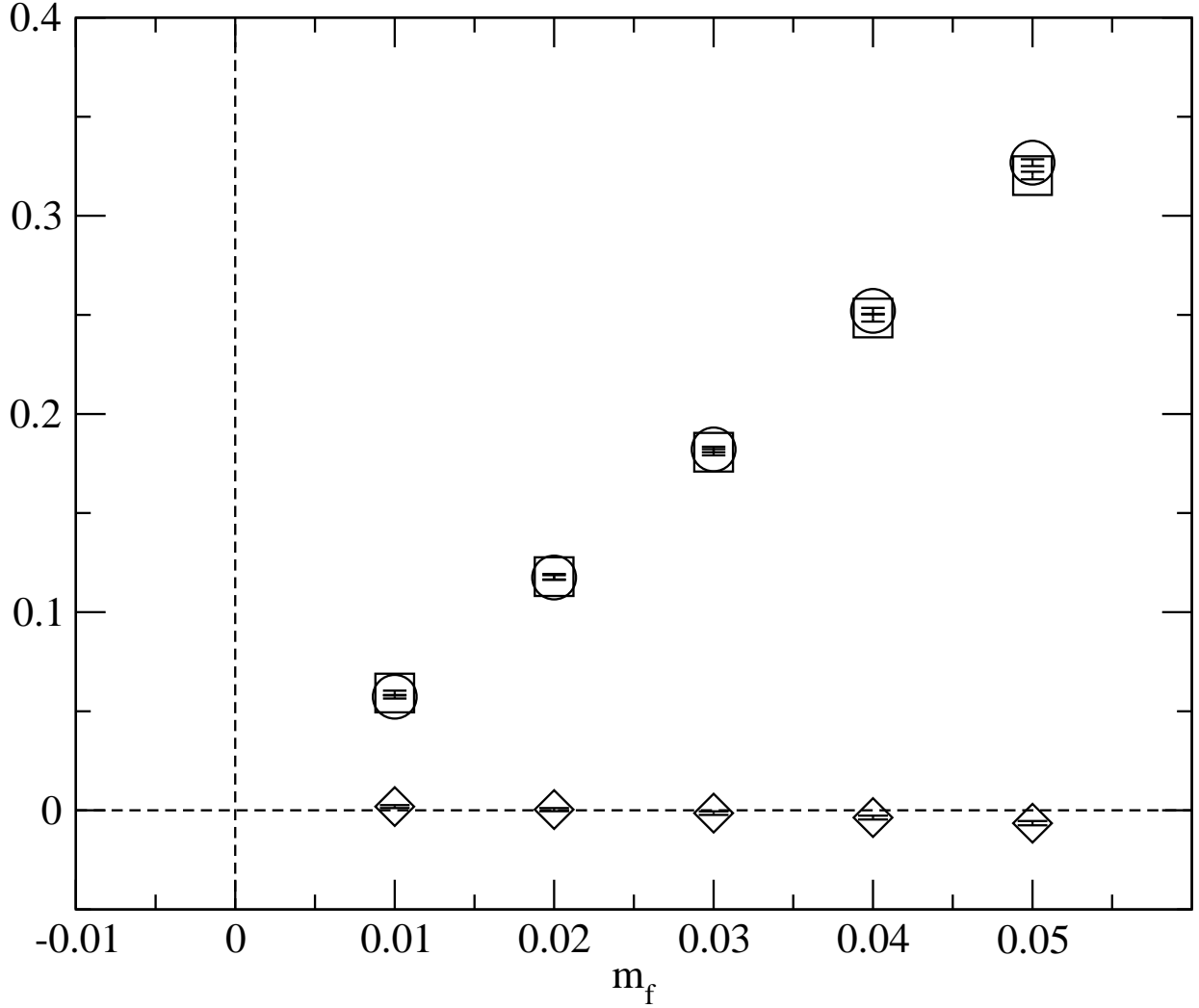}
\caption{The matrix elements $\langle \pi^+ | Q_6 | K^+ \rangle$
($\square$),  $2 m_f \, |\eta_{1,6}| \, \langle \pi^+ | (\bar{s}d)_{\rm
lat} | K^+ \rangle$ ($\circ$) and $\langle \pi^+ | Q_6 | K^+
\rangle_{\rm sub}$ ($\diamond$) showing the noticeable, and very
similar, non-linearity in the first two quantities and the size of the
subtraction for this left-right operator.  The slope of the subtracted
matrix element determines the desired $\alpha^{(8,1)}_{1, \rm lat}$ for
$Q_6$ and is about $30$ times smaller than the slope of the
unsubtracted operator, and of opposite sign.  Note that the subtracted
operator does not vanish at $m_f = -m_{\rm res}$ since the divergent
parts of the operator do not see only the chiral symmetry breaking of
the low energy theory.}

\label{fig:ktopi_Q6_sub_unsub_charm_out}

\end{center}
\end{figure}


\begin{figure}
\begin{center}
\epsfxsize=\hsize
\epsfbox{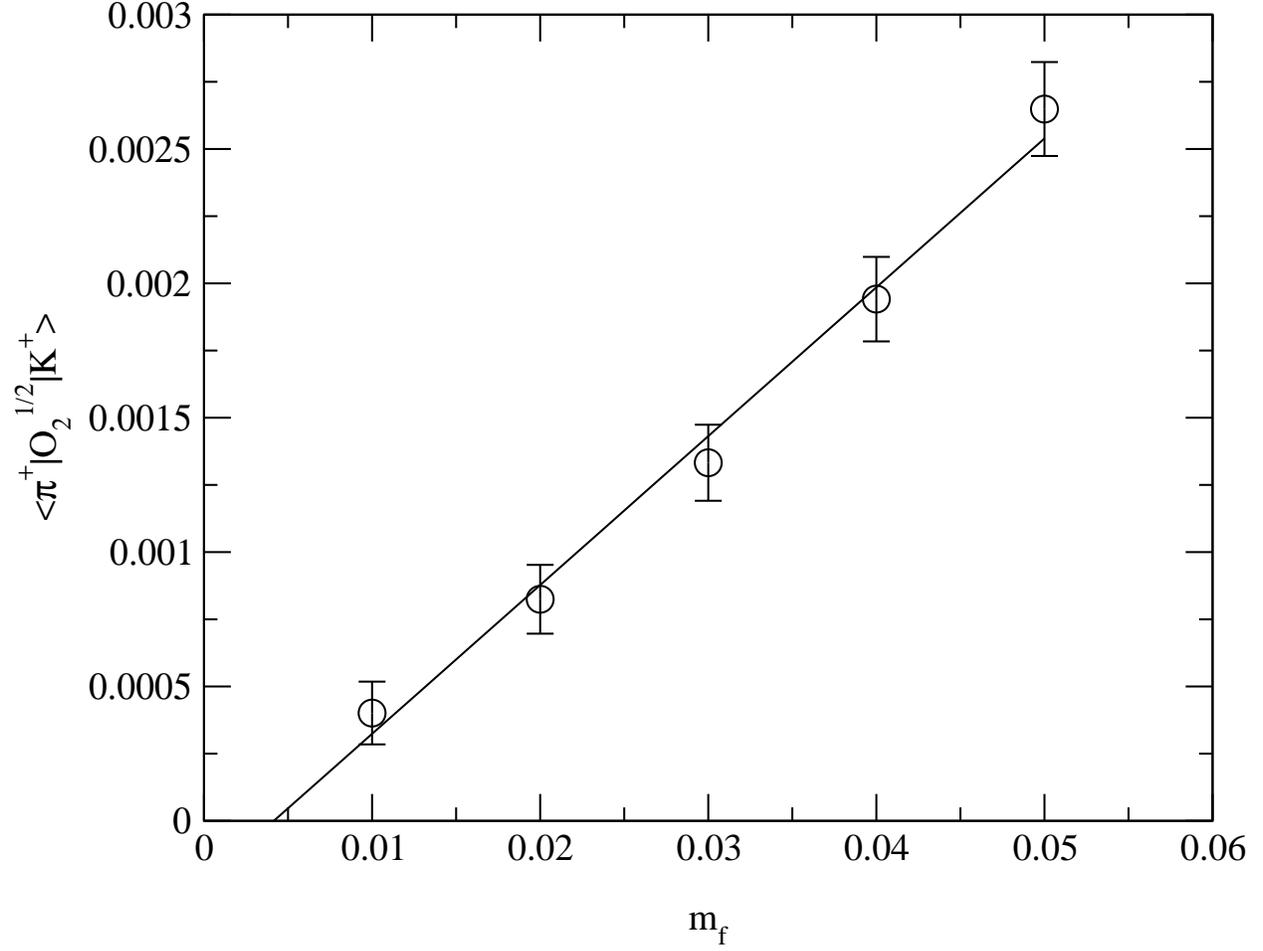}
\caption{The matrix element $\langle \pi^+ | Q^{(1/2)}_{2, \rm lat} |
K^+ \rangle_{\rm sub}$ which has the divergent contribution removed.
Due to the contact term in the Ward-Takahashi identity the matrix
element does not vanish at $m_f = -m_{\rm res}$.  The slope is
related to the matrix elements we seek.}
\label{fig:ktopi_Q2_sub_charm_out}
\end{center}
\end{figure}


\begin{figure}
\begin{center}
\epsfxsize=\hsize
\epsfbox{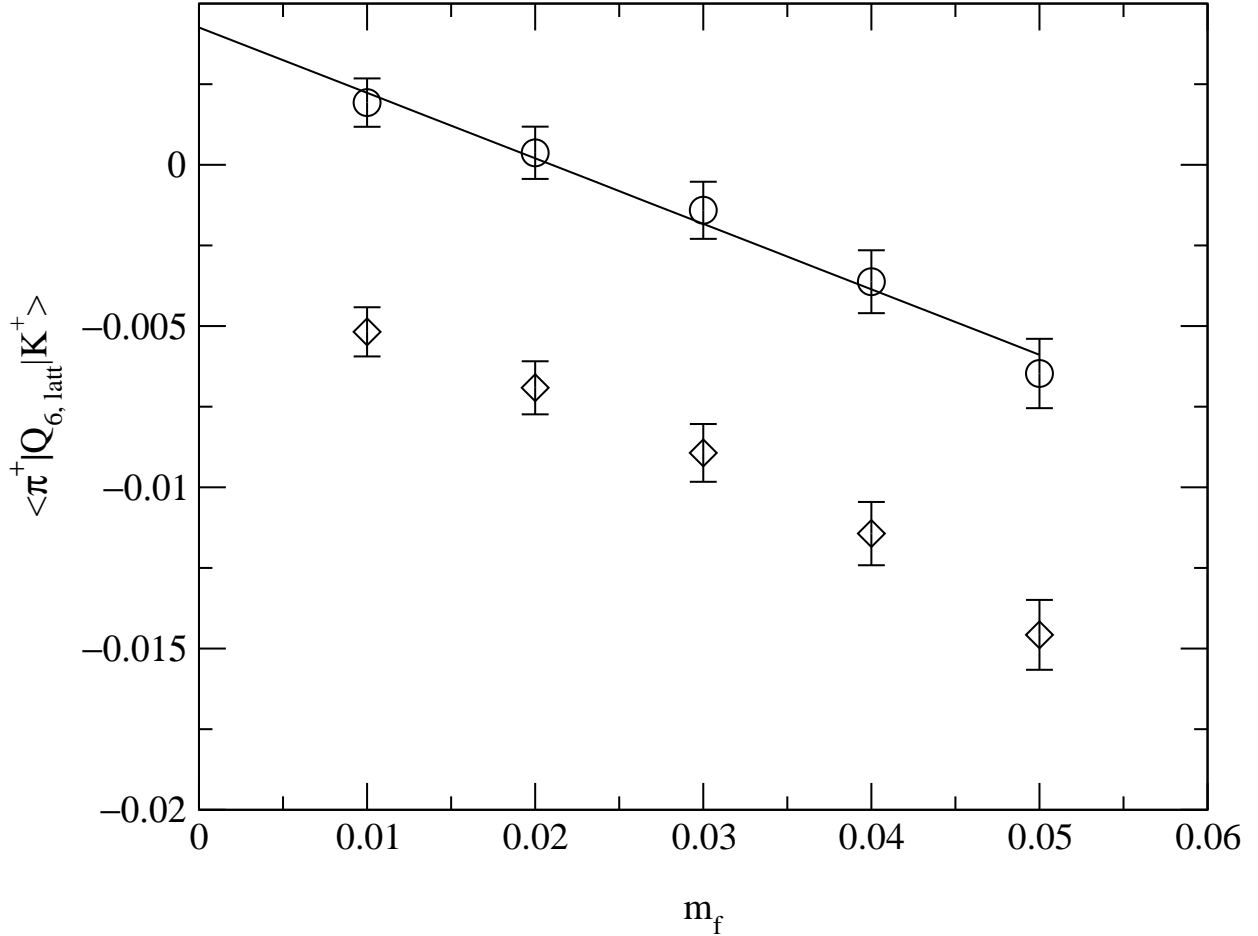}

\caption{The matrix element $\langle \pi^+ | Q^{(1/2)}_{6, \rm lat} |
K^+ \rangle_{\rm sub}$ which has the divergent contribution removed
($\bigcirc$).  The subtraction does not remove the $O(\mres/a^2)$
divergent term, so the matrix element does not vanish at $m_f = 0$.
The line is a linear fit to the data, since the chiral logarithm
corrections are not known, and the slope of this line is related to
physical matrix elements.  From the data, non-linear effects appear
small.  The lower points ($\Diamond$) are the result if the subtraction
in Eq.\ \ref{eq:ktopi_subtraction} has $(m_s + m_d )$ changed to $(m_s
+ m_d + 2\mres)$.  This subtraction will also not exactly remove the
$O(\mres/a^2)$ term, but the two subtractions show that chiral
symmetry breaking from finite $L_s$ is quantitatively
$O(\mres/a^2)$.}

\label{fig:ktopi_Q6_sub_charm_out}
\end{center}
\end{figure}


\begin{figure}
\begin{center}
\epsfxsize=\hsize
\epsfbox{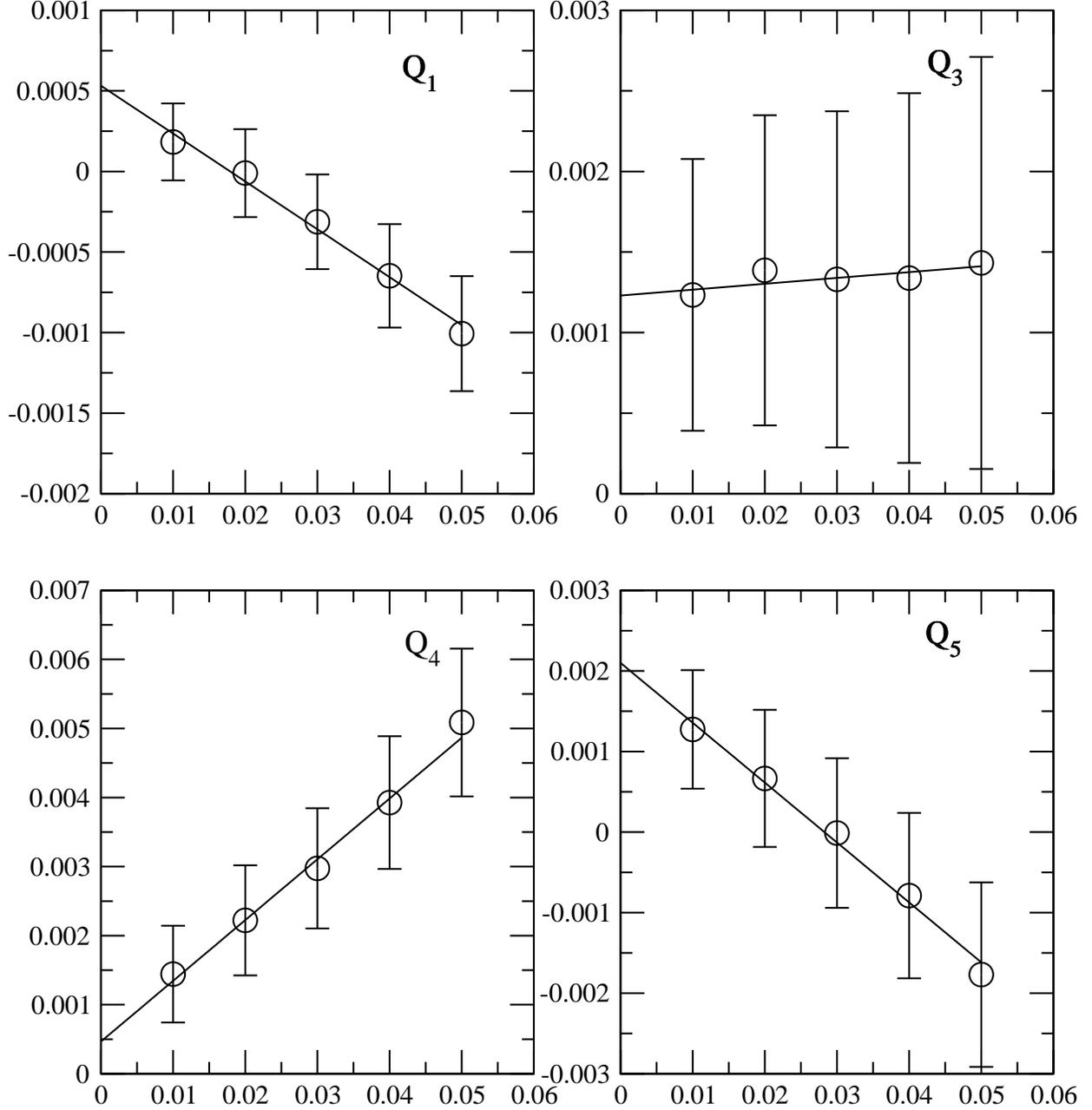}
\vspace{0.2in}
\caption{The matrix element $\langle \pi^+ | Q^{(1/2)}_{i, \rm lat} |
K^+ \rangle_{\rm sub}$, for $i = 1$, 3, 4 and 5, which has the
divergent contribution removed.  Due to the contact term in the
Ward-Takahashi identity the matrix element does not vanish at $m_f =
-m_{\rm res}$.}
\label{fig:ktopi_Q1_Q3_Q4_Q5_sub_charm_out}
\end{center}
\end{figure}


\begin{figure}
\begin{center}
\epsfxsize=\hsize
\epsfbox{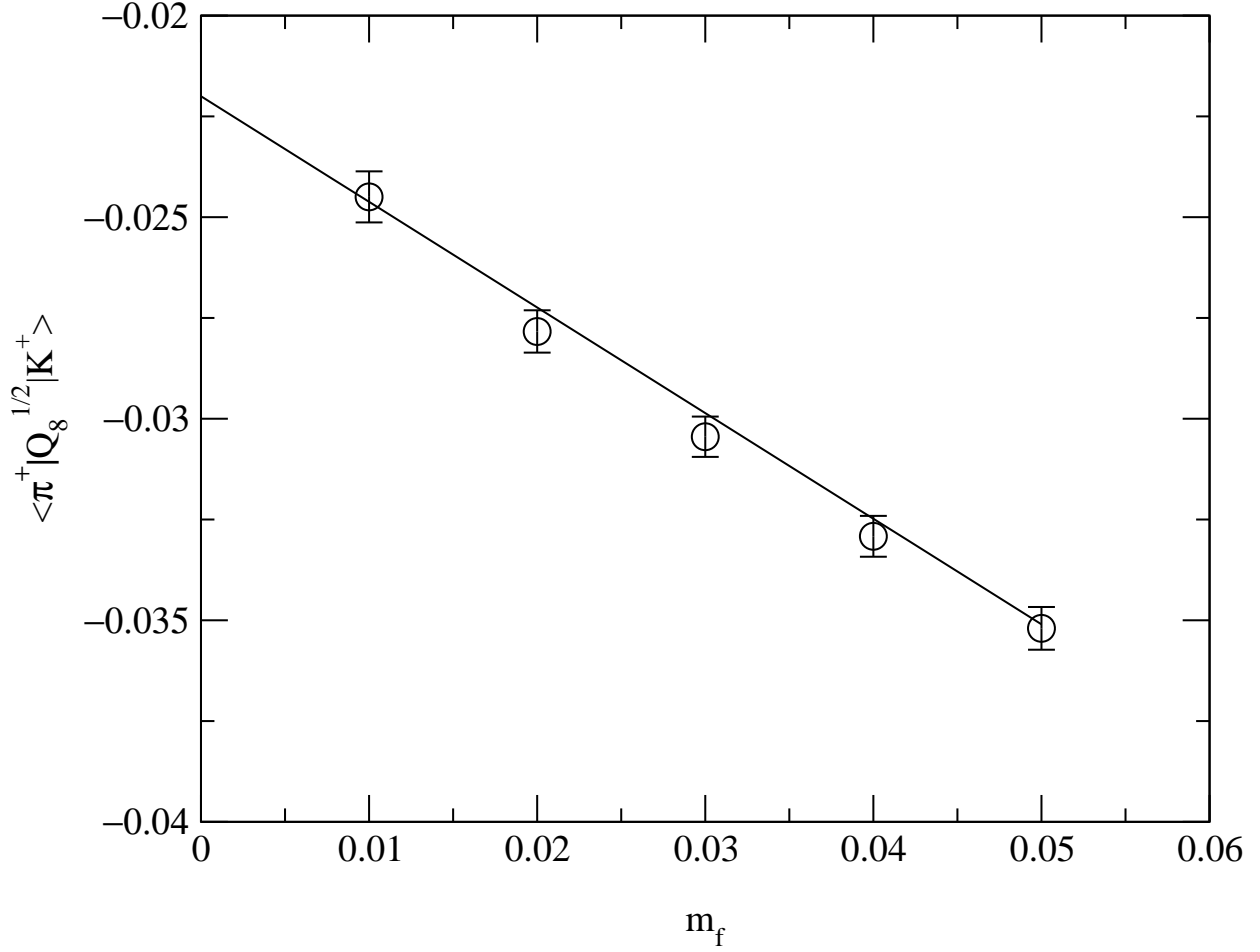}
\caption{The matrix element $\langle \pi^+ | Q^{(1/2)}_{8, \rm lat} |
K^+ \rangle_{\rm sub}$ which has the divergent contribution removed.
Due to the power divergence of this operator, the value
of $m_f$ needed to cancel the chiral symmetry breaking effects of
finite $L_s$ is not precisely known.  Thus we do not know where
to evaluate this matrix element to get $\alpha^{(8,8)}_8$ and must
rely on the $\DIthalf$ amplitude to determine this quantity.}
\label{fig:ktopi_Q8_sub_charm_out}
\end{center}
\end{figure}


\begin{figure}
\begin{center}
\epsfxsize=\hsize
\epsfbox{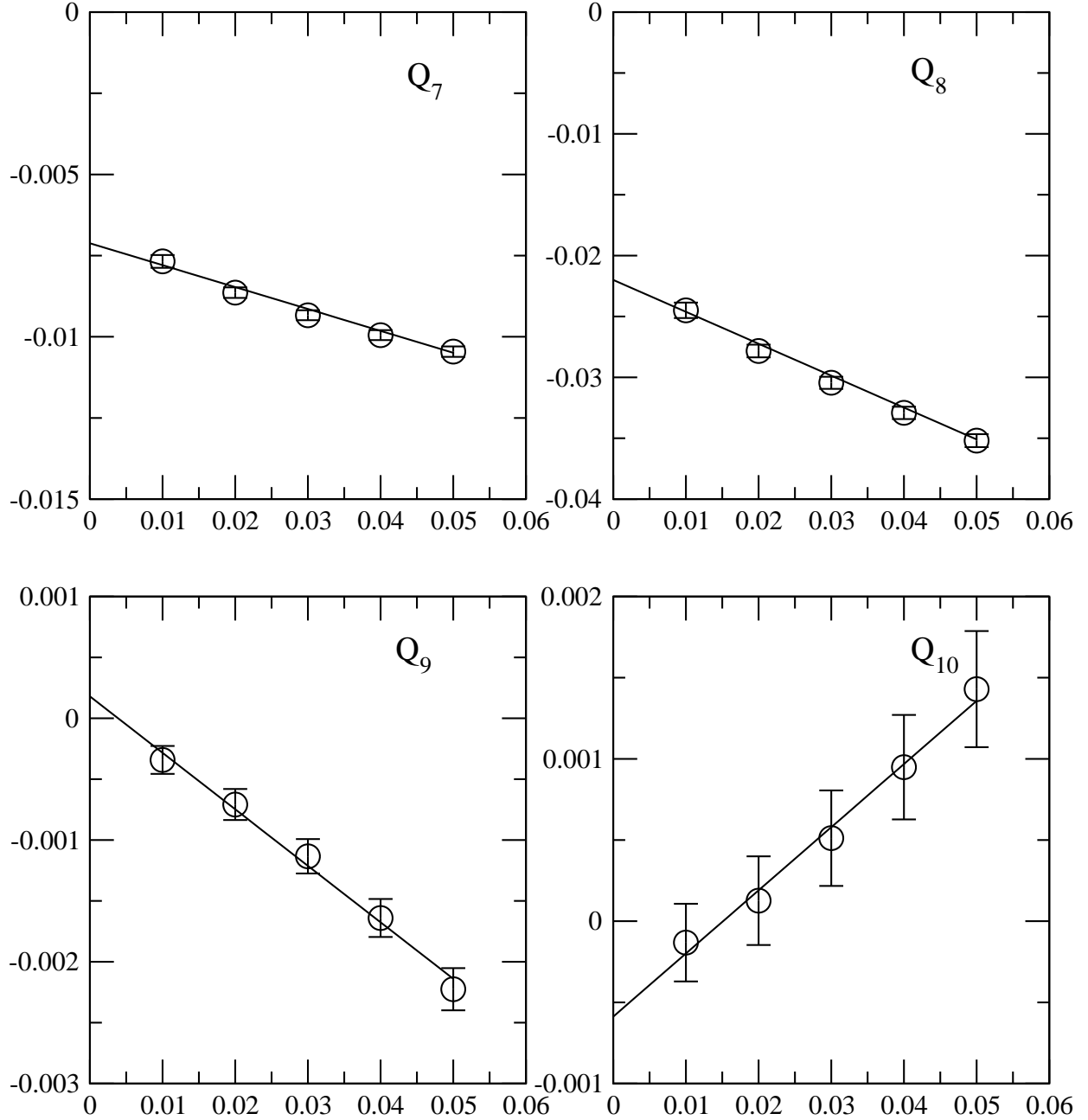}
\vspace{0.2in}
\caption{The matrix element $\langle \pi^+ | Q^{(1/2)}_{i, \rm lat} |
K^+ \rangle_{\rm sub}$, for $i = 7$, 8, 9 and 10, which has the
divergent contribution removed.   For $Q_9$ and $Q_{10}$, the
slope is needed to determine the $K \rightarrow \pi \pi$ matrix
elements.  $Q_7$ and $Q_8$ are shown for completeness.}
\label{fig:ktopi_Q7_Q8_Q9_Q10_sub_charm_out}
\end{center}
\end{figure}


\begin{figure}
\begin{center}
\epsfxsize=\hsize
\epsfbox{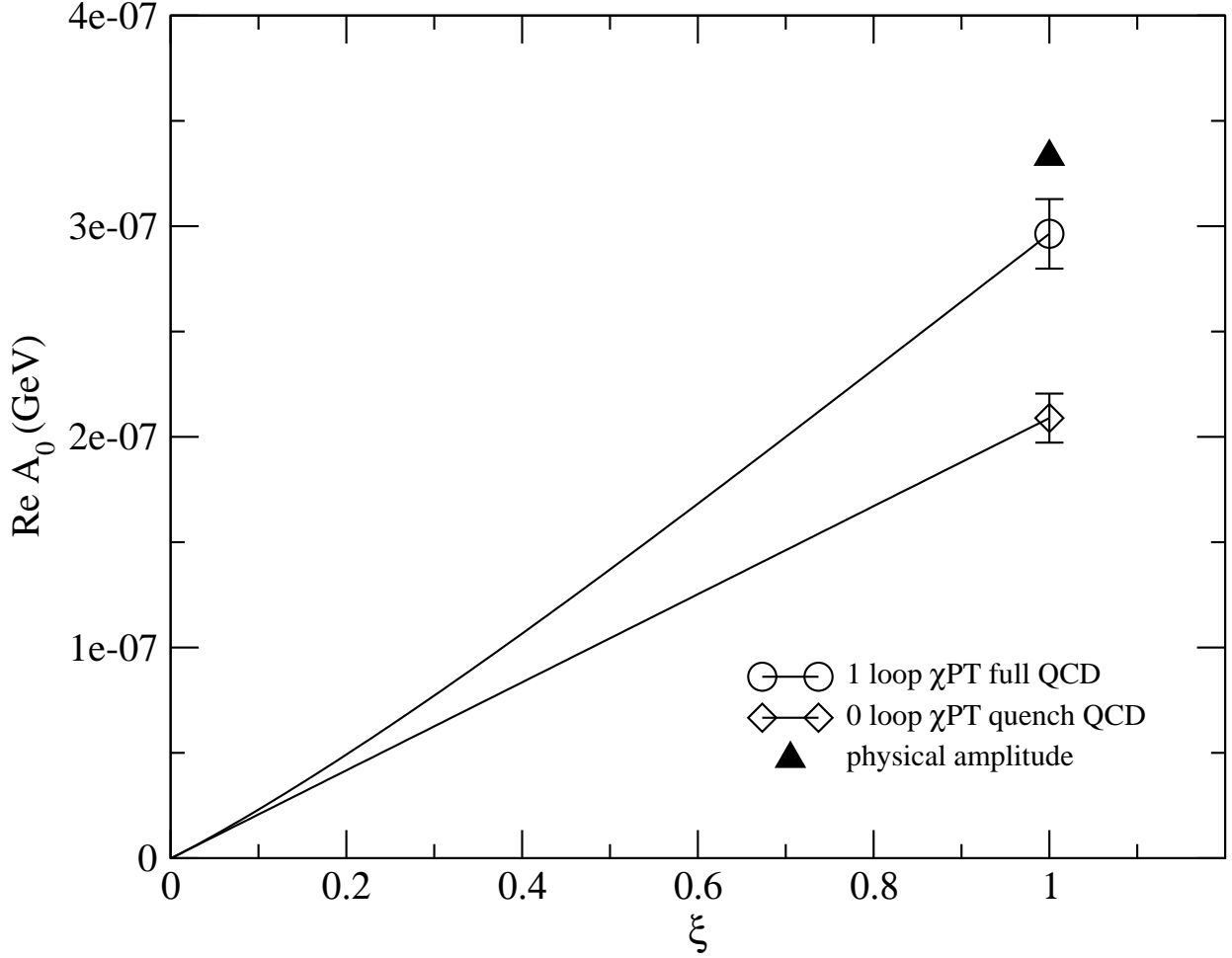}
\caption{Re($A_0$) plotted versus $\xi$, where $\xi$ multiplies the
pseudoscalar masses appearing in Eqs.\ \ref{eq:complete_kpipi} and
\ref{eq:complete_kpipi_log}.  The chiral limit is $\xi = 0$ and the
physical point corresponds to $\xi = 1$.  Two ways of extrapolating to
the physical point are shown:  1) 0-loop chiral perturbation theory in
quenched QCD and 2) 1-loop chiral perturbation theory in full QCD.  The
difference between them gives an indication of the contribution
expected from including all $O(p^4)$ terms in chiral perturbation
theory.  Since all $O(p^4)$ terms are not included in our results, the
close agreement with the experimental value should be regarded as
fortuitous.  The data is for $\mu = 2.13$ GeV.}

\label{fig:Re_A0}
\end{center}
\end{figure}


\begin{figure}
\begin{center}
\epsfxsize=\hsize
\epsfbox{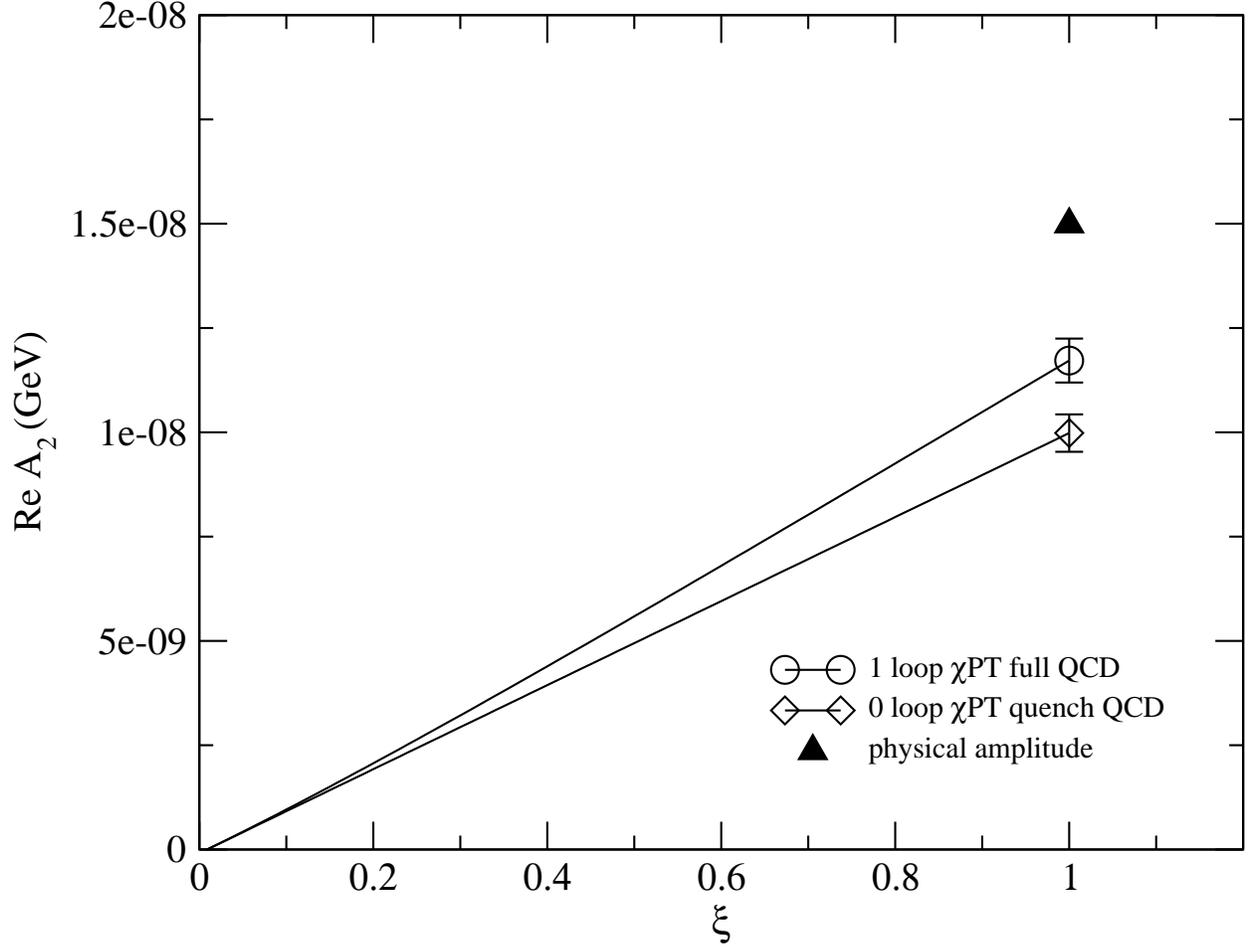}
\caption{As in Fig.\ \ref{fig:Re_A0}, except that Re($A_2$) is plotted
versus $\xi$.  Here the 1-loop chiral perturbation theory extrapolation
in full QCD differs from the experimental result by $\sim$ 18\%.
This is well within the general expectation for higher order
effects in chiral perturbation theory at scales around $m_K$.
The data is for $\mu = 2.13$ GeV.}

\label{fig:Re_A2}
\end{center}
\end{figure}


\begin{figure}
\begin{center}
\epsfxsize=\hsize
\epsfbox{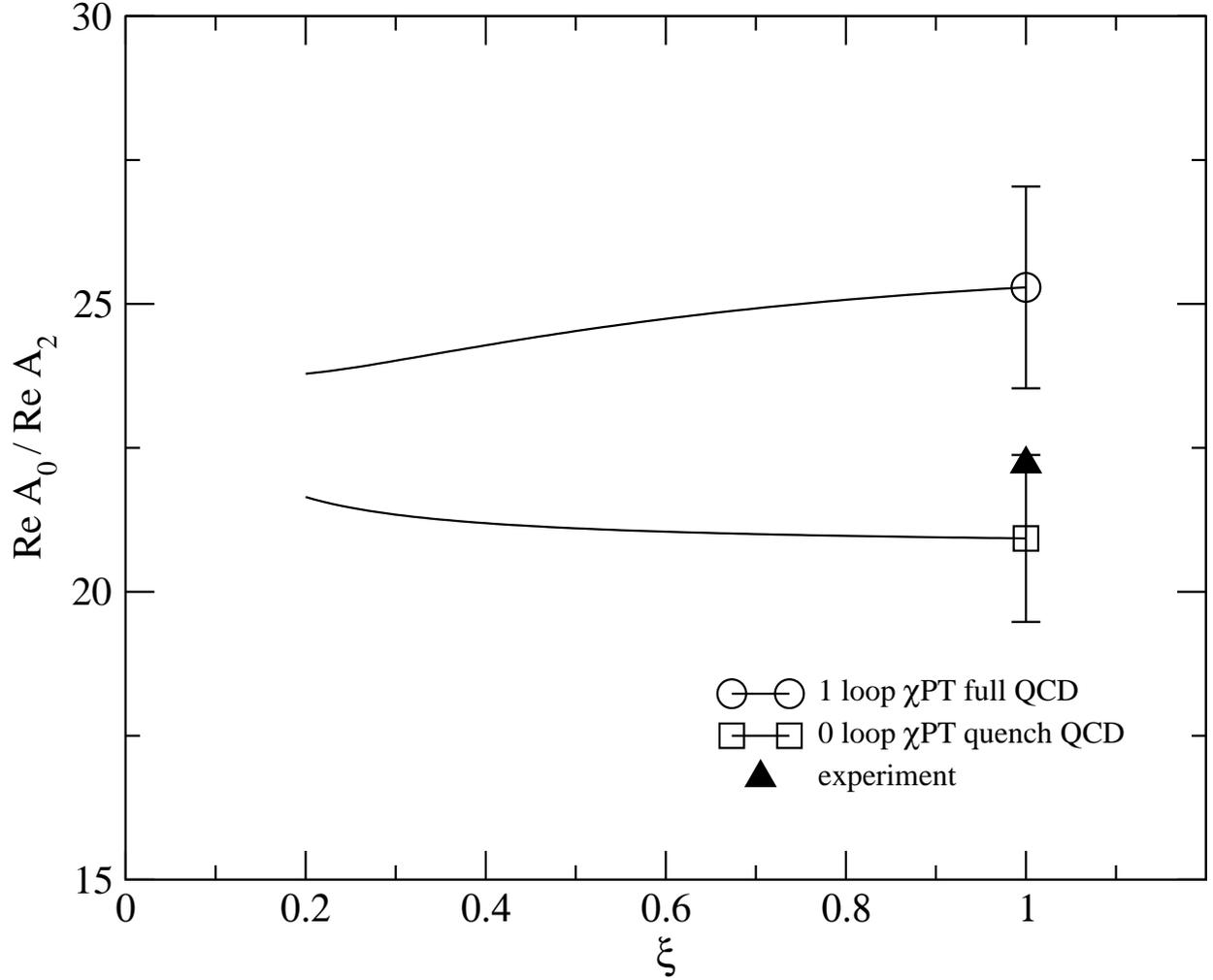}
\caption{As in Fig.\ \ref{fig:Re_A0}, except that Re($A_0$)/ Re($A_2$)
is plotted versus $\xi$.  The two extrapolations are only slightly
different due to the chiral logarithms having coefficients with the
same sign for the dominant operators contributing to Re($A_0$) and
Re($A_2$).  The data is for $\mu = 2.13$ GeV.}

\label{fig:ReA0_over_ReA2}
\end{center}
\end{figure}


\begin{figure}
\begin{center}
\epsfxsize=\hsize
\epsfbox{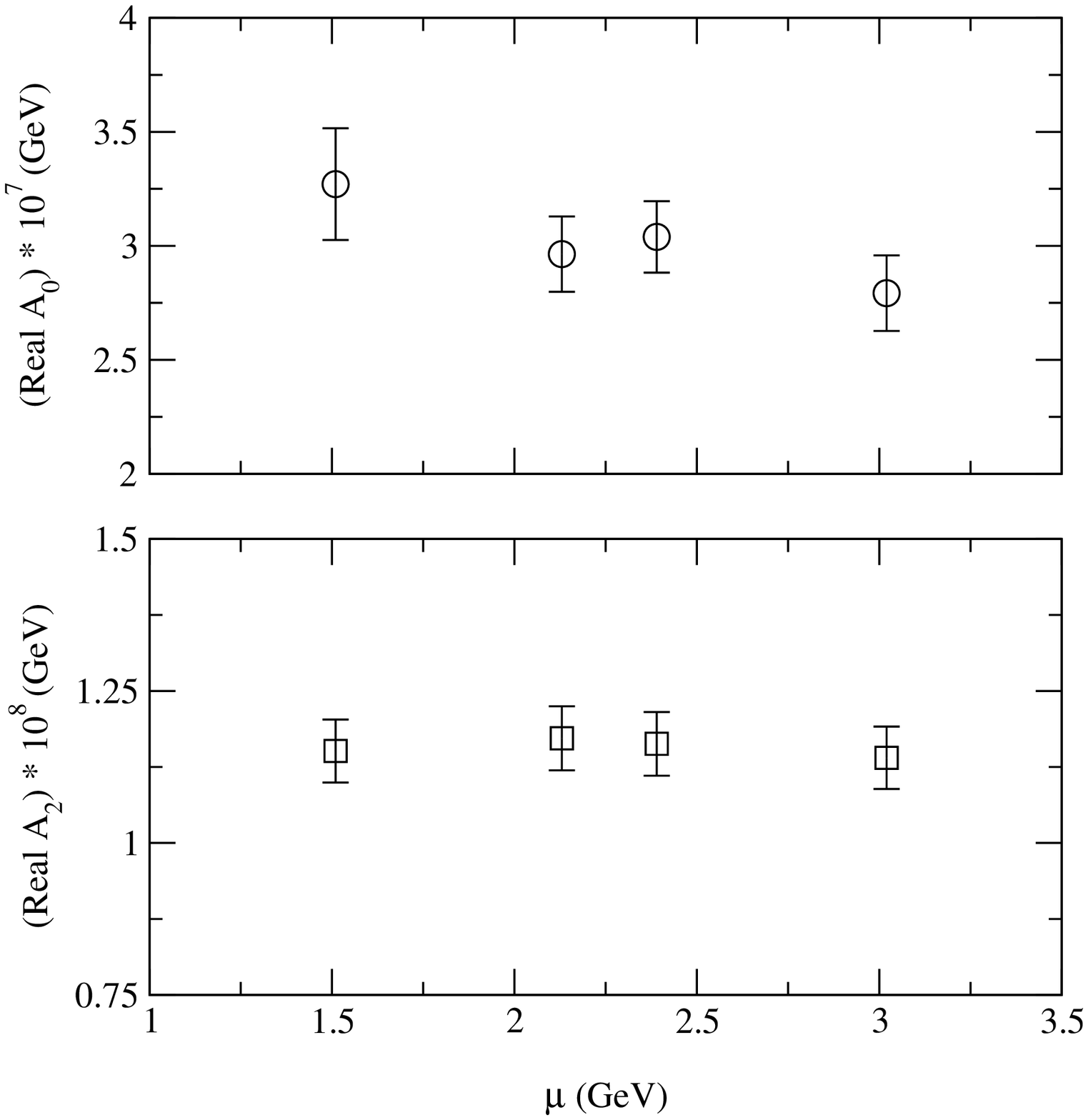}
\caption{A plot of Re($A_0$) (upper panel) and Re($A_2$) (lower panel)
versus $\mu$ for the physical values obtained using 1-loop full QCD
chiral perturbation theory for the extrapolation to the physical kaon
mass.  The results show no statistically significant $\mu$ dependence.
We choose to quote final values with $\mu = 2.13$ GeV.}

\label{fig:ReA0_ReA2_mu_dep}
\end{center}
\end{figure}


\begin{figure}
\begin{center}
\epsfxsize=\hsize
\epsfbox{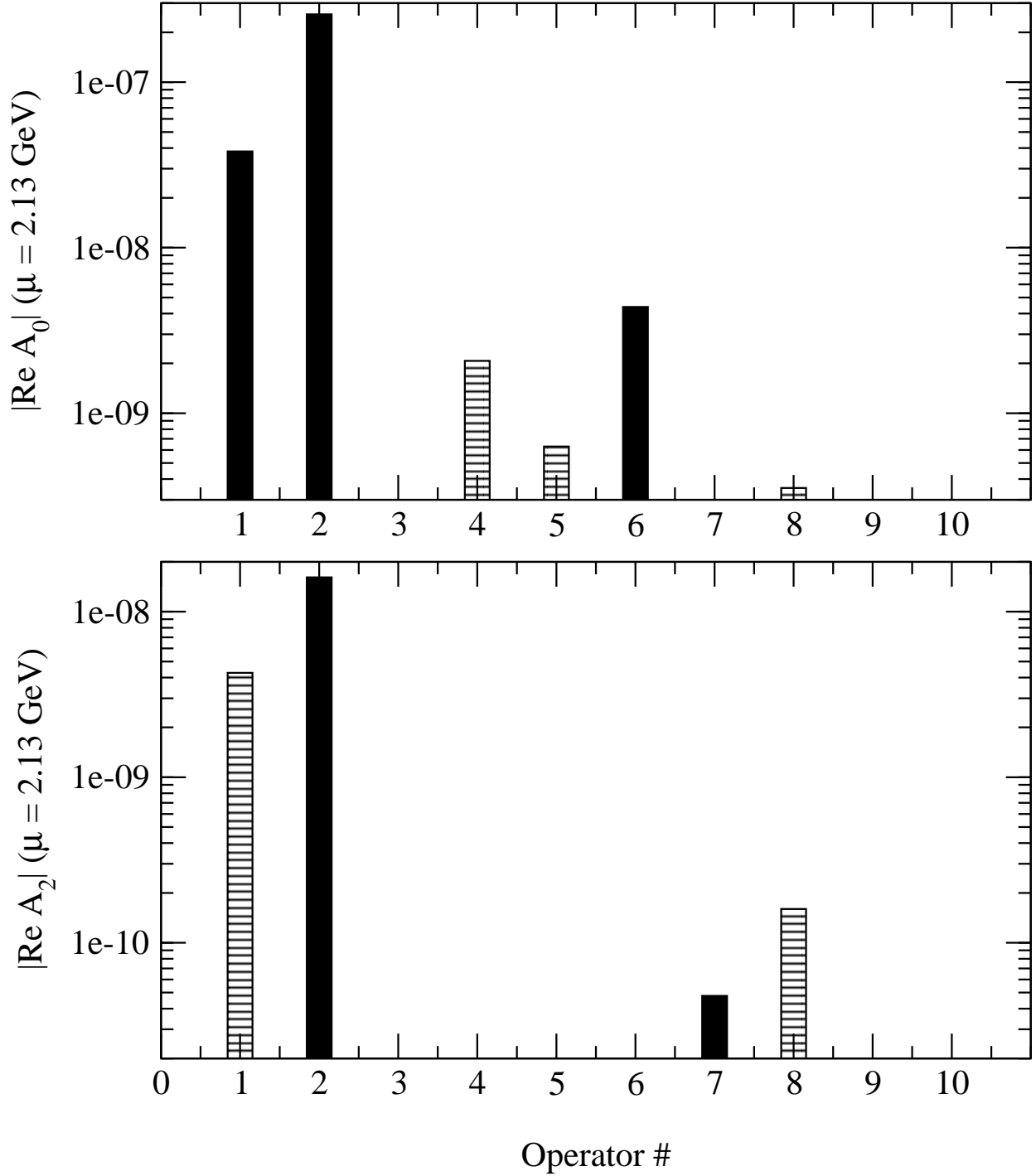}
\caption{A breakdown of the contribution of $Q_{i, \rm cont}$ to
Re($A_0$) (upper panel) and Re($A_2$) (lower panel).  The solid filled
bars in the graph denote positive quantities and the hashed represent
negative quantities.  The data is for $\mu = 2.13$ GeV.}

\label{fig:ReA0_ReA2_break.eps}
\end{center}
\end{figure}


\begin{figure}
\epsfxsize=\hsize
\begin{center}
\epsfbox{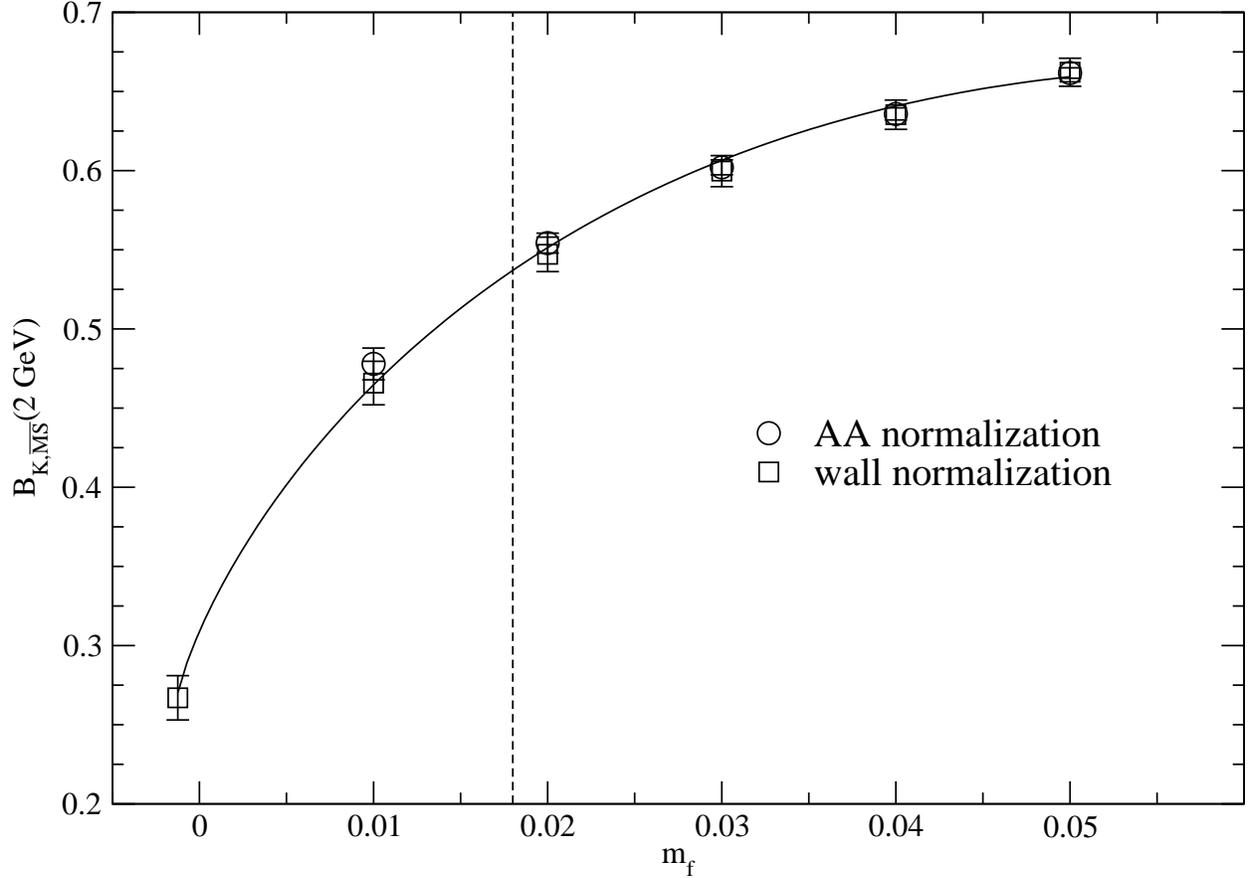}
\end{center}

\caption{Values for $B_{PS,\overline{MS}}(2\,{\rm GeV})$ versus $m_f$.
The points labeled AA normalization ($\bigcirc$) are determined by
normalizing with two-point functions which may introduce zero mode
effects for small quark mass.  The wall normalization points ($\Box$)
determine $B_{PS}$ from a $K \to \pi$ matrix element, where a wall
normalization is used, and zero mode effects should not be introduced
through the normalization.  Some difference can be seen for the smaller
quark masses.  The wall normalization point at $m_f = -\mres$ is the
value of $B_{PS}$ in the chiral limit.  The solid line is the result of
fitting the AA norm points to the form given in Eq.\ \ref{eq:B_PS}.
Since the fit goes below the AA norm point at $m_f = 0.01$, the
extrapolated value agrees with the wall normalization value.  The
dashed line has $m_f = 0.018$ and marks the point where a kaon made of
degenerate quarks has its physical mass.  Using the value of $B_{PS}$
at this point, we find $B_{K,\overline{MS}}(2 \, {\rm GeV}) =
0.532(11)$. }

\label{fig:B_PS}
\end{figure}


\begin{figure}
\begin{center}
\epsfxsize=\hsize
\epsfbox{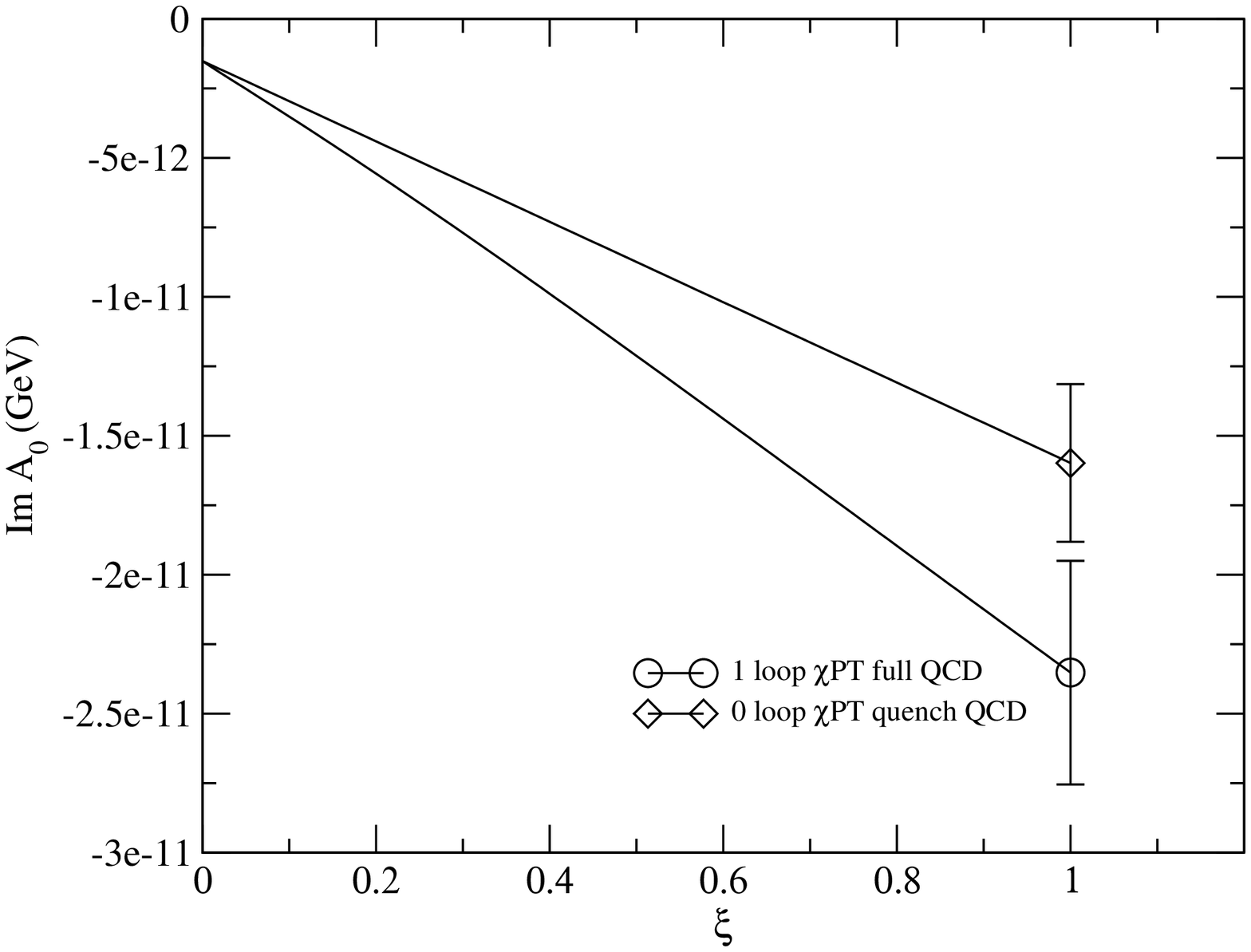}
\caption{As in Fig.\ \ref{fig:Re_A0}, except that Im($A_0$) is
plotted versus $\xi$.  Here a physical value is not directly
known.  The 1-loop chiral perturbation extrapolation in
full QCD is a 47\% correction to the 0-loop extrapolation.
The data is for $\mu = 2.13$ GeV.}

\label{fig:Im_A0}
\end{center}
\end{figure}


\begin{figure}
\begin{center}
\epsfxsize=\hsize
\epsfbox{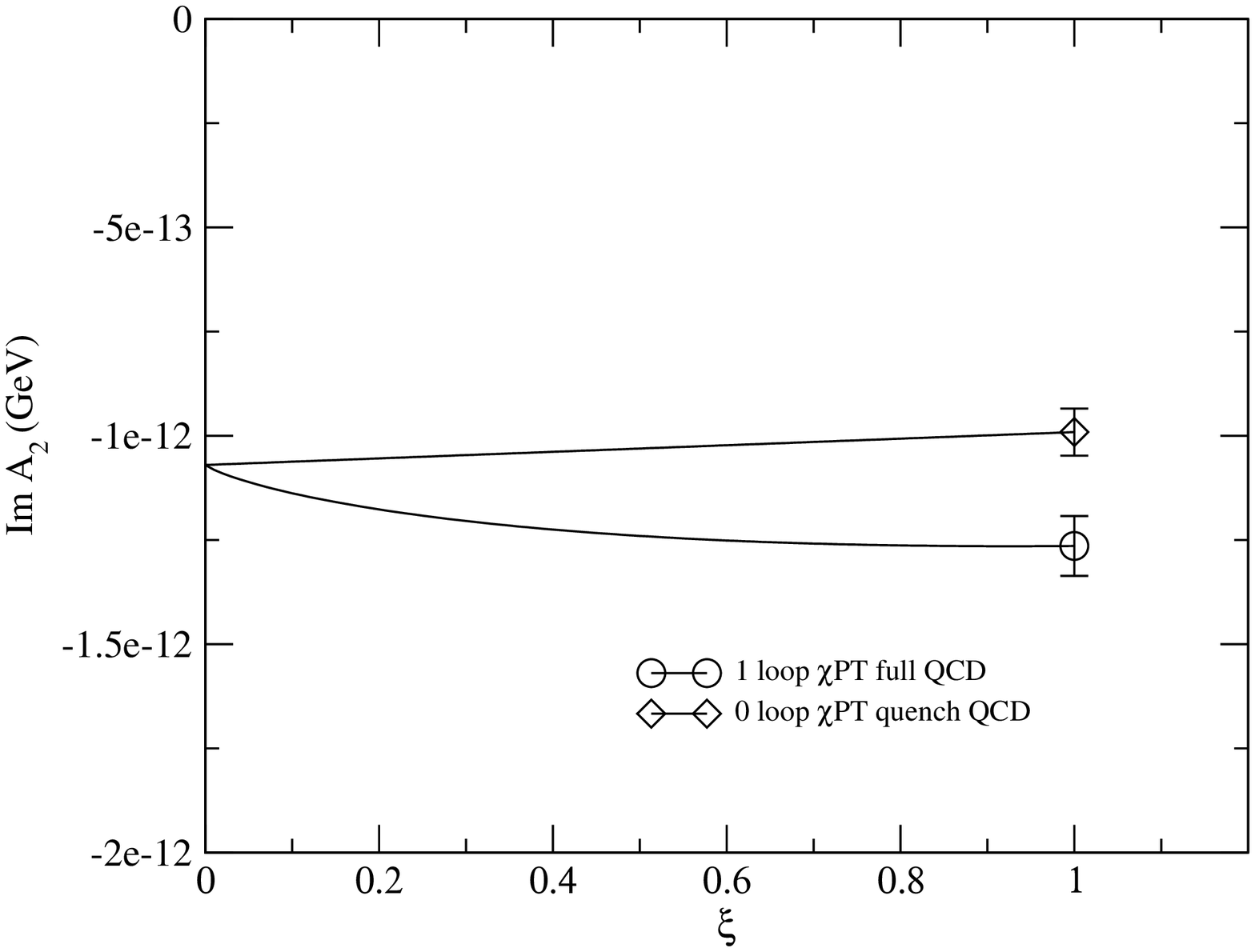}
\caption{As in Fig.\ \ref{fig:Re_A0}, except that Im($A_2$) is
plotted versus $\xi$.  Here a physical value is not directly
known.  The 1-loop chiral perturbation extrapolation in
full QCD is a 28\% correction to the 0-loop extrapolation.
The data is for $\mu = 2.13$ GeV.}

\label{fig:Im_A2}
\end{center}
\end{figure}


\begin{figure}
\begin{center}
\epsfxsize=\hsize
\epsfbox{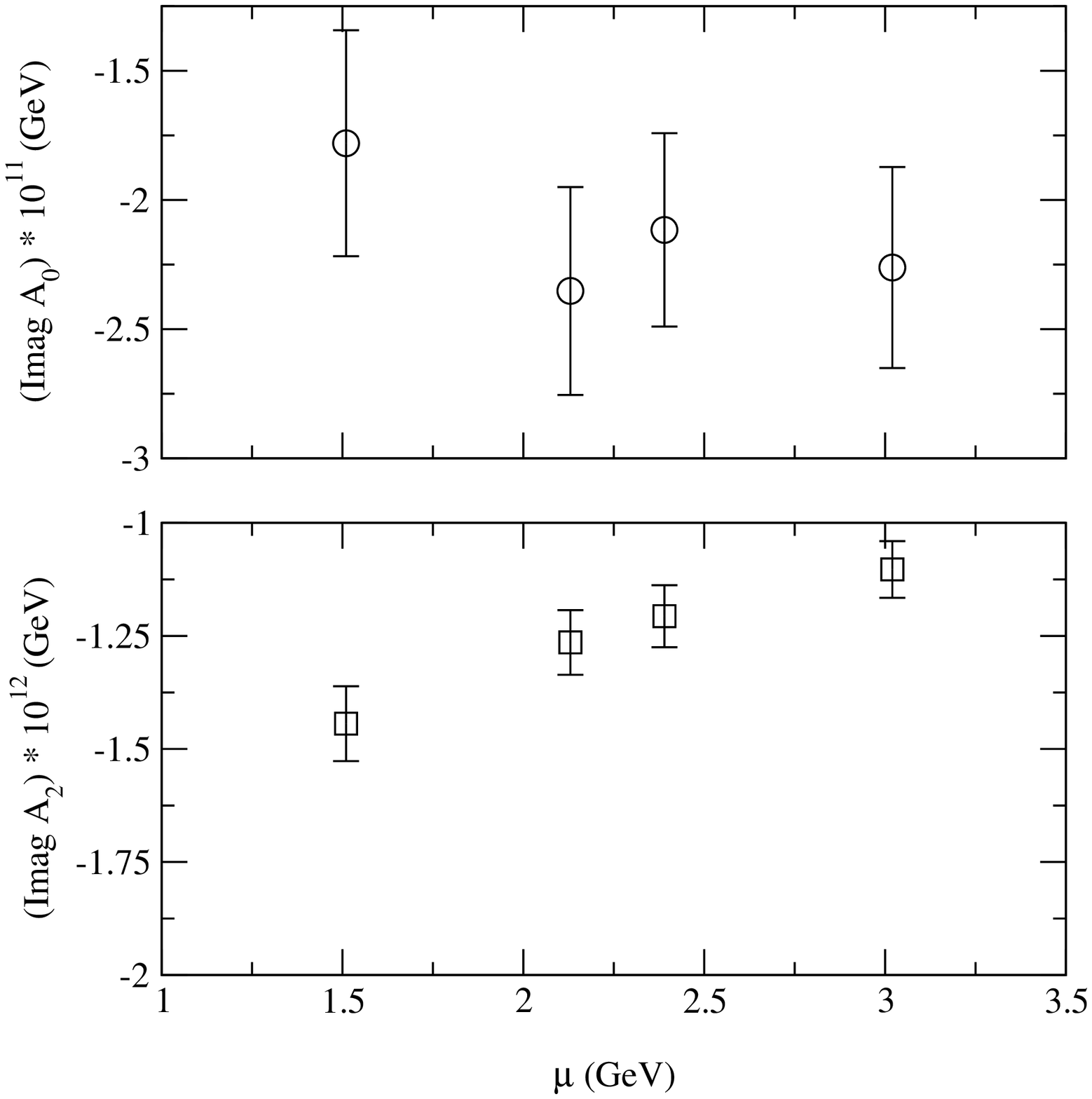}
\caption{A plot of Im($A_0$) (upper panel) and Im($A_2$) (lower panel)
versus $\mu$ for the physical values obtained using 1-loop full QCD
chiral perturbation theory for the extrapolation to the physical kaon
mass.  The results for Im($A_0$) show no statistically significant
$\mu$ dependence, while Im($A_2$) varies by 25\% over this range of
$\mu$.  We choose to quote final values with $\mu = 2.13$ GeV.}

\label{fig:ImA0_ImA2_mu_dep}
\end{center}
\end{figure}


\begin{figure}
\begin{center}
\epsfxsize=\hsize
\epsfbox{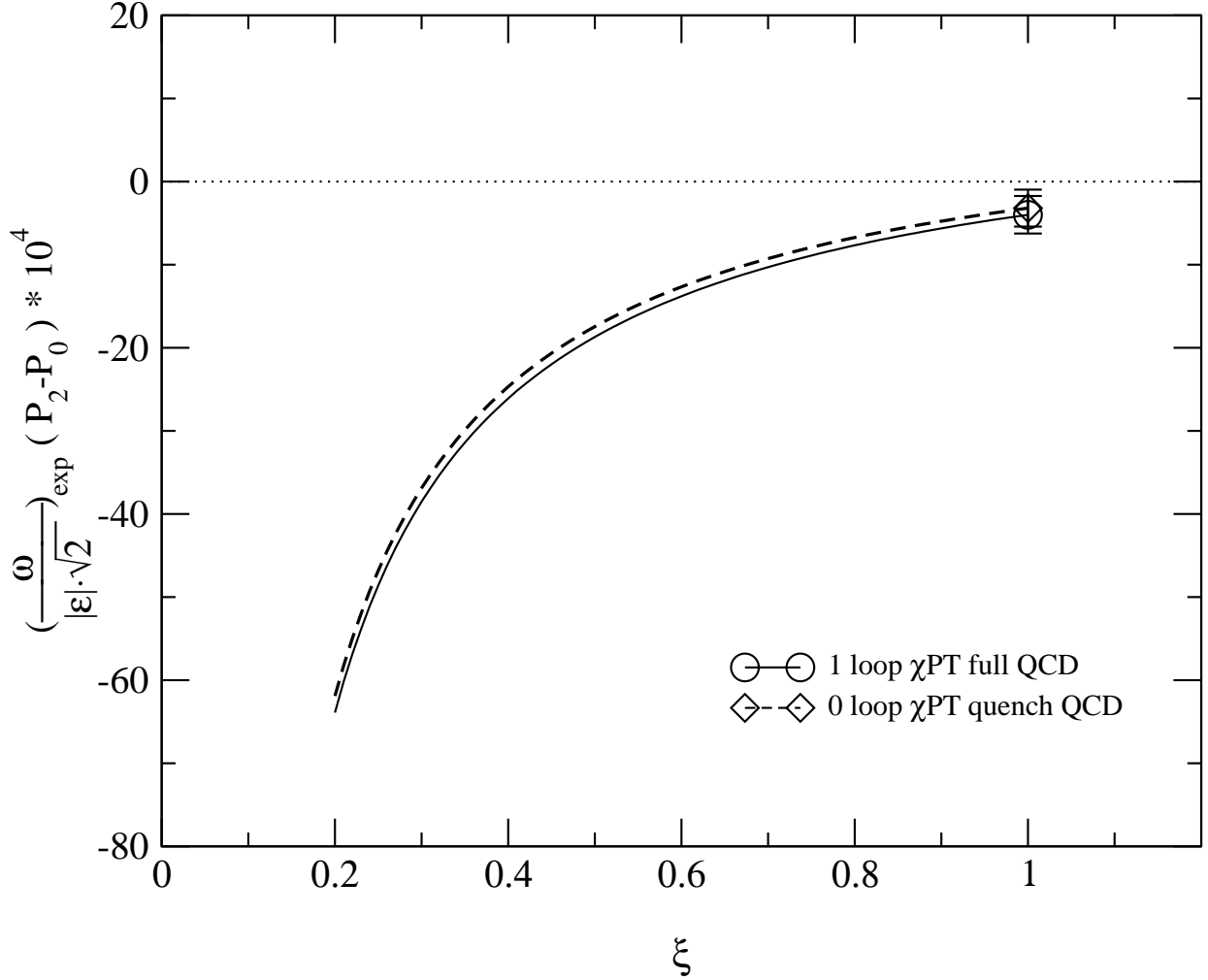}

\caption{A plot of $[\omega/ (\sqrt{2} | \epsilon |)]_{\rm exp}(P_2 -
P_0)$ versus $\xi$, where $\xi = 0$ is the chiral limit and $\xi = 1$
is the physical point.  We only plot points for $\xi \ge 0.2$, since in
the chiral limit only the electroweak (8,8) operators contribute and
$P_2 - P_0 = 0$.  As masses increase from zero, the contributions to
$P_2 - P_0$ of current-current, gluon penguin and electroweak penguin
operators for $\xi < 0.2$ is quite different from the physical world.
As explained in the text, for $0.2 < \xi < 0.5$, the electroweak
penguins continue to dominate by making $|P_2|$ large.  As one
approaches the physical point, the electroweak and gluonic penguins are
cancelling almost completely.  Higher order terms in chiral
perturbation theory could be expected to alter this large
cancellation.  The data is for $\mu = 2.13$ GeV.}

\label{fig:epe_all}
\end{center}
\end{figure}


\begin{figure}
\begin{center}
\epsfxsize=\hsize
\epsfbox{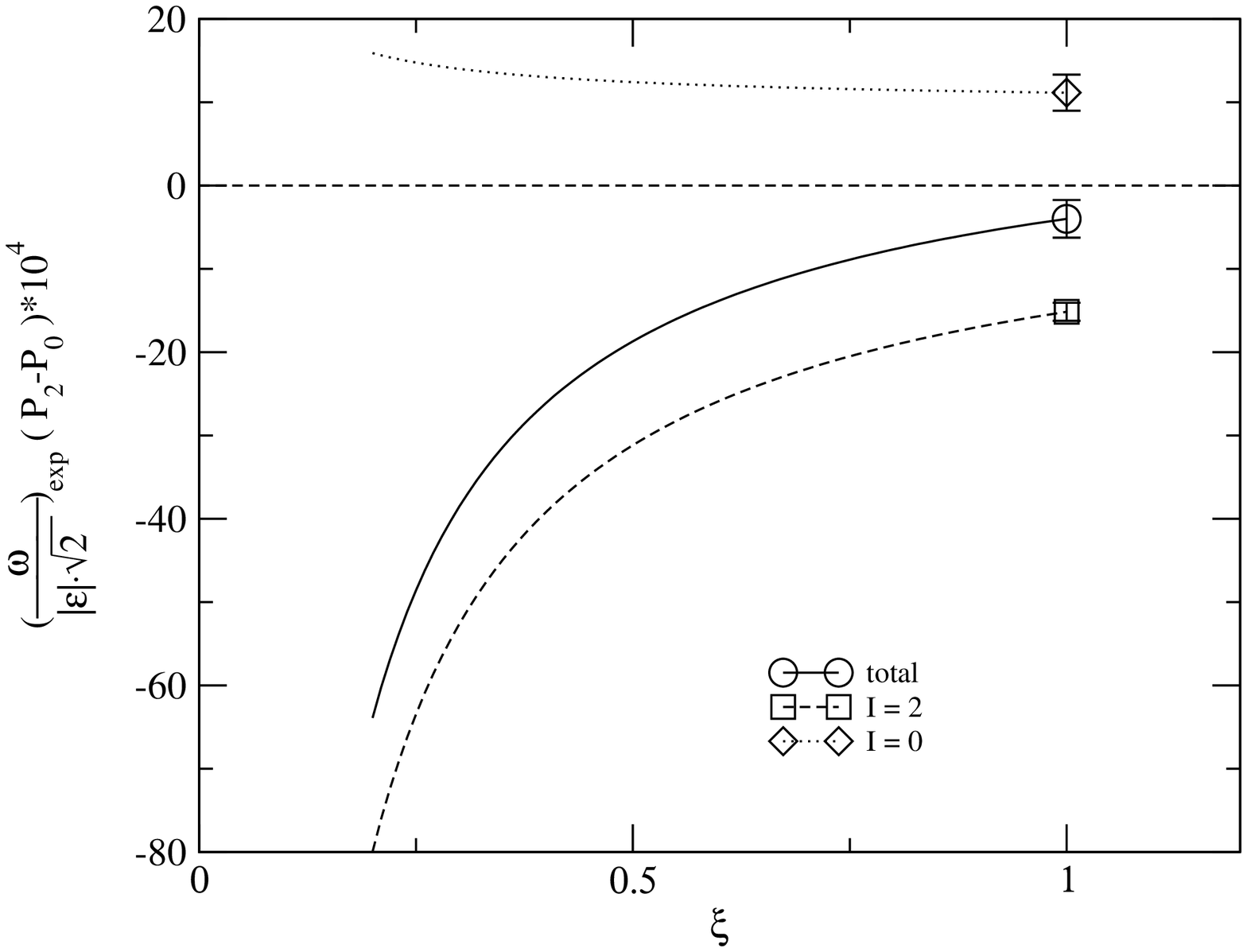}
\caption{The values for $-\repezero$, $\repetwo$ and their sum, using
the 1-loop chiral perturbation theory extrapolation in full QCD, are
plotted versus $\xi$.  The contribution proportional to $P_2$ is going
to zero with increasing $\xi$ due to the increase in Re($A_2$).  $-P_0$
is constant in lowest order chiral perturbation theory, once $\xi$ is
large enough that the electroweak penguins play no role, and has no
chiral logarithm corrections.  At the physical point $\xi = 1$, the two
terms are almost cancelling, producing the small value for $\repe$.
The data is for $\mu = 2.13$ GeV.}

\label{fig:epe_choice2}
\end{center}
\end{figure}


\begin{figure}
\begin{center}
\epsfxsize=\hsize
\epsfbox{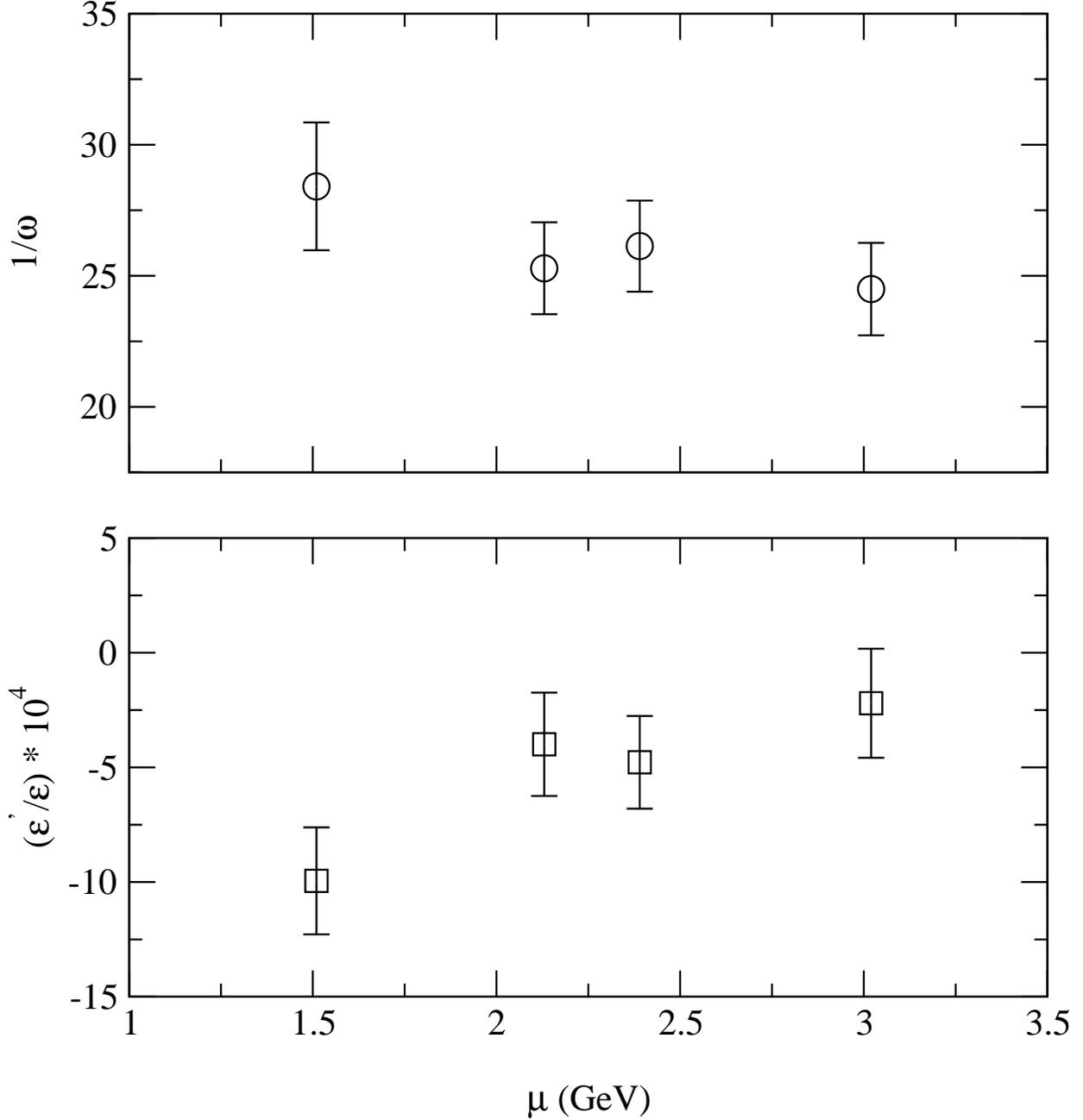}
\caption{A plot of Re($A_0$)/Re($A_2) = 1/\omega$  (upper panel) and
$\repe$ (lower panel) versus $\mu$ for the physical values obtained
using 1-loop full QCD chiral perturbation theory for the extrapolation
to the physical kaon mass.  The results for $1/\omega$ show some $\mu$
dependence beyond the statistical errors.  For $\repe$ the $\mu$
dependence is noticeable, reflecting the visible $\mu$ dependence in
Im($A_2$).  We choose to quote final values with $\mu = 2.13$ GeV.}

\label{fig:omega_epe_mu_dep}
\end{center}
\end{figure}


\begin{figure}
\begin{center}
\epsfxsize=\hsize
\epsfbox{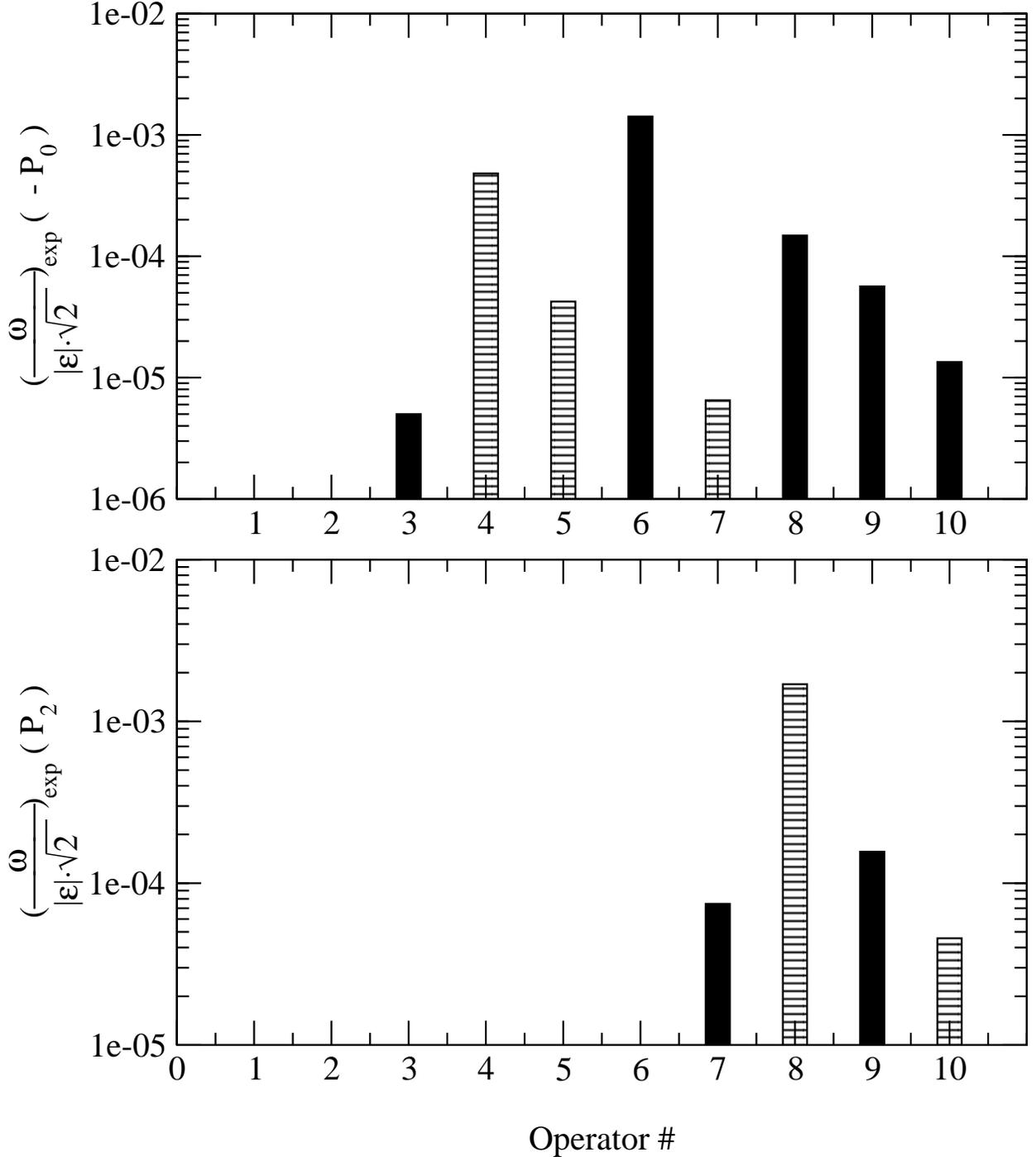}

\caption{A breakdown of the contribution of $Q_{i, \rm cont}$ to the
imaginary amplitudes entering $-\repezero$ (upper panel) and $\repetwo$
(lower panel).  The solid filled bars in the graph denote positive
quantities and the hashed bars represent negative quantities.  The
experimental values for $\omega$ and $|\epsilon|$ are used here and the
data is for $\mu = 2.13$ GeV.}

\label{fig:epe_mom2_break.eps}
\end{center}
\end{figure}

\fi

\end{document}